\documentclass[final,5p,times,twocolumn,sort,compress]{elsarticle}

\usepackage[mathlines,switch]{lineno}
\newif\iflinenos
\linenosfalse
\iflinenos\linenumbers\fi

\usepackage{amsmath}
\usepackage{amsfonts}
\usepackage{amssymb}
\usepackage{bbold}
\usepackage{graphicx}
\graphicspath{{figs/}}
\usepackage{bm}
\usepackage{array}
\usepackage{listings}
\usepackage{color}
\usepackage[normalem]{ulem}
\usepackage[dvipsnames]{xcolor}
\usepackage[colorlinks=true,breaklinks=true]{hyperref}

\definecolor{darkred}{rgb}{0.5,0,0}
\definecolor{darkblue}{rgb}{0,0,0.5}
\definecolor{firebrick}{rgb}{0.75,0.125,0.125}
\definecolor{darkgreen}{rgb}{0,0.5,0}
\hypersetup{urlcolor=darkblue,
	 citecolor=darkgreen,
	 linkcolor=firebrick}

\newcommand{\ie}{{\it i.e.}}

\newcommand{\eg}{{\it e.g.}}

\newcommand{\etc}{{\it etc.}}
\newcommand{\eq}{Eq.}

\newcommand{\fig}{Fig.}

\makeatletter
\DeclareRobustCommand{\_}{%
  \ifmmode\nfss@text{\textunderscore}\else\textunderscore\allowbreak\fi}
\makeatother
\newcommand{\us}{\_}
\providecommand{\Ref}{}
\AtBeginDocument{\renewcommand{\Ref}{Ref.}}
\newcommand{\Refs}{Refs.}
\newcommand{\Tab}{Table}

\newcommand{\equ}[1]{\eq~(\ref{equ:#1})}
\newcommand{\figu}[1]{\fig~\ref{fig:#1}}
\newcommand{\tabl}[1]{\Tab~\ref{tab:#1}}

\newcommand{\magnus}{{\tt Mag}\ensuremath{\nu}{\tt s}}
\definecolor{mygreen}{rgb}{0,0.6,0}
\definecolor{mygray}{rgb}{0.5,0.5,0.5}
\definecolor{mylightgray}{rgb}{0.937,0.937,0.937}
\definecolor{mymauve}{rgb}{0.58,0,0.82}

\lstnewenvironment{pysnip}[1][]{%
  \lstset{numbers=none, frame=none, backgroundcolor={},
          basicstyle=\small\ttfamily,
          xleftmargin=0pt, framexleftmargin=0pt,
          aboveskip=5pt, belowskip=-12pt, #1}}{}

\journal{Computer Physics Communications}

\begin{document}

\begin{frontmatter}

\title{\magnus: neutrino oscillation probabilities for any Hermitian Hamiltonian,\\any number of flavors, and any matter profile}

\author{Mauricio Bustamante\corref{author}}
\ead{mbustamante@nbi.ku.dk}
\cortext[author]{ORCID: \href{http://orcid.org/0000-0001-6923-0865}{0000-0001-6923-0865}.}
\address{Niels Bohr International Academy, Niels Bohr Institute,\\University of Copenhagen, DK-2100 Copenhagen, Denmark}

\date{\today}

\begin{abstract}
Interpreting neutrino oscillation measurements requires computing the probability that a neutrino born with one flavor is detected with another. In vacuum and in matter of constant density, this probability has exact formulas. Where the density varies along the path, as in the Earth, the Sun, or a supernova, it does not. There, the usual approaches either replace the profile by constant-density steps, whose error falls only as the square of their width, or integrate the evolution equation with a general-purpose solver, which must shorten its steps to meet a tighter tolerance. Both become slow at high accuracy, yet an analysis needs the probability at many energies, directions, and parameter values. Each new experiment or model often requires writing and validating a new solver. We present \magnus, an open-source Python code that avoids these limitations. It uses the Magnus expansion, whose error falls as a high power of its step width and whose cost is set by how fast the density varies, not by how many oscillations the path holds. \magnus\ accepts any density profile, any number of flavors, and any Hermitian Hamiltonian, such as those of non-standard interactions, sterile neutrinos, Lorentz-invariance violation, and pseudo-Dirac neutrinos. Where a cheaper route exists, it takes it automatically. For example, where the density varies slowly, as in the Sun, it follows the instantaneous eigenstates and applies the expansion only at resonances. Besides oscillating probabilities, \magnus\ returns phase-averaged ones, which describe solar and astrophysical neutrinos. A typical probability takes a few milliseconds, and batched scans are one to two orders of magnitude cheaper per point. Among the public codes we compared, \magnus\ is the fastest at high accuracy on a varying profile. \magnus\ shifts the effort of studying a new experiment or model from writing a solver to specifying its physics.
\end{abstract}


\end{frontmatter}

\noindent
{\bf Program summary}
\smallskip

\begin{small}

\noindent
{\em Program title:} Mag$\nu$s ({\tt magnuspy}) v1.2.0 \\ \\
\smallskip
{\em Licensing provisions:} GPL-3.0-only \\ \\
\smallskip
{\em Program obtainable from:} \\
\href{https://github.com/mbustama/Magnus}{\tt https://github.com/mbustama/Magnus} \\
\href{https://pypi.org/project/magnuspy/}{\tt https://pypi.org/project/magnuspy/} \\
{\tt pip install magnuspy} \\ \\
\smallskip
{\em Distribution format:} {\tt git} repository; source distribution and built wheel on the Python Package Index (PyPI), installable with {\tt pip} \\ \\
\smallskip
{\em Programming language:} Python 3.10 or later; tested on 3.10, 3.11, 3.12, and 3.13 \\ \\
\smallskip
{\em Classification:} 11.1 General, High Energy Physics and Computing \\ \\
\smallskip
{\em External routines:} {\tt NumPy}, {\tt SciPy}, {\tt joblib}, {\tt Numba}, and {\tt Matplotlib}, all installed automatically \\ \\
\smallskip
{\em Nature of problem:} \\
Computation of neutrino flavor oscillation probabilities for any number of flavors and any Hermitian Hamiltonian, including Hamiltonians that vary along the trajectory, such as those of matter whose density varies continuously, for which no closed form exists. \\ \\
\smallskip
{\em Solution method:}\\
The evolution operator is computed with the Magnus expansion over a chain of slabs, refined until the probabilities converge to a requested tolerance; the result is unitary by construction. Where a request can be answered exactly or more cheaply, the code does so automatically, \eg, by adiabatic transport on smooth profiles or by computing many energies at once. Averaged probabilities are also available. \\ \\
\smallskip
{\em Additional comments, including restrictions, unusual features:}\\
Non-unitary evolution (decay, absorption, decoherence) and nonlinear collective oscillations lie outside the method. Refinement stops once two successive levels agree to within the requested tolerance; this does not guarantee that the answer is that close to the exact one. Probabilities are unitary to round-off at every accuracy setting. \\ \\
{\em Running time:}\\
The cost depends on the requested tolerance, on how rapidly the density varies along the path, and on the number of flavors. For example, a probability through the Earth takes up to a few milliseconds. Batched scans over energy or baseline are one to two orders of magnitude faster per probability.
\end{small}


\setcounter{tocdepth}{3}
\tableofcontents


\section{Introduction}
\label{sec:introduction}

Neutrinos are created and detected in weak interactions as \textit{flavor states} ($\nu_e$, $\nu_\mu$, $\nu_\tau$), but they propagate as superpositions of \emph{propagation states}, the eigenstates of the Hamiltonian, $\mathbb{H}$; in vacuum, these are \textit{mass eigenstates} ($\nu_1$, $\nu_2$, $\nu_3$) of definite mass\ \cite{Pontecorvo:1967fh, Maki:1962mu, Bilenky:2004sn, Fantini:2018itu}. Because the superposition evolves in time, a neutrino created with one flavor has a non-zero probability of being detected with another. The observation of that transition in solar, atmospheric, reactor, and accelerator neutrinos led to the discovery of neutrino mass and of flavor mixing in leptons\ \cite{Kajita:2016cak, McDonald:2016ixn}.

Computing the flavor-transition probability means solving the Schr\"{o}dinger equation
\begin{equation}
 \label{equ:schrodinger}
 i\,\frac{d}{dl}\,\nu_\alpha(l) = \mathbb{H}(l)\,\nu_\alpha(l) \;,
\end{equation}
where $\nu_\alpha$ represents the flavor state of a neutrino born with flavor $\alpha = e$, $\mu$, $\tau$, and $l$ is the distance traveled by the neutrino since its creation along the trajectory. For relativistic neutrinos---the predominant scenario---it stands in for the elapsed time, \ie, $l \approx t$. The Hamiltonian, $\mathbb{H}$, encodes the physics that drives the evolution of the neutrino state: neutrino masses and mixing, energy dependence, interactions with matter, and any new physics.  The probability of a $\nu_\alpha \to \nu_\beta$ transition is $P_{\nu_\alpha \to \nu_\beta}(l) = \vert \nu_\beta^\dagger \nu_\alpha(l) \vert^2$. The probabilities are oscillatory in $l$; hence, flavor transitions are conventionally referred to as ``neutrino oscillations.''

The cost of computing flavor-transition probabilities reflects the complexity of $\mathbb{H}$. Where $\mathbb{H}$ does not depend on $l$, the solution of \equ{schrodinger} is $\nu_\alpha(l) = e^{-i \mathbb{H} l}\,\nu_\alpha(0)$, with $\mathbb{U} \equiv e^{-i \mathbb{H} l}$ the evolution operator. Hence, the oscillation probabilities in vacuum and in matter of constant density have exact solutions\ \cite{Bustamante:2019ggq, NuOscProbExact}. This covers important scenarios such as the propagation of astrophysical neutrinos in the vacuum of space and of natural and human-made neutrinos in the Earth's crust. Once $\mathbb{H}$ depends on $l$---across the density layers of the Earth, inside the Sun, or through a supernova shock---there is, in general, no closed form: \equ{schrodinger} must be integrated along the trajectory. 

For any method that follows the oscillation, the cost of that integration is set by one quantity: the \emph{accumulated phase} along the trajectory,
\begin{equation}
 \label{equ:phase}
 \Phi
 \equiv
 \int_{l_0}^{l_1} \left[\lambda_{\rm max}(l) - \lambda_{\rm min}(l)\right] dl \;,
\end{equation}
where $\lambda_{\rm max}$ and $\lambda_{\rm min}$ are the largest and smallest instantaneous eigenvalues of $\mathbb{H}(l)$. It is the total phase difference the fastest-separating pair of propagation states builds up between production and detection, measured in radians.  In vacuum at three flavors it reduces, \eg, to the familiar $\Delta m^2_{31} L / 2E$, where $\Delta m^2_{31} \equiv m_3^2 - m_1^2$ is the difference between the squared masses of the mass eigenstates $\nu_3$ and $\nu_1$, $L$ is the total path length, and $E$ is the neutrino energy. For a 1-GeV neutrino crossing the Earth along a diameter, $\Phi$ is some $80$ rad; for a 5-MeV neutrino leaving the Sun, some $9 \cdot 10^{5}$ rad (and $2.6 \cdot 10^{4}$ rad on the slower pair, $\nu_2$ and $\nu_1$, the pair that sets the solar $\nu_e$ survival probability); and for a TeV neutrino from $100$~Mpc, some $2 \cdot 10^{16}$ rad.

This paper addresses the case in which the Hamiltonian varies \emph{continuously} along the path, typically through matter of varying density. Then $\mathbb{H}(l)$ does not commute with itself at different positions, so the evolution operator is no longer $e^{-i\int \mathbb{H}\,dl}$. This leaves two standard approaches. One is to replace the matter profile by a staircase: cut the path into slabs of constant density, solve each in closed form, and compose the operators in order. The other is to integrate \equ{schrodinger} with an adaptive differential-equation solver, which adjusts its steps to meet a tolerance. Each has a drawback. The staircase is equivalent to the midpoint rule, so its error falls only as the square of the slab width (Sec.~\ref{sec:order_is_the_slab_product}). The solver must instead resolve the oscillation phase step by step, so its cost grows with $\Phi$. The first drawback comes from the shape of the profile and the second from the size of $\Phi$, not from the physics in the Hamiltonian. 

A better method would have a cost that follows the shape of the profile rather than the phase, and an error that falls faster than the square of the slab width.

\begin{figure*}[t!]
 \centering
 \includegraphics[width=\textwidth]{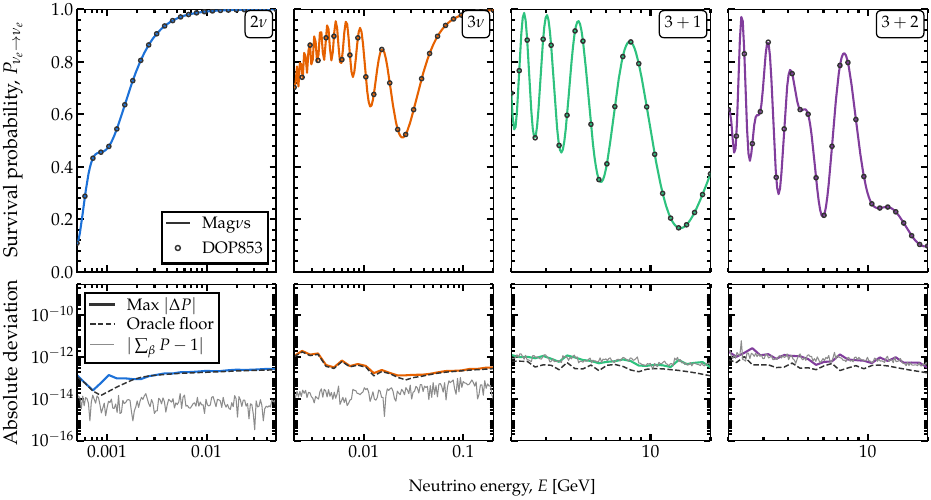}
 \caption{\label{fig:validation}\textbf{Accuracy of neutrino oscillation probabilities computed with \magnus\ at two to five flavors, against an independent solver.} Survival probability $P_{\nu_e \to \nu_e}$ at two, three, four, and five flavors, along an exponentially falling density profile (central density $3 \cdot 10^{3}$~g~cm$^{-3}$, scale height $10$~km, baseline $25$~km). {\it Top}: the probability computed with \magnus\ \cite{Magnus} (solid) and with a {\tt DOP853} integration (eighth-order Runge-Kutta) of the same Hamiltonian. The profile is the same in every panel; the energy window differs, since an eV-scale $\Delta m^2_{41}$ and the standard mass splittings do not resonate at the same energy. {\it Bottom}: the largest deviation between the two over all entries of the probability matrix; the departure of each row of the matrix from summing to one; and the \emph{oracle floor}, the amount by which the {\tt DOP853} result itself changes when its tolerance is tightened from ${\tt rtol} = 10^{-12}$ to $10^{-13}$. Deviations below the floor cannot be resolved by this comparison. The sterile parameters are $\sin^2\theta_{14} = \sin^2\theta_{24} = 0.1$ and $\Delta m^2_{41} = 1$~eV$^2$, with $\sin^2\theta_{15} = \sin^2\theta_{25} = 0.06$ and $\Delta m^2_{51} = 1.7$~eV$^2$ at five flavors. Listing~\ref{lst:validation} shows the code to compute these curves. Every figure in this paper is computed in full in notebook \#28 (\ref{app:notebooks}).}
\end{figure*}


\section{Synopsis}
\label{sec:synopsis}

To address this, we introduce \magnus\ \cite{Magnus}\footnote{Install via {\tt pip install magnuspy}\;.  Tutorials at \Ref~\cite{MagnusDocs}.}, a code that returns oscillation probabilities for any Hamiltonian and any matter profile. \magnus\ is meant to spare its users from writing new code for each new setup or model. A new setup is a different baseline, energy range, or matter profile: a long-baseline beam, an atmospheric oscillogram, a solar or supernova signal. A new model is a different Hamiltonian: non-standard interactions, a sterile state, a long-range force. In \magnus, they are inputs; the probability is the output. We designed it to be as transparent as possible, so that its users can focus on the physics rather than on numerical optimization: \magnus\ chooses its own discretization and method from the Hamiltonian and a requested tolerance.

\magnus\ computes the evolution operator with the \textit{Magnus expansion}\ \cite{Magnus:1954zz, Blanes:2008xlr}: over a segment of the trajectory of the neutrino---within which the Hamiltonian may vary---the expansion writes the operator as a single exponential, $\mathbb{U} = \exp(\Omega)$, where $\Omega$ is a series of integrals of nested commutators of the Hamiltonian. Truncated at order $2n$ in the segment width ($n = 1, \dots, 5$ in \magnus), the series gives an error that falls as the $2n$-th power of that width. The staircase mentioned earlier is its second-order truncation; \magnus\ goes up to order 10.

Three advantages follow, independently of what the Hamiltonian contains: robustness, speed, and flexibility. \magnus\ is \emph{robust}: because the evolution operator is exactly unitary at any order, any tolerance, and any slab count, the probabilities add up to one to within numerical round-off. \magnus\ is \emph{fast} without sacrificing accuracy: its cost follows the profile and not the phase, whereas a differential-equation solver pays for every radian resolved. A scan over neutrino energy or arrival direction is a single batched call, run on compiled {\tt Numba} kernels, rather than thousands of separate ones. And \magnus\ is \emph{flexible}: the Hamiltonian is a callable returning a Hermitian matrix of any size, so five flavors, non-standard interactions, a Lorentz-violating background, and a completely novel neutrino interaction are one and the same call. 

Figure~\ref{fig:validation} shows the survival probability $P_{\nu_e \to \nu_e}$ of a neutrino crossing an exponentially falling density profile, computed with \magnus\ at two, three, four, and five flavors; the last two include one and two sterile states. The deviation from an independent integration of the same Hamiltonian---the {\tt DOP853} eighth-order Runge-Kutta solver---is a few parts in $10^{12}$, at the limit of that integration's own accuracy.

Run as printed, Listing~\ref{lst:validation} takes about half a second on a laptop, and about a second including the start of Python. It asks for ${\tt rtol} = 10^{-12}$ and ${\tt atol} = 10^{-14}$, far beyond what any oscillation analysis requires, only to match the comparison in \figu{validation}. At the default of $10^{-3}$, the listing takes about 0.2~s, most of it one-time start-up. The listing calls four of the wrappers \magnus\ ships for the cases most often met, one per flavor count; the four calls differ only in the energy window and in the mixing parameters each count adds. The wrappers are a convenience: each builds a Hamiltonian internally and hands it to the core function of \magnus. A scenario for which \magnus\ ships no wrapper is the same call with a Hamiltonian of the user's own.

While the Magnus expansion is at the core of \magnus, the code does not rely on the expansion alone. Where the accumulated phase is extreme and the profile varies slowly, as in the Sun, \magnus\ transports the state along the instantaneous eigenstates of $\mathbb{H}$ and reserves the expansion for the narrow windows in which two eigenvalues approach and that transport fails. \magnus\ decides the hand-over from the Hamiltonian itself, with no intervention from the user.

\begin{lstlisting}[language=Python, caption={\textbf{Neutrino oscillation probabilities in \magnus\ at two to five flavors.} The code behind \figu{validation}: the $\nu_e$ survival probability along an exponentially falling density profile at two, three, four (3+1), and five (3+2) flavors, one panel each. Each flavor count has its own wrapper, and the four calls differ only in the energy window and in the sterile mixing. The oscillation parameters are the NuFIT 6.1 best fit~\cite{NuFIT61, Esteban:2024eli}, in normal ordering with Super-Kamiokande atmospheric data. Both sterile cases use $\sin^2\theta_{14} = \sin^2\theta_{24} = 0.1$ and $\Delta m_{41}^2 = 1$~eV$^2$; the 3+2 case adds $\sin^2\theta_{15} = \sin^2\theta_{25} = 0.06$ and $\Delta m_{51}^2 = 1.7$~eV$^2$. All other active-sterile parameters are zero. The tolerances, ${\tt rtol} = 10^{-12}$ and ${\tt atol} = 10^{-14}$, are tightened from the default of $10^{-3}$ to match the comparison.}, label={lst:validation}, float=t]
import numpy as np
import magnus.globaldefs as gd
import magnus.oscprob as oscprob

# Standard oscillation parameters
osc = gd.load_nufit_params('NuFIT 6.1')
# Active-sterile mixing
s14 = s24 = np.sqrt(0.10)
s15 = s25 = np.sqrt(0.06)

def energies(lo, hi):
    return np.logspace(np.log10(lo), np.log10(hi),
                       140)*gd.UNIT_GEV

# Exponentially falling density: 3e3 g/cm^3 at the
# origin, scale height 10 km, 25 km baseline.
# Every mixing parameter left out defaults to zero.
common = dict(L=25.0*gd.UNIT_KM, L0=0.0,
              rho_central=3.e3,
              l_scale=10.0*gd.UNIT_KM,
              density_matter_is_in_g_per_cm3=True,
              nu_i=gd.NUE, nu_f=gd.NUE,
              rtol=1e-12, atol=1e-14,
              magnus_exp_order=8)

P_2nu = oscprob.osc_prob_2nu_matter_exp_density(
    energies(0.0005, 0.05), **common,
    sth=osc['s12'], Dm2=osc['D21'])

P_3nu = oscprob.osc_prob_3nu_matter_exp_density(
    energies(0.002, 0.2), **common, **osc)

P_4nu = oscprob.osc_prob_4nu_matter_exp_density(
    energies(2.0, 20.0), **common, **osc,
    s14=s14, s24=s24, D41=1.0)

P_5nu = oscprob.osc_prob_5nu_matter_exp_density(
    energies(2.0, 20.0), **common, **osc,
    s14=s14, s15=s15, s24=s24, s25=s25,
    D41=1.0, D51=1.7)
\end{lstlisting}

\magnus\ ships ready-made cases for the scenarios most often explored: propagation in vacuum, in matter of constant or exponentially falling density, through the Earth, and through the Sun, on an exponential fit or on any of twelve standard solar models. Each case comes with standard oscillations and with new physics: non-standard neutrino-matter interactions, Lorentz-invariance violation, sterile neutrinos, and pseudo-Dirac neutrinos. The ready-made cases span two to five flavors (three active and up to two sterile); the generic functions take any number. Any other profile, such as a supernova shock front, is a function the user passes in.

Beyond the oscillation probability, \magnus\ returns the \emph{phase-averaged} probability, \ie, the probability with each interference term damped by the spread of its phase across the energy resolution of the measurement. Where a phase runs through many cycles, as for a solar or an astrophysical neutrino, this is the $L/E \to \infty$ limit. Every function can also return the evolution operator $\mathbb{U}$ alongside the probabilities. The probabilities keep only the magnitudes of the transition amplitudes; the operator keeps their phases too. The phases are needed, for example, to compute the probability that a neutrino leaving a star arrives in each mass eigenstate rather than in each flavor.

\magnus\ ships functions that draw commonly used plots: probability against distance or energy, oscillograms, and bi-probability plots. A command-line tool lets the user run \magnus\ as a calculator: it returns the probabilities for one energy and baseline without writing code. The application programming interface (API) is fully documented and publicly available~\cite{MagnusDocs}, along with implementation details and tutorial notebooks.

The rest of the paper is organized as follows. Section~\ref{sec:recap} fixes notation. Section~\ref{sec:method} presents the method. Section~\ref{sec:code} describes the code. Section~\ref{sec:examples} shows it in use. Section~\ref{sec:performance} measures its speed, precision, and reach. Section~\ref{sec:comparison} compares it against two other public codes, with each case checked by a third, independent method. Section~\ref{sec:conclusions} concludes. The appendices contain supporting detail: the Magnus expansion at any order, the predefined oscillation-parameter sets, the test suite, and the notebooks.


\section{Neutrino oscillations: units, notation, conventions}
\label{sec:recap}

In a system of $n \geq 2$ neutrino flavors, $\nu_\alpha$ is the flavor state of a neutrino, represented by an $n$-component column vector, and $\mathbb{H}$ is the $n \times n$ Hamiltonian in flavor space. We use natural units, $c = \hbar = 1$, with energies in eV and distances in eV$^{-1}$, so that $\mathbb{H}L$ is dimensionless. Every length an oscillation function takes is in eV$^{-1}$: baselines, radii, \etc Multiplying by {\tt globaldefs.UNIT\_KM} converts km into eV$^{-1}$; 1~km is about $5.07 \cdot 10^{9}$~eV$^{-1}$. (Earth-geometry helpers, such as {\tt earth.distance\_traveled\_inside\_earth}, work in km instead, so their output must be converted before it is passed on.) Passing a length that is not in eV$^{-1}$ raises no error: a baseline passed in km is read as at most a few mm, yet the call returns a converged, unitary probability. \magnus\ warns when a baseline is short enough to be km left unconverted, but does not halt.

Because neutrinos are relativistic, we identify the elapsed time with the distance traveled and write both as $l$. The state obeys \equ{schrodinger}; its solution defines the evolution operator $\mathbb{U}(l_1, l_0)$  through $\nu_\alpha(l_1) = \mathbb{U}(l_1,l_0)\, \nu_\alpha(l_0)$, with $\mathbb{U}(l_0,l_0) = \mathbb{1}$. The probability that a neutrino born as $\nu_\alpha$ is detected as $\nu_\beta$ is
\begin{equation}
 \label{equ:probability}
 P_{\nu_\alpha \to \nu_\beta}(l_1, l_0)
 =
 \left\lvert \left[\mathbb{U}(l_1,l_0)\right]_{\beta\alpha} \right\rvert^2 \;.
\end{equation}
We take $\mathbb{H}$ Hermitian throughout; $\mathbb{U}$ is then unitary and $\sum_\beta P_{\nu_\alpha \to \nu_\beta} = 1$ for every $\alpha$.

As an example, in vacuum, the Hamiltonian is
\begin{equation}
 \label{equ:h_vacuum}
 \mathbb{H}^{\rm vac}(E)
 =
 \frac{1}{2E}\, \mathbb{R}\, \mathbb{M}^2\, \mathbb{R}^\dagger \;,
\end{equation}
where $\mathbb{M}^2 = {\rm diag}(0, \Delta m^2_{21}, \Delta m^2_{31}, \ldots)$, $\Delta m^2_{ij} \equiv m_i^2 - m_j^2$, $m_i$ and $m_j$ are the masses of the neutrino mass eigenstates $\nu_i$ and $\nu_j$, and $\mathbb{R}$ is the mixing matrix that rotates between the flavor and mass bases.  The matrix is parametrized by the rotation angles and CP-violation phases. At three flavors---the standard case with $\nu_e$, $\nu_\mu$, $\nu_\tau$---it corresponds to the Pontecorvo-Maki-Nakagawa-Sakata (PMNS) mixing matrix, parametrized in terms of three mixing angles, $\theta_{12}$, $\theta_{23}$, and $\theta_{13}$, and one phase, $\delta_{\rm CP}$. 

In \magnus, $\mathbb{R}$ is a product of rotations. Each $\mathbb{R}_{ij}$ rotates by an angle $\theta_{ij}$ in the $(i,j)$ plane: it has $\cos\theta_{ij}$ in entries $(i,i)$ and $(j,j)$, $\sin\theta_{ij}\,e^{-i\delta_{ij}}$ in entry $(i,j)$, $-\sin\theta_{ij}\,e^{i\delta_{ij}}$ in entry $(j,i)$, ones in the remaining diagonal entries, and zeros in all other entries. For instance, at three flavors,
\begin{equation}
 \label{equ:rotation_example}
 \mathbb{R}_{13}
 =
 \begin{pmatrix}
  \cos\theta_{13} & 0 & \sin\theta_{13}\, e^{-i\delta_{13}} \\
  0 & 1 & 0 \\
  -\sin\theta_{13}\, e^{i\delta_{13}} & 0 & \cos\theta_{13}
 \end{pmatrix} \;.
\end{equation}
For $n$ flavors,
\begin{equation}
 \label{equ:mixing_matrix}
 \mathbb{R}
 =
 \begin{cases}
  \mathbb{R}_{12} \;, & n = 2 \;, \\
  \mathbb{R}_{23}\, \mathbb{R}_{13}\, \mathbb{R}_{12} \;, & n = 3 \;, \\
  \mathbb{R}_{34}\, \mathbb{R}_{24}\, \mathbb{R}_{14}\, \mathbb{R}_{23}\, \mathbb{R}_{13}\, \mathbb{R}_{12} \;, & n = 4 \;, \\
  \mathbb{R}_{35}\, \mathbb{R}_{25}\, \mathbb{R}_{15}\, \mathbb{R}_{34}\, \mathbb{R}_{24}\, \mathbb{R}_{14}\, \mathbb{R}_{23}\, \mathbb{R}_{13}\, \mathbb{R}_{12} \;, & n = 5 \;.
 \end{cases}
\end{equation}
At $n = 2$, $\mathbb{R}_{12}$ is a real rotation by the single mixing angle $\theta$. At $n \geq 3$, the phase of $\mathbb{R}_{13}$ is $\delta_{13} \equiv \delta_{\rm CP}$. Each sterile state adds its rotations on the left, and of these only $\mathbb{R}_{14}$, $\mathbb{R}_{24}$, $\mathbb{R}_{15}$, and $\mathbb{R}_{35}$ carry a phase. All other rotations are real. There is no $\mathbb{R}_{45}$: a rotation between the two sterile states only relabels them, since both feel the same potential, and it leaves every probability between active flavors unchanged.

Flavor indices run over $e$, $\mu$ at $n = 2$ (other pairs can be chosen; see Sec.~\ref{sec:two_flavors}); $e$, $\mu$, $\tau$ at $n = 3$; $e$, $\mu$, $\tau$, $s_1$ at $n = 4$; and $e$, $\mu$, $\tau$, $s_1$, $s_2$ at $n = 5$. (\magnus\ accepts any number of flavors, but examples in this paper---save for pseudo-Dirac neutrinos---stop at five.) The sterile states $s_i$ feel no weak interaction; they take part in oscillations only through mixing with active ones.

In matter, neutrinos scatter coherently forward on electrons via charged-current (CC) interactions and on electrons, protons, and neutrons via neutral-current (NC) ones. As a result, the Hamiltonian acquires the Mikheev-Smirnov-Wolfenstein (MSW) term\ \cite{Wolfenstein:1977ue, Mikheev:1986gs}
\begin{equation}
 \label{equ:h_matter}
 \mathbb{H}(E, l)
 =
 \mathbb{H}^{\rm vac}(E)
 +
 V_{\rm CC}(l)\, \mathbb{P}(l) \;,
 \quad
 V_{\rm CC} = \sqrt{2}\, G_F\, n_e(l) \;,
\end{equation}
where $V_{\rm CC}$ is the matter potential, $n_e$ is the electron number density, $G_F$ is the Fermi constant, and $\mathbb{P}$, the projector, is a diagonal matrix. The projector is where the flavor count enters. 

At three flavors, $\mathbb{P} = {\rm diag}(1,0,0)$: the NC potential is common to the three active flavors and cancels from the probabilities. With a sterile state, it no longer cancels, since a sterile neutrino is a gauge singlet and feels neither potential. In this case, $\mathbb{P} = {\rm diag}(1, 0, 0, r/2, \ldots)$ with $r = n_n/n_p$ the neutron-to-proton ratio. The $r/2$ factor is explained as follows. The NC potential of an active flavor is $V_{\rm NC} = -G_F n_n/\sqrt{2}$. Subtracting it from the whole diagonal is what removes it from the active flavors; what it leaves on a sterile state is $-V_{\rm NC}$. With $n_p = n_e$ by charge neutrality, $-V_{\rm NC} = G_F n_n/\sqrt{2} = (n_n/n_p)\,G_F n_e/\sqrt{2} = (r/2)\,V_{\rm CC}$. So the sterile entry is not an interaction of the sterile state but the residue of the subtraction: $1/2$ for isoscalar matter (\ie, matter with $n_n = n_p$). For every flavor count, \magnus\ builds $\mathbb{P}$ in {\tt matter.matter\_potential\_projector}. The wrappers take $r$ as {\tt ratio\_number\_neutrons\_to\_protons}, a number or a function of position, $r(l)$. Earth wrappers use the latter to follow the composition of each layer, so $\mathbb{P}$ varies along the path.

For antineutrinos, the mixing matrix is conjugated \emph{and} the matter potential changes sign. Doing only one of the two returns a plausible but wrong answer, so the wrappers take a {\tt nubar} flag that does both instead of leaving it to the user.


\section{The method}
\label{sec:method}


\subsection{History and context}

The Magnus expansion is due to Magnus\ \cite{Magnus:1954zz}. In numerical analysis, it is one of the standard geometric integrators: methods that preserve a structure of the problem, here unitarity, exactly rather than to within the truncation error\ \cite{Iserles:2000acta, Hairer:2006gni}. \Ref\ \cite{Blanes:2008xlr} reviews it. By default, \magnus\ evaluates it with the Gauss-Legendre \emph{collocation} integrators of \Refs\ \cite{Blanes:2000bit, Blanes:2002opt}: the Hamiltonian is sampled at a few Gauss-Legendre nodes inside each step, and the scheme is built so that those samples deliver the full order. We introduce them below.

The Magnus expansion has been used for neutrino oscillations before. D'Olivo and Oteo brought it to two-flavor evolution in matter\ \cite{DOlivo:1990xs}, Ioannisian and Smirnov developed it for realistic profiles\ \cite{Ioannisian:2008ve}, and Casas \emph{et al.}\ turned it into a practical integrator for three flavors\ \cite{Casas:2016asi}, reporting gains of up to two orders of magnitude in speed over general-purpose solvers.

The Magnus expansion is established; what has been missing is a general-purpose, tested, and documented implementation of it that can be used for a wide range of physics studies with minimal supervision. \magnus\ provides one. It takes an arbitrary Hermitian Hamiltonian at any number of flavors, on a continuous or a layered profile, and chooses its own discretization of the path from a requested tolerance, aligned with any discontinuities. For each request for a probability, it chooses among several ways of computing the answer, suited to different kinds of Hamiltonian (Sec.~\ref{sec:engines}), and reports which one it used. When it cannot confirm that its answer has converged, it warns instead of returning it silently. It implements the Magnus expansion up to ten terms, two quadrature rules, the composition of the path in steps and its adaptive refinement, an exactly unitary matrix exponential, adiabatic transport with Magnus patches, the phase-averaged probability, and batching, which makes a scan over energy or arrival direction one call instead of thousands.


\subsection{Where the method applies and where it does not}
\label{sec:applicability}

The method, and its implementation in \magnus, require only \textbf{that the Hamiltonian be Hermitian}. Hermiticity makes the evolution unitary and the system closed, so that the total probability, summed over all flavors, is conserved. Section~\ref{sec:limitations} lists the cases this excludes, among them open systems such as invisible neutrino decay and decoherence; for those, other codes should be used, such as \texttt{nuSQuIDS}~\cite{Arguelles:2021twb, NuSQuIDS}.

Convergence of the Magnus expansion is not a second requirement on the user: \magnus\ handles it internally. The series, introduced in Sec.~\ref{sec:expansion}, converges absolutely if $\int \lVert \mathbb{H} \rVert_2\, dl < \pi$ over the interval it is applied to\ \cite{Moan:2008conv}, where $\lVert \cdot \rVert_2$ is the spectral norm, which for a Hermitian matrix is its largest eigenvalue in modulus. The condition is sufficient, not necessary, so a particular $\mathbb{H}$ may converge well beyond it; but no constant larger than $\pi$ guarantees convergence for every $\mathbb{H}$. Over a long path the integral exceeds $\pi$, so \magnus\ cuts the path into \emph{slabs}, applies the expansion on each, and composes the results (Sec.~\ref{sec:order_is_the_slab_product}). The number and positions of the slabs follow from the requested tolerance: \magnus\ refines them until successive results agree (Sec.~\ref{sec:refinement}), without user intervention. (The user may instead fix the slabs.) If a slab still exceeds the bound, \magnus\ warns; if the requested tolerance is not reached, because, \eg, \magnus\ reached the maximum number of slabs, it reports that too.


\subsection{The Magnus expansion}
\label{sec:expansion}

Writing $\mathbb{A}(l) \equiv -i\,\mathbb{H}(l)$, \equ{schrodinger} reads $d\mathbb{U}/dl = \mathbb{A}(l)\,\mathbb{U}$. Where $\mathbb{A}$ commutes with itself at different positions (for neutrinos, in vacuum or at uniform density), the solution is $\exp \left( \int \mathbb{A}\,dl \right)$; where it does not, that expression is wrong. The time-ordered exponential that replaces it is the Magnus expansion\ \cite{Magnus:1954zz}:
\begin{equation}
 \label{equ:magnus}
 \mathbb{U}(l, l_0) = \exp\left[\Omega(l, l_0)\right] \;,
 \qquad
 \Omega = \sum_{k \geq 1} \Omega_k \;,
\end{equation}
exactly wherever the series converges (Sec.~\ref{sec:applicability}), with the terms generated by a recursion in the Bernoulli numbers $B_j$ (in the $B_1 = -1/2$ convention)\ \cite{Blanes:2008xlr},
\begin{equation}
 \label{equ:recursion}
 \Omega_1 = \int_{l_0}^{l} \mathbb{A}\,ds \;,
 \quad
 \Omega_k = \sum_{j=1}^{k-1} \frac{B_j}{j!} \int_{l_0}^{l} S_k^{(j)}\, ds \;,
\end{equation}
where $S_k^{(1)} = [\Omega_{k-1}, \mathbb{A}]$ and $S_k^{(j)} = \sum_{m=1}^{k-j} [\Omega_m, S_{k-m}^{(j-1)}]$. The time-ordering means that the earlier-time operators are applied first, \ie, are positioned to the right of later-time operators.

The first three terms read
\begin{align}
 \label{equ:omega123}
 \Omega_1 &= \int_{l_0}^{l}\! ds\, \mathbb{A}(s) \;, \nonumber \\
 \Omega_2 &= \frac{1}{2}\int_{l_0}^{l}\! ds_1 \int_{l_0}^{s_1}\! ds_2\,
 \left[\mathbb{A}(s_1), \mathbb{A}(s_2)\right] \;, \\
 \Omega_3 &= \frac{1}{6}\int_{l_0}^{l}\! ds_1 \int_{l_0}^{s_1}\! ds_2
 \int_{l_0}^{s_2}\! ds_3 \Big(
 \left[\mathbb{A}_1, [\mathbb{A}_2, \mathbb{A}_3]\right] \nonumber \\
 &\hspace{3.1cm} + \left[\mathbb{A}_3, [\mathbb{A}_2, \mathbb{A}_1]\right]
 \Big) \;, \nonumber
\end{align}
with $\mathbb{A}_i \equiv \mathbb{A}(s_i)$. \ref{app:terms} gives the structure at any order. The odd Bernoulli numbers vanish for $j \geq 3$, so whole commutator groups drop out and only $j = 1, 2, 4, 6, \ldots$ contribute. Every term unrolls to a right-nested chain $[\Omega_{m_1}, [\Omega_{m_2}, \ldots [\Omega_{m_j}, \mathbb{A}]\ldots]]$ with $\sum_i m_i = k-1$, so the terms of the $j$-th group are indexed by the compositions of $k-1$ into $j$ positive parts and there are $\binom{k-2}{j-1}$ of them. Their number roughly doubles per order: $\Omega_7$ has $17$ terms and $\Omega_{10}$ has $129$.

In neutrino oscillations, the first term of the Magnus expansion is the one driving the standard slab approximation. Where $\mathbb{H}$ does not depend on position, $\Omega_1 = -i\mathbb{H}\Delta l$ is the whole of the expansion, since every higher term is a nested commutator of $\mathbb{H}$ with itself and vanishes identically. Its exponential is then the exact evolution operator, not a truncation of one. Evaluating it is the problem {\tt NuOscProbExact}\ \cite{NuOscProbExact, Bustamante:2019ggq} solves, for up to four flavors: expanding $\mathbb{H}$ and $e^{-i\mathbb{H}L}$ in the generators of SU($n$) returns the probability from the eigenvalues of $\mathbb{H}$ and traces alone, without ever constructing an eigenvector. The eigenvalues are found in closed form at two and three flavors and numerically at four. Composing one such operator per slab, in the order the neutrino meets them, is the standard slab treatment of a piecewise-constant profile~\cite{Akhmedov:1988kd, Krastev:1989ix, Akhmedov:1998ui}. We take that correspondence further in Sec.~\ref{sec:order_is_the_slab_product}.

\magnus\ implements \equ{recursion} through $\Omega_{10}$. The terms through $\Omega_6$ are written out in the simplified forms of \Ref\ \cite{Blanes:2008xlr}, and those beyond are generated from the recursion. A separate module, {\tt expansionterms}, derives the same terms symbolically, in exact rational arithmetic, at \emph{any} order, and shares no code with the numerical core. The test suite checks the two against each other, term by term through $\Omega_{10}$, and regenerates the hard-coded recursion coefficients (\ref{app:tests}). (Users interested in applying the Magnus expansion to other problems---in magnetic resonance, quantum optics, lattice gauge theory, or condensed matter---may repurpose {\tt expansionterms}.)


\subsection{Truncation is exactly unitary}
\label{sec:unitarity}

Because $\mathbb{H}$ is Hermitian, $\mathbb{A} = -i\mathbb{H}$ is anti-Hermitian, and the commutator of two anti-Hermitian matrices is again anti-Hermitian. Every $\Omega_k$ in \equ{recursion} is a combination, with real coefficients, of nested commutators of $\mathbb{A}$ with itself, so every $\Omega_k$ is anti-Hermitian, and so is any partial sum. Therefore,
\begin{equation}
 \label{equ:unitary}
 \Omega^\dagger = -\Omega
 \quad\Longrightarrow\quad
 \left[\exp(\Omega)\right]^\dagger \exp(\Omega) = \mathbb{1}
\end{equation}
for \emph{every} truncation order and every quadrature accuracy. Truncating early costs accuracy, never unitarity. This is the sense in which the method is a geometric integrator: the exact solution lies in the unitary group, and so does every approximation to it.

In practice, the guarantee is only as good as the routine that computes the exponential. A general-purpose Pad\'e routine knows nothing of the structure of its argument, and returns a matrix that is unitary only to its own error tolerance, so the exactness would be lost at the last step. \magnus\ instead uses a form that is unitary by construction. Because $\Omega$ is anti-Hermitian, $\mathbb{K} \equiv i\Omega$ is Hermitian, so it is diagonalized by a unitary $\mathbb{V}$, and
\begin{align}
 \mathbb{K} &= \mathbb{V}\, {\rm diag}(\lambda_1, \ldots, \lambda_n)\, \mathbb{V}^\dagger \;, \nonumber \\
 \label{equ:expm}
 \exp(\Omega) &= \mathbb{V}\, {\rm diag}\!\left(e^{-i\lambda_1}, \ldots, e^{-i\lambda_n}\right) \mathbb{V}^\dagger \;,
\end{align}
where the eigenvalues $\lambda_1, \ldots, \lambda_n$ of $\mathbb{K}$ are real. The middle factor is then a diagonal of pure phases, and conjugating it by a unitary $\mathbb{V}$ keeps it unitary. Thus, $\exp(\Omega)$ is unitary by construction, not only to the accuracy of an approximation. Section~\ref{sec:exponential} describes how \magnus\ evaluates \equ{expm}: in closed form at two and three flavors, with a compiled eigensolver at four and five, and with a general Hermitian eigendecomposition above five.

In floating-point arithmetic, neither route is unitary to the last bit, because $\mathbb{V}$ is itself computed with rounding. Two measurements show how close it comes, and they measure different things. A single exponential deviates by $\lVert \mathbb{U}^\dagger \mathbb{U} - \mathbb{1}\rVert = 4 \cdot 10^{-16}$ in the median; over a stack of $4096$, the worst reaches $4 \cdot 10^{-15}$ while the median stays put, so a larger workload lengthens the tail without moving the typical case. A probability is built from many such factors, so its deviation grows with their number: across scans whose number of points spans four decades, at two through five flavors, the worst case grows from $3 \cdot 10^{-12}$ to $1.6 \cdot 10^{-11}$. For all practical purposes, \magnus\ preserves unitarity.


\subsection{Two quadrature rules for the Magnus expansion}
\label{sec:quadrature}

Evaluating the Magnus recursion, \equ{recursion}, means sampling $\mathbb{A}$ inside each slab; \magnus\ offers two families of quadrature rule to approximate the integrals in the recursion.

\smallskip

\textit{Collocation integrators (default).---}By default \magnus\ does not evaluate \equ{recursion} as it stands. Each $\Omega_k$ there is a $k$-fold integral over $(k-1)$-fold nested commutators, so the integration and the commutator count both grow steeply with $k$; the sixth-order truncation alone would call for four such integrals. The collocation rules presented below replace them. Each rule, $\Omega_{\rm GL}^{(p)}$, is a quadrature approximation to a \emph{truncation} of the Magnus series in \equ{magnus}, assembled from a few samples of $\mathbb{A}$ at Gauss-Legendre nodes across a slab width, $h$, together with a fixed set of commutators among those samples. Its label is the order $p$ that it attains, not the number of terms that it keeps, \ie,
\begin{equation}
 \label{equ:gl_definition}
 \Omega_{\rm GL}^{(p)}
 =
 \sum_{k=1}^{m_p} \Omega_k + \mathcal{O}\!\left(h^{p+1}\right) \;,
\end{equation}
where $m_p = 1,\, 2,\, 4,\, 6$ for $p = 2,\, 4,\, 6,\, 8$, with $p$ always even. The error in \equ{gl_definition} comes from the next pair of Magnus terms of \equ{magnus}, which is what fixes the order at $p$.  \ref{app:terms} derives the sets of Magnus commutators retained in the collocation integrators and the required evenness of $p$.  (While the construction of collocation integrators is not unique to Gauss-Legendre quadrature, collocation terms written under different quadrature schemes cannot be interchanged or mixed.)

Collocation is economical: for a given truncation of the Magnus series in \equ{magnus}, it evaluates fewer commutators and fewer instances of the Hamiltonian than the series, as we show below. It is the cheapest route to a given accuracy when $\mathbb{H}$ is smooth inside a slab. This is the case, \eg, inside the Earth, where the density is smooth between layer boundaries and \magnus\ puts the slab edges on the boundaries themselves (Sec.~\ref{sec:refinement}).

We use $\Omega_{\rm GL}^{(p)}$ as \equ{magnus} uses $\Omega$. On the $j$-th slab, spanning $l_{j-1}$ to $l_j$ with width $h = l_j - l_{j-1}$, the evolution operator is
\begin{equation}
 \label{equ:slab_exponential}
 \mathbb{U}_j = \exp\left[\Omega_{\rm GL}^{(p)}(l_j, l_{j-1})\right] \;,
\end{equation}
one matrix exponential per slab, with the $\mathbb{U}_j$ time-ordered. As for the Magnus series itself (see Sec.~\ref{sec:unitarity}), truncating the series $\Omega_{\rm GL}^{(p)}$ inside the exponential leaves $\mathbb{U}_j$ exactly unitary.

Following \Refs\ \cite{Blanes:2000bit, Blanes:2002opt}, Magnus orders 2, 4, 6, and 8 are reached from only one, two, three, or four evaluations of $\mathbb{A}$ per slab, taken at the Gauss-Legendre nodes. \magnus\ implements all four. 
(These are \textit{not} the commutator-free family of \Refs\ \cite{Blanes:2006cf, Alvermann:2011cf}, which replaces commutators by products of exponentials of linear combinations of the $\mathbb{A}_i$.) With $\mathbb{A}_j \equiv -i \mathbb{H}_j$ the Hamiltonian at node $j$, the collocation integrators are
\begin{align}
 \label{equ:gl}
 \Omega_{\rm GL}^{(2)} &= h\, \mathbb{A}_1 \;, \nonumber \\
 \Omega_{\rm GL}^{(4)} &= \frac{h}{2}\left(\mathbb{A}_1 + \mathbb{A}_2\right)
 + \frac{\sqrt{3}}{12}\, h^2 \left[\mathbb{A}_2, \mathbb{A}_1\right] \;, \nonumber \\
 \Omega_{\rm GL}^{(6)} &= \alpha_1 + \frac{\alpha_3}{12}
 + \frac{1}{240} \left[-20\,\alpha_1 - \alpha_3 + \mathbb{C}_1,\,
 \alpha_2 + \mathbb{C}_2\right] \;, \nonumber \\
 \Omega_{\rm GL}^{(8)} &= \alpha_1 + \frac{\alpha_3}{12}
 - \frac{7}{120}\,\mathbb{C}_3 + \frac{1}{360}\,\mathbb{C}_6 \;.
\end{align}
Orders 2 and 4 follow \Ref\ \cite{Blanes:2000bit}; orders 6 and 8 are the commutator-optimized forms of \Ref\ \cite{Blanes:2002opt}. The review of \Ref\ \cite{Blanes:2008xlr} reproduces the order-6 form. Orders 4, 6, and 8 need 1, 3, and 6 commutators, the fewest possible at each order\ \cite{Blanes:2002opt}. \tabl{collocation} collects the $\alpha_i$ and the $\mathbb{C}_i$ that build the last two of \equ{gl}.

(We have found no collocation scheme of this form at order 10 or above. Reference\ \cite{Blanes:2000bit} only tabulates the size of the order-10 problem---80 determining equations after time-symmetry, against 27 at order 8; \Refs\ \cite{Blanes:2002opt, Blanes:2008xlr} stop at order 8. The commutator-free family does not reach higher either: \Ref\ \cite{Blanes:2006cf} stops at order 6, and \Ref\ \cite{Alvermann:2011cf} at order 8. Order 8 appears to be where the optimized literature ends. However, given how little order 8 gains over order 6 in \figu{smooth_reach} later, order 10 would rarely be worth reaching for in neutrino oscillation probabilities.)

\begin{table*}[t!]
 {\small
 \begin{tabular}{@{}l l l@{}}
  & Order 6, 3 GL nodes & Order 8, 4 GL nodes \\
 \hline
 $\alpha_1$ & $h\, \mathbb{A}_2$
            & $\frac{h}{2}\left[\left(\frac{1}{2} - \frac{\sqrt{30}}{8}\right)\mathbb{S}_1
              + \left(\frac{1}{2} + \frac{\sqrt{30}}{8}\right)\mathbb{S}_2\right]$ \\[2pt]
 $\alpha_2$ & $\frac{\sqrt{15}}{3}\, h \left(\mathbb{A}_3 - \mathbb{A}_1\right)$
            & $3h\left[(5 - \sqrt{30})\, d_1 \mathbb{R}_1 + (5 + \sqrt{30})\, d_2 \mathbb{R}_2\right]$ \\[2pt]
 $\alpha_3$ & $\frac{10}{3}\, h \left(\mathbb{A}_3 - 2\mathbb{A}_2 + \mathbb{A}_1\right)$
            & $\frac{7\sqrt{30}}{12}\, h \left(\mathbb{S}_1 - \mathbb{S}_2\right)$ \\[2pt]
 $\alpha_4$ & ---
            & $20h\left[(\sqrt{30} - 3)\, d_1 \mathbb{R}_1 - (3 + \sqrt{30})\, d_2 \mathbb{R}_2\right]$ \\
 \hline
 $\mathbb{C}_1$ & $\left[\alpha_1, \alpha_2\right]$
   & $-\frac{1}{28}\left[\alpha_1 + \frac{1}{28}\alpha_3 ,\; \alpha_2 + \frac{3}{28}\alpha_4\right]$ \\[2pt]
 $\mathbb{C}_2$ & $-\frac{1}{60}\left[\alpha_1, 2\alpha_3 + \mathbb{C}_1\right]$
   & $\frac{1}{3}\left[\alpha_1 ,\; -\frac{1}{14}\alpha_3 + \mathbb{C}_1\right]$ \\[2pt]
 $\mathbb{C}_3$ & ---
   & $\left[\alpha_1 + \frac{1}{28}\alpha_3 + \mathbb{C}_1 ,\; \alpha_2 + \frac{3}{28}\alpha_4 + \mathbb{C}_2\right]$ \\[2pt]
 $\mathbb{C}_4$ & --- & $\left[\alpha_2 , \mathbb{C}_1\right]$ \\[2pt]
 $\mathbb{C}_5$ & ---
   & $\left[\alpha_1 + \frac{5}{4}\mathbb{C}_1 ,\; 2\alpha_3 + \mathbb{C}_3 + \frac{1}{2}\mathbb{C}_4\right]$ \\[2pt]
 $\mathbb{C}_6$ & ---
   & $\left[\alpha_1 + \frac{1}{12}\alpha_3 - \frac{7}{3}\mathbb{C}_1 - \frac{1}{6}\mathbb{C}_3 ,\;
      -9\alpha_2 - \frac{9}{4}\alpha_4 + 63\,\mathbb{C}_2 + \mathbb{C}_5\right]$ \\
 \end{tabular}}
 \caption{\label{tab:collocation}\textbf{The Gauss-Legendre collocation integrators.} The quantities entering the collocation integrators of \equ{gl}. The two columns are different objects on different Gauss-Legendre (GL) nodes and cannot be mixed: order 6 samples $\mathbb{A}$ at the slab midpoint and at $(\frac{1}{2} \pm \frac{\sqrt{15}}{10})h$; order 8 at $(\frac{1}{2} \pm v_1)h$ and $(\frac{1}{2} \pm v_2)h$, with $v_1 = \frac{1}{2}\sqrt{(3 + 2\sqrt{6/5})/7}$ and $v_2 = \frac{1}{2}\sqrt{(3 - 2\sqrt{6/5})/7}$. At order 8 the corresponding weights are $w_1 = \frac{1}{2} - \frac{1}{6}\sqrt{5/6}$ and $w_2 = \frac{1}{2} + \frac{1}{6}\sqrt{5/6}$, and we write $d_i \equiv w_i v_i$, $\mathbb{S}_1 = \mathbb{A}_1 + \mathbb{A}_4$, $\mathbb{S}_2 = \mathbb{A}_2 + \mathbb{A}_3$, $\mathbb{R}_1 = \mathbb{A}_4 - \mathbb{A}_1$, $\mathbb{R}_2 = \mathbb{A}_3 - \mathbb{A}_2$. See Sec.~\ref{sec:quadrature} for details.}
\end{table*}

\smallskip

\textit{Cumulative quadrature.---}The alternative rule to collocation is cumulative quadrature. Here, \magnus\ samples $\mathbb{A}$ on a uniform grid of $m$ points inside each slab and integrates with the trapezoid or Simpson rule, selectable; $m$ starts at 100 and the refinement adjusts it together with the number of slabs. The points are $s_i = l_{j-1} + i\,\Delta s$, spaced by $\Delta s = h/(m-1)$. Keeping the first $k$ terms of the series gives
\begin{equation}
 \label{equ:cum_definition}
 \Omega_{\rm cum}^{(k)}
 =
 \sum_{q=1}^{k} \widehat{\Omega}_q \;,
\end{equation}
where each $\widehat{\Omega}_q$ is the Magnus term $\Omega_q$ of \equ{recursion} with its integral replaced by a quadrature sum over the grid, \ie,
\begin{align}
 \label{equ:cum_weights}
 \widehat{\Omega}_1 &= \Delta s \sum_{i=0}^{m-1} w_i\, \mathbb{A}(s_i) \;, \nonumber \\
 \widehat{\Omega}_k &= \sum_{j=1}^{k-1} \frac{B_j}{j!}\, \Delta s \sum_{i=0}^{m-1} w_i\, S_k^{(j)}(s_i) \;,
\end{align}
where the $B_j$ are the Bernoulli numbers of \equ{recursion}, in the same $B_1 = -1/2$ convention, and the $S_k^{(j)}$ are the nested-commutator combinations defined immediately below it. The lower-order terms inside each $S_k^{(j)}(s_i)$ are themselves running integrals, evaluated at every grid point $s_i$, which is what makes the rule cumulative. The $w_i$ are the composite trapezoid or Simpson weights. (For Simpson's rule with an even $m$, the last interval is integrated under a parabola through the last three points; the rule stays fourth-order.) The label $k$ in \equ{cum_definition} counts the terms kept outright, whereas that of the collocation in \equ{gl_definition} counts the error order attained.

The slab evolution operator then follows as in \equ{slab_exponential},
\begin{equation}
 \label{equ:slab_exponential_cum}
 \mathbb{U}_j = \exp\left[\Omega_{\rm cum}^{(k)}(l_j, l_{j-1})\right] \;.
\end{equation}
Thus, the two families of quadrature differ only in how that exponent is built. For the same requested accuracy on a smoothly varying Hamiltonian, the cumulative quadrature is slower than collocation, but it is fully general. This makes it the safer choice if $\mathbb{A}$ has a narrow feature---a kink or a jump \emph{inside} a slab---that the Gauss-Legendre nodes might miss. Also, cumulative quadrature is the only route above order 8 in \magnus. It runs to order 10, the highest \magnus\ implements (Sec.~\ref{sec:expansion}); a request beyond that is refused rather than rounded down.

\smallskip

\textit{Choosing a quadrature.---}In \magnus, the two dials that select which quadrature rule to use are {\tt magnus\_exp\_order} and {\tt integration\_method}, the latter taking {\tt 'gl'} for collocation and {\tt 'simpson'} or {\tt 'trapezoid'} for the cumulative. \tabl{orders} lists what each delivers. 

The two paths interpret {\tt magnus\_exp\_order} differently. On Gauss-Legendre, it names the order the method delivers, so order 6 is a sixth-order method, \ie, its error over the whole trajectory falls as $h^{6}$ in the slab width or, equivalently, as $N_{\rm slabs}^{-6}$ in the number of slabs, since a trajectory of length $L$ cut into equal slabs has $N_{\rm slabs} = L/h$. We derive in $h$ and measure in $N_{\rm slabs}$; the two are equivalent. On Simpson (or trapezoid), {\tt magnus\_exp\_order} names instead the last term $\Omega_k$ keeps. The method delivers an error order that runs ahead of that index (\ref{app:terms} explains why): \eg, order 6 on Simpson keeps $\Omega_5$ and $\Omega_6$ and is an eighth-order method, with the error falling as $h^8$. Those two extra orders are not a virtue of Simpson's rule. They come from $\Omega_5$ and $\Omega_6$, which the Simpson path keeps and the collocation path does not; the collocation path reaches its sixth order from three samples of $\mathbb{A}$ per slab, where Simpson needs a grid of them, making Simpson the more expensive choice. Above order 8, \magnus\ has no collocation scheme, so order 10 runs on Simpson or trapezoid exclusively.

\tabl{orders} shows that every order of convergence, $p$, is even, on both paths. The reason is that the Magnus terms scale in pairs. About the slab midpoint, each $\Omega_k$ carries only odd powers of $h$, which pushes every odd-indexed term up to the power of its even successor---$\Omega_3$ and $\Omega_4$ both at $h^{5}$, $\Omega_5$ and $\Omega_6$ at $h^{7}$, and so on (\ref{app:terms}). As a result, keeping $\Omega_7$ without $\Omega_8$ leaves the error unchanged. A request for an odd-valued {\tt magnus\_exp\_order} on the cumulative path delivers the order of the even request just below it, at the cost of one more term; on the collocation path, no odd scheme exists at all, so an odd-valued request runs the next even one.

\magnus\ defaults to {\tt 'gl'} at order 4. A higher order is not free: the adaptive refinement turns it into fewer slabs, but each slab costs more, so raising the order pays only once the tolerance is tight enough for the saving in slabs to outweigh the extra work in each. Across Earth, solar, and exponential density profiles, order 6 runs $0.96$ to $1.12$ times as fast as order 4 at a requested tolerance ($\texttt{rtol}$) of $10^{-4}$. At a tolerance of $10^{-8}$, it runs $1.08$ to $1.93$ times as fast.  Section~\ref{sec:timing_protocol} details our time-measurement protocol. Going down to order 2 is almost never worthwhile: at a tolerance of $10^{-8}$ on an Earth chord, it needs thousands of slabs where order 6 needs about a hundred.

{\def\arraystretch{1.2}
\begin{table}[t!]
 {\small
 \begin{tabular}{@{}c|ccc@{}}
 & \multicolumn{3}{c@{}}{Order $p$ of convergence delivered, $h^p$} \\
 \cline{2-4}
 & Collocation & \multicolumn{2}{c@{}}{Cumulative quadrature} \\
 {\tt magnus\_exp\_order} & {\tt 'gl'} & {\tt 'simpson'} & {\tt 'trapezoid'} \\
 \hline
 $1$  & $2$          & $2$    & $(2)$  \\
 $2$  & $2$          & $4$    & $(4)$  \\
 $3$  & $4$          & $4$    & $(4)$  \\
 $4$  & $\mathbf{4}$ & $6$    & $(6)$  \\
 $5$  & $6$          & $6$    & $(6)$  \\
 $6$  & $6$          & $8$    & $(8)$  \\
 $7$  & $(8)$        & $(8)$  & $(8)$  \\
 $8$  & $8$          & $10$   & $(10)$ \\
 $9$  & ---          & $(10)$ & $(10)$ \\
 $10$ & ---          & $12$   & $(12)$ \\
 \end{tabular}}
 \caption{\label{tab:orders}\textbf{Order of convergence of each quadrature rule.} Order of convergence, $p$, delivered by each value of {\tt magnus\_exp\_order} on the three quadrature rules of \magnus: the error over a trajectory of length $L$ falls as $h^{p}$ in the slab width, $h$, or equivalently as $N_{\rm slabs}^{-p}$ in the number of slabs, $N_{\rm slabs} = L/h$. On {\tt 'gl'}, {\tt magnus\_exp\_order} is the order itself; an odd value runs the next even scheme, and none is implemented above 8. On {\tt 'simpson'} and {\tt 'trapezoid'}, it is the index of the last $\Omega_k$ kept, and the order delivered is $2\lfloor {\tt magnus\_exp\_order}/2 \rfloor + 2$, one or two ahead of it (\ref{app:terms}). Entries without parentheses are measured, by fitting $\lVert \Delta \mathbb{U} \rVert \propto N_{\rm slabs}^{-p}$ against a tight {\tt DOP853} reference on a smooth problem whose Hamiltonian does not commute with itself at different positions; entries in parentheses are predicted by the two rules above. The default, in bold, is {\tt 'gl'} at order 4.}
\end{table}}


\subsection{Slab composition}
\label{sec:order_is_the_slab_product}

We cut a trajectory into $N_{\rm slabs}$ slabs, evaluate the expansion in each, and compose the slab operators in position order,
\begin{equation}
 \label{equ:composition}
 \mathbb{U}_{\rm tot}
 =
 \mathbb{U}_{N_{\rm slabs}} \cdots \mathbb{U}_2\, \mathbb{U}_1 \;,
\end{equation}
with the slab crossed \emph{first} standing \emph{rightmost}, since each operator acts on the state to its right. Each factor is unitary, so the product is too, whatever their number and whatever the accuracy of each. The order of the factors matters whenever the slab Hamiltonians do not commute, which on a varying profile is almost always. For a real $\mathbb{H}$, reversing the profile transposes $\mathbb{U}_{\rm tot}$: survival probabilities are unchanged, and $P_{\nu_\alpha \to \nu_\beta}$ swaps with $P_{\nu_\beta \to \nu_\alpha}$. With a nonzero CP-violation phase, $\mathbb{H}$ is complex, and reversing the profile can change survival probabilities too.

\smallskip

\textit{Traditional slab approximation.---}Keeping only the first term of the series, $\Omega_1 = \int \mathbb{A}\,dl$, and evaluating its integral at the slab midpoint, $l_j^{\rm mid}$, gives
\begin{equation}
 \label{equ:slab_approx}
 \mathbb{U}_j = \exp\left[-i\,\mathbb{H}(l_j^{\rm mid})\,h\right] \;,
\end{equation}
which tacitly implies a constant Hamiltonian per slab. Composed as in \equ{composition}, this is the traditional slab approximation of \Refs~\cite{Akhmedov:1988kd, Krastev:1989ix, Akhmedov:1998ui}: it is the Magnus expansion truncated at its first term, and its error falls as $h^2$, set by the dropped $\Omega_2$. On the collocation path it is exactly the order-2 scheme of \equ{gl}; the cumulative path at order 1 keeps the same term, with $\mathbb{H}$ integrated across the slab instead of read at its midpoint.

This product was analyzed for a periodically varying density in the seminal work of \Refs\ \cite{Akhmedov:1988kd, Krastev:1989ix}, where it gives rise to parametric resonance, and solved in closed form for the two-layer castle-wall profile in \Ref\ \cite{Akhmedov:1998ui}. For standard oscillations, each slab has the exact solution for uniform matter\ \cite{Ohlsson:1999xb}; for arbitrary standard and non-standard Hamiltonians, {\tt NuOscProbExact}\ \cite{Bustamante:2019ggq, NuOscProbExact} evaluates it in closed form for two, three, and four flavors. It is also what several publicly available oscillation codes do over the Earth: {\tt Prob3++}\ \cite{Prob3pp}, {\tt GLoBES}\ \cite{GLoBES}, {\tt OscProb}\ \cite{OscProb}, {\tt CUDAProb3}\ \cite{Kallenborn:2019ilo}.

Where the medium really is piecewise constant, and the slab edges fall on its jumps, \equ{slab_approx} is exact: each factor is exact, so the product is too, however many slabs there are\ \cite{Bustamante:2019ggq}. The approximation enters only where the density varies \emph{within} a slab. What is dropped there is everything that depends on how $\mathbb{H}$ varies inside a slab: the nested commutators of $\mathbb{H}$ with itself at two or more positions in the slab, from $\Omega_2$ onward, and the departure of $\int \mathbb{H}\,dl$ from its midpoint value. The commutator between neighboring slabs is never dropped, since a product of exponentials crosses one slab and then the next exactly, however far $\mathbb{H}$ jumps between them.  

\smallskip

\textit{Beyond Magnus order 2.---}Magnus orders 4, 6, and 8 put back the in-slab commutators that the traditional, order-2 slab approximation drops. Their errors fall as $h^{4}$, $h^{6}$, and $h^{8}$ in the slab width $h$, against $h^{2}$ at order 2, so halving $h$ divides them by $2^{4}$, $2^{6}$, and $2^{8}$ where the slab approximation gains only $2^{2}$. 

The lower-right panel of \figu{phase_vs_profile} (Sec.~\ref{sec:phase_vs_profile}) shows the gain on an exponentially falling density profile. At about a thousand slabs, the error in the returned probabilities is about $10^{-6}$ at order 2, $10^{-10}$ at order 4, and $10^{-15}$ at order 6, which is at the round-off floor. Order 6 is below $10^{-8}$ already at about a hundred slabs, a level order 2 never reaches on the panel.


\subsection{Tolerance and adaptive refinement}
\label{sec:refinement}

\textit{What tolerance means.---}By default, \magnus\ chooses the number of slabs needed to meet a requested tolerance. It computes the probability matrix, multiplies the number of slabs by $1.5$ ({\tt growth\_factor\_n\_slabs}), recomputes, and stops when two successive levels agree in every entry. Writing $P^{(n)}_{\alpha\beta}$ for the $\nu_\alpha \to \nu_\beta$ probability computed at the $n$-th refinement step, the stopping criterion is
\begin{equation}
 \label{equ:refinement_stop}
 \left\lvert P^{(n)}_{\alpha\beta} - P^{(n-1)}_{\alpha\beta} \right\rvert
 \leq
 {\tt atol} + {\tt rtol}\, \left\lvert P^{(n-1)}_{\alpha\beta} \right\rvert
 \quad \text{for all } \alpha, \beta \;,
\end{equation}
with absolute and relative tolerances ${\tt atol} = {\tt rtol} = 10^{-3}$ by default. A user who wants two consecutive agreements instead of one can ask for them ({\tt strict\_convergence}). We call this sequence of grids the \emph{refinement ladder}. (Normally, each level has 1.5 times as many slabs as the one before. Declared discontinuities (Sec.~\ref{sec:declaring_edges}) can shrink that step, because their edges are present at every level. For example, with 15 declared edges, two successive levels can have 17 and 18 edges, almost the same. Two grids that similar return similar answers whether or not either has converged, so \magnus\ counts an agreement only when the finer grid has at least 25\% more edges than the coarser one.)

\smallskip

\textit{The refinement ladder.---}Every level prior to the one that converges is computed and discarded. On an Earth chord at three flavors, with standard oscillations, reaching a requested accuracy through the ladder takes about three to four times longer than evaluating once at the right slab count, computed in advance. (That count is reported after any converged call to {\tt osc\_prob}, introduced in Sec.~\ref{sec:scope_interface}, through its {\tt convergence\_info} argument. Passing it as {\tt n\_slabs}, with ${\tt rtol} = {\tt atol} = {\tt None}$, makes \magnus\ evaluate once at that count and skip the ladder.)

Four devices keep the ladder short.
\begin{enumerate}
\item \textit{A physics-informed start.} Before refining, \magnus\ estimates the total phase accumulated along the trajectory from $\lVert \Omega_1 \rVert_2$, and starts with enough slabs to hold about $2\pi$, one oscillation cycle, in each. This is deliberately coarser than the convergence guarantee of Sec.~\ref{sec:applicability}, which asks for less than $\pi$: at order 4, slabs this wide already give probabilities accurate to about $10^{-3}$ on smooth profiles, the default tolerance, and the ladder refines from there as the profile requires. The estimate seeds the collocation path only; on the cumulative path, where the accuracy depends jointly on the number of slabs and of points per slab ({\tt n\_tpts\_per\_slab}, which the ladder also grows by $1.5$, {\tt growth\_factor\_n\_tpts\_per\_slab}), the ladder starts from its minimum ({\tt min\_n\_slabs}).
\item \textit{Warm starts.} In a scan over neutrino energy or arrival direction, each point starts from the count at which the previous point converged.
\item \textit{Slab edges on the discontinuities.} A slab that contains a jump of the Hamiltonian, caused, \eg, by a density jump, spoils the convergence whatever order was requested: the expansion assumes a smooth Hamiltonian inside each slab, and no order repairs a jump. The error then falls only about as fast as the slab width itself, and irregularly, depending on where in its slab the jump falls. With an edge placed on every jump, each slab holds a smooth Hamiltonian again, and the error falls as $h^{p}$, as the order promises. The placement of the edges can matter more than the order. This matters most in the Earth, since trajectories through it cross layer boundaries of the Preliminary Reference Earth Model (PREM) density profile. Section~\ref{sec:declaring_edges} describes how \magnus\ places edges on those boundaries itself, how a user declares a jump in any other profile, and how much the placement improves the accuracy.
\item \textit{A user-supplied minimum.} An {\tt n\_slabs} (or {\tt min\_n\_slabs}) passed together with a tolerance is a lower bound on the ladder, not a replacement for it. The ladder starts from whichever is larger: the user's count, or the count estimated from the accumulated phase (item 1 above). This is also the remedy for a profile with features narrower than any slab, turbulent or noisy, which no level of the ladder can see, so no agreement between levels can detect it.
\end{enumerate}

\smallskip

\textit{The refinement cap.---}The ladder does not run without limit. It stops at a maximum number of slabs ({\tt max\_n\_slabs}), set by default to $20\,000$ on the collocation path and $2\,000$ on the cumulative one, whose slabs cost more; it also stops after 50 levels ({\tt max\_num\_loops}), and, on the cumulative path, at 500 points per slab ({\tt max\_n\_tpts\_per\_slab}). If a limit is reached before the tolerance is met, \magnus\ returns the answer from the last level and warns, saying how far from converged it stopped, as a multiple of the tolerance. \magnus\ has no option to silence this warning, because the returned answer is unitary to round-off like every other and can look entirely plausible while being wrong. Section~\ref{sec:diagnostics} discusses how far these warnings can be trusted, and how often an answer comes back outside the tolerance without a warning.


\subsection{The matrix exponential}
\label{sec:exponential}

Every slab ends in a matrix exponential, $\mathbb{U}_j = \exp[\Omega_{\rm GL}^{(p)}]$ or $\mathbb{U}_j = \exp[\Omega_{\rm cum}^{(k)}]$, from \equ{slab_exponential} or \equ{slab_exponential_cum}, and the exponent is anti-Hermitian in either case. Evaluating the exponential is a large part of the cost of a slab. The standard route is to diagonalize the exponent with a library routine, such as {\tt numpy.linalg.eigh}, which returns eigenvalues and eigenvectors, and to exponentiate the eigenvalues. That routine costs about $1~\mu$s per $3 \times 3$ matrix however many matrices it is given at once, because internally it loops over them. What can be done to reduce the cost depends on the size of the matrix.

\smallskip

\textit{Closed-form kernel at two and three flavors.---}The exponential of a $d \times d$ matrix is a polynomial in that matrix of degree $d - 1$ (the Cayley-Hamilton theorem), with coefficients that depend only on the eigenvalues. At these sizes, the eigenvalues have closed forms\ \cite{Zaglauer:1988gz}; no eigenvectors are needed. \magnus\ evaluates this polynomial in a compiled kernel (using {\tt Numba}), as {\tt NuOscProbExact}~\cite{NuOscProbExact, Bustamante:2019ggq} does. Measured against {\tt eigh} under the protocol of Sec.~\ref{sec:timing_protocol}, on stacks of about a hundred to a few thousand slabs, the kernel computes the exponential about 7 times faster at three flavors and 7--13 times at two. End-to-end, an energy scan through the Earth runs about twice as fast.

\smallskip

\textit{Accuracy near degeneracy.---}The closed form also gains accuracy where traditional eigenvector methods lose it. The coefficients contain ratios of the form $(e^{-i\lambda_j} - e^{-i\lambda_k})/(\lambda_j - \lambda_k)$, where $\lambda_j$ and $\lambda_k$ are eigenvalues. A naive evaluation computes this as a small difference divided by a small gap. With $\lambda_j - \lambda_k = 2x$ the same ratio is a phase times $\sin x / x$, which is regular at $x = 0$. This is what \magnus\ evaluates, so a nearly degenerate spectrum needs no special treatment. On random Hermitian Hamiltonians with the two closest eigenvalues pushed together by hand, the error is constant over ten orders of magnitude in the gap.

The closed form fails in one region, at three flavors only. The eigenvalues of a $3 \times 3$ matrix are the roots of its characteristic polynomial, a cubic, and the closed form obtains them through the trigonometric solution of that cubic, which involves the arccosine of a combination of the matrix invariants~\cite{MacFarlane:1968vc}. Where that combination approaches $\pm 1$, which happens when the eigenvalue spectrum is nearly degenerate and the norm of the matrix is large, the arccosine amplifies round-off in its argument, and the eigenvalues lose several digits of accuracy compared to a sixty-digit reference calculation. \magnus\ detects that region and hands such matrices to {\tt eigh}. None of the calculations in this paper reach that region, so all of them use the closed form. The two-flavor closed form has no cubic and no arccosine, and stays accurate down to exact degeneracy.

\smallskip

\textit{Four and five flavors, and beyond.---}For a $4 \times 4$ matrix, the characteristic polynomial is a quartic, whose roots have a closed form too unwieldy to be worth evaluating; for a $5 \times 5$ matrix, it is a quintic, which has no general solution in radicals (the Abel--Ruffini theorem). At four and five flavors, \magnus\ instead diagonalizes numerically, with a compiled iterative (Jacobi) eigensolver that treats all the slabs of a trajectory at once and starts the iteration for each slab from the eigenvectors found for the previous one. If any slab has not converged after 30 iterations, \magnus\ recomputes all of them with {\tt eigh}. Above five flavors, \magnus\ uses {\tt eigh} directly.


\subsection{Adiabatic transport with exact Magnus patches}
\label{sec:hybrid}

The cost of computing a probability with the Magnus expansion grows slowly with the accumulated phase, $\Phi$ in \equ{phase}, but it does grow (Sec.~\ref{sec:phase_vs_profile}). A solar neutrino of a few MeV accumulates a phase of $10^{3}$ to $10^{6}$ radians on its way out of the Sun, and at the upper end of that range the refinement ladder can reach its slab cap ({\tt max\_n\_slabs}) before the tolerance is met. Raising the limit would work, at a cost that grows with the phase. The physics offers a better solution.

Where $\mathbb{H}$ changes slowly compared with the gaps between its eigenvalues---which fails near a \emph{level crossing}, where two eigenvalues approach each other---the neutrino state follows the instantaneous eigenstates of $\mathbb{H}$ as $\mathbb{H}$ changes. This is adiabatic evolution. For standard oscillations in matter, it underlies the adiabatic MSW conversion of solar neutrinos\ \cite{Wolfenstein:1977ue, Mikheev:1986gs, Parke:1986jy, Bethe:1986ej, Kuo:1989qe}, but nothing in it is specific to that Hamiltonian, and \magnus\ applies it to any: standard, with new interactions, or with extra flavors. \magnus\ uses adiabatic evolution where it holds, and the Magnus expansion described earlier where it does not.

\smallskip

\textit{Adiabatic transport.---}At each position $l$ along the neutrino trajectory, we diagonalize the Hamiltonian, $\mathbb{H}(l) = \mathbb{W}(l)\,{\rm diag}(\lambda_k)\,\mathbb{W}^\dagger(l)$, where the $\lambda_k$ are its eigenvalues and $\mathbb{W}(l)$ is the unitary matrix whose columns are its eigenvectors, \ie, the mixing matrix at that position. Adiabatic evolution from $l_0$ to $l_1$ is then
\begin{equation}
 \label{equ:adiabatic}
 \mathbb{U}_{\rm ad}(l_1, l_0)
 =
 \mathbb{W}(l_1)\,{\rm diag}\!\left(e^{-i\Phi_k(l_1)}\right)\mathbb{W}^\dagger(l_0) \;,
\end{equation}
where $\Phi_k = \int_{l_0}^{l_1} \lambda_k\,dl$ is the phase accumulated by the $k$-th eigenstate. It enters through $e^{-i\Phi_k}$, so an error of a few tenths of a radian in $\Phi_k$ matters, however large $\Phi_k$ is: when $\Phi_k$ is $10^{6}$ radians, the integral must be computed to a few parts in $10^{7}$. We integrate with Simpson's rule, which reaches that accuracy on a grid of a few thousand points. (The trapezoid rule does not, and its error is smooth enough to be mistaken for a physical effect.)

Each eigenvector in $\mathbb{W}$ is defined only up to a phase. We fix it along the grid by choosing, at each grid point $l_i$, the phase of $v_k(l_i)$ that makes its overlap with the previous one, $v_k^\dagger(l_{i-1})\, v_k(l_i)$, real and positive. Since $\lVert v_k(l_i) - v_k(l_{i-1}) \rVert^2 = 2 - 2\,{\rm Re}\,[v_k^\dagger(l_{i-1})\, v_k(l_i)]$ for unit vectors, this is the phase that keeps each eigenvector as close as possible to the previous one.

Besides the phase $\Phi_k$ from its eigenvalue, an eigenstate that follows a changing Hamiltonian picks up a second phase, which depends only on how its eigenvector turns along the path: the geometric, or Berry, phase. Because the rule above transports each eigenvector along the grid, the geometric phase is already carried by $\mathbb{W}(l_1)$. For a complex Hamiltonian, it is a genuine phase; for a real one, it reduces to a sign. Finally, \equ{adiabatic} is unitary on any grid, since each of its three factors is.

\smallskip

\textit{Finding where adiabaticity fails.---}Adiabatic evolution fails where the Hamiltonian turns faster than the eigenstates can follow, which happens near a level crossing. We locate those regions with the Hellmann-Feynman identities\ \cite{Feynman:1939zza},
\begin{equation}
 \label{equ:hf}
 \frac{d\lambda_k}{dl} = \langle v_k \left\vert \frac{d\mathbb{H}}{dl} \right\vert v_k \rangle ,
 \;\;
 \gamma_{jk} = \frac{\lvert\langle v_j \lvert \tfrac{d\mathbb{H}}{dl} \rvert v_k \rangle\rvert}
 {(\lambda_k - \lambda_j)^2} \;.
\end{equation}
The second quantity measures how non-adiabatic the pair of levels $(j, k)$ is: small where the levels are far apart compared with how fast $\mathbb{H}$ changes, large where they approach. (For two flavors, $\gamma_{jk} = 1/(2\gamma_{\rm LZ})$, where $\gamma_{\rm LZ} = \lvert \lambda_k - \lambda_j \rvert / (2\lvert d\theta_m/dl \rvert)$, with $\theta_m$ the mixing angle in matter, is the adiabaticity parameter of the Landau-Zener literature.) Both expressions need only the derivative of $\mathbb{H}$, not the derivative of an eigenvector, which would depend on the arbitrary phase above.

We evaluate $\gamma_{jk}$ for every pair of eigenvalues along the trajectory, on a grid of probe points, so a crossing between any pair is found, and several crossings at the same place, as when three eigenvalues approach at once, are all found. Wherever $\gamma_{jk}$ exceeds a threshold, $0.1$ at first, we open a window and widen it until $\gamma_{jk}$ drops below the threshold again, so the window spans the crossing; overlapping windows are merged, and a pair of exactly equal eigenvalues always gets one. Inside the windows, the evolution is computed with the Magnus expansion, as described next; outside them, with \equ{adiabatic}.

\smallskip

\textit{Patching the crossings.---}Across each window, we compute the evolution operator with the Magnus expansion, at order 6 by default, doubling the number of slabs until successive results agree to a tenth of the requested tolerance, or to $10^{-7}$ if that is tighter. The patch has to converge on its own, because the certification described below cannot see its error. The full evolution operator is then the product of the pieces in the order they are crossed,
\begin{equation}
 \label{equ:hybrid_composition}
 \mathbb{U}
 =
 \mathbb{U}_{\rm ad}\,\mathbb{U}_{\rm patch}\, \cdots \,\mathbb{U}_{\rm ad} \;,
\end{equation}
with $\mathbb{U}_{\rm ad}$ from \equ{adiabatic} between windows and $\mathbb{U}_{\rm patch}$ from the Magnus expansion inside them. Multiplying the pieces introduces no approximation of its own: the evolution across two intervals in succession is the product of the evolutions across each, whatever method computed them. The product is unitary because every factor is.

The usual alternative is the Landau--Zener formula\ \cite{Zener:1932ws}, which estimates the probability that a neutrino hops from one eigenstate to the other at a crossing. \magnus\ does not use it, except as a check: for a crossing where the eigenvalues vary linearly, the case the formula describes, the hop probability we compute agrees with it to about $10^{-3}$.

\smallskip

\textit{Certification.---}The procedure above depends on the threshold on $\gamma_{jk}$ and on two grids: the one on which $\mathbb{H}$ is diagonalized and the one on which the crossings are searched for. No choice of these is safe for every Hamiltonian, so \magnus\ does not trust a single pass. It repeats the computation, each time dividing the threshold by three and making both grids twice as fine, until two successive passes agree within the requested tolerance; the result is then reported as \emph{certified}. A window that does not move between two passes holds the same patch in both, so this comparison cannot detect an error in the patch. A result with no window at all is certified only if, in addition, the largest $\gamma_{jk}$ along the trajectory is small enough for adiabatic evolution alone to meet the tolerance. If the passes still disagree when the threshold and both grids have reached their limits ({\tt min\_threshold}, {\tt max\_n\_probe}, {\tt max\_n\_points}), or after twelve passes ({\tt max\_iters}), or if a patch fails to converge, \magnus\ returns the last result, which is unitary to round-off like every other, and reports that it is not certified. (These limits are arguments of {\tt magnus.adiabatic.hybrid\_propagator}; {\tt osc\_prob} uses their defaults.)

We validated the adiabatic transport with Magnus patches, certification included, against a tight {\tt DOP853} integration (order-8 Runge-Kutta) of the evolution equation, on a set of density profiles with zero, one, two separate, and two overlapping crossings, at three, four, and five flavors, with and without CP violation. The method runs 30 to 4\,800 times faster than the integration; its largest error on that set, over certified and uncertified results, is about $10^{-4}$ for a requested tolerance of $10^{-4}$, and $3 \cdot 10^{-3}$ where two resonances lie close enough to merge into one window.

Two ways of failing remain, neither specific to \magnus: any method that evaluates the Hamiltonian on a grid of positions has them. First, a crossing narrower than the spacing of the grid falls between its points and is never seen. Second, two passes can agree with each other and both be wrong, so a certified result can still lie outside the tolerance. Section~\ref{sec:limitations} shows an example of the first.


\subsection{The phase-averaged probability}
\label{sec:averaged}

A neutrino from an astrophysical source arrives with a phase that no experiment resolves. For a TeV neutrino from $100$~Mpc, the accumulated phase is of order $10^{16}$ radians, and none of the quantities that fix it---the distance, the size of the source, the energy of the neutrino---is known to anything near the precision it would take to resolve the oscillations. Whatever the phase is, the measurement averages over many cycles of it. Propagating such a neutrino while resolving the oscillations is expensive and unnecessary; what is needed is the average. In \magnus, the probability functions return it when called with {\tt average=True}. Below, we define what that average is.


\subsubsection{The averaged probability}

We write the Hamiltonian in its eigenbasis, $\mathbb{H} = \mathbb{V}\,{\rm diag}(\lambda_i)\,\mathbb{V}^\dagger$, with $\mathbb{V}$ the matrix of its eigenvectors. The oscillation probability is then a sum of terms, one for each eigenstate pair, each oscillating with the pair's phase difference. Averaging over the phase removes every term with $i \neq j$ and leaves
\begin{equation}
 \label{equ:averaged}
 \left\langle P_{\nu_\alpha \to \nu_\beta} \right\rangle
 =
 \sum_i \lvert \mathbb{V}_{\alpha i}\rvert^2\, \lvert \mathbb{V}_{\beta i}\rvert^2 \;.
\end{equation}
Equation~(\ref{equ:averaged}) is the exact limit of the probability as $L/E \to \infty$. Computing it requires one diagonalization of $\mathbb{H}$.

Three properties of \equ{averaged} are worth noting:
\begin{enumerate}
\item It is symmetric under the exchange of $\alpha$ and $\beta$.
\item In vacuum, it is the same for neutrinos and antineutrinos, since the antineutrino probability is obtained by replacing $\mathbb{V}$ with $\mathbb{V}^\ast$, and \equ{averaged} contains only $\lvert \mathbb{V} \rvert^2$. At three flavors, the CP-violation phase, $\delta_{\rm CP}$, still affects the result through the values of $\lvert \mathbb{V}_{\alpha i} \rvert^2$, but it produces no difference between neutrinos and antineutrinos. In matter, neutrinos and antineutrinos have different averaged probabilities, because the matter potential has opposite signs for the two.
\item It does not depend on the baseline, and it depends on the energy only through the eigenvectors of $\mathbb{H}$. In vacuum, those do not depend on the energy, since the vacuum Hamiltonian is proportional to $1/E$, so the averaged probability is a constant. In matter, the eigenvectors change with the energy, through the competition between the matter potential and the vacuum term, and so does the averaged probability. A non-standard Hamiltonian may have an entirely different energy dependence.
\end{enumerate}
These are properties of the limit. Where part of an interference term survives (see below), none of the three holds in general: the probability depends on the baseline, is not symmetric under the exchange of $\alpha$ and $\beta$, and, in vacuum, differs between neutrinos and antineutrinos when CP is violated.


\subsubsection{Coherent blocks}

Equation~(\ref{equ:averaged}) assumes that the phase between every pair of eigenstates has averaged away. That is a statement about each pair separately. Two eigenvalues close enough that their phase difference stays small over the whole baseline remain coherent, and their cross term survives. To keep those cross terms, \magnus\ groups the eigenvalues into mutually coherent blocks, and sums coherently within each block and incoherently between blocks,
\begin{equation}
 \label{equ:averaged_blocks}
 \left\langle P_{\nu_\alpha \to \nu_\beta} \right\rangle
 =
 \sum_b \left\vert \sum_{i \in b} \mathbb{V}^\ast_{\alpha i}\mathbb{V}_{\beta i} \right\vert^2 \;,
\end{equation}
which reduces to \equ{averaged} when every block contains one eigenvalue. The distinction matters whenever two eigenvalues lie close enough that their phase difference does not average out over the baseline, as for a pseudo-Dirac pair or a sterile state with a small mass splitting. It matters most for a degenerate spectrum: with all eigenvalues equal, nothing oscillates and the probability is the identity, while \equ{averaged} would return a mixture. A pair that is neither coherent nor fully averaged is described by neither equation; its average depends on how the phase is spread in energy, which we compute next.


\subsubsection{The phase average}

Equation~(\ref{equ:averaged}) is the limit in which every interference phase is washed out. A real measurement washes out only the phases it cannot resolve. A detector with a relative energy resolution $\sigma$ averages each phase over the range it covers across that resolution, so a phase that barely changes with energy is not averaged at all. Over 1000~km at 1~GeV, the atmospheric phase is about 6~rad, and \equ{averaged} misses the oscillation entirely. Instead, {\tt average=True} returns the phase average, in which each interference term keeps its phase at the central energy and is damped by the spread of that phase,
\begin{equation}
 \label{equ:phase_average}
 \left\langle P_{\nu_\alpha \to \nu_\beta} \right\rangle
 =
 \sum_{ij} \mathbb{V}^\ast_{\alpha i}\mathbb{V}_{\beta i}\mathbb{V}_{\alpha j}\mathbb{V}^\ast_{\beta j}\,
 e^{-i\phi_{ij}}\, e^{-\sigma^2 \phi_{ij}'^2/2} \;,
\end{equation}
with $\phi_{ij} = (\lambda_i - \lambda_j)L$ and $\phi'_{ij} = d\phi_{ij}/d\ln E$. The damping factor is the average of $e^{-i\phi}$ over a Gaussian spread of width $\sigma$ in $\ln E$, the form familiar from wave packets and finite resolution\ \cite{Kiers:1995zj, Blennow:2005yk, Ohlsson:2000mj}. In vacuum, $\phi' = -\phi$; at $\sigma = 10\%$, a phase of $2\pi$ keeps 82\% of its interference, a phase of 20~rad keeps 14\%, and a phase of 40~rad keeps $3\times10^{-4}$. Therefore, \equ{phase_average} is \equ{averaged} wherever every phase runs through many cycles, the oscillation probability itself where none does, and a smooth interpolation in between. We obtain the slopes from the Hellmann-Feynman identity of \equ{hf}, $d\lambda_i/d\ln E = \langle v_i|\, d\mathbb{H}/d\ln E\, |v_i\rangle$, with $d\mathbb{H}/d\ln E$ from a central difference, so they cost two more evaluations of $\mathbb{H}$ and no diagonalization. The spread is set with {\tt average\_spread}, and is $10\%$ by default.

Only the phases are averaged; the eigenvectors, $\mathbb{V}$, stay at the central energy. This differs from averaging the probability over an energy band, which also averages over the eigenvectors, since they change with energy. In vacuum, the eigenvectors do not change with energy, so the two averages differ only slightly, through the curvature of the phase in energy, which \equ{phase_average} neglects; in matter, they differ more the faster the eigenvectors change. For the averaged $\nu_e$ survival probability of solar neutrinos, computed with the BS2005-AGS,OP solar model at 40 energies from 0.1 to 20~MeV, averaging over a $10\%$ energy band changes 26 of the 40 values by more than $10^{-4}$ relative to the limit of \equ{averaged}. The phase average, \equ{phase_average}, changes none of them, because there is no interference for it to act on. In general, wherever the phase average differs from the limit by less than $10^{-4}$, \magnus\ returns the limit.


\subsubsection{A varying Hamiltonian}

Equations~(\ref{equ:averaged})--(\ref{equ:phase_average}) assume that the Hamiltonian is the same along the whole path. The phase average carries over to a Hamiltonian that varies. As in \equ{phase_average}, an energy offset $u = \delta \ln E$ shifts each instantaneous eigenvalue by $u\, d\lambda_i/d\ln E$ and leaves the eigenvectors unchanged. The probability is averaged over a Gaussian in $u$ of width $\sigma$,
\begin{equation}
 \label{equ:phase_average_varying}
 \left\langle P_{\nu_\alpha \to \nu_\beta} \right\rangle
 =
 \int du\, \frac{e^{-u^2/2\sigma^2}}{\sqrt{2\pi}\,\sigma}\,
 \left[ \mathbb{U}(u)\, \rho_\alpha\, \mathbb{U}^\dagger(u) \right]_{\beta\beta} ,
\end{equation}
where $\mathbb{U}(u)$ is the evolution operator from $l_0$ to $l_1$ under $\mathbb{H}(l) + u\, \mathbb{D}_{\rm diag}(l)$, with $\mathbb{D}_{\rm diag}$ the part of $d\mathbb{H}/d\ln E$ diagonal in the instantaneous eigenbasis, and $\rho_\alpha$ the state the neutrino starts in.

The user sets that state with {\tt average\_initial\_state}. By default, it is {\tt 'flavor'}: the neutrino starts as the flavor state $\nu_\alpha$, $\rho_\alpha = \lvert \nu_\alpha \rangle \langle \nu_\alpha \rvert$, as for a neutrino produced in a medium, \eg, in the Earth or the Sun. In the eigenbasis of $\mathbb{H}(l_0)$, with eigenstates $\lvert v_i \rangle \equiv \lvert v_i(l_0) \rangle$, that state is a coherent superposition, with amplitudes $\mathbb{V}^\ast_{\alpha i}(l_0)$. With {\tt average\_initial\_state='decohered'}, the neutrino starts instead as an incoherent mixture of the same eigenstates, $\rho_\alpha = \sum_i \lvert \mathbb{V}_{\alpha i}(l_0)\rvert^2\, \lvert v_i \rangle \langle v_i \rvert$, as for a neutrino that has lost its coherence before reaching $l_0$, such as one arriving from a distant source (Sec.~\ref{sec:solar_tomography}). Since $\rho_\alpha$ enters \equ{phase_average_varying} linearly, the two differ only by the interference between eigenstates present at production,
\begin{align}
 \label{equ:flavor_vs_decohered}
 \left\langle P_{\nu_\alpha \to \nu_\beta} \right\rangle^{\rm flavor}
 =\;
 & \left\langle P_{\nu_\alpha \to \nu_\beta} \right\rangle^{\rm decoh}
 \\
 & + \sum_{i \neq j} \mathbb{V}^\ast_{\alpha i}(l_0)\, \mathbb{V}_{\alpha j}(l_0)
 \left\langle \left[ \mathbb{U}(u)\, \lvert v_i \rangle \langle v_j \rvert\, \mathbb{U}^\dagger(u) \right]_{\beta\beta} \right\rangle_u ,
 \nonumber
\end{align}
where $\langle \cdot \rangle_u$ is the Gaussian average of \equ{phase_average_varying}. Each term of the sum carries the phase difference accumulated from production onward, and is damped by the spread of that phase, so it vanishes wherever those phases are large. For a constant Hamiltonian, the flavor start gives \equ{phase_average}, and the decohered start gives the limit \equ{averaged} exactly.

\magnus\ evaluates \equ{phase_average_varying} by one of two routes, depending on whether the Hamiltonian varies smoothly or has declared discontinuities.

\smallskip

\textit{Route 1: a smoothly varying Hamiltonian.---}This is the case of the Sun or a supernova. The eigenbases at production and at detection differ, and at a level crossing on the way the neutrino may move from one eigenstate to another. Then, $\mathbb{U}(u)$ is the product of \equ{hybrid_composition} evaluated at the offset $u$,
\begin{equation}
 \label{equ:evolution_offset}
 \mathbb{U}(u)
 =
 \mathbb{U}_{\rm ad}(u)\,\mathbb{U}_{\rm patch}(u) \cdots \mathbb{U}_{\rm ad}(u) \;,
\end{equation}
where each $\mathbb{U}_{\rm ad}(u)$ is \equ{adiabatic} with every phase $\Phi_k$ replaced by $\Phi_k + u\, d\Phi_k/d\ln E$, and each $\mathbb{U}_{\rm patch}(u)$ is the Magnus patch of Sec.~\ref{sec:hybrid} with every eigenvalue shifted in the same way. Across the adiabatic stretches, the average over $u$ has a closed form, as in \equ{phase_average}; across each crossing, the patch is computed at a few values of $u$ and combined by Gauss-Hermite quadrature. No energy is sampled, and the result does not depend on where the windows around the crossings are drawn.

Equation~(\ref{equ:phase_average_varying}) then has two limits in closed form. First, where every phase along the path has averaged away, it reduces, for either initial state, to
\begin{equation}
 \label{equ:averaged_varying}
 \left\langle P_{\nu_\alpha \to \nu_\beta} \right\rangle
 =
 \sum_{ij} \lvert \mathbb{V}_{\alpha i}(l_0)\rvert^2\,
 P^{\rm cross}_{ij}\,
 \lvert \mathbb{V}_{\beta j}(l_1)\rvert^2 \;,
\end{equation}
where $\mathbb{V}(l_0)$ and $\mathbb{V}(l_1)$ are the eigenvector matrices at production and at detection, and $P^{\rm cross}_{ij}$ is the probability that a neutrino produced in eigenstate $i$ arrives in eigenstate $j$: the identity where the evolution is adiabatic, and computed with the Magnus patch across a crossing. This is the standard result for solar and supernova neutrinos\ \cite{Kuo:1989qe, Dighe:1999bi}, here for any number of eigenstates and crossings. It is the path counterpart of the limit \equ{averaged}, to which it reduces for a constant Hamiltonian. Because it adds probabilities, not amplitudes, it drops interference between eigenstates, including where a phase is too small to have averaged away.

Second, on a path with no crossing, $P^{\rm cross}$ is the identity and the average over $u$ can be done in closed form for any phases. For a neutrino that starts decohered, there is no interference to keep, and the result is \equ{averaged_varying} exactly. For a neutrino that starts as a flavor state, it is
\begin{align}
 \left\langle P_{\nu_\alpha \to \nu_\beta} \right\rangle
 =
 \sum_{ij}\,
 & \mathbb{V}^\ast_{\alpha i}(l_0)\,\mathbb{V}_{\alpha j}(l_0)\,
 \mathbb{V}_{\beta i}(l_1)\,\mathbb{V}^\ast_{\beta j}(l_1)
 \label{equ:phase_average_nocross} \\
 & \times e^{-i\Phi_{ij}}\, e^{-\sigma^2 \Phi_{ij}'^2/2} \;,
 \nonumber
\end{align}
where the phases of the eigenvectors in $\mathbb{V}(l_0)$ and $\mathbb{V}(l_1)$ are fixed by transport along the path, as in \equ{adiabatic}; $\Phi_{ij} = \Phi_i - \Phi_j$ is the phase difference accumulated along the path; and $\Phi'_{ij} = d\Phi_{ij}/d\ln E$. For a constant Hamiltonian, \equ{phase_average_nocross} is \equ{phase_average}, and where every $\Phi'_{ij}$ is large, it is \equ{averaged_varying}.

In practice, \magnus\ computes \equ{averaged_varying} first, because it is cheap and is the answer wherever every phase is large, as for a neutrino produced in the solar core. It then computes the phase average: from \equ{phase_average_nocross} if the path has no crossing and the neutrino starts in flavor, and from \equ{phase_average_varying} with the patches if the path has a crossing. It returns the phase average whenever it differs from \equ{averaged_varying} by more than $10^{-4}$, or than the requested tolerance if that is tighter, in absolute terms or relative to the probability. For neutrinos of 10~GeV to 10~TeV that cross the Sun along five chords and arrive decohered (Sec.~\ref{sec:solar_tomography}), \equ{averaged_varying} is off by up to 0.14, while \equ{phase_average_varying} agrees to $4 \times 10^{-5}$ with a brute-force average over $u$.

The crossings are located on a grid of probe points along the path, so a feature narrower than the grid spacing, such as a sharp density front, can fall between two points. To catch it, \magnus\ first checks the profile for such a feature that could also move probability between eigenstates. If it does, the crossings are located with the certified refinement of Sec.~\ref{sec:hybrid} instead. If no refinement resolves the feature, \magnus\ warns that it must be declared (Sec.~\ref{sec:declaring_edges}), which moves the calculation to Route 2.

\smallskip

\textit{Route 2: a profile with declared discontinuities.---}This is the case of the Earth, whose density jumps at the boundaries between layers, or of any profile whose discontinuities the user has declared. At a jump there is no smooth eigenbasis to follow, so $\mathbb{D}_{\rm diag}$, and with it \equ{phase_average_varying}, is not defined. \magnus\ instead computes the probability, by default, at 41 energies  across a window of $\pm 10\%$ around each requested energy, starting from the flavor state, and returns their mean. The probability functions take the half-width of the window as {\tt average\_spread} and the number of energies as {\tt average\_n\_samples}. This is an average over that window, not the phase average, and it depends on the width of the window; \magnus\ warns so, and reports the statistical uncertainty of the mean.

\smallskip

Section~\ref{sec:sun} shows the result on the Sun, and Sec.~\ref{sec:averaging_is_not_estimating} explains why averaging a scan over energy does not converge to the same number.


\section{Code description}
\label{sec:code}


\subsection{Scope and interface}
\label{sec:scope_interface}

\magnus\ is written in Python and requires version 3.10 or later. It depends on {\tt NumPy}\ \cite{Harris:2020xlr}, {\tt SciPy}\ \cite{Virtanen:2019joe}, {\tt joblib}, {\tt Numba}\ \cite{Lam:2015numba}, and {\tt Matplotlib}\ \cite{Hunter:2007ouj}, all installed with it.

\smallskip

\textit{One function, and wrappers around it.---}The central function is {\tt osc\_prob}. It takes a Hamiltonian---either a function ${\tt H\_func}(l)$ of the position that returns a Hermitian matrix of any size, or a constant matrix---and returns the matrix of probabilities $P_{\nu_\alpha \to \nu_\beta}$ for every pair of initial and final flavors (Sec.~\ref{sec:minimal} shows how it is indexed). 

Every other probability function that \magnus\ exposes to the user builds a Hamiltonian for a particular case and passes it to {\tt osc\_prob}. (Three of them evaluate the standard analytic expressions instead; \tabl{wrappers} names them.) For convenience, \magnus\ provides 56 wrappers for common cases, shown in \figu{layers} and \tabl{wrappers}: one for each flavor count (2, 3, 4, or 5), environment (vacuum, constant density, exponential density, the Earth, the Sun), and scenario (standard oscillations, non-standard interactions, Lorentz-invariance violation), except non-standard interactions in vacuum, which need matter. Their names state the flavor count, the environment, and the scenario, which is omitted for standard oscillations: \eg, {\tt osc\_prob\_3nu\_sun}, {\tt osc\_prob\_4nu\_earth\_liv}, and {\tt osc\_prob\_3nu\_matter\_nsi\_constant\_density}. The physical parameters of each wrapper are passed as named arguments: the mixing angle $\theta_{12}$ as {\tt s12}, the matter density as {\tt rho}, the non-standard coupling $\varepsilon_{e\mu}$ as {\tt eps\_em}. Listing~\ref{lst:constant} shows a wrapper and {\tt osc\_prob} computing the same probability.

\smallskip

\textit{Batching energy and baseline scans.---}Every wrapper accepts an array of energies, an array of baselines, or both, and returns the probabilities at every point. Two arrays are paired element by element and must have the same length; a single energy or baseline is paired with every element of the other array. {\tt osc\_prob} itself takes one Hamiltonian and returns one matrix. (Between the two sits {\tt osc\_prob\_energy\_baseline}, which receives the arrays. If the scan has one of the three structures described next, it computes all points in one batched calculation; if not, it calls {\tt osc\_prob} separately for each point. Section~\ref{sec:architecture} shows where it sits among the layers.) \emph{Batching} means computing all the points of a scan together, in shared operations, instead of one at a time, and it makes a scan cheaper than the same probabilities computed individually. 

\magnus\ batches a scan when it has one of three kinds of structure. First, when the Hamiltonian is the sum of a position-independent term and a matter potential multiplied by a fixed matrix, as in \equ{h_matter}, the potential is sampled once along the path and reused for every energy. Second, when the matter potential is constant, the entire scan reduces to the exponential of a single stack of matrices. Third, when several baselines are requested at one energy, the path to each shorter baseline is contained in the path to the longest, so a single propagation along the longest path yields the probabilities at all of them. These are three of the seven computational routes, or \emph{engines}, that \magnus\ tries in a fixed order; Sec.~\ref{sec:engines} describes all of them, and how to see which one answered a call.

\begin{figure*}[t!]
 \centering
 \includegraphics[width=0.98\textwidth]{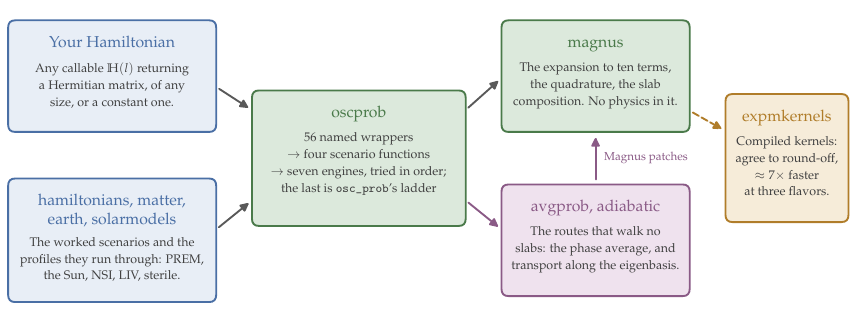}
 \caption{\label{fig:architecture}\textbf{Architecture of \magnus.} The modules of \magnus\ and the path of a request through them. A Hamiltonian enters from the left, either supplied by the user as a function of position or built by {\tt hamiltonians} for one of the implemented scenarios, with the density profile provided by {\tt matter}, {\tt earth}, or {\tt solarmodels}. The request is handled in {\tt oscprob}, which contains the choice of engine of Sec.~\ref{sec:engines} and the refinement ladder of Sec.~\ref{sec:refinement}. From there the computation proceeds along one of two routes: through {\tt magnus}, which evaluates the expansion slab by slab and contains no physics, or through {\tt avgprob} and {\tt adiabatic}, which evaluate no slabs outside the non-adiabatic windows---the first for the phase average, the second for transport along the instantaneous eigenstates---and call {\tt magnus} back to compute the evolution across each window. The dashed arrow to {\tt expmkernels} indicates that this module affects only the speed: a {\tt NumPy} implementation of the same exponential exists behind it, and the test suite verifies that the two agree to round-off.}
\end{figure*}

\smallskip

\textit{Errors and warnings that explain themselves.---}When a call is malformed, \magnus\ raises an error that says what is wrong and what it expected. For instance, requesting an Earth chord by its arrival direction but with the baseline set to {\tt None} raises an error explaining that the arrival direction sets the chord but the baseline may not be {\tt None}. A misspelled keyword raises one that names the closest valid keyword and lists the keywords the wrapper passes on to {\tt osc\_prob}. When a call is well formed but probably not what was meant, \magnus\ warns instead: \eg, when a baseline looks like it was given in kilometers rather than in eV$^{-1}$, or a density does not look like the unit it was declared in. And when the answer itself may be less accurate than requested, the warning says by how much, when it can, and what to change: \eg, how far outside the tolerance the refinement stopped, or that a discontinuity should be declared. 


\subsection{The \magnus\ architecture}
\label{sec:architecture}

Figure~\ref{fig:architecture} shows the modules of \magnus\ and the path of a request through them, from a Hamiltonian on the left to a probability on the right. A request arrives at the {\tt oscprob} module and leaves it either to {\tt magnus}, which evaluates the expansion slab by slab, to {\tt avgprob}, which computes the phase average of the probability, or to {\tt adiabatic}, which transports the neutrino state along the instantaneous eigenstates. The last two return to {\tt magnus} only to cross a non-adiabatic window with a Magnus patch. The route is selected internally by \magnus\ from the form of the request, not by the user; Sec.~\ref{sec:engines} gives the order in which the candidates are tried.

The numerical core is kept separate from the physics: {\tt magnus} imports nothing that concerns neutrinos, and its only dependency within the package is {\tt expmkernels}, the {\tt Numba}-compiled matrix exponential. The module is numerical linear algebra on an arbitrary matrix-valued function $\mathbb{A}(l)$, so it can be tested against the recursion of \equ{recursion} without involving any of the probability code. The physics, the environments, and the numerical core meet in one module, {\tt oscprob}.

\tabl{modules} lists the thirteen modules under {\tt src/magnus/}, together with the {\tt hamiltonians} subpackage, and the responsibility of each. The largest, {\tt oscprob}, holds all the wrappers; to avoid duplicating the same logic in each of them, its functions are organized in four layers, each calling the one below it.

\begin{figure*}[t!]
 \centering
 \includegraphics[width=\textwidth]{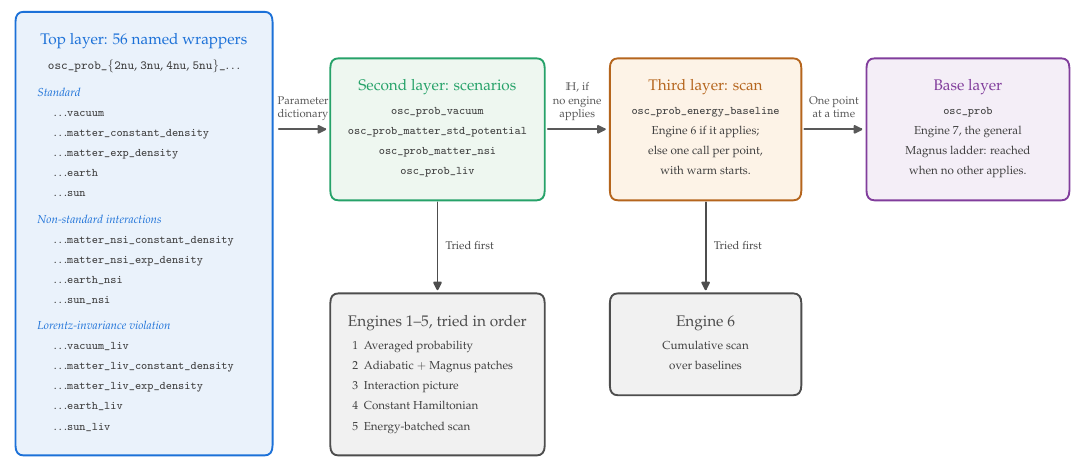}
 \caption{\label{fig:layers}\textbf{The layers of {\tt oscprob}.} The four layers of {\tt oscprob}, and the engines each one tries. A request enters at one of the named wrappers, which convert their named arguments into a parameter dictionary and pass it to the scenario function of the second layer. That function builds the Hamiltonian and tries the first five engines of Sec.~\ref{sec:engines}, in order; if none applies, it passes the Hamiltonian to {\tt osc\_prob\_energy\_baseline}, which tries the cumulative scan over baselines and otherwise calls {\tt osc\_prob} once for every requested point. {\tt osc\_prob} is itself the seventh engine, the general Magnus ladder of Sec.~\ref{sec:refinement}. The wrappers are a product set: the fourteen environment-and-scenario names listed in the left box exist at each of the four flavor counts, giving $4 \times 14 = 56$ in all.}
\end{figure*}

Figure~\ref{fig:layers} shows the four layers:
\begin{enumerate}
\item The top layer consists of the named wrappers, such as {\tt osc\_prob\_3nu\_matter\_constant\_density} in Listing~\ref{lst:constant}. Each converts its named arguments into the parameter dictionary expected by the second layer and calls the function there for its scenario.
\smallskip
\item The second layer consists of four functions, one for each predefined physics scenario shipped with \magnus\ and each applicable to any number of flavors: {\tt osc\_prob\_vacuum}, {\tt osc\us prob\us matter\us std\us potential}, {\tt osc\_prob\_matter\_nsi}, and {\tt osc\_prob\_liv}. Each builds the Hamiltonian for its scenario and tries, in order, the engines that exploit its structure: the phase average, the adiabatic transport with Magnus patches, the interaction-picture expansion, the constant-Hamiltonian engine, and the energy-batched scan. If none of them applies, it passes the Hamiltonian to the third layer.
\smallskip
\item The third layer is {\tt osc\us prob\us energy\us baseline}, which receives the Hamiltonian together with the arrays of energies and baselines. It tries the cumulative scan over baselines; if that does not apply, it calls the base layer once for each point, seeding the refinement of each from the converged slab count of the previous one.
\smallskip
\item The base layer is {\tt osc\_prob}, the general Magnus ladder of Table~\ref{tab:engines}: it computes one probability matrix with the refinement ladder, and it is reached only when no engine above it applies. It can also be called directly with an arbitrary Hamiltonian, as in Listing~\ref{lst:constant}.
\end{enumerate}

The refinement keywords ({\tt n\_slabs}, {\tt max\us n\us slabs}, {\tt t\_breakpoints}, {\tt magnus\_exp\_order}, {\tt integration\us method}, {\tt rtol}, {\tt atol}, {\tt strict\_convergence}, and {\tt n\_jobs}) are declared by {\tt osc\_prob} and by the layers between it and the wrappers, but not by the wrappers, which receive them through {\tt **kwargs}; they are documented in {\tt osc\_prob}.

Extending \magnus\ usually requires only a new Hamiltonian. {\tt osc\_prob} accepts any function that returns a Hermitian matrix and does not depend on how that matrix was constructed, so a new term can be added to the array returned by an existing builder and the result passed on unchanged; Sec.~\ref{sec:building_new_hamiltonian} shows how. A new function in the second layer is needed only for physics that fits none of the four existing scenarios; the documentation\ \cite{MagnusDocs} shows how to write one, and why the shipped scenario functions are faster than a new one unless it dispatches to the same engines.

{\def\arraystretch{1.2}
\begin{table}[t!]
 {\small
 \begin{tabular}{@{}lp{6.1cm}@{}}
 Module & Responsibility \\
 \hline
 {\tt magnus} & Magnus expansion: collocation integrators, cumulative quadrature, slab composition, the unitary exponential \\
 {\tt expmkernels} & Compiled kernels for the exponential: Cayley-Hamilton at $2\times2$ and $3\times3$, batched Jacobi solver at $4\times4$ and $5\times5$ \\
 {\tt expansionterms} & Recursion of \equ{recursion}, in exact rational arithmetic, at any order \\
 {\tt adiabatic} & Adiabatic transport, detection of level crossings, the combined propagator \\
 {\tt avgprob} & Averaged probability \\
 {\tt earth} & PREM profile, geometry \\
 {\tt solarmodels} & Tabulated standard solar models: electron density and composition along a radial ray \\
 {\tt matter} & Density profiles, $V_{\rm CC}$, matter-potential projector \\
 {\tt hamiltonians} & Mixing matrices and the vacuum, matter, NSI, and LIV Hamiltonians \\
 {\tt globaldefs} & Constants, unit conversions, NuFIT sets \\
 {\tt oscprobstd} & Closed-form standard probabilities (validation) \\
 {\tt oscprob} & User-facing probability functions \\
 {\tt plotting} & Pre-packaged figure scripts \\
 {\tt cli} & The {\tt magnus} console script \\
 \end{tabular}}
 \caption{\label{tab:modules}\textbf{Package modules.} The modules of \magnus\ v1.2.0. {\tt magnus} and {\tt hamiltonians} can be used independently of the probability functions.}
\end{table}}


\subsection{The seven engines}
\label{sec:engines}

When a user requests a probability from \magnus, they do not need to select a method to compute it. Behind the interface of Sec.~\ref{sec:scope_interface}, the seven engines of \figu{layers} can answer the request. They are tried in a fixed order, from the most specialized to the most general, and the first one whose conditions the request meets answers it. An engine whose conditions are not met declines, and the request passes to the next one. The general Magnus ladder, which comes last, accepts every request.

\smallskip

\begin{figure*}[t!]
 \centering
 \includegraphics[width=\textwidth]{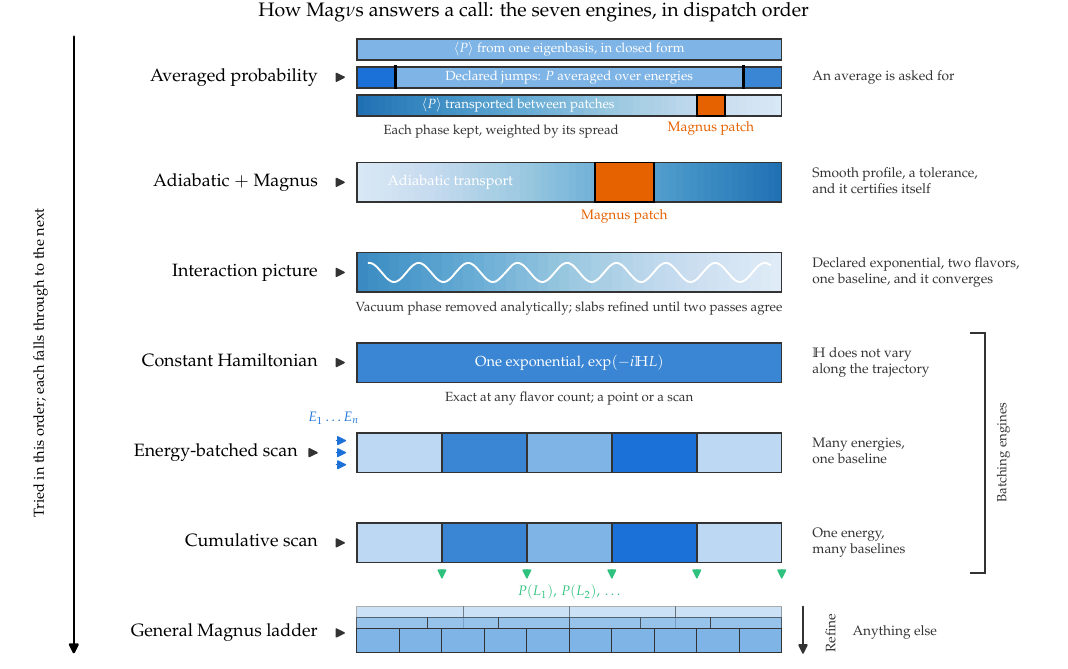}
 \caption{\label{fig:strategies}\textbf{The seven \magnus\ probability engines, in the order they are tried.} A request that does not meet an engine's conditions, summarized on the right, passes to the next, so the general Magnus ladder answers whatever no other engine takes. Each row sketches what its engine does along the trajectory; shading represents the matter density. Averaged probability: the averaged probability from one eigenbasis in closed form; across declared discontinuities, the probability averaged over an energy window; or, where $\mathbb{H}$ varies smoothly, the average transported along the instantaneous eigenstates with a Magnus patch across each non-adiabatic crossing. Adiabatic $+$ Magnus: transport along the instantaneous eigenstates, with a Magnus patch across the resonance. Interaction picture: the vacuum phase removed analytically, leaving only the matter term to integrate over each slab. Constant Hamiltonian: a single exponential. Energy-batched scan: one set of slabs shared by every energy. Cumulative scan: one pass that records the probability at every requested baseline. General Magnus ladder: slabs refined until two successive levels agree. \tabl{engines} gives the conditions of each engine. See Sec.~\ref{sec:engines} for details.}
\end{figure*}

\textit{The engines.---}Figure~\ref{fig:strategies} and \tabl{engines} show the seven engines in the order in which \magnus\ tries them.
\begin{enumerate}
\item \emph{Averaged probability.} This engine answers every request made with {\tt average=True}, by one of three routes chosen by $\mathbb{H}$. When $\mathbb{H}$ does not depend on position, it returns \equ{phase_average}, the probability with each interference term damped by the spread of its phase across the energy resolution. That is the quantity a solar or an astrophysical experiment measures. Where every phase has decohered, it is the limit \equ{averaged}. The engine diagonalizes $\mathbb{H}$ once at each energy and does not divide the path into slabs. When $\mathbb{H}$ varies smoothly along the path, the engine transports along the instantaneous eigenstates, keeping each phase with its weight, with the Magnus patch of Sec.~\ref{sec:hybrid} across every crossing that is not adiabatic. Without such a crossing, the result is \equ{phase_average_nocross} for a neutrino that starts in a flavor state, and \equ{averaged_varying} for one that starts decohered (Sec.~\ref{sec:average_keyword}).

{\def\arraystretch{1.2}
\begin{table*}[t!]
 {\footnotesize
 \begin{tabular}{@{}r@{\hspace{0.8em}}p{2.8cm}p{3.1cm}p{10.5cm}@{}}
 & Engine & Name in {\tt strategy\_info} & Condition \\
 \hline
 1 & Averaged probability & {\tt 'average'} & An averaged probability is requested \\
 2 & Adiabatic $+$ Magnus & {\tt 'hybrid'} & A smooth profile, a requested tolerance, no declared breakpoints, and a certified answer; under {\tt 'auto'}, not a request that the rules of Sec.~\ref{sec:engines} send to another engine \\
 3 & Interaction picture & {\tt 'ip\_exp'} & A profile declared as exponential, at two flavors and one baseline, and a converged iteration \\
 4 & Constant Hamiltonian & {\tt 'constant'} & $\mathbb{H}$ does not vary along the trajectory; any number of flavors, a single point or a scan \\
 5 & Energy-batched scan & {\tt 'separable'} & Many energies at one baseline \\
 6 & Cumulative scan & {\tt 'cumulative'} & Many baselines at one energy \\
 7 & General Magnus ladder & {\tt 'magnus'} & Every other request \\
 \end{tabular}}
 \caption{\label{tab:engines}\textbf{The \magnus\ probability engines and their conditions.} The seven engines, in the order in which they are tried, with the name each reports through {\tt strategy\_info}. A request that fails the condition of a row passes to the next row, so the last is reached whenever no other applies. Since two neighboring engines are different methods, the accuracy can change in a step at the boundary between them; Sec.~\ref{sec:implementation_limits} treats these boundaries. See Sec.~\ref{sec:engines} for details.}
\end{table*}}

When the profile has discontinuities, declared through {\tt t\_breakpoints} or {\tt t\_slab\_edges} (as for the layers of the Earth), $\mathbb{H}$ has no instantaneous eigenstates to transport along. The engine then computes the probability at several energies across a window around the requested one and returns their mean. That mean depends on the width of the window, so it is not the limit returned by the other two routes, and \magnus\ warns whenever it uses this route (Secs.~\ref{sec:average_keyword} and \ref{sec:shock}). Section~\ref{sec:astro} uses this engine for astrophysical neutrinos.
\smallskip
\item \emph{Adiabatic $+$ Magnus.} This engine is the adiabatic transport with Magnus patches of Sec.~\ref{sec:hybrid}. It applies to any Hamiltonian that varies smoothly along the trajectory, at any number of flavors, whether built by a scenario function or supplied by the user. It is the engine for a large accumulated phase: its cost is set by the grid on which $\mathbb{H}$ is diagonalized and by the crossings it patches, not by the phase, which the general ladder would have to resolve slab by slab. It declines when slab edges or breakpoints are supplied, when no tolerance is requested, or when $\mathbb{H}$ is not resolved on the grid on which it searches for crossings; under {\tt strategy='auto'}, it also declines when its certification fails, and it passes on the requests that the default sends to another engine, as described below.
\smallskip
\item \emph{Interaction picture.} At low energy, the vacuum phase is large, and a slab method must resolve every cycle of it, although that part of the evolution is known exactly. Switching to the interaction picture removes this phase. The switch applies to any Hamiltonian that separates into a position-independent part and a position-dependent term of the form $V(l)\,\mathbb{M}$, with $\mathbb{M}$ a fixed matrix: the position-independent part is diagonalized once at each energy, its evolution is factored out analytically, and only the envelope $V(l)$ remains to be integrated. In \magnus, the position-dependent term is the matter term, so the Hamiltonian is $\mathbb{H}(E, l) = \mathbb{H}_0(E) + V_{\rm CC}(l)\,\mathbb{P}$. Here $\mathbb{P}$ is the matter projector (Sec.~\ref{sec:recap}), plus the non-standard-interaction matrix if present, and $\mathbb{H}_0$ collects the vacuum term and, if present, the Lorentz-violating term.

The engine is specific to an exponential envelope, $V_{\rm CC}(l_0 + s) = V_{\rm CC}(l_0)\,e^{-s/l_{\rm scale}}$. The first-order Magnus term needs the integral of the envelope against the vacuum phase, and for an exponential envelope that integral has a closed form. For any other envelope it would have to be computed by quadrature, which would again have to resolve the vacuum phase. The solar density falls close to exponentially over most of the radius, and solar neutrinos, with their large vacuum phase, are the case the engine is for. Over a slab of length $h$ starting at $l_0$, the first-order term is
\begin{equation}
 \label{equ:ip_omega1}
 \big(\tilde{\Omega}_1\big)_{jk}(h)
 = -i\, \big(\tilde{\mathbb{P}}\big)_{jk}\, V_{\rm CC}(l_0)\,
 \frac{e^{(i \Delta_{jk} - 1/l_{\rm scale})h} - 1}{i \Delta_{jk} - 1/l_{\rm scale}} ,
\end{equation}
where a tilde marks a matrix in the eigenbasis of $\mathbb{H}_0$, $\Delta_{jk} = \lambda_j - \lambda_k$ is the gap between two of its eigenvalues, and $l_{\rm scale}$ is the scale height; the expression holds for $j = k$ as well. Equation~(\ref{equ:ip_omega1}) integrates the exponential envelope exactly over the slab: $V_{\rm CC}(l)$ is not replaced by a constant or a straight line within the slab, as a quadrature would do. The only approximation is the truncation of the series at first order. It is accurate wherever the matter term is a small perturbation on the vacuum splitting, and it improves as the energy falls, since $\Delta_{jk}$ grows as $1/E$. The engine doubles the number of slabs until two successive answers agree to the requested tolerance, twice in a row.

Four conditions must hold for this engine to take a request: two flavors, a profile built with {\tt exp\_density\_profile} (the default of the solar wrappers; a tabulated solar model is not one), a single baseline, and no slab edges or breakpoints. Under {\tt strategy='auto'}, it is also skipped when a request at a tolerance of $10^{-6}$ or looser goes to the general ladder, as described below. (The limit of two flavors is empirical: the neglected second-order term is about a thousand times larger at three flavors than at two, and the number of slabs needed to certify an answer is then too large for the engine to be fast.) At an MSW resonance, the matter term is no longer small, the answers stop agreeing before the engine reaches its limit on the number of slabs, and the request passes to the next engine.
\smallskip
\item \emph{Constant Hamiltonian.} When $\mathbb{H}$ does not vary along the trajectory, the Magnus series terminates at its first term: every $\Omega_k$ beyond $\Omega_1$ is a nested commutator of $\mathbb{H}$ with itself, hence zero. The evolution operator is then $\exp(-i\mathbb{H}L)$ exactly, with no discretization at all, and a whole scan over $n_E$ energies is one stacked exponential over an $(n_E, d, d)$ array, with $d$ the number of flavors. The engine applies at any number of flavors, to a single point as well as to a scan, and it accepts a different baseline for each point, which the energy-batched scan below does not. Refinement keywords are accepted and ignored, since they can only ask for a refinement of something already exact; a value that {\tt osc\_prob} would reject is declined, so that the user sees the same error as on any other route.
\smallskip
\item \emph{Energy-batched scan.} A scan over many energies at one baseline, computed point by point, traverses the same profile once per energy although the profile does not depend on it. When $\mathbb{H}$ separates as above (\eg, as the standard, non-standard-interaction, and Lorentz-violating Hamiltonians do), this engine samples the potential once per refinement level and shares it across the batch; the quadrature, the commutators, the exponentials, and the slab products carry the energy as a batch dimension, and energies that have converged drop out of further refinement.
\smallskip
\item \emph{Cumulative scan.} The complementary case is one energy at many baselines, $L_1 < L_2 < \dots < L_N$. The path to each baseline contains the path to the previous one, $\mathbb{U}(0, L_2) = \mathbb{U}(L_1, L_2)\,\mathbb{U}(0, L_1)$, so this engine traverses the profile once and records the running product at each of the $N$ baselines, where computing them independently would traverse it $N$ times. The traversal is chunked and each recorded product is converted to a probability at once, so that peak memory scales with the chunk and not with the number of baselines.
\smallskip
\item \emph{General Magnus ladder.} The last engine is the refinement ladder of Sec.~\ref{sec:refinement}. It assumes only a Hermitian $\mathbb{H}(l)$: it lays slabs along the trajectory, builds the Magnus expansion on each, composes them in order, and raises the slab count until the probabilities computed in two successive stages agree to the requested tolerance. It is reached whenever no engine above it applies.
\end{enumerate}

Comparing the answers of two engines is a useful test of accuracy when the engines use different methods, \eg, the adiabatic engine and the general ladder. It is a weak test when they use the same method: the general ladder, the cumulative scan, and the energy-batched scan all apply the Magnus expansion on slabs, so they tend to err in the same way. A user who doubts a result can test with {\tt cross\_check\_strategies}: it takes the same arguments as the function being checked, computes the request with every engine that applies to it, and reports how far apart the answers are. It is not run by default, since it raises the cost of a call. That said, no engine can see a feature narrower than the grid on which it samples the profile, so all of them can agree and be wrong together; for that, the reference has to come from outside \magnus\ (Sec.~\ref{sec:validation}).

\smallskip

\textit{How a request is assigned to an engine.}---The user does not choose an engine. Under the default, {\tt strategy='auto'}, \magnus\ chooses it from two things: the shape of the request and the tolerance requested.

A request with {\tt average=True} goes to the averaged-probability engine. A Hamiltonian that does not vary along the path goes to the constant-Hamiltonian engine.

For a Hamiltonian that varies smoothly along the path, the tolerance decides. At a tolerance of $10^{-6}$ or looser, which includes the default of $10^{-3}$, a scan over energy goes to the energy-batched engine, a scan over baseline to the cumulative engine, and a single point to the general ladder. A request with a large accumulated phase, as across the Sun, may go to the adiabatic engine instead. At a tighter tolerance, a scan over baseline goes to the cumulative engine. A single point or a scan over energy goes to the general ladder or the energy-batched engine when its accumulated phase is small, and otherwise to the adiabatic engine if that engine can certify its answer. A profile with declared discontinuities never goes to the adiabatic engine.

Since the engine depends on the shape of the request, passing an array instead of looping can change the engine, and with it the cost (Sec.~\ref{sec:cost}). {\tt strategy\_info} reports which engine answered, and the API documentation~\cite{MagnusDocs} gives the full rules.

The keyword {\tt strategy} overrides the default. {\tt 'hybrid'} lets the adiabatic engine answer when it cannot certify its answer. {\tt 'magnus'} excludes the adiabatic and cumulative engines. It is useful for a scan over energy with a large accumulated phase at a tolerance tighter than $10^{-6}$, which {\tt 'auto'} sends to the slower adiabatic engine; it is not useful for a scan over baseline. Under {\tt 'auto'}, the adiabatic engine answers only when it certifies its answer. Hence, the default is as accurate as a certified adiabatic answer or a converged ladder.

Not every entry point tries every engine. The matter scenario functions try the first five, in the order above; {\tt osc\_prob\_energy\_baseline} tries the cumulative scan; and {\tt osc\_prob} is the general ladder itself (Fig.~\ref{fig:layers}). {\tt osc\_prob\_vacuum} needs only the phase average and the constant engine. A Hamiltonian supplied by the user, through {\tt osc\_prob\_earth} or {\tt osc\_prob\_sun}, can reach the phase average, the adiabatic engine, the cumulative scan, and the ladder. A request for the evolution operator, {\tt return\_evolution\_operator=True}, goes to the ladder, the only engine that forms it.

\smallskip

\textit{Seeing which engine answered.}---When an engine declines a request and the next one takes it, the user is usually not told (the exception is the adiabatic engine declining a profile with a jump or a feature too narrow for its sampling grid, which is warned about once per session). This happens on ordinary calls, where a warning each time would be noise. Instead, the user can ask which engine answered once the probability is returned. A dictionary passed as {\tt strategy\_info} to the probability function returns with, among other entries, the engine that answered ({\tt engine}); whether, if it was the adiabatic engine, the answer was certified ({\tt certified}); and the engines that attempted the request and gave up, each with its reason ({\tt declined}). The keyword is accepted by every scenario function and every wrapper, and by {\tt osc\_prob\_earth} and {\tt osc\_prob\_sun}. This makes an answer traceable. A result that moves when a point is added to a scan, or a call that suddenly costs more, is often a change of engine, and the dictionary names it.


\subsection{Example: one request, through the layers}
\label{sec:one_request}

Listing~\ref{lst:layers} follows one request through the four layers of \figu{layers}. Its first block is the whole of what a user sends to \magnus: a call to the wrapper {\tt osc\_prob\_3nu\_earth} for a three-flavor probability at one energy along an Earth chord, with two refinement keywords. The blocks below it are not written by the user; they are the functions inside \magnus\ that the call passes through, in order. No specialized engine takes this request, so it descends all the way to {\tt osc\_prob}.

The wrapper computes the chord through the PREM (see \figu{prem}): the positions where it crosses the boundaries between layers, and the density along it. It collects the oscillation parameters it received by name into a dictionary, and calls the second-layer function for standard oscillations in matter, {\tt osc\_prob\_matter\_std\_potential}, with the flavor count, the density, the energy, the baseline, and the layer boundaries as breakpoints. That function converts the density to the potential $V_{\rm CC}$, builds the Hamiltonian of \equ{h_matter}, and tries the engines that can exploit its structure.

None of engines 1 to 6 of \tabl{engines} takes this request. No average is requested; the density varies along the chord, so $\mathbb{H}$ is not constant; the declared layer boundaries rule out the adiabatic engine; the interaction picture needs two flavors and an exponential profile; and a single energy and a single baseline rule out the energy-batched and cumulative scans. So {\tt osc\_prob\_matter\_std\_potential} passes the Hamiltonian to the third layer, {\tt osc\_prob\_energy\_baseline}, which calls {\tt osc\_prob} once, for the one point. This is engine 7: {\tt osc\_prob} runs the refinement ladder of Sec.~\ref{sec:refinement}, with a slab edge on every layer boundary. The two refinement keywords in the request, {\tt rtol} and {\tt atol}, reach {\tt osc\_prob\_matter\_std\_potential} through the {\tt **kwargs} of the wrapper; from there they are passed down explicitly, layer by layer, to {\tt osc\_prob}.

\begin{lstlisting}[language=Python, caption={\textbf{A probability request through the layers of \magnus.} A request through the four layers of \figu{layers}: one energy along an Earth chord. The first block is what the user writes. The four blocks below it are the functions inside \magnus\ that the call passes through, defined in the module {\tt magnus.oscprob} ({\tt src/magnus/oscprob.py}) and abbreviated to the arguments that carry the request on. See Sec.~\ref{sec:one_request} for details.}, label={lst:layers}, float=tbp]
# ---- What the user writes: the whole request ----
# One energy, one chord, and two refinement keywords
P = osc_prob_3nu_earth(
        3.0*gd.UNIT_GEV, costhz=-0.5,
        L=chord(-0.5)*gd.UNIT_KM, **osc,
        rtol=1e-8, atol=1e-10)

# ---- Inside Magnus: what the call runs through ----
# Top layer: the named wrapper
def osc_prob_3nu_earth(energy, costhz, L,
        s12, s23, s13, dCP, D21, D31, ..., **kwargs):
    edges = earth.prem_layer_edges_along_chord(...)
    rho_func = ...   # PREM density along the chord
    return osc_prob_matter_std_potential(
        3, rho_func, energy, L,
        {'s12': s12, 's23': s23, 's13': s13,
         'dCP': dCP, 'D21': D21, 'D31': D31},
        t_breakpoints=edges, ..., **kwargs)

# Second layer: builds H, tries its engines
def osc_prob_matter_std_potential(num_flavors,
        rho_func, energy, L, osc_params,
        rtol, atol, ..., **kwargs):
    htot = ...   # Vacuum term + V_CC(l) * P
    ...          # Engines 1-5: none applies here
    return osc_prob_energy_baseline(htot, energy,
        L, rtol=rtol, atol=atol, ..., **kwargs)

# Third layer: the scan over (energy, baseline)
def osc_prob_energy_baseline(H_func, energy, L,
        rtol, atol, ..., **kwargs):
    ...          # Engine 6: one baseline, no
    return osc_prob(H_at_energy, L0, L,
                    rtol=rtol, atol=atol, ...)

# Base layer: the refinement ladder, one point
def osc_prob(H_func, l0, l1, rtol, atol, ...):
    ...          # Engine 7; rtol, atol used here
\end{lstlisting}


\subsection{Batching and parallelization}
\label{sec:batching}

In \magnus, a request for many probabilities can be made faster in two ways, batching and parallelization, which act differently and mostly do not combine.

\begin{itemize}
\item \emph{Batching} means passing an array in one call: many energies at one baseline, or many baselines at one energy. A call can also take equal-length arrays of energies and baselines, which it pairs point by point: the first energy with the first baseline, and so on. These points are computed one at a time, unless $\mathbb{H}$ does not vary along the path. To get the probability at every energy for each of several baselines, the user makes one call per baseline, each with the full array of energies.

A scan over energy at one baseline, or over baseline at one energy, can reach one of the batched engines of Sec.~\ref{sec:engines}, which reuse work across the array. The \textit{energy-batched scan} samples the profile once per refinement level and carries the energy as a batch dimension through every slab; the \textit{cumulative scan} traverses the profile once and records the running product at every requested baseline. Both run in a single process (any use of further cores comes from the linear-algebra library under {\tt NumPy}, not from \magnus). The gain is that the profile is not traversed once per point. Where $\mathbb{H}$ does not vary, the batched answers are identical, bit for bit, to those computed one point at a time. Where it varies, both answers meet the requested tolerance but are not identical, since each refines its grid on its own; with the grid fixed, they agree to about $10^{-14}$ at three flavors and $10^{-13}$ at four (Sec.~\ref{sec:validation}).
\item \emph{Parallelization} means computing points in several processes at once, on several cores, as specified by the {\tt n\_jobs} argument. \magnus\ uses the {\tt joblib} library for it: the first point is computed in the calling process, {\tt n\_jobs} processes are started, and the remaining points are divided among them, each computed on the per-point path. Since every point starts its refinement from the first point rather than from its neighbor, a parallel scan agrees with a serial one to the requested tolerance, not bit for bit (Sec.~\ref{sec:cost}).

Section~\ref{sec:cost} measures the gain on an Earth chord where every energy has its own baseline, so that no batched engine applies. On a machine with ten cores (Sec.~\ref{sec:timing_protocol}), ten processes complete a scan of 1\,000 energies 1.9 times as fast as one process does, a scan of 5\,000 energies 2.5 times as fast, and a scan of 20\,000 energies 2.8 times as fast: the gain grows with the size of the scan and stays far from 10.
\end{itemize}

\begin{figure}[t!]
 \centering
 \includegraphics[width=\columnwidth]{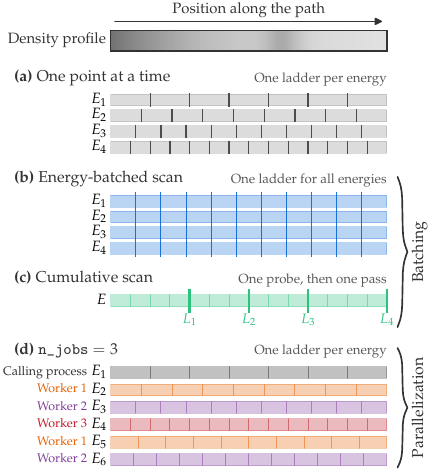}
 \caption{\label{fig:batching}\textbf{Batching against parallelization.} Four ways \magnus\ computes a scan, drawn along one density profile. Each bar is the final slab grid of one refinement ladder, which traverses the profile once per level; its color marks the process that runs it. (a) One point at a time. (b) The energy-batched scan, with one grid shared by every energy. (c) The cumulative scan, at one energy, recording the probability at every requested baseline ($L_1$ to $L_4$) in one pass. (d) Parallelization with {\tt n\_jobs} $= 3$: the first point runs in the calling process and the rest are shared out among three workers. See Sec.~\ref{sec:batching} for details.}
\end{figure}

Figure~\ref{fig:batching} shows why the two differ. Computed one point at a time, a scan of $N$ energies runs $N$ refinement ladders. Batching removes most of that work: the energy-batched scan runs one ladder for all $N$ energies, sampling the profile once per slab, and the cumulative scan covers every baseline in a single pass. Parallelization does not reduce the work: all $N$ ladders still run, but several at a time, on different cores. At best, this divides the run time by the number of processes; in practice, the cost of starting the processes and moving work between them keeps the gain well below that.

Batching over energies and parallelization exclude each other. The energy-batched scan and the constant-Hamiltonian engine answer only at {\tt n\_jobs} $= 1$; any other value sends the scan to the per-point path, computed in parallel, which is slower than the batched path in one process. For a three-flavor scan of 5\,000 energies between 0.5 and 20~GeV along the PREM chord at $\cos\theta_z = -0.9$, with a baseline of 11\,468~km, the batched path takes 0.12~s in one process; ten processes take 0.8~s (Sec.~\ref{sec:cost}). The cumulative scan does not depend on {\tt n\_jobs}: a scan over baselines at one energy is answered in one process regardless.

In practice, for the majority of applications of \magnus, the user should pass arrays and leave {\tt n\_jobs} at its default of 1. Setting {\tt n\_jobs} above 1 pays only for a scan that no batched engine accepts, such as one where every energy has its own baseline; \figu{njobs_scaling} shows the gain in that case.


\subsection{Declaring the structure of a profile}
\label{sec:declaring_edges}

Two keywords let the user place slab edges by hand. They serve similar but different purposes:
\begin{itemize}
\item {\tt t\_breakpoints} names positions that must be slab edges. The ladder of Sec.~\ref{sec:refinement} keeps its own grid, inserts an edge at each named position at every refinement level, then refines as usual to the requested tolerance. It is meant for a profile with a density jump, a kink, or a front at a known position: the user states where the profile is not smooth, and the package chooses the resolution. A tabulated profile, for instance, has a kink at every row of the table (the solar tables are interpolated linearly in the logarithm of the density), and the rows are where to declare breakpoints.

Declaring breakpoints also changes which engine answers. It excludes the adiabatic engine of Sec.~\ref{sec:hybrid}, and with {\tt average=True} it switches to the average over an energy window, a different quantity (Sec.~\ref{sec:average_keyword}). On a scan over baselines, declaring a front is what makes the answer accurate, as Sec.~\ref{sec:shock} shows on the forward shock and contact discontinuity of a supernova ray. At a single point it is not always an improvement, since the general ladder may then be less accurate than the adiabatic engine would have been; \magnus\ says so when it detects an undeclared jump.
\item {\tt t\_slab\_edges} gives the complete grid of positions at which the evolution operator is evaluated, as a list of $[{\rm start}, {\rm end}]$ pairs, one per slab. It replaces the grid of the refinement ladder, so the slabs are fixed and never refined. It is meant for a grid that can be justified in full, with no tolerance requested, so that the call is a single evaluation of the expansion on that grid. The user then takes responsibility for the accuracy: nothing moves or adds an edge. If a tolerance is requested anyway, the ladder has nothing to refine at the default quadrature, repeats the same evaluation, and warns that the tolerance was not achieved. {\tt t\_breakpoints} is ignored when {\tt t\_slab\_edges} is given.

This keyword is intended for piecewise-constant profiles, such as the castle wall of \figu{arrangement}. With one slab per constant piece, the result is exact, because the Hamiltonian is constant within each slab and the Magnus expansion terminates at its first term. For every other profile, {\tt t\_breakpoints} is the keyword to use.
\end{itemize}

\begin{figure}[t!]
 \centering
 \includegraphics[width=\columnwidth]{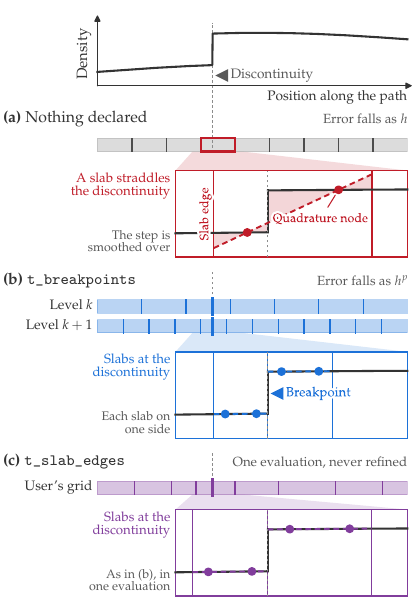}
 \caption{\label{fig:declaring_edges}\textbf{Declaring a density discontinuity.} \textit{Top:} a density profile with one discontinuity. Each enlargement shows the same stretch of path: the true density, the slab edges, the two quadrature nodes of each slab, and the straight line through $\mathbb{H}$ at those nodes, which is how the quadrature sees the slab. The shaded area is the difference from the true density. \textit{(a)} Nothing declared: one slab of the uniform grid straddles the discontinuity. Refining narrows that slab but never removes it, so the error falls only about as fast as the slab width. \textit{(b)} {\tt t\_breakpoints}: the grid changes at each refinement level, but an edge stays at the discontinuity, so the error falls as $h^p$. \textit{(c)} {\tt t\_slab\_edges}: the user's grid is evaluated once, without refinement. It is exact if an edge falls at the discontinuity, and left as is if not. See Sec.~\ref{sec:declaring_edges} for details.}
\end{figure}

Figure~\ref{fig:declaring_edges} shows why a discontinuity must be declared. A slab that straddles a discontinuity spoils the convergence, and neither more slabs nor a higher order repairs it. An edge at the discontinuity removes the problem, because each slab then holds a smooth Hamiltonian. With {\tt t\_breakpoints}, \magnus\ keeps that edge at every level and refines the rest of the grid. With {\tt t\_slab\_edges}, the grid is the user's and is never refined, so it is only as accurate as the edges the user places.

For the most common case, a neutrino crossing the Earth, the user need not declare anything. PREM describes the Earth as concentric layers, with a density that varies smoothly inside each layer and jumps at the boundaries between them. The Earth wrappers find where the trajectory crosses each boundary and pass these positions as {\tt t\_breakpoints} automatically. A core-crossing trajectory ($\cos\theta_z = -0.9$) crosses 16 boundaries. With slabs of equal width, the error stops falling: at orders 2, 4, and 6, it stays between $10^{-3}$ and $10^{-5}$ even with 2\,000 slabs. With an edge at each crossing, the error falls as $h^p$ at every order, so the orders differ only in cost: at 5~GeV, an error of $10^{-8}$ takes about 2\,000 slabs at order 2, 64 at order 4, and 16 at order 6. Over 150 random piecewise-constant profiles, declaring the jumps lowers the median error from about $10^{-3}$ to $10^{-12}$.

Listing~\ref{lst:edges} shows the use of {\tt t\_breakpoints} and {\tt t\_slab\_edges} on the castle-wall profile of \figu{arrangement}. Declaring the 23 density jumps through {\tt t\_breakpoints} gives the exact answer at the first level at which two levels agree, on the grid of the ladder with the jumps added as edges. Giving the 24 slabs through {\tt t\_slab\_edges} gives the same answer in a single evaluation, since the profile is constant within every slab. A grid of twelve slabs, each straddling one jump, is off by $6 \cdot 10^{-4}$; nothing in the package repairs it, because the user owns that grid. Section~\ref{sec:prem} declares the boundaries of PREM, and Sec.~\ref{sec:shock} declares a shock front the same way.

\begin{lstlisting}[language=Python, caption={\textbf{Placing slab edges by hand.} The two ways of placing slab edges by hand, on the castle-wall profile of \figu{arrangement}. {\tt t\_breakpoints} adds the positions of the jumps to the grid of the refinement ladder, which then refines as usual. {\tt t\_slab\_edges} replaces the grid with the given {[}start, end{]} pairs, which stay fixed. The first two calls return the exact result, because the profile is constant within every slab. The third call passes a grid whose slabs straddle the jumps: its result is off by $6 \cdot 10^{-4}$, and nothing corrects it, because the grid is the user's. Section~\ref{sec:shock} declares the two fronts of a supernova shock through {\tt t\_breakpoints} in the same way.}, label={lst:edges}, float=tbp]
import numpy as np
import magnus.oscprob as oscprob
import magnus.globaldefs as gd

osc = gd.load_nufit_params('NuFIT 6.1')
E, L = 2.0*gd.UNIT_GEV, 6000.0*gd.UNIT_KM
# 24 slabs of 250 km, alternating 2 and 8 g/cm^3
edges = np.linspace(0.0, L, 25)
rho = np.where(np.arange(24) % 2 == 0, 2.0, 8.0)

def castle_wall(l):
    k = np.searchsorted(edges, l, side='right') - 1
    return rho[np.clip(k, 0, 23)]

kw = dict(nu_i=gd.NUMU, nu_f=gd.NUE,
          density_matter_is_in_g_per_cm3=True)
prob = oscprob.osc_prob_matter_std_potential

# The jumps declared: the ladder keeps its own grid,
# adds an edge at each of the 23 jumps, then refines
# to the tolerance
P1 = prob(3, castle_wall, E, L, osc,
          t_breakpoints=edges[1:-1],
          rtol=1.0e-8, atol=1.0e-10, **kw)

# The grid given in full, as [start, end] pairs: one
# slab per layer, no refinement, a single evaluation
pairs = np.column_stack([edges[:-1], edges[1:]])
P2 = prob(3, castle_wall, E, L, osc,
          t_slab_edges=pairs, rtol=None, atol=None,
          **kw)

# A grid of 12 slabs, each straddling one jump: the
# user owns this grid, and nothing repairs it
wide = np.column_stack([edges[:-1:2], edges[2::2]])
P3 = prob(3, castle_wall, E, L, osc,
          t_slab_edges=wide, rtol=None, atol=None,
          **kw)

# P1 = P2 = 0.0240046, exact: the profile is constant
# within every slab.  P3 = 0.0246158, off by 6e-4
\end{lstlisting}


\subsection{Documentation and executable examples}
\label{sec:documentation}

\magnus\ is documented at four levels, from overview to detail:
\begin{itemize}
\item \emph{The README} of the repository~\cite{Magnus}: installation, a quick start from Python and from the command line, an overview of what \magnus\ computes, and when it is the right tool.
\item \emph{The documentation site}\ \cite{MagnusDocs}, whose source is kept in the repository beside the code. It presents the method of Sec.~\ref{sec:method} in the form in which it is implemented: one page derives the expansion terms at any order; others cover the adiabatic transport with Magnus patches of Sec.~\ref{sec:hybrid}, the phase average of Sec.~\ref{sec:averaged}, and the engines of Sec.~\ref{sec:engines} with the order in which they are tried. A performance page lists every tuned constant with the set of cases on which it was measured, or states that it was not measured. A diagnostics page explains every warning \magnus\ raises.
\item \emph{The API reference}, part of the documentation site, generated from the docstrings of the source. It documents every public function, with its parameters, units, and examples.
\item \emph{The notebooks}: 29 executable notebooks, listed in \ref{app:notebooks}, work through these topics with examples. One of them, notebook \#28, produces every figure in this paper. A workflow executes all the notebooks whenever the package or the notebooks change in the repository.
\end{itemize}


\subsection{Installation}

The released version of \magnus\ is on the Python Package Index (PyPI)\ \cite{MagnusPyPI}. It is installed with
\begin{verbatim}
pip install magnuspy
\end{verbatim}
This command also installs every dependency \magnus\ needs; for most users, it is the whole installation. \magnus\ requires Python 3.10 or later; it has been tested with up to 3.13.

The PyPI package does not include the notebooks of \ref{app:notebooks} or the test suite of Sec.~\ref{sec:validation} and \ref{app:tests}. The notebooks have saved output and can be read in the repository~\cite{Magnus}. To run or modify them, the user clones the repository:
\begin{verbatim}
git clone https://github.com/mbustama/Magnus.git
\end{verbatim}
The notebooks are in the {\tt notebooks} folder. From the clone, \magnus\ can also be installed directly, which gives access to changes not yet released on PyPI:
\begin{verbatim}
cd Magnus
# The package:
pip install -e .                 
# Plus what the notebooks need:
pip install -e ".[notebooks]"    
# Plus what the tests need:
pip install -e ".[test]"         
\end{verbatim}
Opening the notebooks interactively also requires a Jupyter front end, such as JupyterLab. Only users who contribute to \magnus\ typically need to run the test suite.

Installing \magnus\ in a dedicated virtual environment ({\tt conda} or {\tt venv}) is supported, but not required.


\subsection{Code availability}

We develop \magnus\ in the public GitHub repository\ \cite{Magnus}, which holds the source, the test suite, the notebooks, the documentation source, this paper, release notes, and the change log. This paper describes version 1.2.0. We release \magnus\ under the GNU General Public License v3.0. 


\subsection{Code validation}
\label{sec:validation}

The repository carries about 2\,400 tests, run on Python 3.10 through 3.13, covering 92\% of the statements and branches of the code. \ref{app:tests} groups the tests by what they cover. The validation is performed in layers.

\subsubsection{Against the derivation}

We compare every expansion term the code evaluates, $\Omega_1$ to $\Omega_{10}$, against terms generated independently from the recursion of \ref{app:terms}, on a Hamiltonian chosen so that no term vanishes identically. The two agree to a relative $10^{-11}$ at every order, the round-off of the quadrature that evaluates them.

\subsubsection{Against closed forms}

Two- and three-flavor vacuum oscillations, and two-flavor constant-density matter oscillations, for neutrinos and antineutrinos, agree with their analytic expressions to $10^{-12}$.

\subsubsection{Against an independent solver}

The reference is a {\tt DOP853} integration of the evolution equation (order-8 Runge-Kutta\ \cite{Prince:1981dop, Hairer:1993sod}) at ${\tt rtol} = 10^{-12}$ and ${\tt atol} = 10^{-14}$. Where a test quotes an error close to the accuracy of the reference, the reference is checked by tightening {\tt rtol} to $10^{-13}$. For a constant Hamiltonian, or a piecewise-constant one whose edges are declared, {\tt scipy.linalg.expm} is exact and used instead. On a model Hamiltonian, halving the slab width divides the error by about 4, 16, and 64 at orders 2, 4, and 6 on the default collocation path: the factor $2^{p}$ expected when the global convergence rate equals the order (Sec.~\ref{sec:order_is_the_slab_product}).

\subsubsection{Against itself, where meaningful}

Repeated calls, and a scan over baselines given in any order, are asserted to return identical results, bit for bit, so an optimization that changes an answer fails rather than passing quietly. A parallel run is held to the requested tolerance against a serial one, not to bit identity, since each worker starts its refinement from the first point rather than from its neighbor. The energy-batched scan is held to $10^{-12}$ against the per-point path, with the grid and the tolerances pinned so that the two are arithmetically the same problem; left to refine on its own, it agrees with the per-point path to within the requested tolerance. No Magnus path is ever scored against another as an accuracy measurement: a cross-check between two paths is reported as agreement, never as accuracy.

\subsubsection{Against a population}

Some tests do not check one case at a time. Each draws a set of random density profiles, computes the probability along every one at a requested tolerance of $10^{-3}$, compares each answer with the reference, and then judges the errors as a set: their median, their worst value, and how many are \emph{silent misses}. A silent miss is an answer outside the requested tolerance that carries no warning of any kind. Silent misses matter most, because an inaccurate answer that carries a warning has already been caught by the package. The test passes or fails on these numbers, not on any single profile, because a random profile can always be found on which a method does poorly, whereas a regression in the code shifts the whole set. Two populations run with the test suite. Over 40 random smooth profiles at two and three flavors, the median error is about $10^{-8}$, with one silent miss, 8\% above the tolerance. Over 120 random piecewise-constant profiles at two to five flavors, with the edges deliberately left undeclared, 19 answers fall outside the tolerance, and every one of them comes with a warning.

Larger batteries, run by hand rather than with every test, measured the same quantities on more cases. Over 145 random smooth profiles, the median error was about $6 \cdot 10^{-8}$, with about 4\% silent misses, each within a factor of three of the tolerance. Over 150 random piecewise-constant profiles with their edges declared, the median error was about $10^{-12}$, with no silent miss. Over 164 Earth, solar, vacuum, and constant-density configurations, there was no silent miss.

\subsubsection{Continuous integration}

Every push to the repository runs the above tests on the four supported Python versions, a linter, and a build of the documentation that treats every warning as an error. The notebooks are executed whenever the package or the notebooks change. Before each release, the tests run once more against a regular installation of the package, as a user would have it, and the files uploaded to PyPI are checked for valid metadata.


\subsection{Limitations}
\label{sec:limitations}

The limitations of \magnus\ are of two kinds. Those of the method follow from what every engine assumes: a Hermitian Hamiltonian, fixed before the propagation and sampled on a grid of positions; no implementation would remove them. Those of the implementation follow from choices made in the code, and a later version could remove them.


\subsubsection{Limitations of the method}

\textit{Open systems.}---\magnus\ solves the Schr\"odinger equation for a Hermitian Hamiltonian. Every truncation of the Magnus expansion is anti-Hermitian, so the evolution it produces is unitary at every order; this is the property the method rests on. Decoherence, coupling to a bath, and decay to invisible states\ \cite{Baerwald:2012kc, Valera:2024buc} remove probability from the system or damp the coherence between mass eigenstates. They are described by a density matrix that obeys a non-unitary evolution equation, or by a Hamiltonian with an anti-Hermitian term, neither of which the method admits. {\tt nuSQuIDS}\ \cite{Delgado:2014kpa, NuSQuIDS} is the tool for those problems.

\smallskip

\textit{Collective oscillations.}---In a dense neutrino gas, the Hamiltonian depends on the flavor content of the neutrinos\ \cite{Duan:2010bg, Mirizzi:2015eza}, so the problem is nonlinear and has to be solved self-consistently. In \magnus, the Hamiltonian is fixed before the propagation begins, whether built by the package or supplied by the user, and it does not change with the neutrino state. A self-consistent solution would need an outer iteration that updates $\mathbb{H}$ from the solution and propagates again; \magnus\ could serve as the propagator inside such an iteration, but it does not ship with one.

\smallskip

\textit{A feature narrower than every grid.}---Every engine in \magnus\ samples the Hamiltonian on a grid of positions. A feature narrower than the finest of those grids falls between the samples, so all the engines miss it together, and a cross-check cannot detect the error. We measured this on a Gaussian resonance whose width is $10^{-5}$ of the trajectory: every engine returned an answer wrong by about $10^{-2}$ against a requested tolerance of $10^{-3}$. To catch such features, the scenario functions for standard matter and for non-standard interactions also scan the density profile. When the scan finds variation concentrated between grid points, it warns and names the breakpoints to declare. It catches most such features, but not all. Declaring breakpoints at the feature, either by hand or as named in the warning, brings the error within the requested tolerance. This limitation belongs to every method that samples the Hamiltonian on a grid.


\subsubsection{Limitations of the implementation}
\label{sec:implementation_limits}

\textit{Different eigenbases at the two ends.}---The averaged probability of \equ{averaged_varying} needs the eigenbases at production and at detection. In \magnus, both come from the one Hamiltonian it is given. Some models need two different ones: non-unitary mixing, whose effective form can differ between production and detection\ \cite{Antusch:2014woa, Blennow:2016jkn, Fong:2022oim}, and mixing parameters that run with energy\ \cite{Antusch:2003kp, Antusch:2005gp, Bustamante:2010bf, Babu:2021cxe, Babu:2022non, Ge:2024ibn, Bustamante:2026zst, Bustamante:2026aur}. \magnus\ does not yet support these models.

\smallskip

\textit{A constant Hamiltonian.}---When the Hamiltonian does not depend on position, the evolution operator is a single exponential, which \magnus\ evaluates in closed form (Sec.~\ref{sec:exponential}), to round-off. Most of the cost lies in the dispatch and validation of a general solver, which a constant Hamiltonian does not need. A code built for the constant case alone is cheaper: about sixty times cheaper per point than the three-flavor specialization {\tt NuFast-LBL}~\cite{Denton:2024pzc} (Sec.~\ref{sec:comparison}).

\smallskip

\textit{Density fluctuations at every scale.}---The profile scan described above looks for variation concentrated in one place, so it cannot see variation spread evenly over every scale. We tested this on Kolmogorov density fluctuations of the kind used in supernova studies\ \cite{Loreti:1995ae, Patton:2013dba, Patton:2014lza}, with a $k^{-5/3}$ spectrum spanning 40 to 50~dB. The profile scan found nothing, and the adiabatic engine found the profile resolved on its grid in all 144 configurations tried. The errors were still reported, by the convergence warnings of the refinement ladder. On a turbulent or noisy medium, the user should not rely on the structural diagnostics. Instead, the user should raise the slab-count floor of Sec.~\ref{sec:refinement} ({\tt n\_slabs}) so that several slabs fit into each oscillation length, $2\pi/\Delta_{jk}$, over which the density still fluctuates appreciably: the density modes at those wavelengths are the ones that drive transitions (Sec.~\ref{sec:turbulence}). Section~\ref{sec:jet} propagates through a turbulent envelope of this kind.

\smallskip

\textit{Accuracy at the boundary between two engines.}---Two nearly identical requests can be answered by different engines. A scan over baselines at one energy with a large accumulated phase, for instance, goes to the adiabatic engine up to seven baselines and to the cumulative scan from eight, at the default tolerance. The two are different methods, so the error can change in a step between the two requests, by orders of magnitude in some cases. In the cases measured, the step was toward the more accurate answer. A user who adds one point to a scan and sees the answer move by more than the tolerance is seeing a change of engine, not a fault; {\tt strategy\_info} reports it.


\section{\magnus\ usage and examples}
\label{sec:examples}

\subsection{The full application programming interface (API)}
\label{sec:api}

The examples in this section illustrate the calls to \magnus\ for the cases a user is most likely to encounter, plus a few advanced, longer examples. They are not a complete description of the interface. The complete description is the API reference available at the documentation site\ \cite{MagnusDocs}. It is built from the docstrings of the source code, and includes every function with its description, arguments, default values, and working examples. The notebooks described in \ref{app:notebooks} (especially notebook \#28) contain the calls shown here in full, together with their output.


\subsection{A minimal example in vacuum, step-by-step}
\label{sec:minimal}

\begin{lstlisting}[language=Python, caption={\textbf{A minimal probability calculation.} The simplest complete calculation: the three-flavor vacuum probabilities for a 1-GeV neutrino over 1\,300~km. The oscillation parameters---here and in the rest of Sec.~\ref{sec:examples}---take their defaults, the NuFIT 6.1 best fit with Super-Kamiokande atmospheric data in normal ordering\ \cite{NuFIT61, Esteban:2024eli}. See Sec.~\ref{sec:minimal} for details.}, label={lst:minimal}, float=tbp]
import magnus.oscprob as oscprob
import magnus.globaldefs as gd

# Natural units: energies in eV, lengths in eV^-1
E = 1.0*gd.UNIT_GEV
L = 1300.0*gd.UNIT_KM

P = oscprob.osc_prob_3nu_vacuum(E, L)

print('Pme = %.5f, Pmm = %.5f, Pmt = %.5f'
 % (P[gd.NUMU][gd.NUE],
 P[gd.NUMU][gd.NUMU],
 P[gd.NUMU][gd.NUTAU]))

# Pme = 0.03127, Pmm = 0.39231, Pmt = 0.57643
\end{lstlisting}

Listing~\ref{lst:minimal} is a complete calculation, built in four steps.
\begin{enumerate}
\item \textbf{Import the two modules a typical calculation needs.} {\tt oscprob} holds the probability functions; {\tt globaldefs} holds the physical constants, the unit conversions, and the flavor indices.
\begin{pysnip}
import magnus.oscprob as oscprob
import magnus.globaldefs as gd
\end{pysnip}
\item \textbf{Give the energy and the baseline in natural units.} Every energy crossing the interface is in eV and every length in eV$^{-1}$ (Sec.~\ref{sec:recap}). The factors {\tt UNIT\_GEV} and {\tt UNIT\_KM} convert from the units a user thinks in.
\begin{pysnip}
E = 1.0*gd.UNIT_GEV
L = 1300.0*gd.UNIT_KM
\end{pysnip}
\item \textbf{Call one wrapper.} Its name fixes the flavor count and the environment. The oscillation parameters default to the NuFIT 6.1 best fit in normal ordering; any of them may be overridden by name, \eg, {\tt dCP=0.0}. Nothing in the call selects an engine or a numerical setting: \magnus\ chooses the engine and, where the profile varies, the slab count. (They can still be fixed via arguments; see Secs.~\ref{sec:engines} and~\ref{sec:refinement}.)
\begin{pysnip}
P = oscprob.osc_prob_3nu_vacuum(E, L)
\end{pysnip}
\item \textbf{Read the matrix.} The call returns every flavor pair at once rather than one number, indexed by the constants {\tt NUE}, {\tt NUMU}, and {\tt NUTAU}.
\begin{pysnip}
P[gd.NUMU][gd.NUE]
\end{pysnip}
\end{enumerate}

We index the returned matrix initial flavor first, $P[\alpha][\beta] = P_{\nu_\alpha \to \nu_\beta}$. Each row and each column sums to one, because the evolution is unitary. The corresponding evolution operator element, which the call returns when asked (Sec.~\ref{sec:evolution_operator}), is $\mathbb{U}_{\beta\alpha}$, so $\lvert \mathbb{U}\rvert^2$ comes out indexed the other way:
\begin{pysnip}
# P is indexed initial flavor first
P[gd.NUMU][gd.NUE]           # 0.03127

# The amplitude is indexed the other way
P, U = oscprob.osc_prob_3nu_vacuum(
    E, L, return_evolution_operator=True)
abs(U[gd.NUE][gd.NUMU])**2   # 0.03127
abs(U[gd.NUMU][gd.NUE])**2   # 0.00854
\end{pysnip}

In vacuum, the Hamiltonian does not depend on position, so the Magnus series stops at its first term: the evolution operator is a single exponential, exact to round-off. Listing~\ref{lst:minimal} exercises none of the method of Sec.~\ref{sec:method}. Everything else in this section is the same four steps with a different wrapper. Once the profile varies, the refinement of Sec.~\ref{sec:refinement} engages, at the default {\tt rtol} $=$ {\tt atol} $= 10^{-3}$ unless another tolerance is requested.


\subsection{Choosing the oscillation parameters}
\label{sec:params}

\tabl{parameters} lists every parameter that the shipped wrappers take, at two to five flavors, together with its default. The subsections below show how they are set.


\begin{table*}[t!]
\centering
\footnotesize
\renewcommand{\arraystretch}{1.15}
\begin{tabular}{l l l l l}
 & \textbf{Two flavors} & \textbf{Three flavors} & \textbf{Four flavors (3+1)} & \textbf{Five flavors (3+2)} \\
\hline
\multicolumn{5}{l}{\emph{Standard oscillations, in every wrapper}} \\
Mixing angles & {\tt sth} & {\tt s12}, {\tt s23}, {\tt s13} & as 3$\nu$, $+$ {\tt s14}, {\tt s24}, {\tt s34} & as 3$\nu$, $+$ {\tt s14}, {\tt s15}, {\tt s24}, \\
 & & & & {\tt s25}, {\tt s34}, {\tt s35} \\
CP phases & --- & {\tt dCP} & as 3$\nu$, $+$ {\tt d14}, {\tt d24} & as 3$\nu$, $+$ {\tt d14}, {\tt d15}, {\tt d24}, {\tt d35} \\
Mass splittings [eV$^2$] & {\tt Dm2} & {\tt D21}, {\tt D31} & as 3$\nu$, $+$ {\tt D41} & as 3$\nu$, $+$ {\tt D41}, {\tt D51} \\
Default & required & named set & 3$\nu$: named set; sterile: 0 & 3$\nu$: named set; sterile: 0 \\
\hline
\multicolumn{5}{l}{\emph{Non-standard interactions, in the {\tt \_nsi} wrappers; default 0}} \\
Couplings $\varepsilon_{\alpha\beta}$ & {\tt eps\_aa}, {\tt eps\_ab} & {\tt eps\_ee}, {\tt eps\_em}, {\tt eps\_et}, & as 3$\nu$, $+$ {\tt eps\_es}, {\tt eps\_ms}, & as 3$\nu$, $+$ {\tt eps\_es1}, {\tt eps\_es2}, {\tt eps\_ms1}, \\
 & & {\tt eps\_mm}, {\tt eps\_mt}, {\tt eps\_tt} & {\tt eps\_ts}, {\tt eps\_ss} & {\tt eps\_ms2}, {\tt eps\_ts1}, {\tt eps\_ts2}, {\tt eps\_s1s1}, \\
 & & & & {\tt eps\_s1s2}, {\tt eps\_s2s2} \\
\hline
\multicolumn{5}{l}{\emph{Lorentz-invariance violation, in the {\tt \_liv} wrappers; default 0, except $\Lambda = 1$}} \\
Mixing angles $\xi_{ij}$ & {\tt sxi} & {\tt sxi12}, {\tt sxi23}, {\tt sxi13} & as 3$\nu$, $+$ {\tt sxi14}, {\tt sxi24}, {\tt sxi34} & as 3$\nu$, $+$ {\tt sxi14}, {\tt sxi15}, {\tt sxi24}, \\
 & & & & {\tt sxi25}, {\tt sxi34}, {\tt sxi35} \\
CP phases & --- & {\tt dxiCP} & {\tt dxiCP}, {\tt dxi14}, {\tt dxi24} & {\tt dxiCP}, {\tt dxi14}, {\tt dxi15}, \\
 & & & & {\tt dxi24}, {\tt dxi35} \\
Eigenvalues [eV] & {\tt b1}, {\tt b2} & {\tt b1}--{\tt b3} & {\tt b1}--{\tt b4} & {\tt b1}--{\tt b5} \\
Scale, power & {\tt Lambda}, {\tt n\_liv} & {\tt Lambda}, {\tt n\_liv} & {\tt Lambda}, {\tt n\_liv} & {\tt Lambda}, {\tt n\_liv} \\
\hline
\end{tabular}
\caption{\label{tab:parameters}\textbf{The parameters of the wrapper functions.} Every parameter the wrappers of \tabl{wrappers} take, by flavor count, beyond the energy, the baseline, and the description of the medium. The angles are read in the convention set by {\tt angles} (Sec.~\ref{sec:angle_conventions}); the phases in radians, or in degrees under {\tt angles='deg'}. The standard three-flavor parameters left unset are read from the named set {\tt OSC\_PARAMS\_DEFAULT}, or from another set named through {\tt default\_osc\_params\_set\_name}; the active-sterile ones default to zero, so that a four- or five-flavor call without them returns the three-flavor probabilities. At two flavors, {\tt sth} and {\tt Dm2} are required. The non-standard parameters default to zero, so a {\tt \_nsi} or {\tt \_liv} wrapper called without them returns the standard result; with $n_{\rm LIV} = 0$, a term switched on through the eigenvalues is energy-independent. See Sec.~\ref{sec:params} for details.}
\end{table*}


\subsubsection{Three flavors}

Every {\tt osc\_prob\_3nu\_*} wrapper leaves its six standard parameters unset by default. An unset parameter is read from a named set, {\tt OSC\_PARAMS\_DEFAULT}, which is the NuFIT 6.1 best fit with Super-Kamiokande atmospheric data in normal ordering\ \cite{NuFIT61, Esteban:2024eli}. Naming a different set changes all six at once.

\begin{pysnip}
P = oscprob.osc_prob_3nu_vacuum(E, L,
 default_osc_params_set_name=
  'OSC_PARAMS_NU_FIT_5_2_SK_IO')
# Pme = 0.05166, against 0.03127 by default
\end{pysnip}

Fifty-two NuFIT global-fit sets of best-fit three-flavor mixing parameters are predefined in {\tt globaldefs}, together with {\tt OSC\_PARAMS\_DEFAULT}. Their names in {\tt globaldefs} follow the releases. From NuFIT~4.0 onward, a release splits its fits by whether Super-Kamiokande atmospheric data is included; both are included in {\tt globaldefs}, \eg, {\tt OSC\_PARAMS\_NU\_FIT\_5\_2\_SK\_IO} and {\tt OSC\_PARAMS\_NU\_FIT\_5\_2\_NOSK\_IO}. Earlier releases carry no such split, so their names drop that infix, as in {\tt OSC\_PARAMS\_NU\_FIT\_3\_0\_NO}. Every set can also be loaded by release, ordering, and category through {\tt load\_nufit\_params}, \eg,
\begin{pysnip}
osc = gd.load_nufit_params('NuFIT 5.2',
 ordering='IO', category='without_SK')
P = oscprob.osc_prob_3nu_vacuum(E, L, **osc)
\end{pysnip}
Earlier releases divide their fits in other ways, by reactor-flux treatment before 2.0 and by MINOS event selection in 2.1. The named sets hold one category of each release; passing {\tt category} to the loader reaches the others, \eg,
\begin{pysnip}
# Additional set categories
osc = gd.load_nufit_params('NuFIT 2.1',
 category='LID')   # MINOS event selection
P = oscprob.osc_prob_3nu_vacuum(E, L,
 **osc)
# Pme = 0.05579; 'LEM' gives 0.05558

osc = gd.load_nufit_params('NuFIT 1.3',
 category='huber_fluxes_no_rsbl')
P = oscprob.osc_prob_3nu_vacuum(E, L,
 **osc)
# Pme = 0.04998; 'free_fluxes_rsbl'
# gives 0.04552
\end{pysnip}
\ref{app:params} shows the list of parameter sets in \magnus\ v1.2.0. The predefined sets, and the releases and categories the loader reads, can also be printed from Python, which keeps them current with the installed version:
\begin{pysnip}
# Every predefined set, with its description
sets = gd.OSC_PARAMS_PREDEFINED
for name in sorted(sets):
    desc = sets[name]['description']
    print(name, '|', desc)
# 53 lines, e.g.,
# OSC_PARAMS_NU_FIT_1_0_IO | NuFIT 1.0, IO

# The releases the loader reads, and the
# categories of each
fits = gd.NUFIT_GLOBAL_FITS
for v in fits:
    print(v, list(fits[v]['categories']))
# 18 lines, e.g., NuFIT 2.1 ['LEM', 'LID']
\end{pysnip}

Individual parameter values are passed by name. Anything left unset still comes from the default set, so one parameter can be moved without restating the other five, \eg,
\begin{pysnip}
P = oscprob.osc_prob_3nu_vacuum(E, L,
 dCP=0.0)
# Pme = 0.05098; the other five stay at
# their defaults
\end{pysnip}
The two mechanisms combine. A named set chooses the base, and a parameter passed alongside it replaces that one entry:
\begin{pysnip}
# A named set, with one value moved
P = oscprob.osc_prob_3nu_vacuum(E, L,
 default_osc_params_set_name=
  'OSC_PARAMS_NU_FIT_6_1_SK_IO', dCP=0.0)
# The other five stay at their NuFIT 6.1 IO
# values.  Pme = 0.01800, against
# 0.05241 with the set's own dCP
\end{pysnip}


\subsubsection{Specifying angle conventions}
\label{sec:angle_conventions}

The mixing angles can be stated four ways, which the keyword {\tt angles} selects: {\tt 'sin'} by default, {\tt 'sin2'} for the sines squared that global fits report, {\tt 'rad'} for the angles themselves, or {\tt 'deg'}. With {\tt 'deg'}, the CP phases are read as degrees as well. The keyword has to be given the same value in two places, on {\tt load\_nufit\_params} and on the probability call, since a set loaded under one convention and read under another usually returns a plausible, wrong answer:
\begin{pysnip}
# The convention travels with the numbers
for conv in ('sin', 'sin2', 'rad', 'deg'):
    osc = gd.load_nufit_params('NuFIT 6.1',
     angles=conv)
    P = oscprob.osc_prob_3nu_vacuum(E, L,
     angles=conv, **osc)
# s12 reads 0.5557, 0.3088, 0.5892, 33.759
# and Pme = 0.03127 in all four cases
\end{pysnip}


\subsubsection{Two flavors}
\label{sec:two_flavors}

Two flavors take a different route. A two-flavor system has one mixing angle and one splitting, which no global fit reports as such, so {\tt osc\_prob\_2nu\_*} has no parameter set to fall back on. It does not accept {\tt default\_osc\_params\_set\_name}. Both {\tt sth} and {\tt Dm2} are required arguments; omitting either raises an error. The {\tt angles} keyword applies as it does at three flavors, so {\tt sth} is read as a sine unless another convention is named:
\begin{pysnip}
# Two flavors: both values are required
P = oscprob.osc_prob_2nu_vacuum(E, L,
 sth=0.5557, Dm2=7.537e-5)
# P[0][1] = 0.013089

# The same point, angles in degrees
P = oscprob.osc_prob_2nu_vacuum(E, L,
 sth=33.759, Dm2=7.537e-5, angles='deg')
\end{pysnip}


\subsubsection{Four flavors}
\label{sec:four_flavors}

A fourth, sterile flavor adds three mixing angles, two CP phases, and one mass splitting to the six three-flavor parameters, in the parametrization of Ref.~\cite{Bustamante:2019ggq}. The six three-flavor parameters behave as at three flavors: left unset, they are read from the named set. The six new ones default to zero, so a four-flavor call without them returns the three-flavor probabilities in its upper $3 \times 3$ block. All twelve can be given at once:
\begin{pysnip}
# Four flavors (3+1): twelve parameters
P = oscprob.osc_prob_4nu_vacuum(E, L,
 s12=0.3088, s23=0.4700, s13=0.02248,
 dCP=3.700, D21=7.537e-5, D31=2.511e-3,
 s14=0.10, s24=0.10, s34=0.0,
 d14=0.0, d24=0.0, D41=1.0,
 angles='sin2')
# Pme = 0.02801, Pmm = 0.39958
\end{pysnip}
Here, the angles are given as their sines squared, as global fits report them; the first six values are the NuFIT 6.1 best fit of the default set, rounded. The call returns a $4 \times 4$ matrix, indexed by {\tt NUE}, {\tt NUMU}, {\tt NUTAU}, and {\tt NUS} for the sterile flavor; at five flavors, the two sterile flavors are {\tt NUS1} and {\tt NUS2}.


\subsubsection{Five flavors}
\label{sec:five_flavors}

A fifth flavor adds three further mixing angles, two further CP phases, and a second splitting, for 18 parameters in all:
\begin{pysnip}
# Five flavors (3+2): 18 parameters
P = oscprob.osc_prob_5nu_vacuum(E, L,
 s12=0.3088, s23=0.4700, s13=0.02248,
 dCP=3.700, D21=7.537e-5, D31=2.511e-3,
 s14=0.10, s24=0.10, s34=0.0,
 s15=0.06, s25=0.06, s35=0.0,
 d14=0.0, d15=0.0, d24=0.0, d35=0.0,
 D41=1.0, D51=1.7, angles='sin2')
# Pme = 0.03984, Pmm = 0.40684
\end{pysnip}
As at four flavors, every active-sterile parameter defaults to zero. The mixing matrix has nine angles and five CP phases, {\tt dCP}, {\tt d14}, {\tt d15}, {\tt d24}, and {\tt d35}, where a general $5 \times 5$ unitary matrix has ten angles and six phases. The missing angle and phase belong to a rotation between the two sterile flavors. Since the two sterile states are indistinguishable, that rotation changes no probability among the active flavors, so it is left out.


\subsubsection{Default non-standard parameters}

Non-standard parameters are zero by default (\tabl{parameters}). The couplings of the {\tt \_nsi} wrappers and the Lorentz-violation parameters of the {\tt \_liv} wrappers all default to 0, so a non-standard wrapper called without them returns the standard result. The LIV energy scale {\tt Lambda} defaults to 1 and the power {\tt n\_liv} to 0, so a term switched on through the eigenvalues {\tt b1}, {\tt b2}, \ldots\ is energy-independent unless that power is changed.


\subsection{Batched calls}
\label{sec:batched_calls}

Every wrapper and every scenario function accepts an array for the energy, for the baseline, or for both. The other arguments, such as the mixing parameters and the density, take a single value per call. An array of energies at one baseline returns one probability matrix per energy, stacked along an axis:
\begin{pysnip}
Es = np.logspace(-1, 1, 200)*gd.UNIT_GEV
P = oscprob.osc_prob_3nu_vacuum(Es, L)
P.shape                    # (200, 3, 3)
\end{pysnip}
An array of baselines at one energy works the same way. Here, {\tt E} is still the 1~GeV of Listing~\ref{lst:minimal}:
\begin{pysnip}
Ls = np.linspace(0, 1300, 200)*gd.UNIT_KM
P = oscprob.osc_prob_3nu_vacuum(E, Ls)
P.shape                    # (200, 3, 3)
# P[-1] equals the single call at 1300 km
\end{pysnip}
Arrays for both the energy and the baseline are paired element by element, not crossed into a grid, so they must have the same length; if they do not, \magnus\ raises a {\tt ValueError}. The first two pairs below have the same $L/E$, so, in vacuum, they have the same probabilities, \eg, 0.048968 for $\nu_\mu \to \nu_e$:
\begin{pysnip}
Es = np.array([1.0, 2.0, 5.0])*gd.UNIT_GEV
Ls = np.array([500, 1000, 1300])*gd.UNIT_KM
P = oscprob.osc_prob_3nu_vacuum(Es, Ls)
P.shape          # (3, 3, 3): 3 pairs
P[0, 1, 0], P[1, 1, 0]   # 0.048968 both
\end{pysnip}
To compute a grid of energies and baselines, such as an oscillogram, the user makes one call per baseline, each with the full array of energies.

A batched call is much faster than the same points computed one at a time, because \magnus\ shares the work across them (Sec.~\ref{sec:engines}). Where the Hamiltonian does not vary along the path, as in vacuum or at constant density, the whole scan is one stack of matrix exponentials. Where it varies, the energy-batched engine samples the density profile once for all the energies of a scan, and the cumulative engine computes all the baselines of a scan in a single pass. For scans of 200 points, at two to five flavors, batching is 77--139 times faster than one call per point in vacuum, 108--130 times faster at constant density, and 29--175 times faster along an Earth chord. The gain is smallest where the energies need the most different refinements, as for an eV-scale sterile neutrino along an Earth chord: $1.7 \times$, for 40 energies at four flavors.

In vacuum and at constant density, batched and point-by-point results agree exactly at two and three flavors, and to within $10^{-11}$ at four and five. Along a varying profile, each route refines its answer to the requested tolerance on its own, so the two agree to within about that tolerance, but not exactly.


\subsection{Constant density}
\label{sec:constant_density}

In matter of constant density, the Hamiltonian does not depend on position. Every commutator in \equ{recursion} vanishes, the Magnus series stops at its first term, and the evolution operator is a single matrix exponential. As in vacuum, \magnus\ computes this without exercising any of the method of Sec.~\ref{sec:method}.

\begin{lstlisting}[language=Python, caption={\textbf{Constant density at three and two flavors.} The density is swept from vacuum to about the value at the center of the Earth. The last call returns 200 energies at one density, which the constant-Hamiltonian engine of Sec.~\ref{sec:engines} answers as a single stack of matrix exponentials. See Sec.~\ref{sec:constant_density} for details.}, label={lst:const_density}, float=tbp]
import numpy as np
import magnus.globaldefs as gd
import magnus.oscprob as oscprob

# 'NuFIT 6.1' is the default; re-loaded for clarity only
osc = gd.load_nufit_params('NuFIT 6.1')
E, L = 1.0*gd.UNIT_GEV, 1000.0*gd.UNIT_KM

# Three flavors, one energy and one baseline
for rho in (0.0, 3.0, 8.0, 13.0):
    P = oscprob.osc_prob_3nu_matter_constant_density(
        E, L, rho=rho,
        density_matter_is_in_g_per_cm3=True, **osc)
    print('%5.1f g/cm3  Pme = %.6f'
          % (rho, P[gd.NUMU][gd.NUE]))
#   0.0 g/cm3  Pme = 0.003773
#   3.0 g/cm3  Pme = 0.013589
#   8.0 g/cm3  Pme = 0.050497
#  13.0 g/cm3  Pme = 0.113343

# Two flavors: one angle and one splitting
P = oscprob.osc_prob_2nu_matter_constant_density(
    E, L, rho=3.0, sth=osc['s13'], Dm2=osc['D31'],
    density_matter_is_in_g_per_cm3=True)
# P[0][1] = 0.005766

# Two hundred energies in one call
Es = np.logspace(-1, 1, 200)*gd.UNIT_GEV
P = oscprob.osc_prob_3nu_matter_constant_density(
    Es, L, rho=3.0,
    density_matter_is_in_g_per_cm3=True, **osc)
# P.shape = (200, 3, 3)
\end{lstlisting}

Listing~\ref{lst:const_density} shows the constant-density wrapper at three and two flavors, over a range of densities, and also the whole calculation as one batched call. The number passed as {\tt rho} is read in one of three ways, set by two keywords, both {\tt False} by default:
\begin{itemize}
\item {\tt density\_matter\_is\_in\_g\_per\_cm3=True}, as in Listing~\ref{lst:const_density}, means {\tt rho} is mass density in g~cm$^{-3}$.
\item {\tt density\_is\_of\_number\_of\_electrons=True} means {\tt rho} is the electron number density itself, in eV$^3$.
\item With neither, {\tt rho} is a mass density in natural units, eV$^4$.
\end{itemize}

A mass density and an electron number density differ by the composition of the medium, which enters through a third keyword, {\tt electron\_fraction}, the number of electrons per atomic mass unit of matter, $Y_e = n_e m_u/\rho$, so that $n_e = \rho N_A Y_e$, with $m_u$ the atomic mass unit and $N_A$ Avogadro's number. For a single element, it is the atomic number over the atomic weight, \eg, 26/55.845 for iron. It defaults to 1/2 in the constant-density wrappers. (Inside the Earth, the wrappers instead take $Y_e$ layer by layer from the PREM, as a function of the distance from the center; see Sec.~\ref{sec:prem} and \figu{prem}.) Lowering $Y_e$ to 0.2 at 3~g~cm$^{-3}$ takes $P_{\nu_\mu \to \nu_e}$ from 0.013589 to 0.006726, nearly halving it. It is read only when {\tt rho} is a mass density; a call that supplies the electron number density directly ignores it, since the conversion it governs has already been done by the user.

Four and five flavors take the same form, with the active-sterile parameters added as further keywords. The call {\tt osc\_prob\_4nu\_matter\_constant\_density} with {\tt s14} and {\tt D41} returns a $4 \times 4$ matrix where the three-flavor call returns a $3 \times 3$ one; nothing else about the call changes.

One further keyword, {\tt ratio\_number\_neutrons\_to\_protons}, is the ratio $r = n_n/n_p$ that enters the projector of \equ{h_matter}; since $n_p = n_e$, it is also the number of neutrons per electron. It defaults to 1, the isoscalar value, which is the medium the default $Y_e = 1/2$ already describes. In neutral matter, $Y_e = 1/(1 + r)$. Setting one of the two keywords does not change the other, so both should be set together. The ratio sets the sterile entry of the matter projector of \equ{h_matter}: with $\sin^2\theta_{14} = \sin^2\theta_{24} = 0.1$ and $\Delta m^2_{41} = 1$~eV$^2$, at the energy, baseline, and density of Listing~\ref{lst:const_density}, raising it to 1.5 moves $P_{\nu_\mu \to \nu_\mu}$ from 0.7764 to 0.7587, by 2.3\%. It does not enter the conversion of a mass density into an electron number density, $n_e = \rho N_A Y_e$, which the electron fraction alone sets.


\subsection{Using wrappers vs.~scenario calls vs.~direct calls}
\label{sec:wrapper_vs_direct}

\begin{lstlisting}[language=Python, caption={\textbf{Vacuum and constant density, three ways.} The three-flavor probability, computed three ways. The wrappers take the oscillation parameters as keywords, here the NuFIT 6.1 values, and the density in g~cm$^{-3}$. The scenario function takes the density as a function of position. The hand-built form is the vacuum term divided by the energy plus the potential $V_{\rm CC}$ times the projector of \equ{h_matter}; {\tt VCC\_EARTH\_CRUST} is that potential at 3~g~cm$^{-3}$ and $Y_e = 1/2$. All three return the same probabilities; the times are measured under the protocol of Sec.~\ref{sec:timing_protocol}. See Sec.~\ref{sec:wrapper_vs_direct} for details.}, label={lst:constant}, float=tbp]
import numpy as np
import magnus.globaldefs as gd
import magnus.hamiltonians as ham
import magnus.matter as matter
import magnus.oscprob as oscprob

osc = gd.load_nufit_params('NuFIT 6.1')
E, L = 1.0*gd.UNIT_GEV, 1000.0*gd.UNIT_KM
Es = np.logspace(-1, 1, 200)*gd.UNIT_GEV

# --- 1. With the named wrappers
P_vac = oscprob.osc_prob_3nu_vacuum(E, L, **osc)
P_mat = oscprob.osc_prob_3nu_matter_constant_density(
    E, L, rho=3.0, density_matter_is_in_g_per_cm3=True, **osc)
# Pme = 0.00377306 in vacuum, 0.01358909 in
# matter, in 0.15 ms

P_scan = oscprob.osc_prob_3nu_matter_constant_density(
    Es, L, rho=3.0, density_matter_is_in_g_per_cm3=True, **osc)
# P_scan.shape = (200, 3, 3), in 0.13 ms

# --- 2. With a scenario function
rho_func = lambda l: 3.0*gd.UNIT_G_PER_CM3
P_mat = oscprob.osc_prob_matter_std_potential(
    3, rho_func, E, L, osc_params=osc)
# The same number, to 1e-16, in 3.6 ms

# --- 3. With a Hamiltonian built by hand
H_vac = ham.hamiltonian_3nu_vacuum_energy_independent(
    s12=osc['s12'], s23=osc['s23'], s13=osc['s13'], dCP=osc['dCP'],
    D21=osc['D21'], D31=osc['D31'])
proj = matter.matter_potential_projector(3)

P_vac = oscprob.osc_prob(H_vac/E, 0.0, L)
P_mat = oscprob.osc_prob(
    H_vac/E + gd.VCC_EARTH_CRUST*proj, 0.0, L)
# The same two numbers, in 0.043 ms

P_scan = oscprob.osc_prob_energy_baseline(
    lambda e: H_vac/e + gd.VCC_EARTH_CRUST*proj, Es, L,
    H_func_is_function_only_of_energy=True)
# The same 200 points, in 5.8 ms
\end{lstlisting}

Every wrapper function (\eg, {\tt osc\_prob\_3nu\_vacuum}, {\tt osc\_prob\_3nu\_matter\_constant\_density}) builds a Hamiltonian internally and hands it, through a scenario function, to the engines of Sec.~\ref{sec:engines}, the last of which is {\tt osc\_prob}, the primitive probability function of \magnus. The wrappers are a convenience; a user can do the same by hand, or enter at the scenario layer between the two. Wrappers also contain the most safeguards and validation particular to the cases they are built for; scenario functions contain fewer; and the direct call to {\tt osc\_prob}, the least.  Figure~\ref{fig:layers} shows the wrappers, scenario functions, and their relation to one another and {\tt osc\_prob}.

Listing~\ref{lst:constant} computes the same three-flavor probability all three ways, at one energy and over a scan of two hundred. The three agree: to round-off in matter, bit-for-bit in vacuum. 


\subsubsection{Through a wrapper}
\label{sec:wrappers}

{\def\arraystretch{1.12}
\begin{table*}[t!]
{\footnotesize
\begin{tabular}{@{}p{2.7cm}p{2.5cm}p{5.9cm}p{5.9cm}@{}}
 Environment & Physics & \multicolumn{2}{@{}l@{}}{Wrapper function} \\
 \hline
 Vacuum & Standard & {\tt osc\_prob\_2nu\_vacuum} & {\tt osc\_prob\_3nu\_vacuum} \\
  &  & {\tt osc\_prob\_4nu\_vacuum} & {\tt osc\_prob\_5nu\_vacuum} \\
  & Lorentz violation & {\tt osc\_prob\_2nu\_vacuum\_liv} & {\tt osc\_prob\_3nu\_vacuum\_liv} \\
  &  & {\tt osc\_prob\_4nu\_vacuum\_liv} & {\tt osc\_prob\_5nu\_vacuum\_liv} \\
 Constant density & Standard & {\tt osc\_prob\_2nu\_matter\_constant\_density} & {\tt osc\_prob\_3nu\_matter\_constant\_density} \\
  &  & {\tt osc\_prob\_4nu\_matter\_constant\_density} & {\tt osc\_prob\_5nu\_matter\_constant\_density} \\
  & Non-standard int. & {\tt osc\_prob\_2nu\_matter\_nsi\_constant\_density} & {\tt osc\_prob\_3nu\_matter\_nsi\_constant\_density} \\
  &  & {\tt osc\_prob\_4nu\_matter\_nsi\_constant\_density} & {\tt osc\_prob\_5nu\_matter\_nsi\_constant\_density} \\
  & Lorentz violation & {\tt osc\_prob\_2nu\_matter\_liv\_constant\_density} & {\tt osc\_prob\_3nu\_matter\_liv\_constant\_density} \\
  &  & {\tt osc\_prob\_4nu\_matter\_liv\_constant\_density} & {\tt osc\_prob\_5nu\_matter\_liv\_constant\_density} \\
 Exponential density & Standard & {\tt osc\_prob\_2nu\_matter\_exp\_density} & {\tt osc\_prob\_3nu\_matter\_exp\_density} \\
  &  & {\tt osc\_prob\_4nu\_matter\_exp\_density} & {\tt osc\_prob\_5nu\_matter\_exp\_density} \\
  & Non-standard int. & {\tt osc\_prob\_2nu\_matter\_nsi\_exp\_density} & {\tt osc\_prob\_3nu\_matter\_nsi\_exp\_density} \\
  &  & {\tt osc\_prob\_4nu\_matter\_nsi\_exp\_density} & {\tt osc\_prob\_5nu\_matter\_nsi\_exp\_density} \\
  & Lorentz violation & {\tt osc\_prob\_2nu\_matter\_liv\_exp\_density} & {\tt osc\_prob\_3nu\_matter\_liv\_exp\_density} \\
  &  & {\tt osc\_prob\_4nu\_matter\_liv\_exp\_density} & {\tt osc\_prob\_5nu\_matter\_liv\_exp\_density} \\
 Earth (PREM) & Standard & {\tt osc\_prob\_2nu\_earth} & {\tt osc\_prob\_3nu\_earth} \\
  &  & {\tt osc\_prob\_4nu\_earth} & {\tt osc\_prob\_5nu\_earth} \\
  & Non-standard int. & {\tt osc\_prob\_2nu\_earth\_nsi} & {\tt osc\_prob\_3nu\_earth\_nsi} \\
  &  & {\tt osc\_prob\_4nu\_earth\_nsi} & {\tt osc\_prob\_5nu\_earth\_nsi} \\
  & Lorentz violation & {\tt osc\_prob\_2nu\_earth\_liv} & {\tt osc\_prob\_3nu\_earth\_liv} \\
  &  & {\tt osc\_prob\_4nu\_earth\_liv} & {\tt osc\_prob\_5nu\_earth\_liv} \\
 Sun & Standard & {\tt osc\_prob\_2nu\_sun} & {\tt osc\_prob\_3nu\_sun} \\
  &  & {\tt osc\_prob\_4nu\_sun} & {\tt osc\_prob\_5nu\_sun} \\
  & Non-standard int. & {\tt osc\_prob\_2nu\_sun\_nsi} & {\tt osc\_prob\_3nu\_sun\_nsi} \\
  &  & {\tt osc\_prob\_4nu\_sun\_nsi} & {\tt osc\_prob\_5nu\_sun\_nsi} \\
  & Lorentz violation & {\tt osc\_prob\_2nu\_sun\_liv} & {\tt osc\_prob\_3nu\_sun\_liv} \\
  &  & {\tt osc\_prob\_4nu\_sun\_liv} & {\tt osc\_prob\_5nu\_sun\_liv} \\
\end{tabular}}
\caption{\label{tab:wrappers}\textbf{The probability wrapper functions.} The 56 wrapper functions of \magnus, by environment and by the physics in the Hamiltonian. Each of the fourteen combinations exists at two, three, four, and five flavors. There is no wrapper for non-standard interactions in vacuum, since those are interactions with matter. Every function here builds a Hamiltonian and hands it to the engines of Sec.~\ref{sec:engines}. (Three others, not shown here, do not: {\tt osc\_prob\_2nu\_vacuum\_std}, {\tt osc\_prob\_3nu\_vacuum\_std}, and {\tt osc\_prob\_2nu\_matter\_std} evaluate the standard analytic expressions for the probability instead.) One more, {\tt osc\_prob\_pseudo\_dirac\_vacuum}, covers a pseudo-Dirac spectrum in vacuum (Sec.~\ref{sec:shipped_hamiltonians}). See Sec.~\ref{sec:wrappers} for details.}
\end{table*}}

\tabl{wrappers} lists the 56 wrappers of \magnus. A wrapper is the shortest call whenever the scenario is one \magnus\ ships with. It takes the oscillation parameters as keywords, fills in any left unset from a named set, converts the density from g~cm$^{-3}$, and assembles the Hamiltonian. That last step includes the vacuum term: {\tt osc\_prob\_3nu\_matter\_constant\_density} takes the same six oscillation parameters as {\tt osc\_prob\_3nu\_vacuum} and adds the potential to what it builds from them, which is why {\tt H\_vac} appears only in the third block of Listing~\ref{lst:constant}.

Every wrapper takes its arguments in the same four groups: the energy, the geometry of the environment, the oscillation parameters, and whatever the scenario adds. For instance, for non-standard interactions (NSI) at three flavors in constant density,
\begin{pysnip}
P = oscprob.\
 osc_prob_3nu_matter_nsi_constant_density(
    E, L,            # 1. Energy, baseline
    rho=3.0,         # 2. Density
    density_matter_is_in_g_per_cm3=True,
    **osc,           # 3. Mixing
    eps_em=0.05)     # 4. New physics
# Pme = 0.017126.  Dropping eps_em gives
# 0.013589, the standard wrapper's answer
\end{pysnip}
The fourth group (``new physics'') exists only in the wrappers whose name carries {\tt \_nsi} or {\tt \_liv}; \tabl{parameters} lists its parameters. Every one of them defaults to zero, so one of those wrappers called without them returns the standard result, bit-for-bit identical to what the standard wrapper of the same environment returns. That makes it the control for any new-physics scan. 

Passing {\tt rho} as a scalar makes the Hamiltonian independent of position, so the request reaches the constant-Hamiltonian engine of Sec.~\ref{sec:engines} and the whole scan becomes one stack of matrix exponentials. Two hundred energies cost 0.13~ms, no more than the 0.15~ms of a single energy, so the scan is nearly free.


\subsubsection{Through a scenario function}

{\def\arraystretch{1.2}
\begin{table*}[t!]
{\footnotesize
\begin{tabular}{@{}p{5.0cm}p{4.4cm}p{7.4cm}@{}}
 Scenario function & Covers & Arguments after {\tt num\_flavors} \\
 \hline
 {\tt osc\_prob\_vacuum} & Vacuum & {\tt energy}, {\tt L}, {\tt osc\_params} \\
 {\tt osc\_prob\_matter\_std\_potential} & Standard oscillations in matter & {\tt rho\_func}, {\tt energy}, {\tt L}, {\tt osc\_params} \\
 {\tt osc\_prob\_matter\_nsi} & Non-standard interactions & {\tt rho\_func}, {\tt energy}, {\tt L}, {\tt osc\_params}, {\tt nsi\_params} \\
 {\tt osc\_prob\_liv} & Lorentz violation, in vacuum or matter & {\tt energy}, {\tt L}, {\tt osc\_params}, {\tt liv\_params}; {\tt rho\_func} optional \\
 \hline
 {\tt osc\_prob\_earth} & A chord through PREM & {\tt H\_func}, {\tt energy}, and {\tt costhz} with {\tt L}, or two named locations \\
 {\tt osc\_prob\_sun} & A ray out of the Sun & {\tt H\_func}, {\tt energy}, {\tt L} \\
\end{tabular}}
\caption{\label{tab:scenarios}\textbf{The probability scenario functions.} The scenario functions of \magnus, one per physics scenario. Each takes the flavor count as its first argument, so one function serves every flavor count where \tabl{wrappers} needs four wrappers. \magnus\ ships vacuum Hamiltonians for two through five flavors; past that the user supplies one through {\tt h\_vac\_energy\_indep}, which the same functions then use. Each of the matter functions takes the density as {\tt rho\_func}, a function of position, which is what makes this the layer for a profile that varies. The optional {\tt rho\_func} of {\tt osc\_prob\_liv} is why Lorentz violation is the one non-standard scenario that also exists in vacuum. The refinement keywords of Sec.~\ref{sec:architecture} reach all of them through {\tt **kwargs}. The two functions below the rule are not scenario functions: they take a Hamiltonian the user supplies and apply a geometry to it, so they pair either with a scenario function or with a Hamiltonian built by hand.}
\end{table*}}

\tabl{scenarios} lists the four scenario functions shipped with \magnus\ (plus two other functions at the same level that handle geometry in the Earth and the Sun). The scenario functions sit one layer below the wrappers, one per physics scenario and each applicable at any flavor count. They take the density as a \emph{function of position}, which is what makes them the entry point for a profile that varies; Sec.~\ref{sec:smooth_comparison} calls \magnus\ this way.

Every scenario function takes its arguments in the same groups, with the flavor count first and the physics handed over as dictionaries rather than as keywords. For instance, for NSI at three flavors in constant density,
\begin{pysnip}
rho_func = lambda l: 3.0*gd.UNIT_G_PER_CM3
nsi = dict(eps_ee=0.0, eps_em=0.05,
           eps_et=0.0, eps_mm=0.0,
           eps_mt=0.0, eps_tt=0.0)

P = oscprob.osc_prob_matter_nsi(
    3,               # 1. Flavor count
    rho_func,        # 2. The medium
    E, L,            # 3. Energy, baseline
    osc_params=osc,  # 4. Mixing
    nsi_params=nsi)  # 5. New physics
# Pme = 0.017126, the same number the
# wrapper above returns
\end{pysnip}
Two features differ from the wrapper. First, the flavor count is an argument, not part of the function's name, which is why one scenario function covers what \tabl{wrappers} needs four wrappers for. Second, the dictionaries have no defaults: every coupling has to be present, so the six entries of {\tt nsi\_params} are all written out even though five of them are zero. A wrapper fills those in.

In constant density, scenario functions are the wrong tool. A callable {\tt rho\_func} that returns the same number everywhere is not detected as constant by the scenario function, so the constant-Hamiltonian engine is never reached: a single probability falls through to the general machinery in {\tt osc\_prob}, and a scan to the energy-batched engine. The answer is the same as the wrapper's to round-off, at a far higher cost.


\subsubsection{Through a direct call}

Handing {\tt osc\_prob} a matrix is the only route on which the user has to assemble the Hamiltonian themselves; the wrappers and the scenario functions build it from the parameters they are given. At the same time, it is the cheapest route per probability, about 0.02~ms, because the parameter resolution, input validation, and assembly have already been paid for by the user.

The wrapper and scenario function above both reach the probability from named parameters, building the Hamiltonian internally on the way. At the base layer, the user must perform that construction themselves. For the same case of NSI at three flavors in constant-density matter, this means computing and summing three matrices, \ie,
\begin{pysnip}
H_vac = ham.hamiltonian_3nu_vacuum(
    E,
    s12=osc['s12'], s23=osc['s23'],
    s13=osc['s13'], dCP=osc['dCP'],
    D21=osc['D21'], D31=osc['D31'])

# proj for three flavors is diag(1, 0, 0)
proj = matter.matter_potential_projector(3)

V = gd.VCC_EARTH_CRUST # About 1.145e-13 eV
H_nsi = ham.hamiltonian_3nu_nsi(
    V, 0.0, 0.05, 0.0, 0.0, 0.0, 0.0)

# 1. mixing, 2. matter, 3. new physics
H = H_vac + V*proj + H_nsi

P = oscprob.osc_prob(H, 0.0, L)   # 4. path
# Pme = 0.017126 again
\end{pysnip}
{\tt hamiltonian\_3nu\_nsi} returns the non-standard term alone, $V_{\rm CC}$ times the matrix of couplings, \equ{h_nsi_3nu}, without the standard charged-current term. The standard term is a separate summand, which means that leaving {\tt V*proj} out of the sum does not raise an error: it returns a converged, unitary probability of 0.005962 for a medium with no ordinary matter effect in it. This is unphysical, since NSI coexist with standard interactions; the user must include both.

Assembling the Hamiltonian rarely means writing a matrix from scratch. \magnus\ ships forty-three Hamiltonian builders: vacuum, matter, non-standard interactions, and Lorentz violation at two to five flavors, in position-dependent and position-independent forms, and three for a pseudo-Dirac spectrum. Listing~\ref{lst:constant} uses one of them for the vacuum term and takes the projector from {\tt matter}. Section~\ref{sec:hamiltonians} lists them all, so a direct call is usually a shipped Hamiltonian with something added to it by the user, \eg, a new non-standard contribution.

The direct route does not batch: {\tt osc\_prob} takes one Hamiltonian at one energy and one baseline, and returns one matrix of probabilities. To scan energies with a Hamiltonian built by hand, the user passes it to {\tt osc\_prob\_energy\_baseline}, the third layer of \figu{layers}, which calls {\tt osc\_prob} once per energy. For the 200 energies of Listing~\ref{lst:constant}, that takes 5.8~ms. The wrapper computes the same 200 energies in 0.13~ms, about forty times faster, because it hands them to the constant-Hamiltonian engine as one batch.

Calling {\tt osc\_prob} is the most general route: replacing the scalar {\tt V} by a function of position gives a varying profile, as in Sec.~\ref{sec:hamiltonians}; replacing the whole Hamiltonian matrix is illustrated in Secs.~\ref{sec:building_new_hamiltonian} and~\ref{sec:lri_sun}.


\subsubsection{Which to use}

In brief, a user should use:
\begin{itemize}
\item \textbf{A wrapper} (\tabl{wrappers}), for any scenario \magnus\ already implements (vacuum, constant density, exponential density, the Earth, and the Sun, each at two to five flavors, with standard oscillations, non-standard interactions, or Lorentz violation). It is the shortest call and the one with the most safeguards: it fills in the parameters left unset, converts the density, and builds the Hamiltonian in the form the fastest applicable engine of Sec.~\ref{sec:engines} expects, \eg, a constant density as a number.
\item \textbf{A scenario function} (\tabl{scenarios}), for a density profile that no wrapper provides, given as any function of position, or for more than five flavors, with a vacuum Hamiltonian supplied through {\tt h\_vac\_energy\_indep}. Given the same inputs, it reaches the same engines as a wrapper.
\item \textbf{A direct call} to {\tt osc\_prob}, for a Hamiltonian that no shipped builder produces: a term from new physics beyond non-standard interactions and Lorentz violation, or a matrix that comes from somewhere else entirely.
\end{itemize}


\subsubsection{Returning the evolution operator}
\label{sec:evolution_operator}

Every wrapper and scenario function above returns probabilities; for most purposes, that is the output wanted. But some observables are built from the amplitudes instead. Three examples are the mass-eigenstate content of a neutrino leaving a dense source, which is the factor the source contributes to \equ{averaged_varying}; the flavor composition at a detector after an incoherent mixture of mass eigenstates has crossed the Earth; and the evolution through two media in sequence. None of them can be recovered from a probability matrix, since the phases are gone at that stage. For those cases, every probability function of \magnus, on any of the three routes above and at any flavor count, returns the evolution operator alongside the probabilities when asked via {\tt return\_evolution\_operator}:
\begin{pysnip}
P, U = oscprob.osc_prob_3nu_vacuum(
    E, L, return_evolution_operator=True,
    **osc)
# P is the matrix the call returns without
# the keyword; U is complex and unitary.
# Both of these hold:
np.allclose(P, abs(U.T)**2)
np.allclose(U.conj().T @ U, np.eye(3))
\end{pysnip}
The operator is indexed {\tt U[final, initial]}, so that its moduli squared, transposed, are the probability matrix, indexed the other way. Everything else keeps its meaning: {\tt nu\_i} and {\tt nu\_f} still select one channel of {\tt P} while {\tt U} stays the full operator; arrays of energies or baselines return one operator per point, of shape {\tt (n, d, d)}. {\tt return\_evolution\_operator} is declared by {\tt osc\_prob} and forwarded by every wrapper, as {\tt t\_breakpoints} is.

Inside {\tt osc\_prob}, the request changes two behaviors. First, the refinement ladder compares the operator itself between levels, at the same {\tt rtol} and {\tt atol}, so the operator returned is converged in its phases, not only in its moduli. Second, the specialized engines of Sec.~\ref{sec:engines} stand aside for the call, since only the general Magnus ladder forms the evolution operator; a scan over baselines takes the per-point path instead of the cumulative traversal. Two settings are refused if asked together with {\tt return\_evolution\_operator}, with an error returned naming why: {\tt average=True}, since its routes decohere before any operator is formed; {\tt strategy='hybrid'}, since that engine answers with probabilities only.

The evolution operators returned can be composed. The evolution through two media in sequence is the product of their operators, the later one on the left:
\begin{pysnip}
# Two slabs of constant density in sequence,
# the second denser than the first
H1 = H_vac/E + gd.VCC_EARTH_CRUST*proj
H2 = H_vac/E + 1.5*gd.VCC_EARTH_CRUST*proj
_, U1 = oscprob.osc_prob(H1, 0.0, L,
    return_evolution_operator=True)
_, U2 = oscprob.osc_prob(H2, 0.0, L,
    return_evolution_operator=True)
U = U2 @ U1
P = abs(U.T)**2       # Over both slabs, 2L
\end{pysnip}
The same probabilities come from one call over a two-slab profile with the edge declared through {\tt t\_breakpoints}, as in Listing~\ref{lst:edges}; the product above is what that call assembles internally. Section~\ref{sec:jet} uses the {\tt return\_evolution\_operator} flag and operator composition to first compute the mass-eigenstate content of the neutrino flux that leaves a star, and then its flavor content at a detector on Earth.


\subsubsection{Asking for the averaged probability}
\label{sec:average_keyword}

Every wrapper, scenario function, and {\tt osc\_prob\_energy\_baseline} return the phase average of Sec.~\ref{sec:averaged} when called with {\tt average=True}. Two more keywords configure it: {\tt average\_spread} sets the relative energy spread, $\sigma$, which is 10\% by default, and {\tt average\_initial\_state} sets the state the neutrino starts in, explained below. {\tt osc\_prob} itself, which computes a single point, does not accept {\tt average}. Nor can {\tt average=True} be combined with {\tt return\_evolution\_operator} (Sec.~\ref{sec:evolution_operator}), since the averaged routes never form an evolution operator. On the wrappers and scenario functions, {\tt strategy\_info['engine']} reports {\tt 'average'} when the averaged route answered (Sec.~\ref{sec:which_engine}).

The phase average is not the mean of the probability over an energy range: it damps each interference term by the spread of its phase, but keeps the eigenvectors at the central energy (Sec.~\ref{sec:averaged}). Wherever the limit, \equ{averaged} or \equ{averaged_varying}, agrees with the phase average within $10^{-4}$, or within the requested tolerance if that is tighter, both in absolute terms and relative to the probability, \magnus\ returns the limit. How the phase average is computed follows Sec.~\ref{sec:averaged}.

\smallskip

\textit{A constant Hamiltonian.---}For a Hamiltonian that does not depend on position, the phase average is \equ{phase_average}. It needs only the eigenvectors of $\mathbb{H}$: one diagonalization per energy, no propagation. Over an astrophysical baseline, every interference term is damped, and the result is \equ{averaged}. In the context of Listing~\ref{lst:constant},
\begin{pysnip}
FAR = 1.0e8*gd.UNIT_KM   # Astrophysical L
P = oscprob.osc_prob_3nu_vacuum(
    E, FAR, average=True, **osc)
# P[0, 0] = 0.5481, P[1, 0] = 0.2155: the
# familiar sum_i |U_ei|^2 |U_mi|^2, from the
# eigenvectors alone
\end{pysnip}
At the 1\,000~km of Listing~\ref{lst:constant}, the atmospheric phase is only about 6~rad, and a 10\% spread damps its interference only partly. The result then depends on $\sigma$, and \magnus\ warns about it:
\begin{pysnip}
P = oscprob.osc_prob_3nu_vacuum(
    E, L, average=True, **osc)   # 1000 km
# PhaseAveragingWarning: the phase-averaged
# probability depends on the energy spread
# at 1 of 1 (energy, L) point(s) ...
# P[0, 0] = 0.9844, P[1, 1] = 0.9088;
# without the keyword: 0.9924, 0.9955
\end{pysnip}
A pair of eigenstates whose phase stays small over the baseline, as in the coherent blocks of \equ{averaged_blocks}, keeps its interference almost undamped, because a small phase also changes little across the spread.

\smallskip

\textit{The initial state.---}For a Hamiltonian that varies along the path, the result depends on the state the neutrino starts in, \equ{flavor_vs_decohered}. With {\tt average\_initial\_state='flavor'}, the default, the neutrino starts as a flavor state, as one produced in the medium does, \eg, in a beam, in the atmosphere, in the Earth, or in the Sun. With {\tt 'decohered'}, it starts as an incoherent mixture of the eigenstates at production, as one arriving from a distant source does. The two differ only by the interference present at production, so they agree wherever that interference is damped away, as it is for a neutrino made in the solar core:
\begin{pysnip}
for start in ('flavor', 'decohered'):
    P = oscprob.osc_prob_3nu_sun(
        10.0*gd.UNIT_MEV,
        gd.SUN_RADIUS*gd.UNIT_KM, 0.0,
        nu_i=gd.NUE, nu_f=gd.NUE,
        average=True,
        average_initial_state=start)
    print(start, P)    # 0.297082 both times
\end{pysnip}
Where it is not damped away, the two differ; on the chord of Sec.~\ref{sec:solar_tomography}, they differ by 0.25. For a constant Hamiltonian, {\tt 'decohered'} returns \equ{averaged}.

\smallskip

\textit{Route 1: a smoothly varying Hamiltonian.---}This is the case of the Sun. \magnus\ carries the initial state along the instantaneous eigenstates of the Hamiltonian, applies the Magnus patch of Sec.~\ref{sec:hybrid} across every crossing that is not adiabatic, and reads the result in the flavor basis at detection, as in \equ{evolution_offset}. On a path without such a crossing, the result is \equ{phase_average_nocross} for a flavor start and \equ{averaged_varying} for a decohered start. As in other calls, {\tt rtol} and {\tt atol} set the tolerance: the patches, and propagation along the adiabatic stretches between them, converge to the tighter of the two, $10^{-3}$ by default; if it is tighter than $10^{-4}$, it replaces $10^{-4}$ in the choice between the limit and the phase average. Over chords through the core with impact parameters below $0.15\,R_\odot$, from 30~GeV to 3~TeV, the probability at the default tolerance differs from its value at a tolerance of $10^{-5}$ by at most $10^{-4}$. A discontinuity that is not declared, but could change the probability, is reported with {\tt UnmarkedDiscontinuityWarning}, not treated as smooth.

\smallskip

\textit{Route 2: a profile with declared discontinuities.---}This is the case of the Earth, whose PREM layer boundaries are breakpoints, or of any profile with discontinuities declared through {\tt t\_breakpoints} or {\tt t\_slab\_edges}. Across a jump in the density, the eigenstates change abruptly, so there are none to carry the state along. \magnus\ instead propagates at energies across a window, averages the resulting probabilities, and warns that the result is the mean of the probability over that window, not the phase average. This route starts from a flavor state only; {\tt 'decohered'} raises an error.

\smallskip

\textit{A Hamiltonian built by hand.---}Because {\tt osc\_prob\_energy\_baseline} accepts {\tt average}, a Hamiltonian built by hand is averaged in the same way as a shipped one:
\begin{pysnip}
H = H_vac/E + gd.VCC_EARTH_CRUST*proj
P = oscprob.osc_prob_energy_baseline(
    H, E, FAR, average=True)
# P[0, 0] = 0.8934: the decohered limit,
# now from the eigenvectors in matter
\end{pysnip}

\smallskip

\textit{Where {\tt average} is used.---}Several later sections use {\tt average=True}. In the Sun (Sec.~\ref{sec:sun} and Listing~\ref{lst:sun}), a ray from the center to the surface holds about $1.4 \times 10^{5}$ oscillation cycles at 5~MeV (Sec.~\ref{sec:averaging_is_not_estimating}), far too many to resolve. For the flavor composition of TeV--PeV astrophysical fluxes (Sec.~\ref{sec:astro}), \equ{averaged} is the whole observable. For a long-range force in the Sun (Sec.~\ref{sec:lri_sun}), the Hamiltonian is built by hand. Section~\ref{sec:jet} does not use {\tt average=True}. There, the neutrino crosses the star coherently and decoheres only on its way to Earth, so the average is taken after it leaves the star, from the evolution operator across the star.


\subsection{The Hamiltonians}
\label{sec:hamiltonians}

A direct call to {\tt osc\_prob} takes the Hamiltonian as a matrix, or as a function of position that returns one (Sec.~\ref{sec:wrapper_vs_direct}). \magnus\ ships builders for standard and non-standard terms (Sec.~\ref{sec:shipped_hamiltonians}), and any other term can be written by hand (Sec.~\ref{sec:building_new_hamiltonian}).

{\def\arraystretch{1.15}
\begin{table*}[t!]
{\footnotesize
\begin{tabular}{@{}p{4.3cm}p{6.2cm}p{6.2cm}@{}}
 Returns & \multicolumn{2}{@{}l@{}}{Builder, as {\tt hamiltonian\_}$\ldots$} \\
 \hline
 Vacuum term & {\tt 2nu\_vacuum} & {\tt 3nu\_vacuum} \\
  & {\tt 4nu\_vacuum} & {\tt 5nu\_vacuum} \\
 Same, without the $1/E$ & {\tt 2nu\_vacuum\_energy\_independent} & {\tt 3nu\_vacuum\_energy\_independent} \\
  & {\tt 4nu\_vacuum\_energy\_independent} & {\tt 5nu\_vacuum\_energy\_independent} \\
 Vacuum term at a position & {\tt 2nu\_vacuum\_td} & {\tt 3nu\_vacuum\_td} \\
  & {\tt 4nu\_vacuum\_td} & {\tt 5nu\_vacuum\_td} \\
 Both of the above & {\tt 2nu\_vacuum\_energy\_independent\_td} & {\tt 3nu\_vacuum\_energy\_independent\_td} \\
  & {\tt 4nu\_vacuum\_energy\_independent\_td} & {\tt 5nu\_vacuum\_energy\_independent\_td} \\
 $V_{\rm CC}$ times the projector & {\tt 2nu\_matter} & {\tt 3nu\_matter} \\
  & {\tt 4nu\_matter} & {\tt 5nu\_matter} \\
 Same, from a potential function & {\tt 2nu\_matter\_td} & {\tt 3nu\_matter\_td} \\
  & {\tt 4nu\_matter\_td} & {\tt 5nu\_matter\_td} \\
 Non-standard term alone & {\tt 2nu\_nsi} & {\tt 3nu\_nsi} \\
  & {\tt 4nu\_nsi} & {\tt 5nu\_nsi} \\
 Same, from a potential function & {\tt 2nu\_nsi\_td} & {\tt 3nu\_nsi\_td} \\
  & {\tt 4nu\_nsi\_td} & {\tt 5nu\_nsi\_td} \\
 Lorentz-violating term alone & {\tt 2nu\_liv} & {\tt 3nu\_liv} \\
  & {\tt 4nu\_liv} & {\tt 5nu\_liv} \\
 Same, without the energy factor & {\tt 2nu\_liv\_energy\_independent} & {\tt 3nu\_liv\_energy\_independent} \\
  & {\tt 4nu\_liv\_energy\_independent} & {\tt 5nu\_liv\_energy\_independent} \\
 \hline
 Pseudo-Dirac vacuum term & {\tt pseudo\_dirac\_vacuum} &  \\
 Same, without the $1/E$ & {\tt pseudo\_dirac\_vacuum\_energy\_independent} &  \\
 Its matter term & {\tt pseudo\_dirac\_matter} &  \\
\end{tabular}}
\caption{\label{tab:hamiltonians}\textbf{The Hamiltonian builders.} The $43$ Hamiltonian builders of {\tt magnus.hamiltonians}, each named with the prefix {\tt hamiltonian\_}. The ten variants above the table break exist at two, three, four, and five flavors. The three below the break build a pseudo-Dirac spectrum from the active one, and take the pairs of states to split, so their names carry no number of flavors.}
\end{table*}}


\subsubsection{Shipped Hamiltonians}
\label{sec:shipped_hamiltonians}

\tabl{hamiltonians} lists the 43 Hamiltonian builders shipped in {\tt magnus.hamiltonians}: vacuum, matter, non-standard interactions, and Lorentz violation at two to five flavors, plus three for a pseudo-Dirac spectrum.

A Hamiltonian is a sum of terms, as in \equ{h_matter}. Only the vacuum builders return a complete Hamiltonian; every other builder in \tabl{hamiltonians} returns its own term alone. For instance, at three flavors, the matter builder returns $V_{\rm CC}$ times the projector, which is zero except in the electron entry, {\tt [0][0]}. Added to a vacuum term, it gives \equ{h_matter}. Passed to {\tt osc\_prob} on its own, it returns the probabilities as the identity matrix, since a diagonal Hamiltonian does not mix flavors.

Every builder returns a $d \times d$ {\tt NumPy} array, with $d$ the number of flavors, so terms add with {\tt +} and scale with {\tt *}. The builders take their parameters under the same names as the wrappers, so a three-flavor set loaded with {\tt load\_nufit\_params} can be passed to them with {\tt **}. Every physical parameter must be given, since a builder has no default set to fall back on. The one exception is the neutron-to-proton ratio of the four- and five-flavor matter builders, which is 1 by default.

\smallskip

\textit{Vacuum.---}At three flavors, the vacuum Hamiltonian is
\begin{pysnip}
H = ham.hamiltonian_3nu_vacuum(E, **osc)
H.shape                        # (3, 3)
\end{pysnip}
The call returns \equ{h_vacuum} at three flavors, \ie,
\begin{equation}
 \label{equ:h_vacuum_3nu}
 \mathbb{H}_{3\nu}^{\rm vac}(E)
 =
 \frac{1}{2E}\,
 \mathbb{R}\,
 {\rm diag}\!\left(0,\, \Delta m^2_{21},\, \Delta m^2_{31}\right)
 \mathbb{R}^\dagger \;,
\end{equation}
with $\mathbb{R}$ the PMNS matrix of \equ{mixing_matrix}, built from $\theta_{12}$, $\theta_{23}$, $\theta_{13}$, and $\delta_{\rm CP}$. At two flavors, the builder takes {\tt sth} and {\tt Dm2}, and returns \equ{h_vacuum} with $\mathbb{R} = \mathbb{R}_{12}$ and $\mathbb{M}^2 = {\rm diag}(0, \Delta m^2)$.

At four and five flavors, the builder also takes the mixing angles, phases, and mass-squared splitting of each sterile state, under the same names as in the wrappers:
\begin{pysnip}
st4 = dict(s14=0.1, s24=0.1, s34=0.1,
           d14=0.0, d24=0.0, D41=1.0)
H = ham.hamiltonian_4nu_vacuum(
    E, **osc, **st4)
H.shape                        # (4, 4)

st5 = dict(st4, s15=0.1, s25=0.1, s35=0.1,
           d15=0.0, d35=0.0, D51=1.7)
H = ham.hamiltonian_5nu_vacuum(
    E, **osc, **st5)
H.shape                        # (5, 5)
\end{pysnip}
The two calls return \equ{h_vacuum} at four and five flavors, \ie,
\begin{eqnarray}
 \label{equ:h_vacuum_45nu}
 \mathbb{H}^{\rm vac}(E)
 &=&
 \frac{1}{2E}\,
 \mathbb{R}\,
 \mathbb{M}^2\,
 \mathbb{R}^\dagger \;,
 \nonumber \\
 \mathbb{M}^2_{3+1}
 &=&
 {\rm diag}\!\left(0,\, \Delta m^2_{21},\, \Delta m^2_{31},\, \Delta m^2_{41}\right) ,
 \nonumber \\
 \mathbb{M}^2_{3+2}
 &=&
 {\rm diag}\!\left(0,\, \Delta m^2_{21},\, \Delta m^2_{31},\, \Delta m^2_{41},\, \Delta m^2_{51}\right) ,
\end{eqnarray}
with $\mathbb{R}$ given by \equ{mixing_matrix} at $n = 4$ and 5, respectively.

A builder marked {\tt \_energy\_independent} returns the same matrix without its $1/E$ factor. At one energy, the two builders are interchangeable. Across a scan in energy, the energy-independent one is faster, because the mixing rotation does not depend on the energy and is then computed only once. Over the 200 energies {\tt Es} of Listing~\ref{lst:constant}, rebuilding the Hamiltonian at each energy takes about 4~ms, and building it once and dividing it by each energy takes 0.2~ms:
\begin{pysnip}
# Interchangeable at one energy
H1 = ham.hamiltonian_3nu_vacuum(E, **osc)
H0 = ham.\
 hamiltonian_3nu_vacuum_energy_independent(
    **osc)
H1 - H0/E                     # All zeros

# Over 200 energies: 3.7 ms
Hs = [ham.hamiltonian_3nu_vacuum(e, **osc)
      for e in Es]
# ... and 0.18 ms
Hs = [H0/e for e in Es]
\end{pysnip}

\smallskip

\textit{Matter.---}At three flavors, the matter term is
\begin{pysnip}
V = gd.VCC_EARTH_CRUST         # 1.145e-13 eV
H = ham.hamiltonian_3nu_matter(V)
H.shape                        # (3, 3)
\end{pysnip}
The call returns the matter term of \equ{h_matter} at three flavors, \ie,
\begin{equation}
 \label{equ:h_matter_3nu}
 V_{\rm CC}\, \mathbb{P}
 =
 V_{\rm CC}\,
 {\rm diag}\!\left(1,\, 0,\, 0\right) \;,
\end{equation}
the charged-current potential acting on the electron flavor alone. ({\tt VCC\_EARTH\_CRUST} is $V_{\rm CC}$ at density 3~g~cm$^{-3}$ with $Y_e = 0.5$.) At two flavors, the call is the same; the projector is ${\rm diag}(1, 0)$.

At four and five flavors, the projector also has an entry for each sterile flavor, $r/2$, where $r$ is the neutron-to-proton ratio of the medium (Sec.~\ref{sec:recap}). The builder takes $r$ as a second parameter, 1 by default. In neutral matter, $r = (1 - Y_e)/Y_e$ (Sec.~\ref{sec:constant_density}), so $r$ follows from the electron fraction; for instance, from that of the core of the Earth, which \magnus\ stores in {\tt magnus.earth}:
\begin{pysnip}
import magnus.earth as earth

Ye = earth.Y_E_CORE_PREM       # 0.4656
r = (1.0 - Ye)/Ye              # 1.15

H = ham.hamiltonian_4nu_matter(V, r)
H.shape                        # (4, 4)

H = ham.hamiltonian_5nu_matter(V, r)
H.shape                        # (5, 5)
\end{pysnip}
The builder takes the potential and the ratio independently. The two calls return
\begin{eqnarray}
 \label{equ:h_matter_45nu}
 V_{\rm CC}\, \mathbb{P}_{3+1}
 &=&
 V_{\rm CC}\,
 {\rm diag}\!\left(1,\, 0,\, 0,\, r/2\right) \;,
 \nonumber \\
 V_{\rm CC}\, \mathbb{P}_{3+2}
 &=&
 V_{\rm CC}\,
 {\rm diag}\!\left(1,\, 0,\, 0,\, r/2,\, r/2\right) \;.
\end{eqnarray}

Four of the ten variants in \tabl{hamiltonians} end in {\tt \_td} (for \textit{time-dependent}): the two vacuum forms, {\tt matter}, and {\tt nsi}. Each takes the position along the path, $l$, as its first argument, just as the function of position passed to {\tt osc\_prob} does. The vacuum ones ignore it and return the same matrix at every position. {\tt matter\_td} and {\tt nsi\_td} take the potential as a function of position, {\tt f}, rather than as a number, and evaluate it at $l$; \eg, {\tt matter\_td(l, f)} returns the same matrix as {\tt matter(f(l))}.

\smallskip

\textit{Non-standard interactions.---}At three flavors, the NSI term is
\begin{pysnip}
H = ham.hamiltonian_3nu_nsi(
    V, eps_ee, eps_em, eps_et,
    eps_mm, eps_mt, eps_tt)
H.shape                        # (3, 3)
\end{pysnip}
The call returns the charged-current potential times the matrix of NSI couplings, \ie,
\begin{equation}
 \label{equ:h_nsi_3nu}
 V_{\rm CC}
 \begin{pmatrix}
  \varepsilon_{ee} & \varepsilon_{e\mu} & \varepsilon_{e\tau} \\
  \varepsilon_{e\mu}^\ast & \varepsilon_{\mu\mu} & \varepsilon_{\mu\tau} \\
  \varepsilon_{e\tau}^\ast & \varepsilon_{\mu\tau}^\ast & \varepsilon_{\tau\tau}
 \end{pmatrix} \;.
\end{equation}
The six couplings passed are the diagonal entries and those above it. The builder fills the entries below the diagonal with their complex conjugates, so the matrix is Hermitian even when the off-diagonal couplings are complex. At two flavors, the builder takes two couplings, {\tt eps\_aa} and {\tt eps\_ab}, and returns
\begin{equation}
 \label{equ:h_nsi_2nu}
 V_{\rm CC}
 \begin{pmatrix}
  \varepsilon_{aa} & \varepsilon_{ab} \\
  \varepsilon_{ab}^\ast & 0
 \end{pmatrix} \;,
\end{equation}
where $a$ and $b$ are the two flavors, $e$ and $\mu$ by default (Sec.~\ref{sec:two_flavors}). The lower diagonal entry is zero because only the difference between the two diagonal entries affects probabilities. Adding the same amount to both shifts every energy level of the Hamiltonian equally, which changes only an overall phase of the evolution, so $\varepsilon_{aa}$ stands for the difference $\varepsilon_{aa} - \varepsilon_{bb}$.

At four and five flavors, the matrix of couplings also has rows and columns for the sterile flavors, and the builder takes the entries on and above its diagonal, ten and fifteen of them:
\begin{pysnip}
H = ham.hamiltonian_4nu_nsi(
    V, eps_ee, eps_em, eps_et, eps_es,
    eps_mm, eps_mt, eps_ms, eps_tt,
    eps_ts, eps_ss)
H.shape                        # (4, 4)

H = ham.hamiltonian_5nu_nsi(
    V, eps_ee, eps_em, eps_et, eps_es1,
    eps_es2, eps_mm, eps_mt, eps_ms1,
    eps_ms2, eps_tt, eps_ts1, eps_ts2,
    eps_s1s1, eps_s1s2, eps_s2s2)
H.shape                        # (5, 5)
\end{pysnip}
The two calls return the matrix whose entries are
\begin{equation}
 \label{equ:h_nsi_45nu}
 \left[\mathbb{H}^{\rm NSI}\right]_{\alpha\beta}
 =
 V_{\rm CC}\, \varepsilon_{\alpha\beta} \;,
 \qquad
 \varepsilon_{\beta\alpha} = \varepsilon_{\alpha\beta}^\ast \;,
\end{equation}
with $\alpha, \beta \in \{e, \mu, \tau, s\}$ at four flavors and $\alpha, \beta \in \{e, \mu, \tau, s_1, s_2\}$ at five; \equ{h_nsi_3nu} is the same expression at three flavors.

\smallskip

\textit{Lorentz-invariance violation.---}At three flavors, the Lorentz-invariance-violating (LIV) term is
\begin{pysnip}
H = ham.hamiltonian_3nu_liv(
    E, sxi12, sxi23, sxi13, dxiCP,
    b1, b2, b3, Lambda, n_liv)
H.shape                        # (3, 3)
\end{pysnip}
The call returns
\begin{equation}
 \label{equ:h_liv_3nu}
 \mathbb{H}^{\rm LIV}(E)
 =
 \left(\frac{E}{\Lambda}\right)^{n_{\rm LIV}}
 \mathbb{V}_\xi\,
 {\rm diag}\!\left(b_1,\, b_2,\, b_3\right)
 \mathbb{V}_\xi^\dagger \;.
\end{equation}
Here, $b_1$, $b_2$, and $b_3$ are the eigenvalues of the LIV operator, in eV. The unitary matrix $\mathbb{V}_\xi$ rotates them into the flavor basis; it is built from three angles, $\xi_{12}$, $\xi_{23}$, and $\xi_{13}$, and one phase, $\delta_\xi$, in the same way as the PMNS matrix in \equ{mixing_matrix}. The prefactor sets how the term grows with energy: $\Lambda$ is the energy scale of the operator, in eV, and $n_{\rm LIV}$ is its dimension minus three, so that $n_{\rm LIV} = 0$ gives a term that does not depend on the energy.

The angles $\xi_{ij}$ and the phase $\delta_\xi$ are parameters of the LIV operator alone, independent of the PMNS angles and phase. They fix the basis in which the operator is diagonal, which need be neither the flavor basis nor the mass basis. For instance, with every $\xi_{ij}$ and $\delta_\xi$ set to zero, $\mathbb{V}_\xi$ is the identity and the operator is diagonal in the flavor basis; with them set equal to the PMNS values, it is diagonal in the mass basis.

At four and five flavors, the operator has one eigenvalue per flavor, and $\mathbb{V}_\xi$ is built as $\mathbb{R}$ in \equ{mixing_matrix} at $n = 4$ and $n = 5$, with angles $\xi_{ij}$ in place of $\theta_{ij}$ and phases $\delta_{\xi, ij}$ in place of $\delta_{ij}$:
\begin{pysnip}
H = ham.hamiltonian_4nu_liv(
    E, sxi12, sxi23, sxi13, dxiCP,
    sxi14, dxi14, sxi24, dxi24, sxi34,
    b1, b2, b3, b4, Lambda, n_liv)
H.shape                        # (4, 4)

H = ham.hamiltonian_5nu_liv(
    E, sxi12, sxi23, sxi13, dxiCP,
    sxi14, dxi14, sxi15, dxi15, sxi24,
    dxi24, sxi25, sxi34, sxi35, dxi35,
    b1, b2, b3, b4, b5, Lambda, n_liv)
H.shape                        # (5, 5)
\end{pysnip}
The two calls return
\begin{eqnarray}
 \label{equ:h_liv_45nu}
 \mathbb{H}_{3+1}^{\rm LIV}(E)
 &=&
 \left(\frac{E}{\Lambda}\right)^{n_{\rm LIV}}
 \mathbb{V}_\xi\,
 {\rm diag}\!\left(b_1,\, b_2,\, b_3,\, b_4\right)
 \mathbb{V}_\xi^\dagger \;,
 \nonumber \\
 \mathbb{H}_{3+2}^{\rm LIV}(E)
 &=&
 \left(\frac{E}{\Lambda}\right)^{n_{\rm LIV}}
 \mathbb{V}_\xi\,
 {\rm diag}\!\left(b_1,\, b_2,\, b_3,\, b_4,\, b_5\right)
 \mathbb{V}_\xi^\dagger \;.
\end{eqnarray}
At two flavors, $\mathbb{V}_\xi$ is a single rotation, by the angle $\xi$, and the operator has two eigenvalues.

The builders marked {\tt \_energy\_independent} return these terms without the factor $E^{n_{\rm LIV}}$, which the user then multiplies back in at each energy, as for the vacuum term.

\smallskip

\textit{Pseudo-Dirac.---}A pseudo-Dirac neutrino is a mass state split into a nearly degenerate pair, one active and the other sterile. The pseudo-Dirac builders take the mixing matrix and the mass-squared values of the active sector, and the mass states to split:
\begin{pysnip}
R = ham.pmns_mixing_matrix(
    osc['s12'], osc['s23'], osc['s13'],
    osc['dCP'])
H = ham.hamiltonian_pseudo_dirac_vacuum(
    E, R, [0.0, osc['D21'], osc['D31']],
    {1: 1.0e-18})
H.shape                        # (4, 4)
\end{pysnip}
The last argument lists the mass states to split, each with the splitting of its pair, $\delta m^2_j$, in eV$^2$. The states are counted from 0, so \verb|{1: 1.0e-18}| splits $\nu_2$ by $\delta m^2_2 = 10^{-18}$~eV$^2$. States not listed stay unsplit; with an empty mapping, the builder returns the three-flavor vacuum Hamiltonian, to round-off.

The call returns \equ{h_vacuum} for the enlarged set of states, \ie,
\begin{equation}
 \label{equ:h_pseudo_dirac}
 \mathbb{H}^{\rm vac}_{\rm PD}(E)
 =
 \frac{1}{2E}\,
 \mathbb{R}_{\rm PD}\,
 \mathbb{M}^2_{\rm PD}\,
 \mathbb{R}_{\rm PD}^\dagger \;.
\end{equation}
Each split state $\nu_j$ is replaced by two mass states, $(\nu_j \pm s_j)/\sqrt{2}$, where $s_j$ is its sterile partner, with masses squared $m_j^2$ and $m_j^2 + \delta m^2_j$. For the example above, which splits $\nu_2$, the rows of $\mathbb{R}_{\rm PD}$ are the flavors $e$, $\mu$, $\tau$, and $s_2$, and its columns are $\nu_1$, the two members of the pair, and $\nu_3$:
\begin{equation}
 \label{equ:mixing_pseudo_dirac}
 \mathbb{R}_{\rm PD}
 =
 \begin{pmatrix}
  \mathbb{R}_{e1} & \mathbb{R}_{e2}/\sqrt{2} & \mathbb{R}_{e2}/\sqrt{2} & \mathbb{R}_{e3} \\
  \mathbb{R}_{\mu 1} & \mathbb{R}_{\mu 2}/\sqrt{2} & \mathbb{R}_{\mu 2}/\sqrt{2} & \mathbb{R}_{\mu 3} \\
  \mathbb{R}_{\tau 1} & \mathbb{R}_{\tau 2}/\sqrt{2} & \mathbb{R}_{\tau 2}/\sqrt{2} & \mathbb{R}_{\tau 3} \\
  0 & 1/\sqrt{2} & -1/\sqrt{2} & 0
 \end{pmatrix} ,
\end{equation}
with $\mathbb{R}$ the PMNS matrix, and
\begin{equation}
 \label{equ:mass_pseudo_dirac}
 \mathbb{M}^2_{\rm PD}
 =
 {\rm diag}\!\left(0,\, \Delta m^2_{21},\, \Delta m^2_{21} + \delta m^2_2,\, \Delta m^2_{31}\right) .
\end{equation}
In general, there is one state per active state, plus one per split state. Both matrices are available on their own, from {\tt pseudo\_dirac\_mixing\_matrix} and {\tt pseudo\_dirac\_mass\_squared}.

The matter builder, {\tt hamiltonian\_pseudo\_dirac\_matter}, takes the potential, the number of active states, the same mapping of split states, and the neutron-to-proton ratio, $r$, 1 by default. It returns $V_{\rm CC}$ times the projector in the same enlarged flavor basis. For the example above, which splits $\nu_2$,
\begin{equation}
 \label{equ:matter_pseudo_dirac}
 V_{\rm CC}\, \mathbb{P}_{\rm PD}
 =
 V_{\rm CC}\,
 {\rm diag}\!\left(1,\, 0,\, 0,\, r/2\right) \;,
\end{equation}
with rows $e$, $\mu$, $\tau$, and $s_2$. Since each partner $s_j$ is sterile, its entry is $r/2$, as for the sterile states of \equ{h_matter_45nu}. Split states other than $\nu_2$ add one such entry each. For probabilities in vacuum, {\tt osc\_prob\_pseudo\_dirac\_vacuum} builds this Hamiltonian from the mapping of split states and the standard parameters, or a mixing matrix and masses. With {\tt average=True}, it evaluates the phase average of Sec.~\ref{sec:averaged} from $\delta m^2_j$ directly (through {\tt avgprob.phase\_averaged\_probabilities\_pseudo\_dirac}), and stays accurate to $10^{-16}$ at splittings where the generic route, built on eigenvalue differences, is off by $10^{-4}$ or more.


\subsubsection{Building a new Hamiltonian from scratch}
\label{sec:building_new_hamiltonian}

There are two ways to build a Hamiltonian that \magnus\ does not ship. The first is to add shipped terms, as Sec.~\ref{sec:wrapper_vs_direct} did with a vacuum, a matter, and a non-standard term. With a {\tt \_td} builder, a term can also vary along the path. For instance, with {\tt H\_vac} the vacuum term of that section, a matter potential that grows linearly along the baseline is
\begin{pysnip}
VCC_func = lambda l: V*(1.0 + l/L)

def H_func(l):
    H_mat = ham.hamiltonian_3nu_matter_td(
        l, VCC_func)
    return H_vac + H_mat

P = oscprob.osc_prob(H_func, 0.0, L,
                     rtol=1e-8, atol=1e-8)
# Pme = 0.021309
\end{pysnip}

The second way is to write the matrix directly. {\tt osc\_prob} requires only a function that returns a Hermitian matrix, of any size, so an entirely new term reaches the solver in the same way as a shipped one. For instance, an off-diagonal term between $\nu_e$ and $\nu_\tau$ that grows along the path is
\begin{pysnip}
def H_own(l):
    H = H_vac.copy()
    H[0, 2] += 1.0e-13*(l/L)
    H[2, 0] += 1.0e-13*(l/L)
    return H

P = oscprob.osc_prob(H_own, 0.0, L,
                     rtol=1e-8, atol=1e-8)
# Pme = 0.004865
\end{pysnip}
The Hermiticity of the matrix is the user's to guarantee, since it is what keeps the evolution unitary (Sec.~\ref{sec:applicability}). \magnus\ checks the matrix once, at the start of the path, and raises an error if it is not Hermitian. It does not check later positions, and a matrix that stops being Hermitian along the path returns probabilities that do not add up to one. Every entry of the matrix is in eV, and every position in eV$^{-1}$ (Sec.~\ref{sec:recap}). Section~\ref{sec:lri_sun} shows a complete example.

A custom Hamiltonian may have more than five flavors. The wrappers of \tabl{wrappers}, the builders of \tabl{hamiltonians}, and the compiled exponential kernels of Sec.~\ref{sec:exponential} stop at five. Above five, the user supplies the matrix, either to {\tt osc\_prob} or, as a vacuum term, to a scenario function through {\tt h\_vac\_energy\_indep} (Sec.~\ref{sec:wrapper_vs_direct}), and the exponential of each slab goes through a general Hermitian eigendecomposition, {\tt numpy.linalg.eigh}. Each slab then costs more: on one exponential density profile, at the default tolerance, a probability takes up to two and a half times as long at six flavors as at five, and two to five and a half times as long at eight, depending on the energy. The probabilities remain unitary to within $10^{-13}$.


\subsection{Pre-packaged plotting routines}
\label{sec:plotting}

The {\tt plotting} module of \magnus\ draws several of the standard figures of the field, each with a single call; \tabl{plotting} lists them, and the sections that follow use them. Drawn directly in {\tt matplotlib}, the same figure takes 25--40 lines of grid specification, tick locators, legend keywords, and {\tt savefig} calls.

{\def\arraystretch{1.2}
\begin{table}[t!]
{\footnotesize
\begin{tabular}{@{}p{3.5cm}p{5.0cm}@{}}
 Function ({\tt plot\_}) & Draws \\
 \hline
 {\tt probability\_vs\_energy} & Probability against energy \\
 {\tt probability\_vs\_baseline} & Probability against baseline \\
 {\tt curves} & Any swept variable, optional error subpanel \\
 {\tt curves\_stacked} & One panel per case, over a shared abscissa \\
 {\tt oscillogram} & Probability over energy and arrival direction \\
 {\tt biprobability} & Antineutrino against neutrino probability \\
 {\tt probability\_with\_average} & Oscillating probability, with its average \\
 {\tt probability\_with\_profile} & Density profile on top, probabilities along it below \\
\end{tabular}}
\caption{\label{tab:plotting}\textbf{The plotting functions.} The plotting functions of {\tt magnus.plotting}, each named with the prefix {\tt plot\_}. Each returns the {\tt matplotlib} figure it built and its axes. See Sec.~\ref{sec:plotting} for details.}
\end{table}}

Most of these functions take the abscissa first, then a list of curves, each a dictionary with its data, {\tt y}, and its {\tt label}. For instance, the probability of $\nu_\mu \to \nu_e$ against energy, in vacuum and at constant density, with {\tt P\_scan} the 200-energy scan of Listing~\ref{lst:constant}, is drawn by
\begin{pysnip}
import magnus.plotting as plotting

P_vac = oscprob.osc_prob_3nu_vacuum(Es, L)
fig, ax = plotting.\
    plot_probability_vs_energy(
    Es/gd.UNIT_GEV,
    [dict(y=P_vac[:, gd.NUMU, gd.NUE],
          label='Vacuum'),
     dict(y=P_scan[:, gd.NUMU, gd.NUE],
          label='Matter')],
    nu_i=gd.NUMU, nu_f=gd.NUE)
\end{pysnip}
The flavor indices {\tt nu\_i} and {\tt nu\_f} only label the vertical axis, here $P_{\nu_\mu \to \nu_e}$; the curves are the data passed. The API documentation~\cite{MagnusDocs} gives the arguments of every {\tt plotting} function.

Each figure comes out in a preset style, so a set of figures is consistent without any styling passed. Since each function returns the {\tt matplotlib} figure and its axes, the style can be changed afterwards. For instance, the figure above is resized, given a grid, and saved by
\begin{pysnip}
fig.set_size_inches(6, 3)
ax.grid(True)
fig.savefig('numu_to_nue.pdf')
\end{pysnip}


\subsection{Probability vs.~energy and distance}
\label{sec:prob_vs}

Two plots appear in almost every study of neutrino oscillations: the probability against the distance traveled, at a fixed energy, and against the neutrino energy, at a fixed distance. \magnus\ computes each with a single call: given a list of distances and one energy, or a list of energies and one distance, it returns the probabilities at every entry of the list.

\begin{figure*}[t!]
 \centering
 \includegraphics[width=\textwidth]{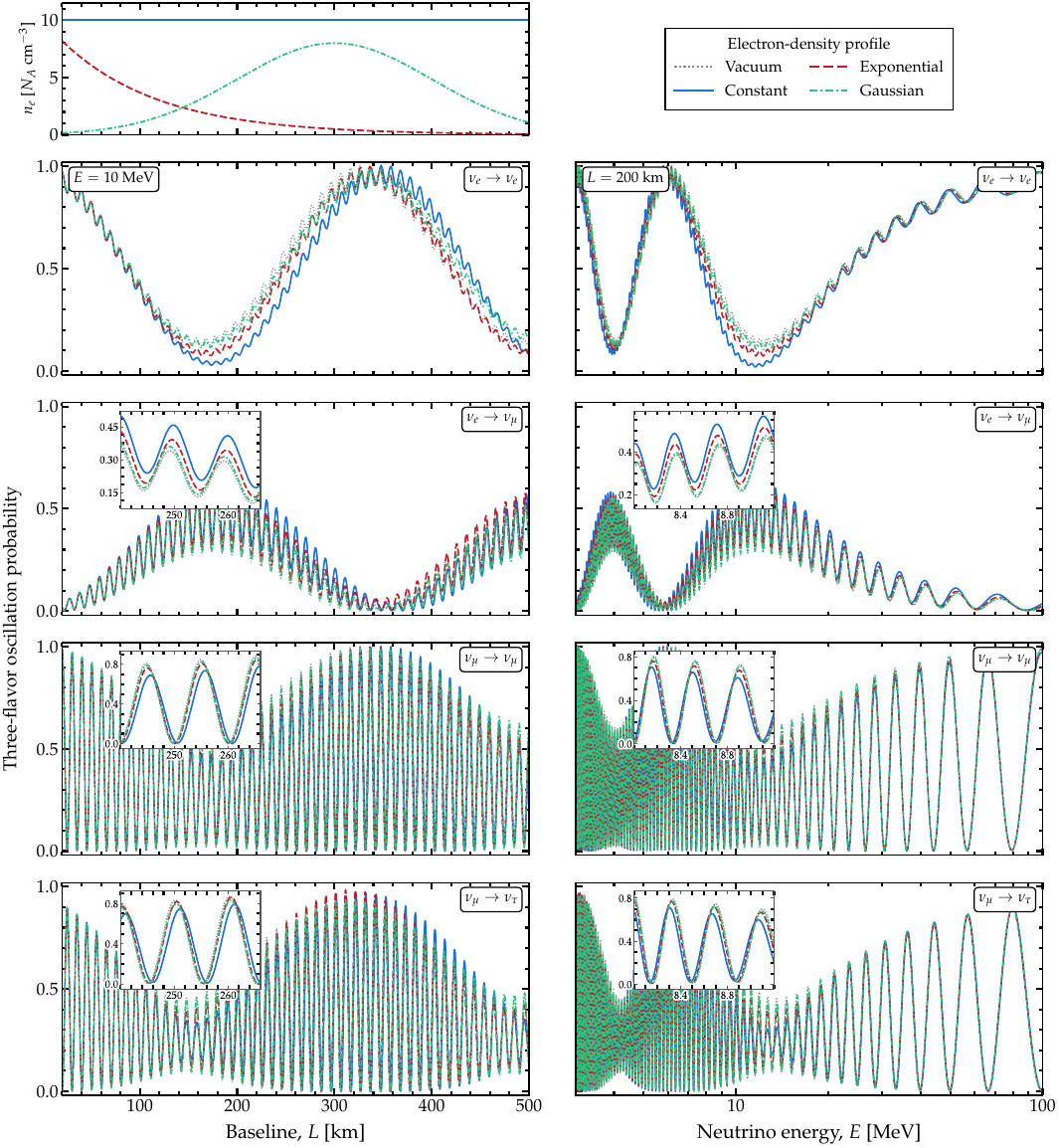}
 \caption{\label{fig:prob_vs}\textbf{Probability against distance and against energy.} Oscillation probabilities of a neutrino that travels through matter, against the distance traveled ({\it left}) and against the neutrino energy ({\it right}). Three matter profiles are drawn, plus the vacuum as a reference. {\it Top left}: the profiles, as the electron number density at each point along the path. The constant profile holds 10~$N_A$ electrons per cm$^3$, with $N_A$ Avogadro's number. The exponential profile starts at that value and halves every 69~km. The Gaussian profile peaks at 8~$N_A$ electrons per cm$^3$, 300~km from the start; its width is 100~km. {\it Rows}: the survival of a $\nu_e$, its conversion into a $\nu_\mu$, the survival of a $\nu_\mu$, and its conversion into a $\nu_\tau$. In the left column the energy is 10~MeV; in the right column the distance is 200~km. In the lower three rows, an inset enlarges a window where the four curves separate. Listing~\ref{lst:prob_vs} computes the curves. See notebook \#28 and Sec.~\ref{sec:prob_vs} for details.}
\end{figure*}

Figure~\ref{fig:prob_vs} shows both plots at three flavors, in vacuum and for three illustrative electron-density profiles: constant, exponentially falling, and a Gaussian bump near the middle of the path. Of the three, the constant profile moves the probabilities furthest from the vacuum ones. Along the path (left column), the exponential profile stays close to the constant one near the start, where its density is still high, then moves back toward the vacuum curves as its density falls. The Gaussian profile does the reverse: it stays close to the vacuum curves until the bump, then moves away from them.

Listing~\ref{lst:prob_vs} computes the probabilities. A constant profile is passed as a single number, and a varying one as a function of the distance. Each column takes four calls, one per profile and one in vacuum, and each call receives the list of distances, or of energies. For constant and exponential profiles, \magnus\ also has wrappers---not used in Listing~\ref{lst:prob_vs}---which take the density and, for the exponential, its scale length, instead of a profile function. They return the same probabilities as the listing; \eg, for $\nu_e \to \nu_e$ at 500~km, the last distance of the left column:
\begin{pysnip}
P = oscprob.osc_prob_3nu_matter_exp_density(
 10.0*MEV, 500.0*KM, L0=0.0,
 rho_central=10.0*NA_CM3,
 l_scale=100.0*KM, **osc,
 nu_i=gd.NUE, nu_f=gd.NUE, **KW)
# 0.082635, as in the listing

P = oscprob.\
 osc_prob_3nu_matter_constant_density(
 10.0*MEV, 500.0*KM, 10.0*NA_CM3, **osc,
 nu_i=gd.NUE, nu_f=gd.NUE, **KW)
# 0.071325, as in the listing
\end{pysnip}

\begin{lstlisting}[language=Python, caption={\textbf{Probability against distance and against energy.} The curves of \figu{prob_vs}: the probability at 5\,000 distances for one energy, then at 3\,000 energies for one distance, for three matter profiles and for vacuum. Each scan is a single call. The constant profile is a number; the two varying ones are functions of the distance traveled. See Sec.~\ref{sec:prob_vs} for details.}, label={lst:prob_vs}, float=tbp]
import numpy as np
import magnus.oscprob as oscprob
import magnus.globaldefs as gd

osc = gd.load_nufit_params('NuFIT 6.1')
KM, MEV = gd.UNIT_KM, gd.UNIT_MEV
# One mole of electrons per cubic centimeter, in the
# natural units that magnus works in
NA_CM3 = gd.N_AV/gd.CONV_CM_TO_INV_EV**3

# Give a constant profile as a single number, and
# a varying one as a function of the distance
# traveled, returning the electron number density
# there.  The distance arrives in eV^-1, like
# every length in magnus
ne_constant = 10.0*NA_CM3

def ne_exponential(l):
    return 10.0*NA_CM3*np.exp(-l/(100.0*KM))

def ne_gaussian(l):
    return 8.0*NA_CM3*np.exp(
     -(l - 300.0*KM)**2/(2*(100.0*KM)**2))

# The first keyword says that the functions return
# an electron number density, not a mass density
KW = dict(density_is_of_number_of_electrons=True,
          rtol=1e-6, atol=1e-6)
PROFILES = (('constant', ne_constant),
            ('exponential', ne_exponential),
            ('gaussian', ne_gaussian))

# Left column: 5,000 distances at one energy.  The
# list of distances goes where a single one would
L = np.linspace(20.0, 500.0, 5000)*KM
P_vs_L = {name: oscprob.osc_prob_matter_std_potential(
 3, ne, 10.0*MEV, L, osc, L0=0.0, **KW)
 for name, ne in PROFILES}

# Right column: 3,000 energies at one distance
E = np.logspace(np.log10(3.0), 2.0, 3000)*MEV
P_vs_E = {name: oscprob.osc_prob_matter_std_potential(
 3, ne, E, 200.0*KM, osc, L0=0.0, **KW)
 for name, ne in PROFILES}

# The vacuum curves, for reference
P_vac_L = oscprob.osc_prob_3nu_vacuum(
 10.0*MEV, L, **osc)
P_vac_E = oscprob.osc_prob_3nu_vacuum(
 E, 200.0*KM, **osc)
\end{lstlisting}

In Listing~\ref{lst:prob_vs}, \magnus\ computes the two columns of \figu{prob_vs} with different engines. The left column is a scan over distance at one energy. Through a varying profile, \magnus\ walks the profile once and reads off the probability at each distance; this is the cumulative engine of Sec.~\ref{sec:engines}, and it computes the whole left column in a few hundredths of a second. The right column is a scan over energy at one distance, which cannot reuse a single walk, since the Hamiltonian changes with the energy. Instead, the energy-batched engine of Sec.~\ref{sec:engines} samples the potential once and evolves all the energies together. It applies to any Hamiltonian made of an energy-dependent part plus the potential times a fixed matrix, such as the standard, non-standard-interaction, and Lorentz-violating Hamiltonians, and it computes the whole right column in under a second. \magnus\ picks each engine by itself (Sec.~\ref{sec:engines}); {\tt strategy\_info} reports which one answered:
\begin{pysnip}
info = {}
P = oscprob.osc_prob_matter_std_potential(
 3, ne_exponential, E, 200.0*KM, osc,
 L0=0.0, strategy_info=info, **KW)
info['engine']   # 'separable'
\end{pysnip}
Both columns agree to within $2 \cdot 10^{-8}$ with the same calculation at the tighter tolerances {\tt rtol=1e-8} and {\tt atol=1e-10}.


\subsection{Layered matter}
\label{sec:arrangement}

\begin{figure}[t!]
 \centering
 \includegraphics[width=\columnwidth]{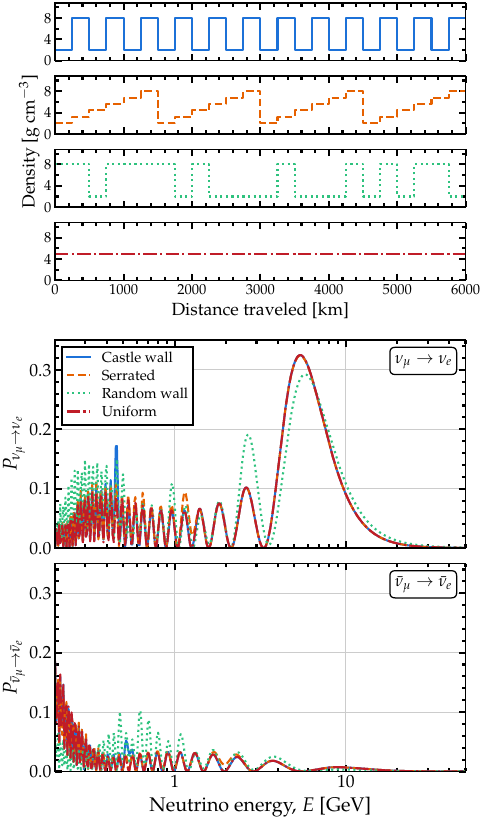}
 \caption{\label{fig:arrangement}\textbf{Same mean density, different probabilities.} Appearance probability along four piecewise-constant density profiles, for neutrinos ({\it middle}) and for antineutrinos ({\it bottom}), computed with \magnus\ at three flavors. {\it Top}: the four profiles, each made of twenty-four slabs of 250~km with a mean density of 5~g~cm$^{-3}$. The castle wall alternates slabs of 2 and 8~g~cm$^{-3}$; the random wall holds the same slabs in a random order; the serrated profile repeats four times a ramp from 2 to 8~g~cm$^{-3}$ in six steps; the uniform profile is 5~g~cm$^{-3}$ throughout. The slab edges are declared, so each curve is exact: it is the product of the twenty-four slab exponentials. Listing~\ref{lst:arrangement} computes the castle-wall curves. See notebook \#28 and Sec.~\ref{sec:arrangement} for details.}
\end{figure}

Figure~\ref{fig:arrangement} shows the appearance probability along four piecewise-constant density profiles, for neutrinos and for antineutrinos. All four are made of twenty-four slabs of 250~km and have the same mean density, 5~g~cm$^{-3}$; they differ only in how the density is distributed along the path. Yet the probabilities differ: against the uniform profile, the random wall departs by up to 0.10 in $P_{\nu_\mu \to \nu_e}$ and 0.12 in $P_{\bar{\nu}_\mu \to \bar{\nu}_e}$.

The largest feature of the neutrino panel is a broad peak near 5.4~GeV, enhanced by the matter resonance. At these energies, the oscillation length in matter is thousands of kilometers, many slabs long, so the neutrino responds mainly to the mean density: the castle wall, the serrated profile, and the uniform profile peak at the same energy and height, and the random wall only slightly later and lower. Antineutrinos have no such peak: for them the matter potential has the opposite sign, and, in the normal mass ordering used here, there is no resonance.

The castle wall adds a narrow peak at 0.45~GeV, where its $P_{\nu_\mu \to \nu_e} = 0.18$, more than twice that of the uniform profile. Its densities alternate every 250~km, so the profile repeats every 500~km, close to the oscillation lengths in matter at 0.45~GeV, which are 460 to 515~km in slabs of either density. With the period of the profile matched to the oscillation length, the effect of each pair of slabs adds to that of the previous pair instead of averaging away. This is the parametric resonance of Sec.~\ref{sec:order_is_the_slab_product}\ \cite{Akhmedov:1988kd, Krastev:1989ix}, on the castle-wall profile that \Ref\ \cite{Akhmedov:1998ui} solved in closed form. Antineutrinos show the same peak at 0.52~GeV, a third as tall: the sign of the matter potential changes the oscillation length in matter, and with it the energy at which that length matches the period of the profile.

Listing~\ref{lst:arrangement} computes the castle-wall curves. The two calls differ only in {\tt nubar}, used for the first time in a listing: it conjugates the mixing matrix and reverses the sign of the potential in one step (Sec.~\ref{sec:recap}). The slab edges are passed through {\tt t\_breakpoints}, so no slab straddles a density jump. The Hamiltonian is constant inside every slab, the Magnus expansion terminates at its first term, and the result is exact: it equals the product of the 24 exponentials, to round-off. The eight curves of the figure (four profiles, for neutrinos and antineutrinos) take 0.08~s, since each call goes to the energy-batched engine of Sec.~\ref{sec:engines}, which samples the profile once for all energies.

\begin{lstlisting}[language=Python, caption={\textbf{Neutrinos and antineutrinos through a castle wall.} The castle-wall curves of \figu{arrangement}. The slab edges are passed through {\tt t\_breakpoints}, so that no slab of the expansion straddles a jump; within each slab the Hamiltonian is constant, and the result is the exact product of the slab exponentials. The two calls differ only in {\tt nubar}. The other three profiles of the figure differ from this one only in the array {\tt rho}. See Sec.~\ref{sec:arrangement} for details.}, label={lst:arrangement}, float=tbp]
import numpy as np
import magnus.oscprob as oscprob
import magnus.globaldefs as gd

osc = gd.load_nufit_params('NuFIT 6.1')
# 24 slabs of 250 km, alternating 2 and 8 g/cm^3
n, width = 24, 250.0*gd.UNIT_KM
rho = np.where(np.arange(n) % 2 == 0, 2.0, 8.0)
edges = np.arange(n + 1)*width

def castle_wall(l):
    """The density at position l, in g/cm^3."""
    k = np.searchsorted(edges, l, side='right') - 1
    return rho[np.clip(k, 0, n - 1)]

E = np.logspace(-0.7, 1.7, 400)*gd.UNIT_GEV
# The slab edges are declared, so no slab straddles a
# jump; within each, one term of the expansion is exact
P, Pbar = [oscprob.osc_prob_matter_std_potential(
    3, castle_wall, E, n*width, osc,
    t_breakpoints=edges[1:-1], nubar=nubar,
    nu_i=gd.NUMU, nu_f=gd.NUE,
    density_matter_is_in_g_per_cm3=True,
    rtol=1.0e-8, atol=1.0e-10)
    for nubar in (False, True)]
\end{lstlisting}


\subsection{Bi-probability plot}
\label{sec:biprobability}

\begin{figure}[t!]
 \centering
 \includegraphics[width=\columnwidth]{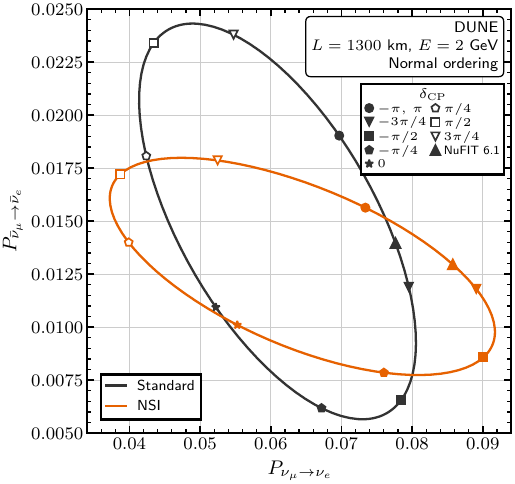}
 \caption{\label{fig:biprobability}\textbf{Bi-probability plot at the DUNE baseline.} Antineutrino against neutrino appearance probability, as $\delta_{\rm CP}$ varies, at 2~GeV, over the 1300-km baseline of DUNE, through matter of constant density 3~g~cm$^{-3}$. The two curves are for standard oscillations and for oscillations with non-standard interactions (NSI), with the couplings of \figu{bsm}: $\varepsilon_{ee} = 0.10$, $\varepsilon_{e\mu} = 0.05$, and $\varepsilon_{\mu\tau} = 0.03$. Markers show selected values of $\delta_{\rm CP}$ and the NuFIT 6.1 best fit. Listing~\ref{lst:biprobability} computes the curves. See notebooks \#05 and \#28, and Sec.~\ref{sec:biprobability} for details.}
\end{figure}

Figure~\ref{fig:biprobability} shows a bi-probability plot\ \cite{Minakata:2001qm}: the antineutrino appearance probability against the neutrino one, at one energy and baseline. Each value of $\delta_{\rm CP}$ gives one point, so taking the phase through its range traces a closed curve. The figure draws two curves, for standard oscillations and for oscillations with non-standard interactions, and they cross at four points. At each crossing, one measurement of the two probabilities fits both cases; \eg, at $P_{\nu_\mu \to \nu_e} = 0.080$ and $P_{\bar{\nu}_\mu \to \bar{\nu}_e} = 0.008$, it fits the standard case with $\delta_{\rm CP} = -0.58\pi$ and the non-standard one with $\delta_{\rm CP} = -0.30\pi$. A new interaction can then pass for a shift in the phase. Separating the two takes a second measurement, at another energy or baseline.

Listing~\ref{lst:biprobability} computes the two curves. Both come from the same loop over $\delta_{\rm CP}$: one calls the standard wrapper, the other the wrapper with non-standard interactions. Any two hypotheses that a measurement might confuse can be compared in the same way, by changing the wrapper or its parameters; notebook \#05, \eg, compares the two mass orderings.

\begin{lstlisting}[language=Python, caption={\textbf{Bi-probability curves at the DUNE baseline.} The two curves of \figu{biprobability}. The function {\tt locus} takes a wrapper once around $\delta_{\rm CP}$, computing the neutrino and the antineutrino probability at each of 181 values. The oscillation parameters are the default set, NuFIT 6.1. No tolerance is passed: the Hamiltonian is constant, so the Magnus expansion ends at its first term, and the result is exact. The curves are drawn twice: first with {\tt plotting.plot\_biprobability}, in the preset style of Sec.~\ref{sec:plotting}, then with {\tt matplotlib} directly. See notebooks \#05 and \#28, and Sec.~\ref{sec:biprobability} for details.}, label={lst:biprobability}, float=tbp]
import numpy as np
import matplotlib.pyplot as plt
import magnus.oscprob as oscprob
import magnus.globaldefs as gd
import magnus.plotting as plotting

# DUNE: 2 GeV, 1300 km; a crust-like 3 g/cm^3
E, L, rho = 2.0*gd.UNIT_GEV, 1300.0*gd.UNIT_KM, 3.0
kw = dict(nu_i=gd.NUMU, nu_f=gd.NUE,
          density_matter_is_in_g_per_cm3=True)
eps = dict(eps_ee=0.10, eps_em=0.05, eps_mt=0.03)
std = oscprob.osc_prob_3nu_matter_constant_density
nsi = oscprob.osc_prob_3nu_matter_nsi_constant_density

def locus(wrapper, **couplings):
    """(P_mu e, P_mubar ebar) once around delta_CP."""
    out = []
    for dcp in np.linspace(-np.pi, np.pi, 181):
        P = [wrapper(E, L, rho, nubar=nubar, dCP=dcp,
                     **couplings, **kw)
             for nubar in (False, True)]
        out.append(P)
    return np.array(out)

P_std = locus(std)
P_nsi = locus(nsi, **eps)

# With the shipped plotter
fig, ax = plotting.plot_biprobability(
    [P_std[:, 0], P_nsi[:, 0]],
    [P_std[:, 1], P_nsi[:, 1]],
    labels=['Standard', 'NSI'])

# With matplotlib directly
fig, ax = plt.subplots()
for P, label in ((P_std, 'Standard'), (P_nsi, 'NSI')):
    ax.plot(P[:, 0], P[:, 1], label=label)
ax.set_xlabel(r'$P_{\nu_\mu \to \nu_e}$')
ax.set_ylabel(r'$P_{\bar\nu_\mu \to \bar\nu_e}$')
ax.legend()
\end{lstlisting}


\subsection{The Earth}
\label{sec:prem}

\begin{figure}[t!]
 \centering
 \includegraphics[width=\columnwidth]{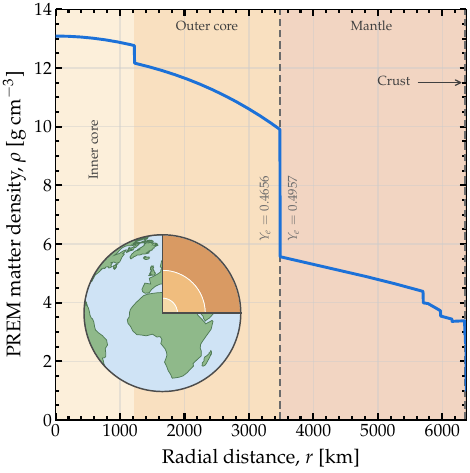}
 \caption{\label{fig:prem}\textbf{The Earth\textquoteright s density profile.} The PREM density profile\ \cite{Dziewonski:1981xy} against radial distance, with a cut-away globe showing the same layers in the same colors. The electron fraction changes at three radii, marked by the dashed lines: from 0.4656 to 0.4957 at the core-mantle boundary, at 3\,480~km; to 0.4952 at 6\,346.6~km; and to 0.5551 in the ocean, at 6\,368~km. The last two are too close together to separate here. \magnus\ places a slab edge wherever a chord crosses one of the nine boundaries between PREM's ten shells, so that no slab straddles a density jump. See Sec.~\ref{sec:prem} for details.}
\end{figure}

Figure~\ref{fig:prem} shows the density profile \magnus\ uses for the Earth, the Preliminary Reference Earth Model (PREM)\ \cite{Dziewonski:1981xy}, built from seismic data. It treats the Earth as a spherically symmetric body of radius $R_\oplus = 6\,371$~km, made of ten concentric shells. The density falls from 13.1~g~cm$^{-3}$ at the center to 1.02~g~cm$^{-3}$ in the ocean at the surface, smoothly inside each shell and in jumps between shells. The largest jump is at the core-mantle boundary, 3\,480~km from the center, where the density drops from 9.90 to 5.57~g~cm$^{-3}$. The electron fraction, $Y_e$, reflects the chemical composition, so it changes with radius too, but only at three of the nine boundaries: \magnus\ uses 0.4656 in the core, 0.4957 in the mantle, 0.4952 in the crust, and 0.5551 in the ocean. The {\tt earth} module stores these as {\tt Y\_E\_CORE\_PREM}, {\tt Y\_E\_MANTLE\_PREM}, {\tt Y\_E\_CRUST\_PREM}, and {\tt Y\_E\_OCEAN\_PREM}. Every Earth wrapper takes {\tt electron\_fraction\_core}, {\tt electron\_fraction\_mantle}, {\tt electron\_fraction\_crust}, and {\tt electron\_fraction\_ocean} to change them.

The trajectory of a neutrino through the Earth is set by its zenith angle at the detector, $\theta_z$. By default, the detector sits on the surface; Sec.~\ref{sec:underground} shows how to place sources and detectors underground. A neutrino arriving from straight below has $\cos\theta_z = -1$ and crosses an Earth diameter. One arriving along the horizon has $\cos\theta_z = 0$ and does not travel inside the Earth at all. In between, the chord comes closest to the center at a radius of $R_\oplus \sqrt{1 - \cos^2\theta_z}$, so a trajectory closer to horizontal samples only the outer shells.

For a given zenith angle, {\tt earth} returns the chord length, \ie, the baseline:
\begin{pysnip}
import magnus.earth as earth

costhz = -0.9
L = earth.distance_traveled_inside_earth(
    costhz)                    # 11467.8 km
\end{pysnip}
A chord meets every PREM boundary it reaches twice, once going in and once coming out. For instance, at $\cos\theta_z = -0.9$, the chord comes within 2\,777~km of the center, so eight of the nine boundaries lie above that point, and the chord crosses them sixteen times:
\begin{pysnip}
edges = earth.prem_layer_edges_along_chord(
    costhz)
len(edges)                     # 16
\end{pysnip}
Every {\tt osc\_prob\_*\_earth} wrapper computes these positions and passes them on as {\tt t\_breakpoints}, so a slab edge falls on every crossing, at every level of refinement and at any tolerance.

An Earth wrapper takes the trajectory as a zenith angle, {\tt costhz}, and a baseline, {\tt L}. From {\tt costhz}, the wrapper finds the density and the electron fraction at each point of the chord, and the slab edges. {\tt L} is in the natural units of \magnus, so the chord length from {\tt earth}, in kilometers, is converted. At three flavors, for a 10-GeV neutrino, $P_{\nu_\mu \to \nu_e}$ is
\begin{pysnip}
import magnus.oscprob as oscprob
import magnus.globaldefs as gd

P = oscprob.osc_prob_3nu_earth(
    10.0*gd.UNIT_GEV, costhz=costhz,
    L=L*gd.UNIT_KM)
P[gd.NUMU][gd.NUE]             # 0.124242
\end{pysnip}
By default, the electron fraction changes from layer to layer, as above. Passing {\tt electron\_fraction} replaces the four values with a single one for the whole Earth. At 5~GeV along this chord, the two choices differ by 0.07 in $P_{\nu_\mu \to \nu_e}$:
\begin{pysnip}
kw = dict(costhz=costhz, L=L*gd.UNIT_KM,
          nu_i=gd.NUMU, nu_f=gd.NUE)
E = 5.0*gd.UNIT_GEV

oscprob.osc_prob_3nu_earth(E, **kw) # 0.0849
oscprob.osc_prob_3nu_earth(E, **kw, 
    electron_fraction=0.5)          # 0.0129
\end{pysnip}


\subsubsection{Neutrinos \& antineutrinos, 2--5 flavors}
\label{sec:nu_nubar}

\begin{figure}[t!]
 \centering
 \includegraphics[width=\columnwidth]{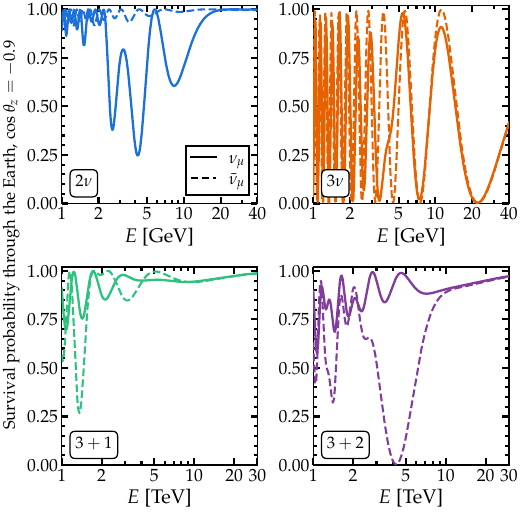}
 \caption{\label{fig:nu_nubar}\textbf{Neutrinos and antineutrinos through the Earth.} Survival probability of $\nu_\mu$ and $\bar{\nu}_\mu$ along a PREM chord at $\cos\theta_z = -0.9$, computed with \magnus. {\it Top}: two flavors, the $(\nu_e, \nu_\mu)$ reduction of the 1--3 sector, and three flavors, both from 1 to 40~GeV. {\it Bottom}: $3+1$ and $3+2$ from 1 to 30~TeV, where an eV-scale splitting places its resonance, with $\sin^2\theta_{14} = \sin^2\theta_{24} = 0.10$ and $\Delta m^2_{41} = 1$~eV$^2$, joined in the $3+2$ panel by $\sin^2\theta_{15} = \sin^2\theta_{25} = 0.06$ and $\Delta m^2_{51} = 1.7$~eV$^2$. Listing~\ref{lst:nu_nubar} computes all four pairs. See notebooks \#15 and \#28, and Sec.~\ref{sec:nu_nubar} for details.}
\end{figure}

Figure~\ref{fig:nu_nubar} shows $\nu_\mu$ and $\bar{\nu}_\mu$ survival probabilities along the same chord as before, at $\cos \theta_z = -0.9$, at two to five flavors. Listing~\ref{lst:nu_nubar} computes all eight curves. One dictionary carries the chord, the channel, and the tolerances through every call, so the four flavor counts differ only in the wrapper, the energy range, and the sterile mixing that 3+1 and 3+2 add.

\begin{lstlisting}[language=Python, caption={\textbf{Neutrinos and antineutrinos through the Earth.} The $\nu_\mu$ and $\bar{\nu}_\mu$ curves of \figu{nu_nubar}. {\tt both\_signs} calls a wrapper twice, with {\tt nubar=False} for the neutrino and {\tt nubar=True} for the antineutrino. See notebooks \#15 and \#28, and Sec.~\ref{sec:nu_nubar} for details.}, label={lst:nu_nubar}, float=t]
import numpy as np
import magnus.oscprob as oscprob
import magnus.earth as earth
import magnus.globaldefs as gd

costhz = -0.9
L = earth.distance_traveled_inside_earth(costhz) \
 *gd.UNIT_KM
kw = dict(costhz=costhz, L=L, nu_i=gd.NUMU,
          nu_f=gd.NUMU, rtol=1.0e-8, atol=1.0e-10)

def both_signs(wrapper, E, **params):
    """Survival probability, for a nu then a nubar."""
    return [wrapper(E, nubar=nubar, **params, **kw)
            for nubar in (False, True)]

# Two flavors is the (nu_e, nu_mu) reduction of the 1-3
# sector, so it takes that one angle and splitting by
# name.  Every wrapper above two flavors defaults to
# the same NuFIT 6.1 set, so it needs none passed.
osc = gd.load_nufit_params('NuFIT 6.1')
E = np.logspace(0, np.log10(40), 260)*gd.UNIT_GEV
P2, P2bar = both_signs(oscprob.osc_prob_2nu_earth, E,
                       sth=osc['s13'], Dm2=osc['D31'])
P3, P3bar = both_signs(oscprob.osc_prob_3nu_earth, E)

# The sterile pairs run three decades higher, where an
# eV-scale splitting places its resonance
E_TEV = np.logspace(0, np.log10(30), 200)*gd.UNIT_TEV
s4 = dict(s14=np.sqrt(0.10), s24=np.sqrt(0.10), s34=0.0,
          D41=1.0)
s5 = dict(**s4, s15=np.sqrt(0.06), s25=np.sqrt(0.06),
          s35=0.0, D51=1.7)
P4, P4bar = both_signs(oscprob.osc_prob_4nu_earth,
                       E_TEV, **s4)
P5, P5bar = both_signs(oscprob.osc_prob_5nu_earth,
                       E_TEV, **s5)
\end{lstlisting}

In the top row, the ordinary matter resonance depletes the $\nu_\mu$ and leaves the $\bar{\nu}_\mu$ almost unaffected, because the matter potential enters their Hamiltonians with opposite signs. The two-flavor panel shows this cleanly. A $(\nu_e, \nu_\mu)$ system has no $\theta_{23}$ oscillation, so the only structure left is the resonance: the $\nu_\mu$ survival probability dips to 0.25 at 4~GeV, while the $\bar{\nu}_\mu$ one stays above 0.9. The three-flavor panel restores the $\theta_{23}$ oscillation, which is near-maximal. Both curves then swing between zero and one several times, and the resonance is no longer easy to see. It still shows as a difference between the $\nu_\mu$ and $\bar{\nu}_\mu$ curves, which oscillate out of step: below 5~GeV they differ by up to 0.86, and the difference shrinks as the energy rises, until the two curves coincide above 20~GeV.

In the bottom row, for 3+1 and 3+2, it is the other way around: the resonance depletes the $\bar{\nu}_\mu$ and leaves the $\nu_\mu$ almost unaffected. A sterile state feels no matter potential at all. The active states all feel the same neutral-current potential, $V_{\rm NC}$, which is common to them and cancels out of their oscillations; the $\nu_e$ also feels the charged-current potential, $V_{\rm CC}$, since only it scatters off electrons that way. Therefore, measured from the active states, each sterile state carries $-V_{\rm NC}$. A resonance needs a positive potential on the flavor that is mostly the lighter of the two mixing mass states, or a negative one on the flavor that is mostly the heavier. At two and three flavors, $V_{\rm CC}$ is positive for neutrinos and acts on the $\nu_e$, which in the normal mass ordering is mostly the lighter state, so the $\nu_\mu$ resonates and is depleted. At 3+1 and 3+2, the potential that matters is the one on the sterile state, $-V_{\rm NC}$. For neutrinos, it is positive, but it acts on the heavier state, since $\Delta m^2_{41} > 0$ makes the sterile state mostly the heavier one, so there is no resonance. For antineutrinos, the potential changes sign, so the $\bar{\nu}_\mu$ resonates instead.

At 3+1, the resonance shows as a dip of the $\bar{\nu}_\mu$ survival probability to 0.26 near 1.4~TeV, while the $\nu_\mu$ one falls no lower than 0.70. Searches for a sterile state in atmospheric neutrinos look for exactly this dip\ \cite{IceCube:2016rnb, IceCube:2020phf}. At 3+2, the second sterile state adds a second dip, at three times the energy, deeper than the first (nearly to zero) and about three times as wide. Above 10~TeV, both curves come back together.


\subsubsection{Probabilities between two locations}
\label{sec:named_baselines}

\begin{figure*}[t!]
 \centering
 \includegraphics[width=\textwidth]{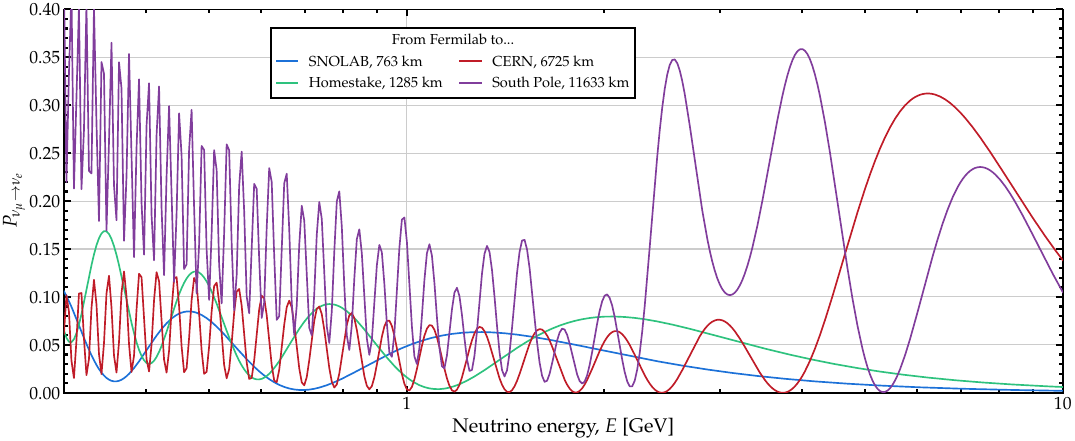}
 \caption{\label{fig:named_baselines}\textbf{Four chords from Fermilab.} Appearance probability $P_{\nu_\mu \to \nu_e}$ against energy for a beam sent from Fermilab to four named sites, computed with \magnus\ along the PREM chord joining each pair of locations, at the layered composition of Sec.~\ref{sec:prem}. The legend gives the chord length of each. The two shortest show the pattern of a beam experiment: a first oscillation maximum at 1 to 2~GeV, with faster oscillations below it. The two longest reach the matter resonance, which is why the probability along them is several times larger. Listing~\ref{lst:named} computes the curves. See notebooks \#04 and \#28, and Sec.~\ref{sec:named_baselines} for details.}
\end{figure*}

\begin{table}[b!]
{\normalsize
\begin{tabular}{@{}p{2.6cm}p{2.5cm}p{2.7cm}@{}}
 Site name & Latitude & Longitude \\
 \hline
 {\tt baikal} & $51^\circ\,45'\,54''$ & $104^\circ\,24'\,54''$ \\
 {\tt cern} & $46^\circ\,14'\,1.8''$ & $6^\circ\,3'\,11.4''$ \\
 {\tt desy} & $53^\circ\,34'\,19.79''$ & $9^\circ\,52'\,27.59''$ \\
 {\tt ess} & $55^\circ\,44'\,6''$ & $13^\circ\,15'\,5.04''$ \\
 {\tt fermilab} & $41^\circ\,49'\,55''$ & $-88^\circ\,15'\,26''$ \\
 {\tt gran\_sasso} & $42^\circ\,25'\,15.8''$ & $13^\circ\,30'\,58.43''$ \\
 {\tt homestake} & $44^\circ\,21'\,5.76''$ & $-103^\circ\,45'\,4.68''$ \\
 {\tt kamioka} & $36^\circ\,25'\,50.05''$ & $137^\circ\,18'\,41.15''$ \\
 {\tt km3net\_arca} & $36^\circ\,16'\,0''$ & $16^\circ\,6'\,0''$ \\
 {\tt km3net\_orca} & $42^\circ\,48'\,0''$ & $6^\circ\,2'\,0''$ \\
 {\tt north\_pole} & $90^\circ\,0'\,0''$ & $0^\circ\,0'\,0''$ \\
 {\tt pyhaasalmi} & $63^\circ\,39'\,31''$ & $26^\circ\,2'\,28''$ \\
 {\tt snolab} & $46^\circ\,28'\,18''$ & $-81^\circ\,11'\,12''$ \\
 {\tt south\_pole} & $-90^\circ\,0'\,0''$ & $0^\circ\,0'\,0''$ \\
 {\tt tokai} & $36^\circ\,27'\,59''$ & $140^\circ\,36'\,24''$ \\
\end{tabular}
}
\caption{\label{tab:sites}\textbf{Named locations on the Earth\textquoteright s surface.} The fifteen locations {\tt magnus.earth} carries by name, with the coordinates it stores for each. A name passed as {\tt loc\_ini} or {\tt loc\_fin} stands for the pair beside it. Coordinates are in degrees, minutes and seconds, which is the form the module stores and the one {\tt loc\_ini} and {\tt loc\_fin} accept; {\tt earth.dms\_to\_decimal} returns the same coordinate in decimal degrees.}
\end{table}

\textit{Locations given as coordinates.---}The examples so far set the trajectory by its zenith angle. A long-baseline experiment is described instead by where the beam is made and where it is detected, and the Earth wrappers accept that form too. Passing {\tt loc\_ini} and {\tt loc\_fin} replaces {\tt costhz} and {\tt L}: the chord joining the two sites is the baseline, and its direction fixes which PREM shells it crosses. Each location is a (latitude, longitude) pair, each stated in degrees, minutes, and seconds. North and east are positive; south and west are negative, with the sign read from the first non-zero part:
\begin{pysnip}
# (degree, minute, second), north and east
# positive, south and west negative
fnal = ((41, 49, 55), (-88, 15, 26))
hs = ((44, 21, 5.76), (-103, 45, 4.68))
P = oscprob.osc_prob_3nu_earth(
    2.0*gd.UNIT_GEV, loc_ini=fnal,
    loc_fin=hs, nu_i=gd.NUMU, nu_f=gd.NUE)
P                              # 0.0797091
\end{pysnip}

Naming two locations fixes the baseline at the chord between them; {\tt costhz} or {\tt L} passed alongside is ignored. Like any full chord, it reads the same from both ends, since it meets every radius twice, so \magnus\ evaluates the Hamiltonian over the first half and mirrors the rest. Section~\ref{sec:cost} quantifies the saving.

\smallskip

\textit{Named sites.---}Figure~\ref{fig:named_baselines} shows the appearance probability $P_{\nu_\mu \to \nu_e}$ along four chords, each running from Fermilab to a named site, and Listing~\ref{lst:named} computes them. Table~\ref{tab:sites} lists the fifteen locations the {\tt earth} module carries by name. Passing a name is the same as passing the coordinates beside it:
\begin{pysnip}
P = oscprob.osc_prob_3nu_earth(
    2.0*gd.UNIT_GEV,
    loc_ini='fermilab', loc_fin='homestake',
    nu_i=gd.NUMU, nu_f=gd.NUE)
P                              # 0.0797091
\end{pysnip}
This call and the one with coordinates above return the same probability, 0.0797091; Listing~\ref{lst:named} uses names. The four chords reach very different depths. The one to SNOLAB stays inside the crust, 11~km down at its deepest. The one to Homestake, the DUNE baseline, reaches 32~km. The one to CERN reaches 960~km, into the lower mantle, and the one to the South Pole reaches 3\,770~km, into the outer core.

\begin{lstlisting}[language=Python, caption={\textbf{Baselines between named sites.} The four curves of \figu{named_baselines}. Two site names replace {\tt costhz} and {\tt L}: the wrapper takes the chord between them as the baseline and its direction as the trajectory. {\tt chord\_length\_inside\_earth} returns the chord length. Site names are in \tabl{sites}. See notebooks \#04 and \#28, and Sec.~\ref{sec:named_baselines} for details.}, label={lst:named}, float=b]
import numpy as np
import magnus.oscprob as oscprob
import magnus.earth as earth
import magnus.globaldefs as gd

E = np.logspace(np.log10(0.3), 1.0, 400)*gd.UNIT_GEV
sites = ('snolab', 'homestake', 'cern', 'south_pole')

# Two site names replace costhz and L: the chord
# between them is the baseline
P = {site: oscprob.osc_prob_3nu_earth(
         E, loc_ini='fermilab', loc_fin=site,
         nu_i=gd.NUMU, nu_f=gd.NUE,
         rtol=1.0e-8, atol=1.0e-10)
     for site in sites}

# The chord itself, in km, for a legend
a = earth.loc_coords_dms['fermilab']
b = earth.loc_coords_dms['homestake']
L_km = earth.chord_length_inside_earth(
    a['lat'], a['lon'], b['lat'], b['lon'])
print('%.0f km' % L_km)   # 1285
\end{lstlisting}


\subsubsection{New physics}
\label{sec:bsm}

\begin{figure}[t!]
 \centering
 \includegraphics[width=\columnwidth]{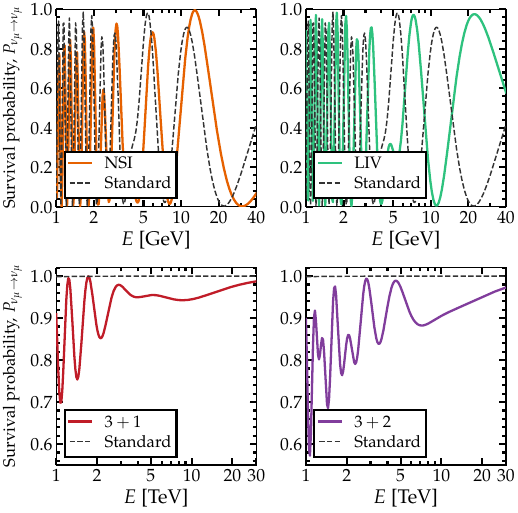}
 \caption{\label{fig:bsm}\textbf{New physics along an Earth chord.} Muon-neutrino survival probability $P_{\nu_\mu \to \nu_\mu}$ along a PREM chord at $\cos\theta_z = -0.9$, computed with \magnus\ for four departures from the standard three-flavor case. {\it Top left}: non-standard interactions with $\varepsilon_{ee} = 0.10$, $\varepsilon_{e\mu} = 0.05$ and $\varepsilon_{\mu\tau} = 0.03$. {\it Top right}: an isotropic, energy-independent Lorentz-violating term, with eigenvalues $b_1 = b_2 = 0$ and $b_3 = 5.4 \cdot 10^{-14}$~eV, the value that advances the phase by $\pi$ over this chord. {\it Bottom}: the 3+1 and 3+2 systems of \figu{nu_nubar}, where the eV-scale splittings, $\Delta m_{41}^2$ and $\Delta m_{51}^2$, place the matter resonance at the TeV scale. Listing~\ref{lst:bsm} computes the curves. See notebooks \#07, \#08, \#09, and \#28, and Sec.~\ref{sec:bsm} for details.}
\end{figure}

\begin{lstlisting}[language=Python, caption={\textbf{New physics along an Earth chord.} The five scenarios of \figu{bsm}, in six calls: the standard case is computed twice, once for each energy range. Every call shares the chord, the channel and the tolerances, differing only in the wrapper and in the parameters of the extra term. See notebooks \#07, \#08, \#09, and \#28, and Sec.~\ref{sec:bsm} for details.}, label={lst:bsm}, float=b]
import numpy as np
import magnus.oscprob as oscprob
import magnus.earth as earth
import magnus.globaldefs as gd

costhz = -0.9
L = earth.distance_traveled_inside_earth(costhz) \
 *gd.UNIT_KM
kw = dict(costhz=costhz, L=L, nu_i=gd.NUMU,
          nu_f=gd.NUMU, rtol=1.0e-8, atol=1.0e-10)
E = np.logspace(0, np.log10(40), 260)*gd.UNIT_GEV
# Three decades up is where an eV-scale splitting
# places its matter resonance
E_TEV = np.logspace(0, np.log10(30), 200)*gd.UNIT_TEV

eps = dict(eps_ee=0.10, eps_em=0.05, eps_mt=0.03)
liv = dict(b1=0.0, b2=0.0, b3=np.pi/L, Lambda=1.0,
           sxi12=0.0, sxi23=1.0/np.sqrt(2.0),
           sxi13=0.0, dxiCP=0.0, n_liv=0)
s4 = dict(s14=np.sqrt(0.10), s24=np.sqrt(0.10),
          s34=0.0, D41=1.0)
s5 = dict(**s4, s15=np.sqrt(0.06),
          s25=np.sqrt(0.06), s35=0.0, D51=1.7)

# The same call each time, a different extra term
P_std = oscprob.osc_prob_3nu_earth(E, **kw)
P_nsi = oscprob.osc_prob_3nu_earth_nsi(E, **eps, **kw)
P_liv = oscprob.osc_prob_3nu_earth_liv(E, **liv, **kw)
P_std_tev = oscprob.osc_prob_3nu_earth(E_TEV, **kw)
P_3p1 = oscprob.osc_prob_4nu_earth(E_TEV, **s4, **kw)
P_3p2 = oscprob.osc_prob_5nu_earth(E_TEV, **s5, **kw)
\end{lstlisting}

Figure~\ref{fig:bsm} shows four departures from the standard three-flavor case along one deep Earth chord: non-standard interactions\ \cite{Farzan:2017xzy}, Lorentz-invariance violation\ \cite{Kostelecky:2003xn, Kostelecky:2011gq}, one sterile state\ \cite{Gariazzo:2015rra, Dentler:2018sju}, and two sterile states\ \cite{Sorel:2003hf}. Each adds a different term to the Hamiltonian. As Listing~\ref{lst:bsm} shows, the calls differ only in the wrapper and in the parameters of the new term.

\begin{figure}[t!]
 \centering
 \includegraphics[width=\columnwidth]{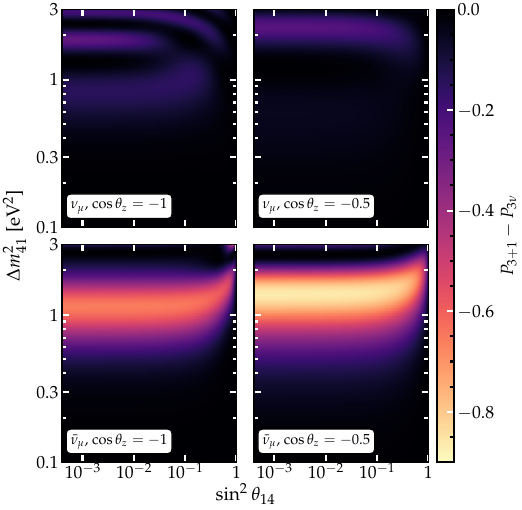}
 \caption{\label{fig:sterile_scan}\textbf{A scan of the sterile-state parameters.} Change in the muon-neutrino survival probability when one sterile state is added, $P_{3+1} - P_{3\nu}$, as a function of $\Delta m_{41}^2$ and $\sin^2 \theta_{14}$, at a fixed energy of 5~TeV. {\it Top row}: $\nu_\mu$. {\it Bottom row}: $\bar{\nu}_\mu$. {\it Left column}: a core-crossing chord, $\cos\theta_z = -1$. {\it Right column}: a shallower one, $\cos\theta_z = -0.5$. The other sterile parameters are held at $\sin^2\theta_{24} = 0.10$ and $\theta_{34} = 0$. Each panel holds $80 \times 80$ probabilities. See notebooks \#07 and \#28, and Sec.~\ref{sec:bsm} for details.}
\end{figure}

Figure~\ref{fig:sterile_scan} scans the sterile parameters themselves. The wrapper takes $\Delta m^2_{41}$ and $\theta_{14}$ like any other argument, so the map is a double loop over the two, and Listing~\ref{lst:sterile_scan} computes all of it. The map shows the reversal of Sec.~\ref{sec:nu_nubar} across parameter space. The $\nu_\mu$ survival probability changes by at most 0.27. The $\bar{\nu}_\mu$ panels carry a resonance band near $\Delta m^2_{41} = 1$~eV$^2$ that removes up to 0.87 of the flux. The band is strongest at small $\theta_{14}$ and fades as $\sin^2\theta_{14}$ approaches 1, because the $\nu_\mu$ reaches the sterile state through $\lvert \mathbb{R}_{\mu 4} \rvert^2 = \cos^2\theta_{14} \sin^2\theta_{24}$, with $\theta_{24}$ fixed.

\begin{lstlisting}[language=Python, caption={\textbf{A scan of the sterile-state parameters.} The four panels of \figu{sterile_scan}, one call of {\tt panel} each, for one chord ({\tt costhz}) and one sign ({\tt nubar}). In each, the three-flavor probability is computed once and subtracted from the 3+1 one at every point of the grid. {\tt S14} holds $\sin\theta_{14}$; the figure plots its square. See notebooks \#07 and \#28, and Sec.~\ref{sec:bsm} for details.}, label={lst:sterile_scan}, float=tbp]
import numpy as np
import magnus.oscprob as oscprob
import magnus.earth as earth
import magnus.globaldefs as gd

E = 5.0*gd.UNIT_TEV
D41 = np.logspace(np.log10(0.1), np.log10(3.0), 80)
S14 = np.logspace(np.log10(0.02), 0.0, 80)
s24 = np.sqrt(0.10)

def panel(costhz, nubar):
    """One chord and one sign, over the whole plane."""
    L = earth.distance_traveled_inside_earth(costhz) \
     *gd.UNIT_KM
    kw = dict(costhz=costhz, L=L, nu_i=gd.NUMU,
              nu_f=gd.NUMU, nubar=nubar,
              rtol=1.0e-8, atol=1.0e-10)
    # Standard 3nu oscillations
    std = oscprob.osc_prob_3nu_earth(E, **kw)
    return np.array([[oscprob.osc_prob_4nu_earth(
        E, s14=s, s24=s24, s34=0.0, D41=d, **kw)
        - std for s in S14] for d in D41])

# Change between 3nu and 3+1 survival probability
dP = {(cz, nb): panel(cz, nb)
      for cz in (-1.0, -0.5) for nb in (False, True)}
\end{lstlisting}


\subsubsection{Earth oscillograms}
\label{sec:earth_oscillograms}

\begin{figure}[t!]
 \centering
 \includegraphics[width=\columnwidth]{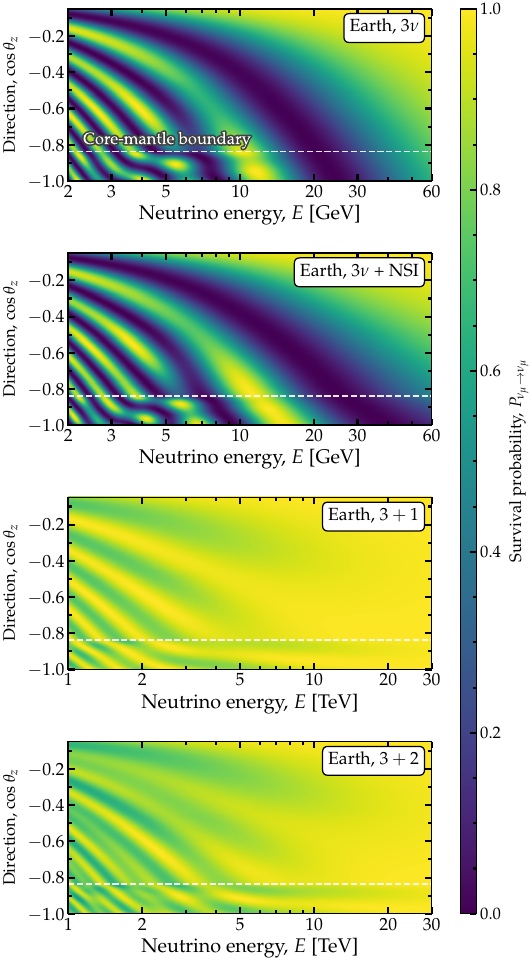}
 \caption{\label{fig:oscillogram}\textbf{Oscillograms through the Earth.} Oscillogram: the $\nu_\mu$ survival probability, $P_{\nu_\mu \to \nu_\mu}$, over neutrino energy and arrival direction, computed with \magnus\ along the PREM profile\ \cite{Dziewonski:1981xy}. {\it Top to bottom}: standard three flavors; three flavors with the non-standard couplings of \figu{bsm}; 3+1 and 3+2 with the sterile parameters of \figu{nu_nubar}. The dashed line in each panel marks $\cos\theta_z = -0.838$, the direction whose chord just grazes the outer core; steeper trajectories enter the core, where the density nearly doubles. Listing~\ref{lst:oscillogram} computes the oscillograms. See notebooks \#06 and \#28, and Sec.~\ref{sec:earth_oscillograms} for details.}
\end{figure}

Figure~\ref{fig:oscillogram} shows oscillograms of $\nu_\mu$ survival through the Earth, \ie, $P_{\nu_\mu \to \nu_\mu}$ across energy and arrival direction, for the standard case, with non-standard interactions, and with one and two sterile states. Listing~\ref{lst:oscillogram} computes them.

\begin{lstlisting}[language=Python, caption={\textbf{Oscillograms through the Earth.} The four panels of \figu{oscillogram}. One wrapper call per arrival direction returns an energy sweep, with the chord length taken from the direction alone. Only the wrapper and its extra parameters change between panels. Each panel is drawn twice: first by {\tt plotting.plot\_oscillogram}, then by {\tt matplotlib} directly. Both put $\cos\theta_z$ on the horizontal axis; \figu{oscillogram} puts it on the vertical one. See Sec.~\ref{sec:earth_oscillograms} for details.}, label={lst:oscillogram}, float=tbp]
import numpy as np
import matplotlib.pyplot as plt
import magnus.oscprob as oscprob
import magnus.earth as earth
import magnus.globaldefs as gd
import magnus.plotting as plotting

chord = earth.distance_traveled_inside_earth
CZ = np.linspace(-1.0, -0.05, 170)
E_GEV = np.logspace(np.log10(2.0),
                    np.log10(60.0), 200)*gd.UNIT_GEV
E_TEV = np.logspace(0.0,
                    np.log10(30.0), 200)*gd.UNIT_TEV
kw = dict(nu_i=gd.NUMU, nu_f=gd.NUMU,
          rtol=1.0e-8, atol=1.0e-10)

eps = dict(eps_ee=0.10, eps_em=0.05, eps_mt=0.03)
s4 = dict(s14=np.sqrt(0.10), s24=np.sqrt(0.10),
          s34=0.0, D41=1.0)
s5 = dict(**s4, s15=np.sqrt(0.06),
          s25=np.sqrt(0.06), s35=0.0, D51=1.7)

def oscillogram(wrapper, E, **params):
    """One panel: an energy sweep per direction."""
    return np.array([wrapper(E, costhz=c,
        L=chord(c)*gd.UNIT_KM, **params, **kw)
        for c in CZ]).T

PANELS = [(oscprob.osc_prob_3nu_earth, E_GEV, {}),
          (oscprob.osc_prob_3nu_earth_nsi, E_GEV, eps),
          (oscprob.osc_prob_4nu_earth, E_TEV, s4),
          (oscprob.osc_prob_5nu_earth, E_TEV, s5)]
P = [oscillogram(w, E, **p) for w, E, p in PANELS]

# With the shipped plotter, one call per panel
for (_, E, _), grid in zip(PANELS, P):
    fig, ax = plotting.plot_oscillogram(
        CZ, np.log10(E/gd.UNIT_GEV), grid,
        nu_i=gd.NUMU, nu_f=gd.NUMU)

# Or with matplotlib directly, one axes per panel
for (_, E, _), grid in zip(PANELS, P):
    fig, ax = plt.subplots()
    im = ax.pcolormesh(CZ, np.log10(E/gd.UNIT_GEV),
                       grid, shading='gouraud')
    ax.set_xlabel(r'$\cos\theta_z$')
    ax.set_ylabel(r'$\log_{10}(E/{\rm GeV})$')
    fig.colorbar(im, ax=ax,
                 label=r'$P_{\nu_\mu \to \nu_\mu}$')
\end{lstlisting}

Figures~\ref{fig:nu_nubar} and \ref{fig:bsm} hold the arrival direction fixed; an oscillogram varies it. The direction sets two things: the length of the chord and the densities along it. From $\cos\theta_z = -0.05$ to $-1$, the chord grows twentyfold, from 640 to 12\,742~km. The densities along it grow far less, until the chord reaches the core-mantle boundary, at $\cos\theta_z = -0.838$.

The bands in each panel come from two effects, which differ in their tilt. Oscillation bands follow a fixed value of $L/E$: lower in a panel, the chord is longer, so the same band lies at a higher energy, and the band runs diagonally. Resonance bands follow the density instead, which changes little across the mantle, so they run nearly vertically. Below the core-mantle boundary, the chord crosses mantle, core, and mantle again, and the density nearly doubles in the core: a short version of the castle-wall profile of Sec.~\ref{sec:arrangement}. Where the length of each crossing matches the oscillation length in matter, the effects of the crossings add up, which gives this strip a pattern of its own.

The two sterile panels in \figu{oscillogram} show the $\nu_\mu$, which has no sterile resonance: that resonance is in the $\bar{\nu}_\mu$ channel (Sec.~\ref{sec:nu_nubar}). All the bands in these panels are oscillation bands. For the $\nu_\mu$, matter suppresses the mixing with the sterile state, more so at higher energy, since the vacuum term, $\Delta m^2_{41}/2E$, falls with energy while the potential does not. 


\subsubsection{Underground sources and detectors}
\label{sec:underground}

By default, both ends of an Earth trajectory sit on the surface. Detectors, and some sources, sit underground, at depths from 100~m to 2--3~km. Passing {\tt source\_depth} or {\tt detector\_depth} to a wrapper puts either end underground. The zenith angle keeps its meaning, since it is measured at the detector. The depth then fixes the baseline: with a nonzero depth, passing {\tt L}, {\tt loc\_ini}, or {\tt loc\_fin} raises an error. A buried end also breaks the symmetry of the chord, which no longer reads the same from both ends, so \magnus\ does not mirror it (Sec.~\ref{sec:cost}).

How much burying the detector changes the probability depends on zenith angle. Along a long chord, at $\cos \theta_z = -0.8$, the last 2~km are a small part of it and barely change the probability:
\begin{pysnip}
E, costhz = 10.0*gd.UNIT_GEV, -0.8
kw = dict(nu_i=gd.NUMU, nu_f=gd.NUMU)

# On the surface, the baseline is given
L = earth.distance_traveled_inside_earth(
    costhz)*gd.UNIT_KM
oscprob.osc_prob_3nu_earth(E, costhz=costhz,
    L=L, **kw)
                               # 0.911587

# 2 km down, the depth fixes the baseline
oscprob.osc_prob_3nu_earth(
    E, costhz=costhz,
    detector_depth=2.0*gd.UNIT_KM, **kw)
                               # 0.911600
\end{pysnip}
The two differ by $10^{-5}$. In contrast, a neutrino arriving horizontally, at $\cos \theta_z = 0$, does not cross the Earth to reach a detector on the surface, but it does to reach one 2~km down:
\begin{pysnip}
E, costhz = 10.0*gd.UNIT_GEV, 0.0
kw = dict(nu_i=gd.NUMU, nu_f=gd.NUMU)

# Horizontal: no path at all at the surface
L = earth.distance_traveled_inside_earth(
    costhz)*gd.UNIT_KM
oscprob.osc_prob_3nu_earth(
    E, costhz=costhz, L=L, **kw)
                               # 1.000000

# 2 km down it crosses 160 km of rock
oscprob.osc_prob_3nu_earth(
    E, costhz=costhz,
    detector_depth=2.0*gd.UNIT_KM, **kw)
                               # 0.997560
\end{pysnip}
Here the probabilities differ by $2 \cdot 10^{-3}$. A nonzero {\tt source\_depth} acts in the same way.

PREM puts 3~km of ocean at the top of the Earth. That describes KM3NeT in the Mediterranean and Baikal-GVD in a lake, but not IceCube, under ice, or SNOLAB and Super-Kamiokande, under rock. Two keywords replace that medium: {\tt density\_matter\_ocean} sets the density of the ocean shell, and {\tt electron\_fraction\_ocean} its electron fraction. How much the replacement matters depends on how much of the path lies within 3~km of the surface. A steep trajectory crosses the shell and leaves it behind. A near-horizontal one can stay inside it for its whole length. There, ice instead of water lowers the probability by 0.1\%, and rock raises it by 2\%. Ice keeps the electron fraction of water, so only its density changes; rock also takes the electron fraction of the crust, 0.4952:
\begin{pysnip}
# PREM puts an ocean in the outer 3 km
L = earth.distance_traveled_inside_earth(
    -0.02)*gd.UNIT_KM
kw = dict(costhz=-0.02, L=L, nu_i=gd.NUMU,
          nu_f=gd.NUE)
E = 1.0*gd.UNIT_GEV
oscprob.osc_prob_3nu_earth(E, **kw)
                               # 0.020930
# Under ice, as at IceCube
oscprob.osc_prob_3nu_earth(E, **kw,
    density_matter_ocean=0.92) # 0.020900
# Under rock, as at SNOLAB
oscprob.osc_prob_3nu_earth(E, **kw,
    density_matter_ocean=2.65,
    electron_fraction_ocean=0.4952)
                               # 0.021325
\end{pysnip}


\subsubsection{Using a custom Earth density profile}
\label{sec:custom_earth}

The Earth wrappers take their density from PREM, and only the density of the outermost shell can be changed by keyword (Sec.~\ref{sec:underground}). A profile of one's own takes a call one layer down, to the scenario function {\tt osc\_prob\_matter\_std\_potential}, which the wrappers themselves call. The profile is passed as a function of position, and the positions where its density jumps as slab edges.

Listing~\ref{lst:coarse_earth} replaces PREM's ten shells with four wider ones: the inner core, the outer core, the mantle, and the crust with the ocean. Each holds a constant density, the mass-weighted mean of the PREM region it replaces, and PREM's electron fraction for that region (for the outermost shell, the crust's). Along the chord at $\cos \theta_z = -0.9$, at 10~GeV, the four-shell Earth gives $P_{\nu_\mu \to \nu_\mu} = 0.784$, against 0.806 for PREM. The electron fraction matters as much: with $Y_e = 0.5$ in every shell, the four-shell Earth gives 0.818.

The four-shell Earth has three boundaries, at radii of 1\,221.5, 3\,480, and 6\,346.6~km. The chord at $\cos \theta_z = -0.9$ crosses only the outer two, each twice, which makes four slab edges, against sixteen for PREM. The listing passes those four positions as {\tt t\_breakpoints}, as the Earth wrappers do for PREM.

\begin{lstlisting}[language=Python, caption={\textbf{A custom Earth profile.} A four-shell Earth in place of PREM. The profile is a function of position that returns an electron density; naming the radii where it jumps keeps every slab inside one shell. See Sec.~\ref{sec:custom_earth} for details.}, label={lst:coarse_earth}, float=tbp]
import numpy as np
import magnus.oscprob as oscprob
import magnus.earth as earth
import magnus.matter as matter
import magnus.globaldefs as gd

# Four shells for PREM's ten, each at the mass-weighted
# mean density of the region it replaces
EDGES = [1221.5, 3480.0, 6346.6]       # km
RHO = [12.894, 10.901, 4.476, 2.520]   # g/cm^3
YE = [0.4656, 0.4656, 0.4957, 0.4952]  # PREM's own

costhz = -0.9
r_of = earth.earth_radial_distance_from_depth
L = earth.distance_traveled_inside_earth(costhz) \
 *gd.UNIT_KM
osc = gd.load_nufit_params('NuFIT 6.1')

def ne_coarse(l):
    """Electron density [eV^3] at position l."""
    r = r_of(costhz, l/gd.UNIT_KM)
    i = int(np.searchsorted(EDGES, r))
    return matter.num_density_e_func(
        r, lambda _: RHO[i], electron_fraction=YE[i],
        ratio_number_neutrons_to_protons=
            (1.0 - YE[i])/YE[i],
        density_matter_is_in_g_per_cm3=True)

# Those four edges are PREM boundaries, so their
# crossings come straight out of the helper
cross = earth.prem_layer_edges_along_chord(costhz)
keep = cross[np.isclose(
    r_of(costhz, cross)[:, None], EDGES).any(axis=1)]

E = 10.0*gd.UNIT_GEV
kw = dict(nu_i=gd.NUMU, nu_f=gd.NUMU,
          rtol=1.0e-8, atol=1.0e-10)
P = oscprob.osc_prob_matter_std_potential(
    3, ne_coarse, E, L, osc_params=osc, L0=0.0,
    t_breakpoints=keep*gd.UNIT_KM,
    density_is_of_number_of_electrons=True, **kw)
P                               # 0.784445
\end{lstlisting}


\subsection{Strategy report after a call}
\label{sec:which_engine}

Every wrapper and scenario function takes a dictionary as {\tt strategy\_info} and fills it in place with a report on how the probability was computed. Three functions do not take it, because they evaluate a closed-form expression and use no engine: {\tt osc\_prob\_2nu\_vacuum\_std}, {\tt osc\_prob\_3nu\_vacuum\_std}, and {\tt osc\_prob\_2nu\_matter\_std}. Nor does the direct call {\tt osc\_prob}, which reports on its refinement ladder through {\tt convergence\_info} instead (Sec.~\ref{sec:refinement}).

The report has seven entries: {\tt 'engine'}, the engine that answered, named as in \tabl{engines}; {\tt 'family'}, the group of engines that share its method, and so can err in the same way (Sec.~\ref{sec:engines}); {\tt 'certified'}, whether the hybrid engine certified its answer, or {\tt None} if another engine answered; {\tt 'declined'}, the engines that attempted the request and gave up, each with its reason; {\tt 'trace'}, every engine tried, in order, with its details; {\tt 'hidden\_feature'}, the result of a scan of the profile for a feature narrower than the grids on which the engines sample it, or {\tt None} if the call declares where the profile changes, as on the Earth, or if the density is constant; and {\tt 'sampling'}, described below. For instance,
\begin{pysnip}
L = earth.distance_traveled_inside_earth(
    -0.9)*gd.UNIT_KM
info = {}
P = oscprob.osc_prob_3nu_earth(
    10.0*gd.UNIT_GEV, costhz=-0.9, L=L,
    nu_i=gd.NUMU, nu_f=gd.NUMU,
    strategy_info=info)
sorted(info)
# ['certified', 'declined', 'engine',
#  'family', 'hidden_feature', 'sampling',
#  'trace']
info['engine']                 # 'magnus'
info['family']          # 'magnus-ladder'
\end{pysnip}
This single energy went to the general Magnus ladder.

The {\tt 'sampling'} entry shows how many points a scan along the trajectory needs to resolve the oscillation. It gives the shortest oscillation length, how many oscillations of that length fit in the trajectory, and how many points a scan needs to sample them twice per cycle, as the Nyquist criterion requires. A scan with fewer points is aliased: each value is correct, but a curve drawn through them is not the probability. For this call,
\begin{pysnip}
s = info['sampling']
s['oscillation_length']/gd.UNIT_KM
                               # 3227.2
s['cycles_over_trajectory']    # 3.6
s['nyquist_points']            # 9
\end{pysnip}

On the same chord, an array of energies, a phase average, and a constant density each go to a different engine:
\begin{pysnip}
kw = dict(costhz=-0.9, L=L, nu_i=gd.NUMU,
          nu_f=gd.NUMU, strategy_info=info)
E = np.linspace(5.0, 15.0, 3)*gd.UNIT_GEV
oscprob.osc_prob_3nu_earth(E, **kw)
info['engine']              # 'separable'

oscprob.osc_prob_3nu_earth(
    10.0*gd.UNIT_GEV, **kw, average=True)
info['engine']                # 'average'

oscprob.osc_prob_3nu_matter_constant_density(
    10.0*gd.UNIT_GEV, L, 4.0,
    density_matter_is_in_g_per_cm3=True,
    nu_i=gd.NUMU, nu_f=gd.NUMU,
    strategy_info=info)
info['engine']               # 'constant'
\end{pysnip}

When an engine declines a request, {\tt 'declined'} gives the reason. In this example, along an Earth diameter (12\,742~km), the density falls as $3\,e^{-l/(50~{\rm m})}$~g~cm$^{-3}$, with $l$ the distance traveled. That is narrower than the spacing of the grid on which the hybrid engine probes the profile, so the hybrid engine declines, with a warning, and the interaction-picture engine answers:
\begin{pysnip}
info = {}
P = oscprob.osc_prob_2nu_matter_exp_density(
    5.0*gd.UNIT_MEV, 12742.0*gd.UNIT_KM,
    0.0, 3.0, 0.05*gd.UNIT_KM, sth=0.5,
    Dm2=2.5e-3,
    density_matter_is_in_g_per_cm3=True,
    nu_i=0, nu_f=1, strategy_info=info)
info['declined']
# [('hybrid', 'the profile is not
#   resolved at the probe scale')]
info['engine']                # 'ip_exp'
\end{pysnip}
With a fall-off length of 50~km instead of 50~m, the hybrid engine resolves the profile and certifies its own answer:
\begin{pysnip}
# 50.0*gd.UNIT_KM, not 0.05*gd.UNIT_KM
info['engine']                # 'hybrid'
info['certified']             # True
info['declined']              # []
\end{pysnip}


\subsection{The Sun}
\label{sec:sun}

\begin{figure}[t!]
 \centering
 \includegraphics[width=\columnwidth]{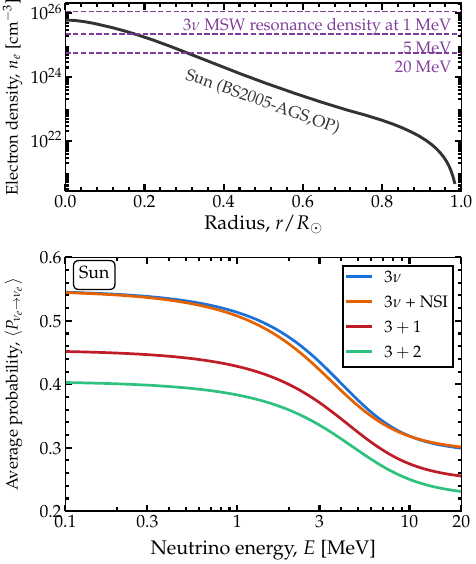}
 \caption{\label{fig:solar}\textbf{The averaged solar survival probability.} {\it Top}: the electron number density along a radial ray of the BS2005-AGS,OP solar model\ \cite{Bahcall:2004pz}. The marked densities are those of the MSW resonance, $\sqrt{2} G_F n_e = \Delta m^2_{21}\cos 2\theta_{12}/2E$, at three energies: a 20-MeV neutrino meets it a third of the way out; a 1-MeV neutrino never does. {\it Bottom}: the phase-averaged survival probability, \equ{averaged_varying}, computed with {\tt average=True} for four Hamiltonians on the same profile. The two sterile cases lie lowest, $3+2$ below $3+1$, because each extra state takes a share of the flux that does not return. Listing~\ref{lst:sun} generates the data. See notebooks \#10, \#13, and \#28, and Sec.~\ref{sec:sun} for details.}
\end{figure}

Figure~\ref{fig:solar} shows the electron density of the BS2005-AGS,OP solar model\ \cite{Bahcall:2004pz}, and the averaged survival probability it produces for four Hamiltonians. Along a radial ray, the density falls smoothly by five orders of magnitude. On the way out, a 5-MeV neutrino accumulates a phase of $3 \cdot 10^4$~radians between the first two eigenstates, and of $9 \cdot 10^5$~radians between the first and third. The survival probability at the surface thus runs through a full oscillation each time the production point moves by about 160~km, or the energy by a few parts in $10^4$. Yet, neutrinos are produced over tens of thousands of kilometers, and no detector resolves energy that finely. What a solar experiment measures is the phase-averaged probability of Sec.~\ref{sec:averaged}.

The energy dependence of the probability in \figu{solar} is set by where the resonance density sits relative to the density at production. The resonance density falls with energy as $1/E$. At 20~MeV, the profile falls to the resonance density a third of the way out. A neutrino born at the center, at ten times that density, is mostly the matter eigenstate that turns into $\nu_2$ on the way out, so it leaves the Sun as $\nu_2$, and its average is $\lvert \mathbb{R}_{e2} \rvert^2 \approx 0.30$. At 1~MeV, the resonance density lies above the central density, so the neutrino is born below it and never crosses it. Matter then only lowers the average a little, to 0.51, from the vacuum value, 0.55, which the curve nearly reaches at 0.1~MeV.

The probability in \figu{solar} shows the passage between the two regimes, over an energy range that spans the spectrum of neutrinos from $^8$B decay. Along the ray, the density changes slowly compared with the local oscillation length, so a neutrino stays in the eigenstate it was born in: $P^{\rm cross}$ in \equ{averaged_varying} is the identity, at every energy. What remains are the two eigenbases at the ends of the ray. The one at the surface is the vacuum mixing matrix, the same for any profile. The one at the other end is set by the density where the neutrino is born.

Listing~\ref{lst:sun} computes the four curves of \figu{solar}. The BS2005-AGS,OP table is one of the twelve standard solar models that ship with \magnus\ (Sec.~\ref{sec:solar_model_comparison}); every Sun wrapper takes it by name through {\tt density\_profile}. Without that keyword, a wrapper uses its default, {\tt 'exp'}, an exponential fit to the solar density (Sec.~\ref{sec:solar_model_comparison}). For the sterile states, it also reads the neutron-to-proton ratio from the table (Sec.~\ref{sec:solar_model_comparison}). 

The only difference from an ordinary probability call is {\tt average=True} (Sec.~\ref{sec:average_keyword}). With it, \magnus\ does not propagate the phase along the ray. It searches the ray for crossings that are not adiabatic, which it would bridge with the Magnus patch of Sec.~\ref{sec:hybrid}, and combines the eigenbases at the two ends. On the Sun, it finds no such crossing for any of the four Hamiltonians, so the result depends only on those two eigenbases. The neutrino starts in a flavor state, the default of {\tt average\_initial\_state}, so the result is \equ{phase_average_nocross}. Its interference terms carry the phases above, which damp them to nothing, so the result equals \equ{averaged_varying} to within $5 \cdot 10^{-15}$.

In Listing~\ref{lst:sun}, only the wrapper changes between the four probability calls, and with it the Hamiltonian. The cost of a call follows the number of pairs of levels that the engine tests for adiabaticity along the ray, not the content of the Hamiltonian: three pairs at three flavors, six at $3+1$, and ten at $3+2$. The non-standard interactions change the matter term but add no levels, so they cost the same as the standard case. Sterile states add levels, and so do pseudo-Dirac partners, one per split pair (Sec.~\ref{sec:shipped_hamiltonians}). The four curves take about 0.5~s, 0.5~s, 0.9~s, and 1.2~s, respectively, over ninety energies.

\begin{lstlisting}[language=Python, caption={\textbf{Averaged solar probabilities.} The four averaged solar probabilities of \figu{solar}, runnable as it stands: the BS2005-AGS,OP table\ \cite{Bahcall:2004pz} ships with \magnus, and every Sun wrapper takes it by name through {\tt density\_profile}. The same keywords serve the four wrappers. The keyword {\tt average=True} is the only difference from an instantaneous call: it returns \equ{averaged_varying} from the eigenbases at the ends of the ray, without propagating the phase. The four curves take about three seconds. Naming {\tt 'B16-GS98'} instead gives its curve in \figu{solar_models}. See notebooks \#10, \#13, and \#28, and Sec.~\ref{sec:sun} for details.}, label={lst:sun}, float=tbp]
import numpy as np
import magnus.oscprob as oscprob
import magnus.globaldefs as gd
import magnus.solarmodels as solarmodels

# The table ships with Magnus. Its columns are read
# only to plot the density (top panel of the figure)
model = 'BS05-AGS-OP'
r = solarmodels.load_solar_model(model)['r_over_r_sun']
ne_sun = solarmodels.electron_density_profile(model)
ne = ne_sun(r*gd.SUN_RADIUS*gd.UNIT_KM)

# The solar spectrum, 0.1 to 20 MeV
E = np.logspace(-1.0, np.log10(20.0), 90)*gd.UNIT_MEV
osc = gd.load_nufit_params('NuFIT 6.1')

# Shared by the four calls: the model by name, from
# the center to its last row, nu_e to nu_e, the average
R = solarmodels.table_edge(model)
run = dict(nu_i=gd.NUE, nu_f=gd.NUE, average=True,
 density_profile=model)

# Standard three-flavor oscillations
P3 = oscprob.osc_prob_3nu_sun(E, R, 0.0, **osc, **run)

# The same, with non-standard interactions
eps = dict(eps_ee=0.10, eps_em=0.05, eps_mt=0.03)
P3_nsi = oscprob.osc_prob_3nu_sun_nsi(
 E, R, 0.0, **osc, **eps, **run)

# One and two sterile states.  n_n/n_p is read from
# the table; passing ratio_number_neutrons_to_protons
# overrides it
ster = dict(s14=0.10**0.5, s24=0.10**0.5, D41=1.0)
P4 = oscprob.osc_prob_4nu_sun(
 E, R, 0.0, **osc, **ster, **run)
P5 = oscprob.osc_prob_5nu_sun(
 E, R, 0.0, **osc, **ster, s15=0.06**0.5,
 s25=0.06**0.5, D51=1.7, **run)

# P3 = 0.5449 at 0.1 MeV, 0.2993 at 20 MeV
\end{lstlisting}

Without the {\tt average} keyword, the same call returns the instantaneous probability at the surface. That answer is correct, even if it is not what \figu{solar} shows. With an oscillation length of about 160~km, the probability swings across most of the interval [0,1] within a hundred kilometers of the surface:
\begin{pysnip}
dL = np.array([0, 50, 100])*gd.UNIT_KM
P = oscprob.osc_prob_3nu_sun(
 5.0*gd.UNIT_MEV, R - dL, 0.0, **osc,
 nu_i=gd.NUE, nu_f=gd.NUE,
 density_profile=model)
# P = 0.666, 0.092, 0.259 in about 2 s
\end{pysnip}
These values are not numerical noise. The hybrid engine of Sec.~\ref{sec:hybrid} computes and certifies them: two successive passes, on finer grids, agree to the requested tolerance, $10^{-3}$ by default.


\subsubsection{Varying the production point}
\label{sec:sun_production_point}

\begin{figure}[t!]
 \centering
 \includegraphics[width=\columnwidth]{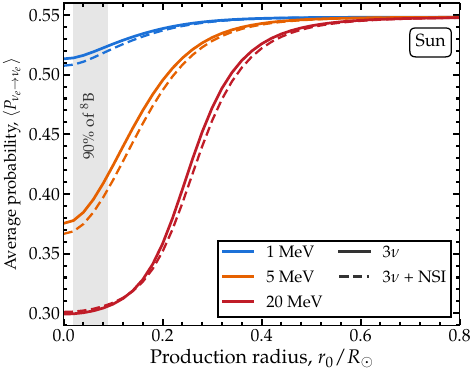}
 \caption{\label{fig:solar_production}\textbf{Dependence on the production point in the Sun.} Phase-averaged survival probability of a solar neutrino as a function of where it is produced, along the radial ray of \figu{solar}, at three energies, with and without the non-standard interactions (NSI) of \figu{solar}. The density is from the BS2005-AGS,OP solar model~\cite{Bahcall:2004pz}. At 5 and 20~MeV, each curve rises from its value at the center to the vacuum value, 0.55, as the production point moves out past the radius where the resonance density of \figu{solar} meets the profile: a sixth of the way out at 5~MeV, a third at 20~MeV. At 1~MeV, the resonance density lies above the central density, and the curve rises only from 0.51. The band marks where $90\%$ of the $^8$B neutrinos are made, in the same model. See notebooks \#10, \#13, and \#28, and Sec.~\ref{sec:sun_production_point} for details.}
\end{figure}

Section~\ref{sec:sun} placed the neutrino at the center of the Sun. Figure~\ref{fig:solar_production} moves its production point instead, $r_0$, at fixed energy. At 20~MeV, a neutrino born at the center leaves as $\nu_2$, with an average of 0.30. The farther out it is born, the lower the density at production, and the closer the average comes to the vacuum value, 0.55, which it nearly reaches by $R_\odot/2$. The rise passes through the radius where the resonance density of \figu{solar} meets the profile, about $R_\odot/3$; a neutrino born there has an average of $\cos^4\theta_{13}/2 \approx 0.48$. As the energy drops, the rise moves inward, to about $R_\odot/6$ at 5~MeV. At 1~MeV, the resonance density lies above the central density, and the curve rises only mildly, from 0.51 to 0.55. Each point of the figure is one call with {\tt average=True}, as in Listing~\ref{lst:sun}, with a different {\tt L0}:
\begin{pysnip}
P = oscprob.osc_prob_3nu_sun(
 5.0*gd.UNIT_MEV, R,
 0.05*gd.SUN_RADIUS*gd.UNIT_KM,
 **osc, **run)
# P = 0.389, against 0.375 from the center
\end{pysnip}
At 5~MeV, moving the production point from the center to $0.05\,R_\odot$, in the middle of where the $^8$B neutrinos are made, raises the average by 0.014.

The $^8$B neutrinos are made between $0.02\,R_\odot$ and $0.09\,R_\odot$ (the band in \figu{solar_production}). The density is lowest at the outer edge of the band, and even there it exceeds the resonance density of a neutrino with more than 2.6~MeV. So a $^8$B neutrino with more than 2.6~MeV is born at a density above its resonance density, wherever in the band it is made, just like one born at the center. That is why \figu{solar}, computed for a neutrino born at the center, holds for the $^8$B neutrinos at high energy: at 20~MeV, the average is 0.301 from $0.05\,R_\odot$, against 0.299 from the center. The non-standard interactions shift the curves by up to 0.012.


\subsubsection{Probability averaging}
\label{sec:averaging_is_not_estimating}

In principle, the averaged probability could be estimated from the instantaneous one, as its mean over a window of baselines at the end of the ray. On the Sun, that estimate is close, but biased. The window must span many oscillation lengths for the mean to wash out the oscillation, yet across such a window the density changes, and so does the averaged probability, which depends on the density at the end of the ray. In notebook \#13, for a 5-MeV neutrino at two flavors, on a ray from the center through the inner tenth of the solar radius, the mean over the last 6, 12, 24, and 48 oscillation lengths lies between 0.596 and 0.605, while the true averaged probability at the end of the ray is 0.596. The bias is small, but it grows as the window widens, from below $10^{-6}$ over 6 oscillation lengths to 0.009 over 48.

A scan of the instantaneous probability along the ray has a second problem: aliasing. A scan needs at least two points per cycle of the oscillation it samples. With fewer, each value is still correct, but a curve drawn through them is not the probability. On the Sun, the fastest oscillation of a 5-MeV neutrino, between the first and third eigenstates, has a length of 4.9~km, so a scan of the last 100~km in steps of 50~km is aliased. Passing {\tt strategy\_info} reports it:
\begin{pysnip}
dL = np.array([0, 50, 100])*gd.UNIT_KM
info = {}
P = oscprob.osc_prob_3nu_sun(
 5.0*gd.UNIT_MEV, R - dL, 0.0, **osc,
 nu_i=gd.NUE, nu_f=gd.NUE,
 density_profile=model,
 strategy_info=info)
s = info['sampling']
s['cycles_per_step']          # 10.1
s['aliased']                  # True
s['nyquist_points']           # 277028
\end{pysnip}
Ten cycles fit between neighboring points. To resolve every cycle from the center to the surface, a scan would need $2.8 \cdot 10^5$ points. \magnus\ reports aliasing, but does not warn about it, since nearly every realistic scan of a solar ray is aliased.

The keyword {\tt average=True} avoids both problems: it takes no window and runs no scan. It evaluates the limit directly from the eigenbases at the two ends of the ray (Sec.~\ref{sec:averaged}). At two flavors, on an adiabatic passage, the averaged survival probability of a neutrino started decohered has the textbook form
\begin{equation}
 \label{equ:two_flavor_adiabatic}
 \langle P^{2\nu}_{\nu_e \to \nu_e} \rangle_{\rm approx}
 =
 \tfrac12 + \tfrac12\cos2\theta_m(l_0)\cos2\theta_m(l_1) \;,
\end{equation}
with $\theta_m$ the mixing angle in matter at the two ends of the ray. Notebook \#13, with {\tt average\_initial\_state='decohered'}, reproduces this expression to within $4 \cdot 10^{-16}$ across 1--20~MeV.


\subsubsection{Comparing solar models}
\label{sec:solar_model_comparison}

{\def\arraystretch{1.2}
\begin{table}[t!]
 {\small
 \begin{tabular}{@{}lp{5.9cm}@{}}
 {\tt density\_profile} & Solar density profile \\
 \hline
 {\tt 'exp'} & Exponential fit~\cite{Giunti:2007ry}; the default \\
 {\tt 'BP2000'} & BP2000~\cite{Bahcall:2000nu}; to $0.95\,R_\odot$ \\
 {\tt 'BP04'} & BP04~\cite{Bahcall:2004fg}; to $0.95\,R_\odot$ \\
 {\tt 'BS05-OP'} & BS2005, GS98 composition~\cite{Bahcall:2004pz}; to $0.98\,R_\odot$ \\
 {\tt 'BS05-AGS-OP'} & BS2005, AGS05 composition~\cite{Bahcall:2004pz}; to $0.98\,R_\odot$ \\
 {\tt 'B16-GS98'} & B16, GS98 composition~\cite{Vinyoles:2016djt} \\
 {\tt 'B16-AGSS09met'} & B16, AGSS09met composition~\cite{Vinyoles:2016djt} \\
 {\tt 'B23-GS98'} & B23, GS98 composition~\cite{Herrera:2023b23} \\
 {\tt 'B23-AGSS09'} & B23, AGSS09 composition~\cite{Herrera:2023b23} \\
 {\tt 'B23-C11'} & B23, C11 composition~\cite{Herrera:2023b23} \\
 {\tt 'B23-AAG21'} & B23, AAG21 composition~\cite{Herrera:2023b23} \\
 {\tt 'B23-MB22m'} & B23, MB22 meteoritic composition~\cite{Herrera:2023b23} \\
 {\tt 'B23-MB22p'} & B23, MB22 photospheric composition~\cite{Herrera:2023b23} \\
 \end{tabular}}
 \caption{\label{tab:solar_models}\textbf{Solar density profiles.} The solar density profiles that every Sun wrapper takes through {\tt density\_profile}. The tabulated standard solar models ship with \magnus. The B16 and B23 tables reach the surface. Past the last row of the older models, the density continues along the slope of the last tabulated interval, unless {\tt stop\_at\_table\_edge=True}, which returns {\tt NaN} there, with a warning. With sterile states, $n_n/n_p$ is read from the same table, unless overridden with {\tt ratio\_number\_neutrons\_to\_protons}. The solar compositions are GS98~\cite{Grevesse:1998bj}, AGS05~\cite{Asplund:2004eu}, AGSS09~\cite{Asplund:2009fu}, C11~\cite{Caffau:2010qc}, AAG21~\cite{Asplund:2021aag}, and MB22~\cite{Magg:2022rxb}. See notebook \#13 and Sec.~\ref{sec:solar_model_comparison} for details.}
\end{table}}

\begin{figure}[t!]
 \centering
 \includegraphics[width=\columnwidth]{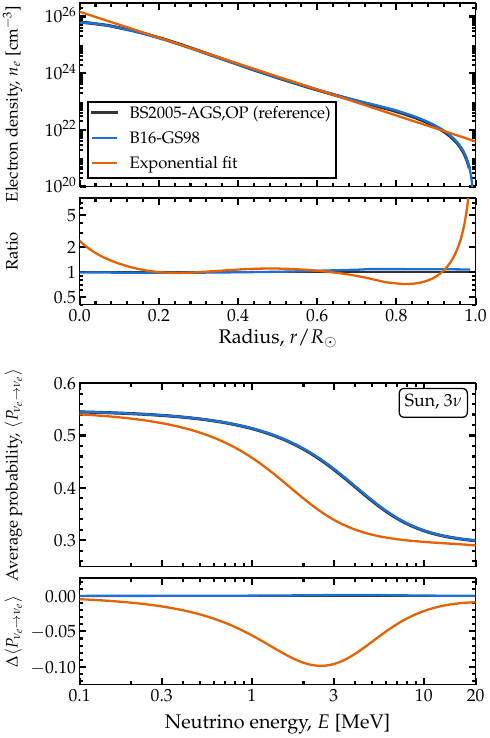}
 \caption{\label{fig:solar_models}\textbf{Three solar models compared.} {\it Top}: the tabulated BS2005-AGS,OP\ \cite{Bahcall:2004pz} and B16-GS98\ \cite{Vinyoles:2016djt} standard solar models, and the exponential fit that the Sun wrappers use by default, $n_e = 245\,N_A\,e^{-10.54\,r/R_\odot}$~cm$^{-3}$\ \cite{Giunti:2007ry}, with $N_A$ the Avogadro number; underneath, the ratio of each to BS2005-AGS,OP. The two tables agree to within about 2\% over the inner half of the Sun and 9\% over the outer half. The fit is 2.4 times too dense at the center and several times too dense in the outermost layers. {\it Bottom}: $\langle P_{\nu_e \to \nu_e}\rangle$ at three flavors on each profile; underneath, the difference from the BS2005-AGS,OP result. The two tables give the same probability to within $1.3 \cdot 10^{-3}$; the fit is low by up to 0.1, at 2.5~MeV. See notebooks \#10, \#13, and \#28, and Sec.~\ref{sec:solar_model_comparison} for details.}
\end{figure}

\tabl{solar_models} lists the solar density profiles that every Sun wrapper accepts through {\tt density\_profile}. Besides an exponential profile, which is the default, \magnus\ ships twelve tabulated standard solar models from five series published between 2001 and 2023, several of them in more than one solar composition. Each is selected by name, which is not case-sensitive, \eg, {\tt density\_profile='B16-GS98'}.

From the tabulated mass density, $\rho$, and hydrogen mass fraction, $X$, the wrapper builds the electron number density, $n_e = \rho N_A\,[X/m_{\rm H} + 2(1 - X)/m_{\rm He}]$, with $m_{\rm H}$ and $m_{\rm He}$ the atomic masses of $^1$H and $^4$He in atomic mass units (everything heavier than hydrogen counted as $^4$He), and interpolates it log-linearly in radius. Below the first tabulated radius, it holds the density at its first value, since the core is flat. Past the last one, which lies at $0.95\,R_\odot$ for BP2000 and BP04 and at $0.98\,R_\odot$ for BS2005, it continues the density along the logarithmic slope of the last tabulated interval. Passing {\tt stop\_at\_table\_edge=True} returns {\tt NaN} there instead, with a warning. With sterile states, it also reads $n_n/n_p = [2(1-X)/m_{\rm He}]/[X/m_{\rm H} + 2(1-X)/m_{\rm He}]$ from the same table. In notebook \#13, we compare all twelve models: they give the same averaged $\nu_e$ survival probability to within $2.4 \cdot 10^{-3}$. Most of that spread is between series of models; the six compositions of the B23 series differ by at most $2 \cdot 10^{-4}$.

Figure~\ref{fig:solar_models} compares the probabilities computed with three choices of solar density profile. Two are tabulated standard solar models: BS2005-AGS,OP\ \cite{Bahcall:2004pz}, used throughout this section, and the newer B16-GS98\ \cite{Vinyoles:2016djt}, with updated nuclear rates, equation of state, and opacities. The third is the exponential fit that textbooks quote\ \cite{Giunti:2007ry} and that the Sun wrappers use by default. The densities of the two solar models are within about 2\% of each other over the inner half of the Sun and 9\% over the outer half. The exponential fit is 2.4 times too dense at the center, where the real profile flattens, and several times too dense again near the surface, where the real profile plunges.

The two tabulated models give the same averaged probability to within $1.3 \cdot 10^{-3}$ at every energy, far below the precision of any solar measurement. The exponential fit gives a probability lower by up to 0.1, at 2.5~MeV. The reason is in Sec.~\ref{sec:sun}: on an adiabatic passage, the averaged probability depends only on the eigenbases at the two ends of the ray. At the surface, the eigenbasis is that of vacuum, whatever the profile. At the center, it is set by the central density, where the fit is 2.4 times too dense. The probability falls from 0.55 at low energy to 0.30 at high energy, and the fall happens near the resonance energy, which is inversely proportional to the density. With the fit, the fall happens at energies about 2.4 times lower. The values at low and high energy are nearly the same for every profile, so the difference is largest in between, at a few MeV.


\subsubsection{Comparing against approximations}
\label{sec:sun_compare_approx}

\begin{figure}[t!]
 \centering
 \includegraphics[width=\columnwidth]{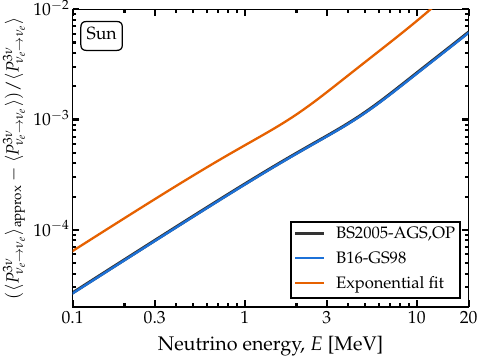}
 \caption{\label{fig:solar_approx}\textbf{Error of the two-flavor solar approximation.} Relative error of the two-flavor approximation, $\langle P^{3\nu}_{\nu_e \to \nu_e} \rangle_{\rm approx}$ of \equ{two_flavor_reduction}, to the averaged solar survival probability $\langle P^{3\nu}_{\nu_e \to \nu_e} \rangle$ computed with \magnus, on the three solar density profiles of \figu{solar_models}. The approximation overestimates the probability at every energy, by an amount that grows in proportion to the energy. On the two tabulated models, it reaches 0.6\% at 20~MeV. On the exponential fit, whose central density is 2.4 times higher, it is 2.2 to 3.4 times larger, and its curve leaves the panel above 12~MeV. See notebook \#28 and Sec.~\ref{sec:sun_compare_approx} for details.}
\end{figure}

Solar analyses rarely solve the three-flavor problem. The standard approximation\ \cite{Kuo:1989qe} reduces it to two flavors in two steps. First, the 1--2 sector is solved on its own, as a two-flavor problem with $\theta_{12}$ and $\Delta m^2_{21}$, on the electron density multiplied by $\cos^2\theta_{13}$. Second, $\theta_{13}$ is folded back in through
\begin{equation}
 \label{equ:two_flavor_reduction}
 \langle P^{3\nu}_{\nu_e \to \nu_e} \rangle_{\rm approx}
 =
 \sin^4\theta_{13} + \cos^4\theta_{13}\,\langle P^{2\nu}_{\nu_e \to \nu_e} \rangle_{\rm approx} \;.
\end{equation}
Here, $\langle P^{2\nu}_{\nu_e \to \nu_e} \rangle_{\rm approx}$ is \equ{two_flavor_adiabatic}, with $\theta_m$ computed from $\theta_{12}$, $\Delta m^2_{21}$, and the electron density multiplied by $\cos^2\theta_{13}$. The approximation rests on $\Delta m^2_{31}$ being large compared with $\Delta m^2_{21}$ and with the matter potential, so that the third mass state decouples. In vacuum, the approximation is exact. In matter, it drops terms of relative size $2 E V_{\rm CC}/\Delta m^2_{31}$, about 0.1 at the center of the Sun for a 20-MeV neutrino, multiplied by the admixture of the decoupled state, $\sin^2\theta_{13} \approx 0.02$.

Figure~\ref{fig:solar_approx} evaluates them on the three density profiles of \figu{solar_models}. The three-flavor result, $\langle P^{3\nu}_{\nu_e \to \nu_e} \rangle$, is computed with \magnus: {\tt P3} of Listing~\ref{lst:sun} for the BS2005-AGS,OP table, and the same call on the other two profiles. The two-flavor result needs the density multiplied by $\cos^2\theta_{13}$, which a wrapper cannot provide, since it takes the solar model by name. So it goes through the scenario function, which accepts any profile, here {\tt ne\_sun} of Listing~\ref{lst:sun} multiplied by $\cos^2\theta_{13}$:
\begin{pysnip}
# The 1-2 sector on cos^2 th13 * n_e
c13sq = 1.0 - osc['s13']**2   # cos^2 th13
P2 = oscprob.osc_prob_matter_std_potential(
 2, lambda l: c13sq*ne_sun(l), E, R,
 dict(sth=osc['s12'], Dm2=osc['D21']),
 L0=0.0, nu_i=gd.NUE, nu_f=gd.NUE, average=True,
 density_is_of_number_of_electrons=True)
# Fold theta_13 back in
P_approx = osc['s13']**4 + c13sq**2*P2
(P_approx - P3)/P3   # 6e-3 at 20 MeV
\end{pysnip}
The {\tt P2} equals \equ{two_flavor_adiabatic} to within $4 \cdot 10^{-16}$ at every energy.

The approximation overestimates the probability at every energy, on every profile, and its relative error grows in proportion to the energy, as the size of the dropped terms does. On the two tabulated models, which give nearly the same probability (Sec.~\ref{sec:solar_model_comparison}), the error is about $3 \cdot 10^{-5}$ at 0.1~MeV, $3 \cdot 10^{-4}$ at 1~MeV, and 0.6\% at 20~MeV. On the exponential fit, it is 2.2 to 3.4 times larger, 2\% at 20~MeV, because the fit's central density is 2.4 times higher and the dropped terms grow with it.


\subsection{A supernova shock front}
\label{sec:shock}

\begin{figure*}[t!]
 \centering
 \includegraphics[width=\textwidth]{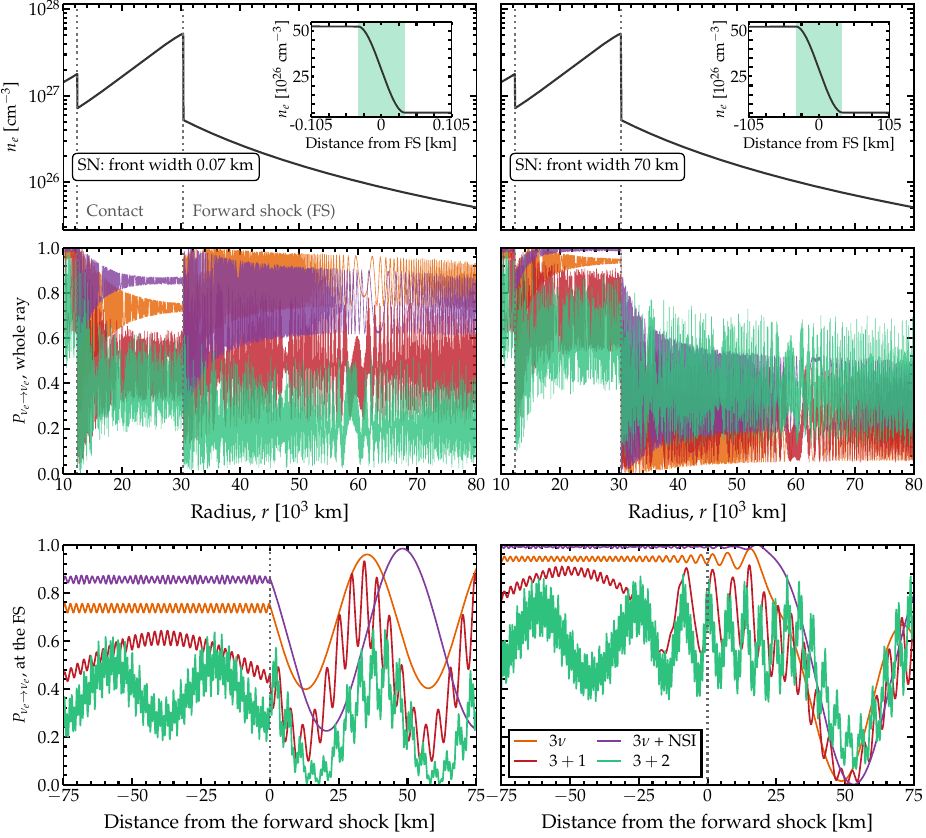}
 \caption{\label{fig:shock}\textbf{A supernova shock front.} Electron density along a supernova ray, and the survival probability of a 15-MeV $\nu_e$ along it. {\it Top}: the density from $10^{4}$ to $8 \cdot 10^{4}$~km, with the contact discontinuity at 12\,348~km, where the density jumps by a factor of 2.5, and the forward shock at 30\,323~km, where it jumps tenfold. The columns differ only in the width of both fronts: 0.07~km on the left, 70~km on the right. The insets enlarge the forward shock, with its width shaded. {\it Center}: the probability along the ray, computed with \magnus\ at ${\tt rtol} = 10^{-8}$ for four Hamiltonians. The ray holds thousands of oscillation lengths of the pair split by $\Delta m^2_{31}$, so the panel shows the envelope of the oscillation. The two columns agree bitwise up to the contact discontinuity. {\it Bottom}: a window of $\pm$75~km around the forward shock, where the oscillation is resolved. Two flavors are not drawn: with $\Delta m^2_{21}$ alone, the probability stays above 0.95. The profile follows \Refs\ \cite{Schirato:2002tg, Fogli:2003dw}, with the radii of a simulation snapshot\ \cite{Kneller:2014oea}. Listing~\ref{lst:shock} computes the left column. See notebooks \#14 and \#28, and Sec.~\ref{sec:shock} for details.}
\end{figure*}

Figure~\ref{fig:shock} shows the survival probability of a 15-MeV $\nu_e$ along a supernova ray crossed by a shock, for two widths of the shock fronts, 0.07 and 70~km. The forward shock is the blast wave of the explosion. It propagates outward through an envelope whose density falls as $r^{-2.4}$, and compresses the matter behind it tenfold. Behind it lies the contact discontinuity, which separates the shocked envelope from the ejecta; across it, the density drops by a factor of 2.5, while the pressure is continuous. An outgoing neutrino thus crosses two density jumps: first the contact discontinuity, then the forward shock. Between them lies the shocked shell, thinned by a rarefaction. Because a shock alters the adiabaticity of the level crossings, it leaves a signature in the neutrino signal\ \cite{Schirato:2002tg, Fogli:2003dw}.

At three flavors, only the pair of eigenstates split by $\Delta m^2_{31}$ is affected. The resonance density of the pair split by $\Delta m^2_{21}$ is 7 times lower than the lowest density on the ray, so that pair never reaches resonance; accordingly, the two-flavor survival probability stays above 0.95. At 15~MeV, the resonance density of the pair split by $\Delta m^2_{31}$ lies between the densities on either side of the forward shock, so the resonance is crossed at the front, where the oscillation length is about 50~km.

The 0.07-km front is crossed suddenly, before the state can follow the eigenstates. Across the 70-km front, which is comparable to the oscillation length, the state partly follows them. As a result, beyond the forward shock, the mean survival probability at three flavors is 0.84 for the sharp fronts and 0.20 for the wide ones. The other Hamiltonians shift by different amounts; for 3+2, the shift is reversed, from 0.22 to 0.38.

Listing~\ref{lst:shock} computes the curves of the left column. It declares both fronts through {\tt t\_breakpoints}, as the positions where each begins and ends, so that every level of the refinement places slab edges there. Without the breakpoints, a 0.07-km front falls inside a slab at every level, and a slab that contains a jump degrades the quadrature regardless of the number of slabs. The three-flavor scan is then wrong beyond the forward shock by up to 0.25, and \magnus\ warns:
\begin{pysnip}
P = oscprob.osc_prob_matter_std_potential(
 3, ne_shock, 15.0*gd.UNIT_MEV, Ls, osc,
 L0=10000.0*gd.UNIT_KM, rtol=1.0e-8,
 atol=1.0e-10, nu_i=gd.NUE, nu_f=gd.NUE,
 density_is_of_number_of_electrons=True)
# UnmarkedDiscontinuityWarning: the
# Hamiltonian is discontinuous at the scale
# of the grid this scan builds, and no
# t_breakpoints were given.  A slab
# straddling a density jump degrades the
# quadrature no matter how many slabs are
# used; pass t_breakpoints at the
# discontinuities.
\end{pysnip}
With or without breakpoints, the scan also raises two tolerance warnings, because at ${\tt rtol} = 10^{-8}$ the refinement reaches its cap of 20\,000 slabs before certifying the result. With the fronts declared, the cap is immaterial: raising it tenfold changes the three-flavor and NSI curves by about $10^{-5}$, and the $3+1$ and $3+2$ curves by up to $4 \cdot 10^{-3}$, below the resolution of \figu{shock}. The scans go to the cumulative engine, as the last line of the three-flavor call confirms; declared breakpoints would keep the hybrid engine of Sec.~\ref{sec:hybrid} out in any case. Notebook \#14 checks the result against an independent integration, and Sec.~\ref{sec:shock_comparison} measures the cost of the front width.

\begin{lstlisting}[language=Python, caption={\textbf{Through a supernova shock front.} The left column of \figu{shock}: for each Hamiltonian, one scan of 4\,000 baselines at one energy. {\tt t\_breakpoints} declares where each front begins and ends, so every refinement level puts slab edges there. Notebook \#14 builds {\tt ne\_shock}, the electron density of \figu{shock}, and the oscillation parameters, {\tt osc}. See Sec.~\ref{sec:shock} for details.}, label={lst:shock}, float=tbp]
import numpy as np
import magnus.oscprob as oscprob
import magnus.globaldefs as gd

# Both fronts, 0.07 km wide, centered on the contact
# discontinuity and on the forward shock: declare
# where each begins and ends
w = 0.07*gd.UNIT_KM
edges = [r + s*w/2 for r in (12348.0*gd.UNIT_KM,
 30323.0*gd.UNIT_KM) for s in (-1.0, 1.0)]

# 4000 baselines from 10,200 to 80,000 km, at one
# energy: the cumulative engine answers
Ls = np.linspace(10200.0, 80000.0, 4000)*gd.UNIT_KM
run = dict(L0=10000.0*gd.UNIT_KM, t_breakpoints=edges,
 rtol=1.0e-8, atol=1.0e-10, nu_i=gd.NUE, nu_f=gd.NUE,
 density_is_of_number_of_electrons=True)
E = 15.0*gd.UNIT_MEV

# Standard three-flavor oscillations
info = {}
P3 = oscprob.osc_prob_matter_std_potential(
 3, ne_shock, E, Ls, osc, strategy_info=info, **run)
info['engine']   # 'cumulative'

# The same, with non-standard interactions
eps = dict(eps_ee=0.10, eps_em=0.05, eps_et=0.0,
 eps_mm=0.0, eps_mt=0.03, eps_tt=0.0)
P3_nsi = oscprob.osc_prob_matter_nsi(
 3, ne_shock, E, Ls, osc, eps, **run)

# One and two sterile states
osc4 = dict(osc, s14=0.10**0.5, s24=0.10**0.5,
 s34=0.0, D41=1.0, d14=0.0, d24=0.0)
P4 = oscprob.osc_prob_matter_std_potential(
 4, ne_shock, E, Ls, osc4, **run)
osc5 = dict(osc4, s15=0.06**0.5, s25=0.06**0.5,
 s35=0.0, D51=1.7, d15=0.0, d35=0.0)
P5 = oscprob.osc_prob_matter_std_potential(
 5, ne_shock, E, Ls, osc5, **run)
\end{lstlisting}

A detector measures the survival probability averaged over its energy resolution. The average washes out two phases: the one accumulated after the forward shock, thousands of cycles by the end of the ray at 15~MeV and far more on the way to Earth, and the one accumulated between the two fronts. The latter arises because the neutrino can change eigenstate at either front, so the paths through the shocked shell interfere; the interference completes a cycle every 15~keV. On the Sun, \equ{averaged_varying} required only the eigenbases at the ends of the ray, because the passage was adiabatic and $P^{\rm cross}$ was the identity. Here, the fronts are crossed suddenly, $P^{\rm cross}$ depends on their width, and the amplitudes of the two paths add before the modulus is taken.

\begin{figure}[t!]
 \centering
 \includegraphics[width=\columnwidth]{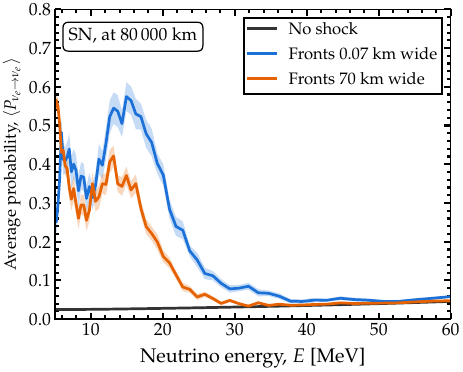}
 \caption{\label{fig:shock_energy}\textbf{The shock signature against energy.} Averaged $\nu_e$ survival probability at the end of the ray of \figu{shock}, at 80\,000~km, without a shock and with the two front widths of that figure. With a shock, each point is the mean over a $\pm$10\% energy window, which is what {\tt average=True} returns when fronts are declared; the bands are the standard errors of those means. Without a shock, the passage is adiabatic, and the probability stays at a few percent. With one, it rises between 5 and about 25~MeV, where the resonance of the pair split by $\Delta m^2_{31}$ lies at or near the forward shock. Above 30~MeV, the resonance lies in undisturbed matter, and the probability falls back toward the no-shock value. See notebooks \#14 and \#28, and Sec.~\ref{sec:shock} for details.}
\end{figure}

Without declared fronts, {\tt average=True} resolves the 70-km fronts: it returns 0.37, against $0.35 \pm 0.04$ with the fronts declared. It cannot resolve the 0.07-km fronts: it treats them as smooth, returns 0.04, close to the few percent obtained without a shock (\figu{shock_energy}), instead of 0.56, and raises {\tt UnmarkedDiscontinuityWarning}. With the fronts declared, the call returns the correct average:
\begin{pysnip}
P = oscprob.osc_prob_matter_std_potential(
 3, ne_shock, 15.0*gd.UNIT_MEV, Ls[-1], osc,
 L0=10000.0*gd.UNIT_KM, t_breakpoints=edges,
 average=True, nu_i=gd.NUE, nu_f=gd.NUE,
 density_is_of_number_of_electrons=True)
# PhaseAveragingWarning: average=True on a
# profile with discontinuities has no closed
# form, so the probability was propagated
# across an energy window of +/-10.0% and
# averaged over 41 samples.  This is the
# average over that window, not the L/E ->
# infinity limit, and it depends on the
# window: the largest standard error of the
# mean here is 4.47e-02.
# Pass average_spread to set the half-width
# of the window to the resolution of the
# measurement, and average_n_samples to
# reduce the error, which falls as the
# inverse square root of the number of
# samples.
# P = 0.56
\end{pysnip}
The warning reports the two defaults of this route, named constants of {\tt magnus.avgprob}: a window of $\pm$10\% around the energy, typical of the energy resolution of a detector, and 41 samples, spaced evenly in $1/E$, and hence in phase. 

The keywords {\tt average\_spread} and {\tt average\_n\_samples} change these values:
\begin{pysnip}
P = oscprob.osc_prob_matter_std_potential(
 3, ne_shock, 15.0*gd.UNIT_MEV, Ls[-1], osc,
 L0=10000.0*gd.UNIT_KM, t_breakpoints=edges,
 average=True, average_spread=0.05,
 average_n_samples=161, nu_i=gd.NUE,
 nu_f=gd.NUE,
 density_is_of_number_of_electrons=True)
# P = 0.584
\end{pysnip}
Unlike the limit returned on a smooth profile, this average depends on the width of the window, because the averaged probability itself varies with energy across it (\figu{shock_energy}). The width is best set to the energy resolution of the detector being modeled. The number of samples sets the statistical error of the average, which falls as the inverse square root of that number, while each sample costs a full propagation. The warning reports that error, so the number of samples needed for a target error $\delta$ follows directly: $N\,(\sigma/\delta)^2$, with $N$ the samples used and $\sigma$ the reported error. Here, 161 samples halve the error of the default, from 0.04 to 0.02, and take four times as long; an error of 0.01 needs about 800 samples. As in Listing~\ref{lst:shock}, each propagation reaches the slab cap and raises the tolerance warnings.

\begin{figure*}[t!]
 \centering
 \includegraphics[width=\textwidth]{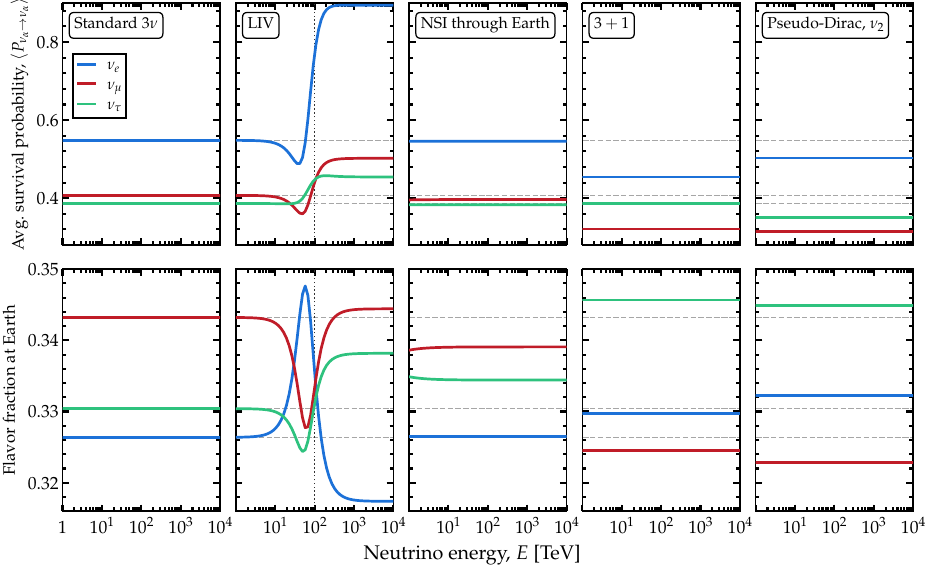}
 \caption{\label{fig:astro}\textbf{Flavor composition of an astrophysical flux.} Averaged probabilities, \equ{averaged}, for five Hamiltonians, from a source at 100~Mpc. \emph{Top}: survival probabilities, $\langle P_{\nu_\alpha \to \nu_\alpha}\rangle$. \emph{Bottom}: flavor fractions at Earth from a pion-decay source, $1:2:0$; the standard case is repeated in each panel for comparison. Lorentz-invariance violation: the operator of \equ{h_liv_3nu} with $n_{\rm LIV} = 1$, $\Lambda = 1$~eV, $b_1 = b_2 = 0$, $b_3 = 1.26 \cdot 10^{-31}$~eV, and $\mathbb{V}_\xi$ the PMNS matrix with $\delta_\xi = 0$, so that the new term equals $\Delta m^2_{31}/2E$ at 100~TeV (vertical line). Non-standard interactions: $\varepsilon_{ee} = 0.10$, $\varepsilon_{e\mu} = 0.05$, and $\varepsilon_{\mu\tau} = 0.03$, across the core of the Earth, $\cos\theta_z = -1$. Sterile state: $\sin^2\theta_{14} = \sin^2\theta_{24} = 0.1$, $\theta_{34} = 0$, and $\Delta m^2_{41} = 1$~eV$^2$. Pseudo-Dirac: $\nu_2$ split by $\delta m^2 = 10^{-13}$~eV$^2$. For the last two, the fractions are renormalized to the three active flavors. See Listing~\ref{lst:astro}, notebooks \#10, \#13, and \#28, and Sec.~\ref{sec:astro} for details.}
\end{figure*}

Figure~\ref{fig:shock_energy} shows the effect of the shock on the energy spectrum. Without a shock, the passage is adiabatic, a $\nu_e$ leaves as $\nu_3$, and the survival probability is a few percent at all energies. With a shock, it rises to tens of percent between 5 and about 25~MeV. The window is set by where the resonance density, which falls as $1/E$, sits on the profile. Between 1.8 and 18~MeV, the resonance density lies within the jump at the forward shock, so the resonance is crossed at the front. Above 18~MeV, it lies beyond the front, in undisturbed matter, where the crossing is adiabatic; the jump still changes the mixing angle, but less at higher energy, so the effect fades. Between 5.3 and 13~MeV, the resonance density also lies within the jump at the contact discontinuity. The resonance is then crossed at both fronts, and the result depends on the phase accumulated between them. Across most of the window, the sharp fronts give the larger probability: the oscillation length at the resonance, 20 to 60~km between 6 and 18~MeV, is comparable to the 70-km ramp, so the state partly follows the wide fronts. Near 5~MeV, the wide fronts give the larger probability.


\subsection{High-energy astrophysical neutrinos}
\label{sec:astro}

Figure~\ref{fig:astro} shows the flavor composition of a TeV--PeV astrophysical flux at Earth and the average flavor-transition probabilities behind it, for five Hamiltonians: standard three-flavor, Lorentz-invariance violation (LIV), non-standard interactions (NSI) through the Earth, $3+1$ flavors, and pseudo-Dirac neutrinos with the $\nu_2$ mass eigenstate split. The top row shows the survival probabilities $\langle P_{\nu_\alpha \to \nu_\alpha}\rangle$; the bottom row, the flavor fractions reaching Earth from a source that produces neutrinos through the full pion decay chain, described next.

\smallskip

\textit{How the neutrinos are produced.}---In cosmic accelerators---active galaxies, gamma-ray bursts, superluminous supernovae---protons may interact with ambient matter and radiation to produce pions\ \cite{Margolis:1977wt, Stecker:1978ah, Kelner:2006tc}. The pions decay via $\pi^+ \to \mu^+ + \nu_\mu$, followed by $\mu^+ \to \bar{\nu}_\mu + \nu_e + e^+$, and their charge conjugates, so that the flux leaving the source carries flavor ratios $(f_e, f_\mu, f_\tau)_{\rm S} = \left(\frac{1}{3}, \frac{2}{3}, 0\right)$, with $f_{\alpha, {\rm S}}$ the fraction of $\nu_\alpha + \bar{\nu}_\alpha$ in the total. Neutrino telescopes cannot ordinarily tell $\nu$ from $\bar\nu$, so $\nu_\alpha$ below means $\nu_\alpha + \bar{\nu}_\alpha$. For illustration, we adopt that full pion-decay composition---the nominal expectation---though there are other possibilities~(see, \eg, \Ref~\cite{Bustamante:2015waa}).

\smallskip

\textit{What is observable.}---Over Mpc--Gpc baselines, the oscillation length is minute against the distance; neither the distance, the size of the production region, nor the energy resolution of a detector is known to anything approaching the precision the phase would demand for the oscillations to be resolved. As a result, a neutrino telescope is sensitive only to the average flavor-transition probabilities $\langle P_{\nu_\alpha \to \nu_\beta}\rangle$ of \equ{averaged}.

\begin{figure*}[t!]
 \centering
 \includegraphics[width=\textwidth]{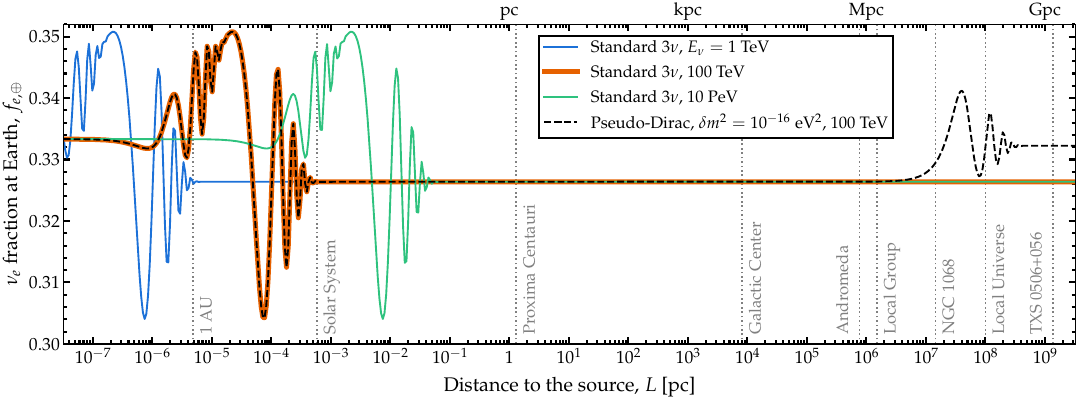}
 \caption{\label{fig:astro_baseline}\textbf{The averaged flavor composition settles once the source is far enough.} Fraction of $\nu_e$ at Earth, $f_{e,\oplus}$, of a flux produced by pion decay, $\left(\frac{1}{3}:\frac{2}{3}:0\right)_{\rm S}$, against the distance to the source, computed with {\tt average=True}, \ie, the phase average of \equ{phase_average} at the default 10\% energy spread. Curves are for standard three-flavor oscillations in vacuum at 1~TeV, 100~TeV, and 10~PeV, and for a pseudo-Dirac spectrum, with $\nu_2$ split by $\delta m^2 = 10^{-16}$~eV$^2$, at 100~TeV; the latter lies on the 100-TeV standard curve out to about 8~Mpc. Close to the source, the oscillations are only partly averaged, and the curves depend on the energy spread. Farther out, every phase has averaged away and the composition settles, at $f_{e,\oplus} = 0.326$ under standard oscillations, at a distance that grows with the energy. The pseudo-Dirac pair, with its much smaller splitting, settles a second time, at $f_{e,\oplus} = 0.332$, beyond about 250~Mpc. See Sec.~\ref{sec:astro} and notebook \#28 for details.}
\end{figure*}

With {\tt average=True}, \magnus\ computes this probability in closed form from the eigenvectors of $\mathbb{H}$, without propagating a phase. (It returns the phase average of \equ{phase_average}, which over these distances equals \equ{averaged}, since every phase runs through many cycles.) Under standard oscillations in vacuum, $\mathbb{V}$ in \equ{averaged} is the PMNS matrix, $\mathbb{R}$, and the probability is the familiar $\sum_i \lvert \mathbb{R}_{\alpha i}\rvert^2 \lvert \mathbb{R}_{\beta i}\rvert^2$. The flavor ratios at Earth are then
\begin{equation}
 \label{equ:flavor_ratio_earth}
 f_{\alpha, \oplus}
 =
 \sum_\beta \langle P_{\nu_\beta \to \nu_\alpha}\rangle\, f_{\beta, {\rm S}} \;.
\end{equation}
The flavor composition is a rich probe of astrophysics and fundamental physics; see, \eg, \Refs~\cite{Esmaili:2009dz, Barenboim:2003jm, Beacom:2003nh, Kachelriess:2006ksy, Lipari:2007su, Hummer:2010ai, Bustamante:2015waa, Arguelles:2015dca, Rasmussen:2017ert, Bustamante:2019sdb, Ackermann:2019cxh, Ackermann:2019ows, Arguelles:2019rbn, Bustamante:2020bxp, Song:2020nfh, Liu:2023flr}. With NuFIT~6.1 best-fit values of the mixing parameters, the flavor fractions at Earth are nearly equal, $(0.33, 0.34, 0.33)_\oplus$: oscillations over astrophysical distances create a $\nu_\tau$ component even when the source produces none. Run backward, the measured flavor composition can constrain the sources\ \cite{Bustamante:2019sdb, Song:2020nfh, Liu:2023flr, IceCube:2025ole}.

If the vacuum Hamiltonian is extended, \eg, by mixing with sterile states or by a new-physics term such as Lorentz-invariance violation, the averaged probability keeps the same form, $\sum_i \lvert \mathbb{V}_{\alpha i}\rvert^2 \lvert \mathbb{V}_{\beta i}\rvert^2$, with $\mathbb{V}$ the eigenvectors of the full Hamiltonian\ \cite{Ahlers:2018yom, Ahlers:2020miq}. Matter, which the neutrinos meet only when they cross the Earth, is treated differently; we return to it below.

Listing~\ref{lst:astro} computes the five cases of \figu{astro}. The standard, Lorentz-violating, and $3+1$ cases take one wrapper call each, with {\tt average=True}, over all sixty energies at once. Each call takes the distance to the source, which may be surprising: the averaged probability is usually written without a baseline, because it is the limit of \equ{averaged}, $L/E \to \infty$, in which every interference term has averaged away. However, \magnus\ does not assume that limit. With {\tt average=True}, it returns the phase average of \equ{phase_average}, which damps each interference term according to the size of its phase, $\Delta m^2_{ij} L/2E$, and so needs $L$. Where every phase is large, the two coincide: at the default energy spread of 10\%, a phase of 40~rad keeps $3 \cdot 10^{-4}$ of its interference (Sec.~\ref{sec:averaged}), and \magnus\ then returns \equ{averaged} itself. For the mass splittings of standard oscillations, this holds at the distance of any real source: the smallest phase, from $\Delta m^2_{21}$ at 10~PeV, the highest energy of \figu{astro}, exceeds 40~rad beyond about 0.07~pc. Hence, every astrophysical baseline gives the same result, to the last digit; for instance, continuing Listing~\ref{lst:astro},
\begin{pysnip}
# One megaparsec, in natural units
Mpc = 3.0857e19*gd.UNIT_KM
# The standard case at 1 Mpc and at 1 Gpc
P1 = oscprob.osc_prob_3nu_vacuum(E,
 1.0*Mpc, average=True, **osc)
P2 = oscprob.osc_prob_3nu_vacuum(E,
 1.0e3*Mpc, average=True, **osc)
# Largest difference over all energies
# and channels: none
np.abs(P1 - P2).max()        # 0.0
\end{pysnip}
The same holds for the Lorentz-violating and $3+1$ cases of \figu{astro}. The baseline matters only when some phase is not large: over a short baseline, or for a splitting far smaller than $\Delta m^2_{21}$, as in the pseudo-Dirac case below. Over $10^8$~km, less than an astronomical unit, the $\Delta m^2_{21}$ phase is too small to average away at most energies of the figure. The wrapper warns, and returns the phase average over the 10\% energy spread, which depends on that spread:
\begin{pysnip}
P = oscprob.osc_prob_3nu_vacuum(E,
 1.0e8*gd.UNIT_KM, average=True, **osc)
# PhaseAveragingWarning: depends on
# the energy spread at 46 of 60 points
at_earth(P[-1]).round(3)
# [0.333 0.666 0.001]
\end{pysnip}
At 10~PeV, the composition is still that of the source: over this distance, the oscillation has barely begun.

Figure~\ref{fig:astro_baseline} follows the composition from $10^6$~km out to Gpc distances. Under standard oscillations, $f_{e,\oplus}$ settles, to within $10^{-3}$, beyond about 1~AU at 1~TeV, $4 \cdot 10^{-4}$~pc at 100~TeV, and 0.04~pc at 10~PeV: the phases grow as $L/E$, so the higher the energy, the farther out the composition settles. Every known or suspected source of high-energy astrophysical neutrinos lies far beyond that. A pseudo-Dirac pair adds a phase of its own, $\delta m^2 L/2E$, set by a much smaller splitting. With $\delta m^2 = 10^{-16}$~eV$^2$, at 100~TeV, the composition leaves the standard value at about 8~Mpc and settles on a new one, $f_{e,\oplus} = 0.332$, beyond about 250~Mpc, which are the distances of real sources. For such a spectrum, the baseline passed must be the actual one; we return to it at the end of this section.

\begin{figure}[t!]
 \centering
 \includegraphics[width=\columnwidth]{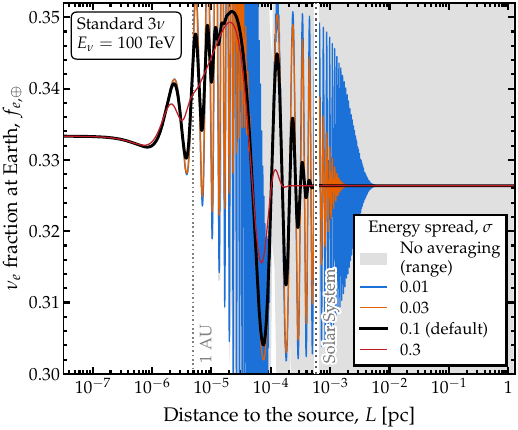}
 \caption{\label{fig:astro_spread}\textbf{The energy spread sets where the averaged composition settles, not its value.} Fraction of $\nu_e$ at Earth, $f_{e,\oplus}$, from a pion-decay source, against the distance to the source, for standard three-flavor oscillations in vacuum at 100~TeV. The curves are the phase average of \equ{phase_average} for relative energy spreads $\sigma = 0.01$, 0.03, 0.1 (the default of {\tt average\_spread}), and 0.3. The distance beyond which each curve settles scales as $1/\sigma$; all of them settle at the same value, $f_{e,\oplus} = 0.326$. The gray band is the range swept by the unaveraged $f_{e,\oplus}$, which never settles. Dotted lines mark 1~AU and the heliopause, at 120~AU. See Sec.~\ref{sec:astro} and notebook \#28 for details.}
\end{figure}

Figure~\ref{fig:astro_spread} shows the $\nu_e$ fraction at Earth, $f_{e,\oplus}$, from a pion-decay source, against the distance to the source, for standard three-flavor oscillations in vacuum at 100~TeV. The curves are computed with {\tt average=True} for four relative energy spreads, $\sigma = 0.01$, 0.03, 0.1 (the default of {\tt average\_spread}), and 0.3. The gray band is the range swept by $f_{e,\oplus}$ computed without averaging. The spread sets how strongly each interference term is damped [\equ{phase_average}], and so the distance at which the composition stops changing: as the spread grows from 1\% to 30\%, the distance beyond which $f_{e,\oplus}$ settles, to within $10^{-3}$, shrinks as $1/\sigma$, from $4 \cdot 10^{-3}$~pc to $1.4 \cdot 10^{-4}$~pc. The value it settles to, 0.326, is the same for every spread. Without averaging, the composition never settles: it keeps oscillating at every distance. Hence, the exact value of the spread passed with {\tt average\_spread} matters only for a source close enough that the phases have not yet averaged away; for any realistic astrophysical source of high-energy neutrinos, the result does not depend on it.

In \figu{astro}, only the matter case propagates a probability, and only across the Earth; the other four cases, and the vacuum leg of the matter case, are closed-form averages. The flux decoheres in the vacuum mass basis long before it reaches the Earth, and then crosses the Earth \emph{coherently}. Therefore, the two legs combine as probability matrices, not as amplitudes, and the flavor ratios at Earth are
\begin{equation}
 \label{equ:astro_earth_leg}
 f_{\beta,\oplus}
 =
 \sum_{\alpha,\gamma} f_{\alpha, {\rm S}}\,
 \langle P_{\nu_\alpha \to \nu_\gamma}\rangle\,
 P^{\oplus}_{\nu_\gamma \to \nu_\beta} \;,
\end{equation}
where the first factor is the averaged probability in vacuum and the second is an ordinary propagation through the density profile of the Earth. In Listing~\ref{lst:astro}, this is the product {\tt P\_std @ P\_in}. (The Earth call starts refinement at 32 slabs only to avoid a harmless convergence warning on the coarser default grid.)

\smallskip

\begin{lstlisting}[language=Python, caption={\textbf{Flavor composition of astrophysical neutrinos.} The five cases of \figu{astro}: the phase-averaged matrix $\langle P_{\nu_\alpha \to \nu_\beta}\rangle$ of \equ{averaged} for each Hamiltonian, over the sixty energies of the figure, contracted with the pion-decay source fractions. The standard, Lorentz-violating, and $3+1$ cases are one wrapper call each with {\tt average=True}; the NSI case composes the averaged vacuum matrix with one propagation through the core; the pseudo-Dirac case is one call to {\tt osc\_prob\_pseudo\_dirac\_vacuum}, which forms each pair's phase from its splitting. {\tt f\_earth} holds the five compositions. See Sec.~\ref{sec:astro} for details.}, label={lst:astro}, float=t]
import numpy as np
import magnus.oscprob as oscprob
import magnus.earth as earth
import magnus.globaldefs as gd

osc = gd.load_nufit_params('NuFIT 6.1')
E = np.logspace(3, 7, 60)*gd.UNIT_GEV   # 1 TeV-10 PeV
f_src = np.array([1/3, 2/3, 0.0])       # pion decay
# A source 100 Mpc away: far enough for every phase
# to average away at every energy.  The average
# depends on the baseline through the phases
L_src = 100.0*3.0857e19*gd.UNIT_KM

def at_earth(P):
    # The source fractions, zero for a sterile state,
    # contracted with the averaged matrix; the active
    # fractions renormalized to one
    f = np.zeros(len(P)); f[:3] = f_src
    f = (f @ P)[:3]
    return f/f.sum()

# Standard: the averaged matrix at each energy
P_std = oscprob.osc_prob_3nu_vacuum(E, L_src,
 average=True, **osc)                   # (60, 3, 3)

# LIV: the n = 1 operator, aligned with the PMNS
# angles, sized to equal the vacuum term at 100 TeV
b3 = osc['D31']/(2*(100.0*gd.UNIT_TEV)**2)
P_liv = oscprob.osc_prob_3nu_vacuum_liv(E, L_src,
 average=True, sxi12=osc['s12'], sxi23=osc['s23'],
 sxi13=osc['s13'], dxiCP=0.0, b1=0.0, b2=0.0,
 b3=b3, Lambda=1.0, n_liv=1, **osc)

# NSI through the Earth: the flux arrives decohered,
# then crosses the Earth coherently, so the two legs
# compose as probability matrices.  The ladder starts
# at 32 slabs: the matter term alone winds more than
# pi across a coarser one
eps = dict(eps_ee=0.10, eps_em=0.05, eps_mt=0.03)
L_in = earth.distance_traveled_inside_earth(-1.0)
P_in = oscprob.osc_prob_3nu_earth_nsi(E, costhz=-1.0,
 L=L_in*gd.UNIT_KM, n_slabs=32, rtol=1e-6, atol=1e-8,
 **osc, **eps)
P_nsi = np.asarray(P_std) @ np.asarray(P_in)

# 3+1: one sterile state, sin^2 of both active-
# sterile angles 0.1, Dm41^2 = 1 eV^2
P_4 = oscprob.osc_prob_4nu_vacuum(E, L_src,
 average=True, s14=np.sqrt(0.1), s24=np.sqrt(0.1),
 D41=1.0, **osc)

# Pseudo-Dirac: the second mass state is paired
# with a partner split by 1e-13 eV^2
P_pd = oscprob.osc_prob_pseudo_dirac_vacuum(E,
 L_src, {1: 1.0e-13}, average=True, **osc)

cases = dict(std=P_std, liv=P_liv, nsi=P_nsi,
             four=P_4, pd=P_pd)
f_earth = {k: np.array([at_earth(P) for P in v])
           for k, v in cases.items()}
\end{lstlisting}

\textit{What varies with energy, and what does not.}---Under standard oscillations, the composition does not depend on the energy at all: \equ{averaged} depends only on the eigenvectors, and in vacuum they do not depend on the energy (Sec.~\ref{sec:averaged}). Three of the four departures from standard oscillations are also flat in energy, or nearly so:
\begin{itemize}
 \item
  In the $3+1$ case, the composition at Earth moves from $(0.326, 0.343, 0.330)_\oplus$ to $(0.330, 0.325, 0.346)_\oplus$. The sterile state takes away 13.3\% of the flux, which was purely active at the source; these triplets are the three active fractions, rescaled to add to one. The Hamiltonian is a vacuum one, so the composition does not depend on the energy.
 \item
  In the pseudo-Dirac case, the composition moves to $(0.332, 0.323, 0.345)_\oplus$, and does not depend on the energy either, for the same reason. At $\delta m^2 = 10^{-13}$~eV$^2$, the pair has decohered at every energy shown (see below).
 \item
  With NSI through the Earth, the fractions move by less than 0.005, and vary by less than 0.001 across the energies shown: above a TeV, the matter term dominates the Hamiltonian inside the Earth, and it does not depend on the energy. Above about 100~TeV, a neutrino crossing the core is likely to interact on the way and never reach the detector, an effect \magnus\ does not account for.
\end{itemize}

The Lorentz-violating case is different: its term grows with the energy, as $E$, while the vacuum term falls, as $1/E$, so the eigenvectors change with the energy. The vacuum term sets them well below 100~TeV, and the new term well above. Across the crossover, $f_{e, \oplus}$ rises from 0.33 to 0.35 and falls to 0.32, and $f_{\mu, \oplus}$ dips to 0.33 and recovers; the $\nu_e$ survival probability reaches 0.89. Above the crossover, the composition settles close to the standard one, because $\mathbb{V}_\xi$ is the PMNS matrix with $\delta_\xi = 0$. A measurement of the flavor composition at several energies could detect such an energy dependence\ \cite{Liu:2023flr, Telalovic:2025xor}.

\begin{figure}[t!]
 \centering
 \includegraphics[width=\columnwidth]{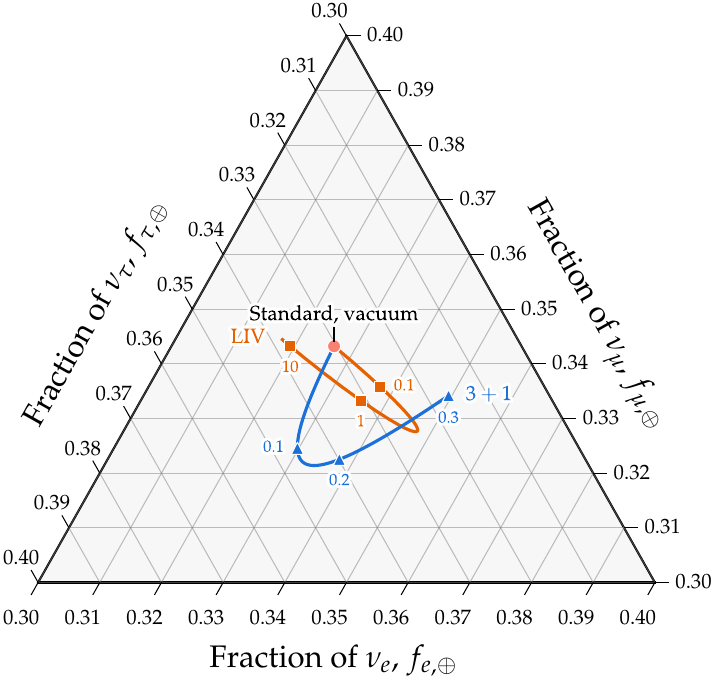}
 \caption{\label{fig:astro_ternary}\textbf{The flavor triangle of high-energy astrophysical neutrinos.} Flavor composition at Earth of a pion-decay flux of astrophysical neutrinos at 100~TeV. The flavor composition at the source is $\left(\frac{1}{3}:\frac{2}{3}:0\right)_{\rm S}$. Results are for standard, three-flavor oscillations in vacuum, for Lorentz-invariance violation (LIV), and for active-sterile mixing under 3+1 flavors. The standard expectation is $(0.326, 0.343, 0.330)_\oplus$. Along the LIV curve, the eigenvalue $b_3$ of the $n = 1$ LIV operator of \figu{astro} grows from zero; the marks are where the new term is 0.1, 1, and 10 times the vacuum one. At the value of $b_3$ used in \figu{astro}, these ratios occur at 32, 100, and 316~TeV. Along the 3+1 curve, the two active-sterile mixing angles grow together from zero; the marks are $\sin^2\theta_{14} = \sin^2\theta_{24} = 0.1$, 0.2, and 0.3, with flavor fractions renormalized to the active flavors. See Sec.~\ref{sec:astro} and notebooks \#10, \#13, and \#28 for details.}
\end{figure}

\smallskip

\textit{The flavor triangle.}---Figure~\ref{fig:astro_ternary} shows how the flavor composition at Earth depends on the new-physics parameters themselves, at a fixed energy of 100~TeV, for two of the departures of \figu{astro}, plotted on the flavor triangle. Along the LIV curve, the eigenvalue $b_3$ grows from zero: the composition leaves the standard point, swings out where the new term is comparable to the vacuum one, and settles where the new term alone puts it, close to the standard point. Along the $3+1$ curve, the two active-sterile mixing angles grow together from zero, and the composition moves steadily away from the standard point\ \cite{Ahlers:2020miq}. Over these ranges, every fraction stays within 0.03 of its standard value, so the triangle is drawn over 0.30 to 0.40 on each axis.

\smallskip

\textit{Pseudo-Dirac neutrinos.}---In a pseudo-Dirac spectrum, a mass eigenstate is split into two nearly degenerate states, one active and one sterile\ \cite{Wolfenstein:1981kw, Petcov:1982ya} (see Sec.~\ref{sec:shipped_hamiltonians}). In the fifth panel of \figu{astro}, only $\nu_2$ is split, by $\delta m^2 = 10^{-13}$~eV$^2$. Once the two members of the pair have decohered, half of the $\nu_2$ component of the flux is in the sterile member, which a detector does not see: about 19\% of the flux. Because only $\nu_2$ loses flux, the flavor fractions change. If all three mass eigenstates were split, every probability between active flavors would be halved, since in the pseudo-Dirac limit each active state mixes maximally with its partner [\equ{mixing_pseudo_dirac}]; the active fractions, which are ratios, would stay standard. Reference~\cite{Beacom:2003eu} proposed using the flavor composition to detect pseudo-Dirac splittings.

In vacuum, \magnus\ computes the averaged probability of a pseudo-Dirac spectrum with a dedicated function, {\tt osc\_prob\_pseudo\_dirac\_vacuum}. It takes the same mapping of split states as the Hamiltonian builders of Sec.~\ref{sec:shipped_hamiltonians}, from the index of each split mass state to its splitting, together with the standard oscillation parameters, which default to NuFIT~6.1. Listing~\ref{lst:astro} computes the fifth panel of \figu{astro} with it. At a single energy, continuing the listing,
\begin{pysnip}
# nu_2 split by 1e-13 eV^2, at 100 TeV
# from 100 Mpc, as in the fifth panel
E1 = 100.0*gd.UNIT_TEV
pdv = oscprob.osc_prob_pseudo_dirac_vacuum
P = pdv(E1, L_src, {1: 1.0e-13},
 average=True)
P.shape            # (4, 4): e, mu, tau, s_2
at_earth(P).round(3)  # [0.332 0.323 0.345]
\end{pysnip}

Whether the two members of the pair count as one state or as two depends on the phase [\equ{phase}] that the pair accumulates over the baseline, $\delta m^2 L/2E$. At 100~Mpc and 100~TeV, for $\delta m^2$ below about $10^{-18}$~eV$^2$, this phase is under 0.1~rad: the pair stays coherent and acts as a single state, as in \equ{averaged_blocks}, so the fractions are those of an unsplit spectrum. For $\delta m^2$ above about $5 \cdot 10^{-16}$~eV$^2$, the phase exceeds 40~rad: the interference between the two members has averaged away, and \equ{averaged} counts them as separate states. The fifth panel of \figu{astro}, at $10^{-13}$~eV$^2$, is well inside this regime. Between the two regimes, part of the interference survives, so the result depends on the energy spread, and \magnus\ warns so. No switch between the regimes is needed: the phase average of \equ{phase_average} damps the interference of the pair according to the spread of its phase, so it moves smoothly from one regime to the other:
\begin{pysnip}
# The composition at Earth for three
# splittings of nu_2, at 100 TeV and
# 100 Mpc, from coherent to decohered
for dm2 in (1.0e-19, 1.0e-17, 1.0e-13):
    P = pdv(E1, L_src, {1: dm2},
     average=True)
    print(at_earth(P).round(3))
# [0.326 0.343 0.33 ]   Coherent: standard
# [0.328 0.338 0.334]   Between: warns
# [0.332 0.323 0.345]   Decohered
\end{pysnip}

The function builds the eigensystem of the pseudo-Dirac spectrum directly, using each given $\delta m^2$, so the pair phase is exact however small the splitting. A diagonalization of the Hamiltonian (which the general route below relies on) resolves eigenvalues only to about $10^{-16}$ of the largest one, so it would obtain a small splitting with a large relative error. This matters exactly where pseudo-Dirac neutrinos are observable: splittings between the two regimes, $10^{-18}$--$10^{-15}$~eV$^2$ over astrophysical distances, are only $10^{-16}$--$10^{-12}$ of $\Delta m^2_{31}$. For a phase of a few radians, the error in the probabilities is below $10^{-7}$ for $\delta m^2 \gtrsim 10^{-10}\,\Delta m^2_{31}$, but reaches $10^{-2}$ at $4 \cdot 10^{-16}\,\Delta m^2_{31}$, as large as the shift in flavor fractions that the splitting itself produces.

In matter, there is no dedicated function, and the general route applies: build the Hamiltonian from its vacuum and matter parts, {\tt hamiltonian\_pseudo\_dirac\_vacuum} and {\tt hamiltonian\_pseudo\_dirac\_matter} (Sec.~\ref{sec:shipped_hamiltonians}), and pass it to {\tt osc\_prob\_energy\_baseline}. The precision problem does not arise here: the matter potential is not the same for the active member of a pair and for its sterile partner, so it separates the two by far more than $\delta m^2/2E$, and no eigenvalue difference comes near the resolution of the diagonalization. For an astrophysical flux, matter enters only when the neutrinos cross the Earth on their way to the detector. The probability for that crossing is multiplied by the averaged probability in vacuum, as matrices, as in \equ{astro_earth_leg}. Below, the pseudo-Dirac flux of the fifth panel of \figu{astro} crosses the Earth along a chord through the core. The matter part takes, at each position, the charged-current potential and the neutron-to-proton ratio of the local layer, which sets the entry of the sterile partner:
\begin{pysnip}
import magnus.hamiltonians as ham
import magnus.matter as matter

# Inputs of the vacuum part: the PMNS
# matrix and the three m^2 [eV^2]
U = ham.pmns_mixing_matrix(osc['s12'],
 osc['s23'], osc['s13'], osc['dCP'])
m2 = np.array([0.0, osc['D21'],
 osc['D31']])
pairs = {1: 1.0e-13}

# PREM along the chord through the core:
# at position l, V_CC and the neutron-
# to-proton ratio of the local layer
rad = earth.earth_radial_distance_from_depth
rho = earth.density_matter_func_prem
yef = earth.electron_fraction_func_prem
def prem(l):
    r = rad(-1.0, l/gd.UNIT_KM)
    ye = yef(r)
    ne = matter.num_density_e_func(r, rho,
     electron_fraction=ye,
     density_matter_is_in_g_per_cm3=True)
    return np.sqrt(2.0)*gd.GF*ne, (1-ye)/ye

# The Hamiltonian, of energy and position:
# vacuum part plus matter part
hpd = ham.hamiltonian_pseudo_dirac_vacuum
hpdm = ham.hamiltonian_pseudo_dirac_matter
def H(E, l):
    V, r_np = prem(l)
    return (hpd(E, U, m2, pairs)
     + hpdm(V, 3, pairs, r_np))

# Through the core, with the PREM layer
# boundaries declared as breakpoints
edges = earth.prem_layer_edges_along_chord(
 -1.0)*gd.UNIT_KM
P_core = oscprob.osc_prob_energy_baseline(
 H, E, L_in*gd.UNIT_KM, t_breakpoints=edges,
 rtol=1e-6, atol=1e-8)

# The averaged vacuum leg, then the Earth
P_both = (np.asarray(P_pd)
 @ np.asarray(P_core))
at_earth(P_both[-1]).round(3)
# [0.332 0.323 0.345]
\end{pysnip}
Crossing the core changes the composition by at most $3 \cdot 10^{-5}$, at 1~TeV, and by less than $10^{-6}$ above 10~TeV: at these energies, the matter potential exceeds the vacuum term by orders of magnitude, and it is diagonal in flavor, so crossing the Earth changes no flavor. It is the off-diagonal couplings of NSI that make the Earth leg matter in the NSI panel of \figu{astro}.


\subsection{Assorted long examples}


\subsubsection{A cavity in the Earth's crust}
\label{sec:cavity}

\begin{figure}[t!]
 \centering
 \includegraphics[width=\columnwidth]{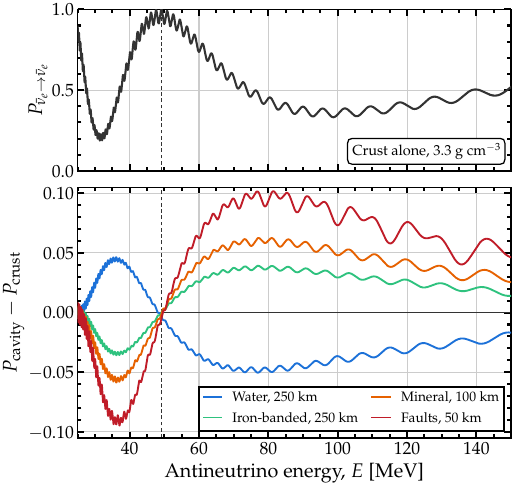}
 \caption{\label{fig:cavity}\textbf{A cavity in the Earth\textquoteright s crust.} Effect of a cavity in the Earth's crust on the $\bar{\nu}_e$ survival probability over a baseline of $1\,500$~km, computed with \magnus, after \Ref\ \cite{Arguelles:2012nw}. {\it Top}: the crust alone, uniform at 3.3~g~cm$^{-3}$ with $Y_e = 0.5$. {\it Bottom}: the change that four cavities make to it, each centered on the baseline, with the density and width its label gives. The water cavity carries $Y_e = 0.555$, the other three 0.5. The dashed line marks 49~MeV, where the crust curve peaks. The beam of \Ref\ \cite{Arguelles:2012nw} spans 5--150~MeV; below 25~MeV, the oscillation is too rapid to draw. Listing~\ref{lst:cavity} computes every curve. See notebook \#28 and Sec.~\ref{sec:cavity} for details.}
\end{figure}

\begin{lstlisting}[language=Python, caption={\textbf{A cavity in the Earth's crust.} A cavity search, after \Ref\ \cite{Arguelles:2012nw}: the $\bar{\nu}_e$ survival probability over 1\,500~km of Earth's crust, with and without a cavity of different density centered on the baseline. A cavity is a density profile with two walls, whose positions are declared as breakpoints so that the refinement ladder runs inside the cavity. See notebook \#28 and Sec.~\ref{sec:cavity} for details.}, label={lst:cavity}, float=tbp]
import numpy as np
import magnus.oscprob as oscprob
import magnus.matter as matter
import magnus.globaldefs as gd

L0 = 1500.0                     # km, source to detector
RHO_CRUST, YE_CRUST = 3.3, 0.5  # Near PREM's crust
E = np.linspace(25.0, 150.0, 2500)*gd.UNIT_MEV
osc = gd.load_nufit_params('NuFIT 6.1')
kw = dict(osc_params=osc, L0=0.0, nu_i=gd.NUE,
          nu_f=gd.NUE, nubar=True, rtol=1.0e-8,
          atol=1.0e-10,
          density_is_of_number_of_electrons=True)

def n_e(rho, ye):
    """Electron density [eV^3] of uniform matter."""
    return matter.num_density_e_func(
        0.0, lambda _: rho, electron_fraction=ye,
        ratio_number_neutrons_to_protons=(1.0-ye)/ye,
        density_matter_is_in_g_per_cm3=True)

NE_CRUST = n_e(RHO_CRUST, YE_CRUST)

def cavity(rho, ye, w):
    """Crust holding a cavity of width w, centered."""
    d = (L0 - w)/2.0
    ne_in = n_e(rho, ye)

    def profile(l):
        x = np.asarray(l, dtype=float)/gd.UNIT_KM
        return NE_CRUST + (ne_in - NE_CRUST) \
            *((x >= d) & (x <= d + w))

    return profile, np.array([d, d + w])*gd.UNIT_KM

# An empty crust needs no profile, only a number
P0 = oscprob.osc_prob_matter_std_potential(
    3, NE_CRUST, E, L0*gd.UNIT_KM, **kw)

# Water, iron, a mineral deposit, a zone of faults:
# (rho, Y_e, w) for each
CASES = [(1.0, 0.555, 250.0), (5.0, 0.5, 250.0),
         (10.0, 0.5, 100.0), (25.0, 0.5, 50.0)]
dP = []
for rho, ye, w in CASES:
    profile, walls = cavity(rho, ye, w)
    # The two walls are density jumps: declare them
    P = oscprob.osc_prob_matter_std_potential(
        3, profile, E, L0*gd.UNIT_KM,
        t_breakpoints=walls, **kw)
    dP.append(P - P0)
\end{lstlisting}

A region of anomalous density along the path of a neutrino through an otherwise uniform medium changes the oscillation probability measured at the far end. Using neutrinos to explore the interior of the Earth was proposed decades ago~\cite{DeRujula:1983ya, Ermilova:1986ph}; oscillations in matter offer one way to do it (see, \eg, \Ref~\cite{Winter:2006vg}). Reference~\cite{Arguelles:2012nw} applied this to a search for cavities in the Earth's crust. We use \magnus\ to implement the setup of \Ref\ \cite{Arguelles:2012nw}: a low-energy $\bar{\nu}_e$ beam crossing $1\,500$~km of crust, with the survival probability measured from 5 to 150~MeV. The example is illustrative; we do not claim that such an experiment is feasible.

Listing~\ref{lst:cavity} sets this up. The crust is uniform at 3.3~g~cm$^{-3}$ with $Y_e = 0.5$, a round value close to the 0.4952 of the PREM crust. A cavity is a slab of different density, centered on the baseline. We show four cavities, with the densities and labels of \Ref\ \cite{Arguelles:2012nw}: water at 1~g~cm$^{-3}$ with $Y_e = 0.555$, an iron-banded formation at 5~g~cm$^{-3}$, a mineral deposit at 10~g~cm$^{-3}$, and a zone of seismic faults at 25~g~cm$^{-3}$; the last two are denser than any natural rock, and serve to show how the effect grows with density. Their widths along the path range from 250 to 50~km, the densest being the narrowest. Each wall of a cavity is a density jump, so we pass the positions of both walls to the scenario function via {\tt t\_breakpoints}; this places a slab edge on each wall, so that no slab straddles a jump (Sec.~\ref{sec:declaring_edges}).

Figure~\ref{fig:cavity} shows the resulting probabilities. Without a cavity, the survival probability peaks at 0.9997 near 49~MeV, where the oscillation driven by $\Delta m^2_{21}$ completes half a cycle (the peak would reach exactly 1 if $\theta_{13}$ were zero). A cavity shifts this peak along the energy axis, leaving its shape almost unchanged. A denser cavity adds column density and pushes the peak to higher energy; a lighter one removes column density and pulls it to lower energy. Consequently, just below the peak of the crust curve, a denser cavity lowers the probability, and just above it, raises it; a lighter cavity does the opposite. Every curve of the change crosses zero between the peak of the crust curve and that of the cavity curve, between 48.2 and 49.5~MeV.

\begin{figure}[t!]
 \centering
 \includegraphics[width=\columnwidth]{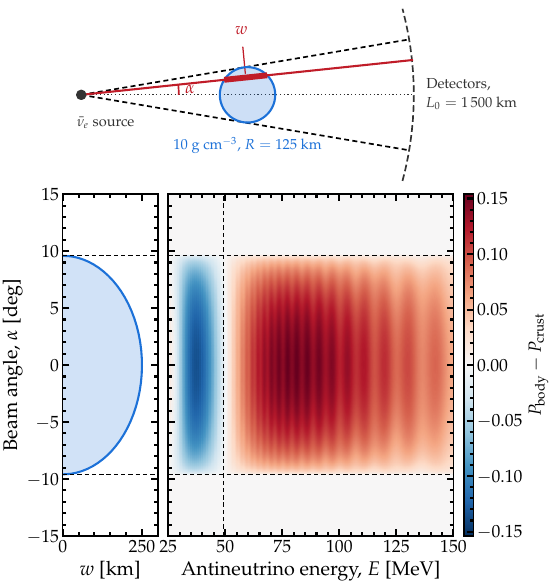}
 \caption{\label{fig:cavity_sweep}\textbf{A beam swept across a buried body.} {\it Top}: the geometry. The beam leaves the source at an angle $\alpha$ and reaches a detector on the dashed arc, $1\,500$~km away. The two dashed straight lines are the beams tangent to the body, at $\alpha = \pm 9.59^\circ$. The beam crosses a width $w$ of the body. {\it Bottom left}: width against angle, \ie, the cavity's silhouette. {\it Bottom right}: change in probability over energy and angle. The dashed vertical line marks the $49$~MeV of \figu{cavity}. Listing~\ref{lst:cavity_sweep} computes the map. See notebook \#28 and Sec.~\ref{sec:cavity} for details.}
\end{figure}

Figure~\ref{fig:cavity_sweep} sweeps the beam across a buried body. The source stays fixed and the baseline keeps its length of $1\,500$~km; only the direction of the beam changes, by an angle $\alpha$. The body is a sphere of radius 125~km at the density of the mineral deposit, 10~g~cm$^{-3}$, centered 750~km from the source along the baseline at $\alpha = 0$. (\Ref\ \cite{Arguelles:2012nw} uses elliptical cavities, but a sphere makes the same point.) The beam crosses the body only for $\lvert \alpha \rvert < 9.59^\circ$; the width it crosses is largest, 250~km, at $\alpha = 0$, and falls to zero at the edges of that band.

The change in probability follows that width: it is largest near $\alpha = 0$ and fades toward the edges of the band. Outside the band, the beam misses the body, the profile is the uniform crust, and the change is exactly zero. The sweep is what separates a large, light body from a small, dense one. Along a single direction, the probability depends mainly on the excess column density, \ie, the density contrast times the width, so density and width cannot be told apart; the angular width of the band fixes the size of the body independently. Listing~\ref{lst:cavity_sweep} computes the map.

\begin{lstlisting}[language=Python, caption={\textbf{A beam swept across a buried body.} The map of \figu{cavity_sweep}, which reuses {\tt n\_e}, {\tt NE\_CRUST}, {\tt L0}, and {\tt kw} from Listing~\ref{lst:cavity}. For each beam angle, {\tt crossing} finds where the chord enters and leaves the body. Each call uses these two points twice: as the edges of the body in the density profile, and as {\tt t\_breakpoints}. Angles at which the beam misses the body are skipped, since there the probability equals the crust-only {\tt P0}. A cavity is a density profile with two walls, whose positions are declared as breakpoints, so that no slab straddles a wall. See Sec.~\ref{sec:cavity} for details.}, label={lst:cavity_sweep}, float=tbp]
import numpy as np

BODY_R, BODY_D0 = 125.0, 750.0   # km: size, position
NE_BODY = n_e(10.0, 0.5)         # a mineral deposit
E = np.linspace(25.0, 150.0, 400)*gd.UNIT_MEV
ALPHA = np.linspace(-15.0, 15.0, 220)

def crossing(alpha_deg):
    """Entry and exit points in the body [km]."""
    a = np.radians(alpha_deg)
    miss = abs(BODY_D0*np.sin(a))
    if miss >= BODY_R:
        return None              # the beam misses it
    half = np.sqrt(BODY_R**2 - miss**2)
    mid = BODY_D0*np.cos(a)
    return mid - half, mid + half

P0 = oscprob.osc_prob_matter_std_potential(
    3, NE_CRUST, E, L0*gd.UNIT_KM, **kw)

dP = np.zeros((len(E), len(ALPHA)))
for j, alpha in enumerate(ALPHA):
    seg = crossing(alpha)
    if seg is None:
        continue                 # no body, no change
    lo, hi = seg

    def body(l, lo=lo, hi=hi):
        x = np.asarray(l, dtype=float)/gd.UNIT_KM
        return NE_CRUST + (NE_BODY - NE_CRUST) \
            *((x >= lo) & (x <= hi))

    # The walls move with the angle, so every call
    # declares its own pair
    dP[:, j] = oscprob.osc_prob_matter_std_potential(
        3, body, E, L0*gd.UNIT_KM,
        t_breakpoints=np.array(seg)*gd.UNIT_KM,
        **kw) - P0
\end{lstlisting}


\subsubsection{Geoneutrinos}
\label{sec:geoneutrinos}

\begin{figure}[t!]
 \centering
 \includegraphics[width=\columnwidth]{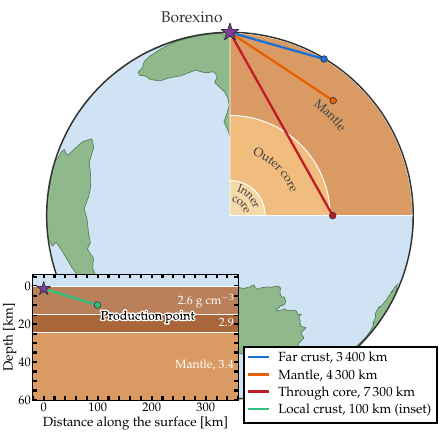}
 \caption{\label{fig:geoneutrinos}\textbf{Geoneutrino geometry at Borexino.} The geometry of a geoneutrino measurement at Borexino, at Gran Sasso, 1.4~km underground. A quarter of the Earth is cut away to expose the PREM layers of \figu{prem}. Three chords reach production points beyond the local crust: in the far crust, 20~km deep and 3\,400~km away; in the mantle, 1\,000~km deep and 4\,300~km away; and at the base of the mantle, 2\,800~km deep and 7\,300~km away, on a path through the outer core. The inset shows the local crust, to scale in distance and stretched in depth, with the crust layers of PREM and a production point 10~km deep and 100~km away. See notebook \#28 and Sec.~\ref{sec:geoneutrinos} for details.}
\end{figure}

Geoneutrinos are the $\bar{\nu}_e$ emitted in the decay chains of uranium-238 and thorium-232 inside the Earth\ \cite{Bellini:2013wsa}. Inverse beta decay can detect only those above 1.8~MeV. The most energetic geoneutrinos, from the uranium chain, reach 3.3~MeV, so the detectable spectrum spans 1.8--3.3~MeV. The sources are the crust and the mantle. Uranium and thorium are lithophile elements: they concentrate in silicate rock, and are expected to be absent from the metallic core. The core still enters the calculation, though, as the medium that neutrinos from the far side of the mantle cross. For a detector at or near the surface, the production points lie from a few km to a full Earth diameter away. As an example, we take Borexino~\cite{Borexino:2008gab, Borexino:2019gps} as the detector, at Gran Sasso, 1.4~km underground.

Figure~\ref{fig:geoneutrinos} shows the setup: four production points illustrate the kinds of path a geoneutrino takes to Borexino. One lies in the local crust, 10~km deep and 100~km away. One lies in the far crust, 20~km deep and 3\,400~km away, on a path that dips into the upper mantle. One lies in the mantle, 1\,000~km deep and 4\,300~km away. The last lies at the base of the mantle, 2\,800~km deep and 7\,300~km away, on a path through the outer core.

\begin{figure}[t!]
 \centering
 \includegraphics[width=\columnwidth]{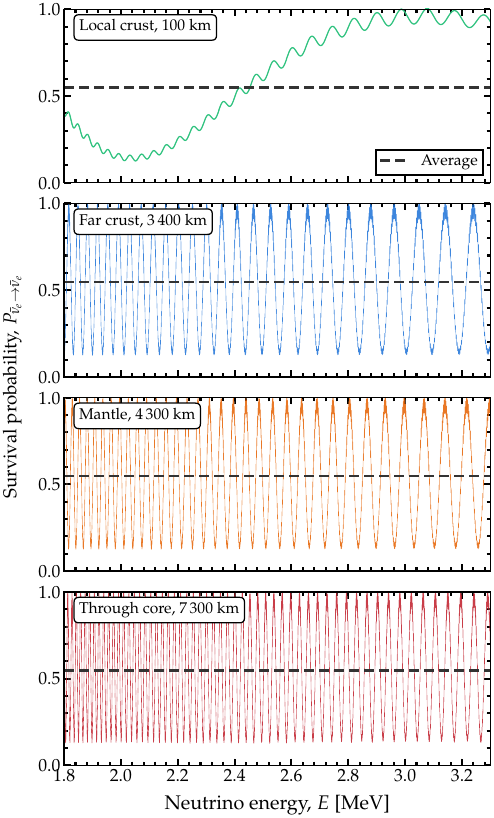}
 \caption{\label{fig:geoneutrino_energy}\textbf{Geoneutrino survival against energy.} Survival probability of $\bar{\nu}_e$ geoneutrinos reaching Borexino, from the four production points of \figu{geoneutrinos}. In every panel, the fast oscillation of the pair split by $\Delta m^2_{31}$ is resolved, riding as a fine ripple on the slower oscillation of the pair split by $\Delta m^2_{21}$. This is visible only in the near-production case (\textit{top panel}). In the three distant-production cases (\textit{bottom three panels}), the line marks the phase average in vacuum, 0.548; matter raises it by 0.3--0.8\%, less than the width of the line. Listing~\ref{lst:geoneutrinos} computes the four panels. See notebook \#28 and Sec.~\ref{sec:geoneutrinos} for details.}
\end{figure}

\begin{figure}[t!]
 \centering
 \includegraphics[width=\columnwidth]{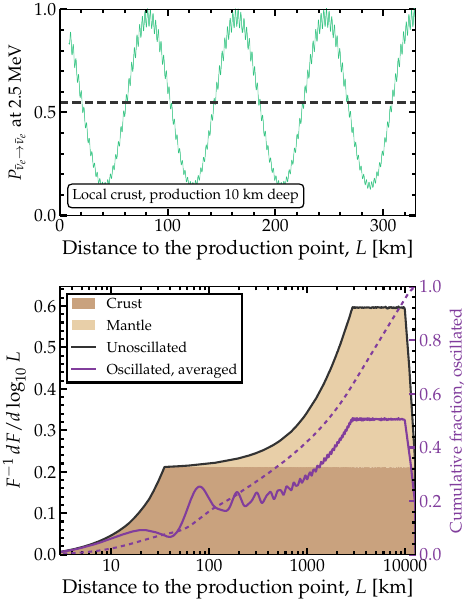}
 \caption{\label{fig:geoneutrino_flux}\textbf{Where the geoneutrino flux comes from.} Where the detectable geoneutrino flux at Borexino comes from and what oscillations leave of it. {\it Top}: the survival probability at 2.5~MeV against the distance to a production point 10~km deep, across the local crust, with the pair split by $\Delta m^2_{31}$ resolved; the horizontal line is the phase average. {\it Bottom}: the distribution of the flux of \equ{geoflux} in the distance to the production point, per unit $\log_{10} L$ and as a fraction of the total, for a spherically symmetric Earth with a 35-km crust holding 7~TW of radiogenic power, a mantle with the abundances of the geochemical model of Ref.~\cite{Bellini:2013wsa}, and no uranium or thorium in the core. The unoscillated curve has $\langle P_{\bar\nu_e \to \bar\nu_e} \rangle = 1$. The oscillated curve folds in the survival probability averaged over the detectable window with equal weight in energy; its cumulative fraction is on the right axis. See notebook \#28 and Sec.~\ref{sec:geoneutrinos} for details.}
\end{figure}

Figure~\ref{fig:geoneutrino_energy} shows the survival probability against energy for these production points. Two oscillations appear: a slow one, from the pair split by $\Delta m^2_{21}$, and a fast one, from the pairs split by $\Delta m^2_{31}$. From the local crust, the slow oscillation completes less than one cycle across the window. From the three distant points, both complete tens to thousands of cycles, far more than a detector resolves in energy, so a measurement of a distant reservoir sees only the average probability. Within a few hundred kilometers of the detector, the slow oscillation does not average out across the window. For such nearby points, the full probability must be used instead of the average; \figu{geoneutrino_flux} uses it out to 331~km, where it differs from the average by 0.01.

For the three distant production points, \figu{geoneutrino_energy} marks the phase-averaged probability in vacuum, 0.548. Matter changes it little. The passage through the Earth is adiabatic: even the largest jump in electron density, at the core-mantle boundary, changes the mixing angle in matter by only about a percent at these energies, so the neutrino stays in the eigenstate in which it was produced. The average then depends only on the eigenstates at the two ends of the path [\equ{averaged_varying}, with $P^{\rm cross}$ the identity; Sec.~\ref{sec:averaged}]. At both ends, the matter potential is at most 2\% of $\Delta m^2_{21}/2E$, so those eigenstates are nearly the vacuum ones. As a result, matter raises the average by only 0.3--0.8\%, less than the width of the line in \figu{geoneutrino_energy}, and the vacuum value serves our calculation. The snippet below checks this along the mantle chord, with the density of PREM and the electron fraction of the mantle, using {\tt average=True} (Sec.~\ref{sec:average_keyword}):
\begin{pysnip}
from magnus.earth import (
 distance_traveled_inside_earth as chord,
 earth_radial_distance_from_depth as radius,
 density_matter_func_prem as prem,
 Y_E_MANTLE_PREM)

osc = gd.load_nufit_params('NuFIT 6.1')

# The mantle chord: cosine of the zenith
# angle at the detector, and the depths of
# the production point and of Borexino [km]
c = -0.552
geo = dict(source_depth=1000.0,
           detector_depth=1.4)

# Length of the chord [km]
L = chord(c, **geo)

# Density along the chord [g/cm^3]: PREM,
# with its ocean layer replaced by rock
def rho(l):
    # Radius at distance l along the chord
    r = radius(c, l/gd.UNIT_KM, **geo)
    return prem(r,
                density_matter_ocean=2.65)

E = 2.5*gd.UNIT_MEV

# Phase-averaged survival probability in
# matter, at electron fraction of the mantle
P_m = oscprob.osc_prob_matter_std_potential(
 3, rho, E, L*gd.UNIT_KM, osc,
 average=True, nubar=True,
 nu_i=gd.NUE, nu_f=gd.NUE,
 electron_fraction=Y_E_MANTLE_PREM,
 density_matter_is_in_g_per_cm3=True)
# 0.551

# The same, in vacuum
P_v = oscprob.osc_prob_3nu_vacuum(E,
 L*gd.UNIT_KM, average=True, nubar=True,
 nu_i=gd.NUE, nu_f=gd.NUE)
# 0.548
\end{pysnip}

Listing~\ref{lst:geoneutrinos} computes the panels of \figu{geoneutrino_energy} with the Earth wrapper, {\tt osc\_prob\_3nu\_earth}. Each production point enters through its depth and the cosine of its zenith angle at the detector. Every chord crosses PREM boundaries---six on the far-crust chord, nine on the chord through the core---and the wrapper internally places a slab edge on each. Since Gran Sasso is a continental site, {\tt density\_matter\_ocean} replaces the ocean layer of PREM with rock.

(When a source is buried, the wrapper automatically places it on the far side of the chord, where the chord first reaches the source depth. However, the local-crust production point, 10~km deep and 100~km away, lies instead on the near side: along its chord, a depth of 10~km is first reached about 1\,100~km from the detector. To overcome this, the last call in Listing~\ref{lst:geoneutrinos} swaps the two ends of the path: it treats the detector as the source, 1.4~km deep, and the production point as the detector, 10~km deep, with the zenith angle measured at the production point. The result is unchanged, because the survival probability is the same along a path and along its reverse.)

\begin{lstlisting}[language=Python, caption={\textbf{Geoneutrino survival against energy.} The four panels of \figu{geoneutrino_energy}: the survival probability of a geoneutrino reaching Borexino from the far crust, the mantle, the base of the mantle, and the local crust, across the detectable window, on one energy grid fine enough to resolve the pair split by $\Delta m^2_{31}$ on the longest chord. The call declares the PREM boundaries on each chord by itself. The ocean layer of PREM is replaced by rock. The local production point is run in reverse, from the detector. See notebook \#28 and Sec.~\ref{sec:geoneutrinos} for details.}, label={lst:geoneutrinos}, float=tbp]
import numpy as np
import magnus.oscprob as oscprob
import magnus.globaldefs as gd

# Twelve energies per cycle of the pair split by
# Dm31^2 on the longest chord, 7,300 km
E = 1/np.linspace(1/1.8, 1/3.3, 22500)*gd.UNIT_MEV

# Borexino, 1.4 km underground.  A production point
# enters as its depth and the cosine of its zenith
# angle at the detector; the call declares the PREM
# boundaries on the chord by itself.  The oscillation
# parameters are the defaults, NuFIT 6.1.
points = {'far crust': (20.0, -0.272),    # 3,400 km
          'mantle': (1000.0, -0.552),     # 4,300 km
          'core': (2800.0, -0.872)}       # 7,300 km
P = {}
for name, (depth, costhz) in points.items():
    P[name] = oscprob.osc_prob_3nu_earth(E,
     costhz=costhz, source_depth=depth*gd.UNIT_KM,
     detector_depth=1.4*gd.UNIT_KM, nubar=True,
     nu_i=gd.NUE, nu_f=gd.NUE,
     density_matter_ocean=2.65)

# The local point, 10 km deep and 100 km away,
# run in reverse, from the detector (see text)
P['local'] = oscprob.osc_prob_3nu_earth(E,
 costhz=0.078, source_depth=1.4*gd.UNIT_KM,
 detector_depth=10.0*gd.UNIT_KM, nubar=True,
 nu_i=gd.NUE, nu_f=gd.NUE,
 density_matter_ocean=2.65)
\end{lstlisting}

Figure~\ref{fig:geoneutrino_flux} shows how the detectable flux is distributed over the distance to the production point and how much of it survives the oscillations. A flux measurement weighs the curves of \figu{geoneutrino_energy} by that distribution. The Earth model chosen for the figure is the simplest one that carries both reservoirs: a spherically symmetric crust 35~km thick holding 7~TW of radiogenic power at a thorium-to-uranium ratio of 4.5, a mantle with the abundances of the geochemical model of Ref.~\cite{Bellini:2013wsa}, no uranium or thorium in the core, and the density of PREM throughout. The flux at the detector, located at $\mathbf{r}_{\rm det}$, is
\begin{equation}
 F = \int_\oplus d^3r \, \frac{\varepsilon(\mathbf{r})}{4 \pi L^2} \, \langle P_{\bar\nu_e \to \bar\nu_e} (L) \rangle \,,
 \label{equ:geoflux}
\end{equation}
where $L \equiv \lvert \mathbf{r} - \mathbf{r}_{\rm det} \rvert$ and $\varepsilon(\mathbf{r})$ is the number of detectable $\bar{\nu}_e$ emitted per unit volume and time at position $\mathbf{r}$, fixed by the density of PREM and the abundances above. The survival probability from that point, $\langle P_{\bar\nu_e \to \bar\nu_e}(L) \rangle$, is the mean over the detectable window, 1.8--3.3~MeV. In spherical coordinates centered on the detector, $d^3r = L^2 \, dL \, d\Omega$ and the $L^2$ cancels: the production points between $L$ and $L + dL$ contribute $dL / 4\pi$ times the emissivity integrated over the directions in which that shell lies inside the Earth. The lower panel of \figu{geoneutrino_flux} shows this distribution per unit $\log_{10} L$, normalized to the total flux without oscillations, so that the area under a curve between two distances is the share of the flux from between them. Without oscillations, the crust supplies 61\% of the detectable flux, and the region within 350~km of the detector supplies 30\%. Out to 331~km, the full probability must be used instead of the average.

The oscillated curve is the unoscillated one multiplied by $\langle P_{\bar\nu_e \to \bar\nu_e}(L) \rangle$, the survival probability averaged over the detectable window. Up to 331~km, we compute the probability in full, at each distance on the grid, with the last call of Listing~\ref{lst:geoneutrinos}, and then average it over the window. Beyond 331~km, we use the average in vacuum instead, which differs from the full result by 0.01 at 331~km and by less farther out. The upper panel of \figu{geoneutrino_flux} shows the full probability at 2.5~MeV: the slow oscillation has a period of 82~km, and the fast one, with a period of 2.5~km, appears as a ripple on it. Close to the detector, the window holds too few cycles of the slow oscillation to average it out, so the oscillated curve in the lower panel ripples out to a few hundred kilometers; farther out, it is smooth. Over the whole Earth, oscillations leave 54\% of the detectable flux.

Both curves in the lower panel of \figu{geoneutrino_flux} flatten at large distances, for a geometric reason. The production points at a distance $L$ from the detector lie on a sphere of radius $L$ centered on it. The part of this sphere at a distance between $r$ and $r + dr$ from the center of the Earth has area $2 \pi L r \, dr / R_\oplus$. In \equ{geoflux}, this area is divided by $4 \pi L^2$, and on the logarithmic axis of the panel it is multiplied by $L$, so the result does not depend on $L$. Hence, a layer of the Earth contributes the same flux per decade of distance as long as the sphere reaches every depth of that layer. For the crust, this holds from 35~km out to a full Earth diameter; for the mantle, from about 2\,900 to 9\,900~km. The total stays flat over that range because the core, which holds no uranium or thorium in this model, adds nothing to it.


\subsubsection{Neutrino tomography of the Sun}
\label{sec:solar_tomography}

\begin{figure}[t!]
 \centering
 \includegraphics[width=\columnwidth]{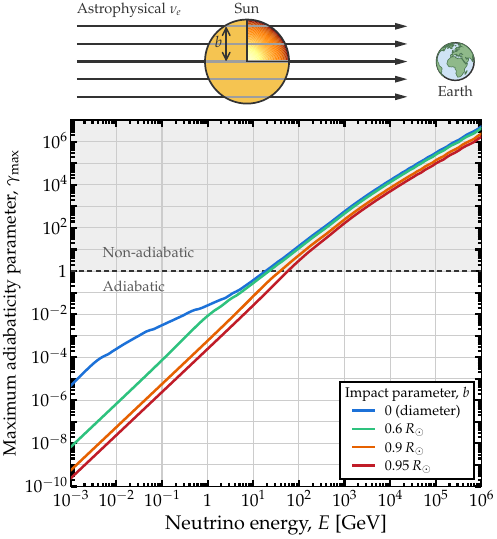}
 \caption{\label{fig:solar_adiabaticity}\textbf{Adiabaticity along a solar chord.} {\it Top}: the setup. An isotropic flux of $\nu_e$ crosses the Sun and continues to Earth, where it may oscillate. The impact parameter $b$ is marked on one ray. {\it Bottom:} Adiabaticity of a neutrino trajectory through the Sun against energy, computed with \magnus\ on the B16-GS98 solar density profile\ \cite{Vinyoles:2016djt}, tabulated up to the surface, for four impact parameters, each defining a chord through the Sun. The parameter $\gamma_{\rm max}$ is the adiabaticity parameter of \equ{hf}, evaluated at its maximum along the chord and over the three pairs of levels. See notebook \#28 and Sec.~\ref{sec:solar_tomography} for details.}
\end{figure}

Neutrinos made outside the Solar System arrive from all directions, so a share of them crosses the Sun on its way to Earth. At around 10~MeV, they come mainly from the diffuse supernova neutrino background; from TeV to PeV, they are part of the high-energy astrophysical flux. Below, we compute what that does to their oscillations, using the density profile of the B16-GS98 solar model\ \cite{Vinyoles:2016djt} of Table~\ref{tab:solar_models}. 

At the energies of this section, up to 1~PeV, a neutrino from a source outside the Solar System has traveled far enough for its oscillations to average out on the way (\figu{astro_baseline}), so it reaches the Sun as an incoherent mixture of $\nu_1$, $\nu_2$, and $\nu_3$. The quantity to compute is the phase-averaged probability, propagated through the varying profile of the Sun as described in Sec.~\ref{sec:averaged}. Where the passage is adiabatic, the neutrino stays in the instantaneous eigenstates of $\mathbb{H}$; where it is not, \magnus\ crosses the region with a Magnus patch. Where every phase has averaged away, the result is \equ{averaged_varying}, with $P^{\rm cross}$ the level-crossing matrix between the eigenbases at entry and exit. Both ends of the path lie in vacuum, where the eigenvectors of $\mathbb{H}$ form the vacuum mixing matrix, so $\mathbb{V}(l_0) = \mathbb{V}(l_1) = \mathbb{R}$ in \equ{averaged_varying}. If the passage is adiabatic, $P^{\rm cross}$ is the identity, and \equ{averaged_varying} reduces to the vacuum average of \equ{averaged}: the Sun leaves no trace. Whether the passage is adiabatic depends on the energy, through \equ{hf}.

This may seem at odds with the MSW effect, which changes solar neutrinos strongly (Sec.~\ref{sec:sun}). The difference lies in where the path starts. A solar neutrino is born deep inside the Sun, at high density, as an eigenstate of $\mathbb{H}$ in matter; an adiabatic passage outward turns it into a single vacuum mass eigenstate, which is a different state. A neutrino crossing the Sun from outside starts and ends in vacuum, so an adiabatic passage returns it to the same mass eigenstate it entered as.

Figure~\ref{fig:solar_adiabaticity} shows how adiabatic the passage is, from MeV to PeV, at three flavors, for four impact parameters. The adiabaticity parameter $\gamma_{jk}(l)$ of \equ{hf} is defined for each pair of eigenvalues of $\mathbb{H}$, 1-2, 1-3, and 2-3, at each point $l$ of the path; it is large where the two eigenvalues come close. The figure plots its largest value along the chord and over the three pairs, $\gamma_{\rm max}$. Where $\gamma_{\rm max} > 1$, the passage is not adiabatic: somewhere along the chord, the neutrino moves between eigenstates of $\mathbb{H}$. The {\tt adiabatic} module of \magnus\ computes $\gamma_{\rm max}$:
\begin{pysnip}
import numpy as np
import magnus.adiabatic as adiabatic
import magnus.hamiltonians as ham
import magnus.matter as matter
import magnus.globaldefs as gd
import magnus.solarmodels as solarmodels

osc = gd.load_nufit_params('NuFIT 6.1')
R = gd.SUN_RADIUS*gd.UNIT_KM
prof = solarmodels.electron_density_profile
ne_sun = prof('B16-GS98')
b = 0.3*R                # impact parameter
half = np.sqrt(R**2 - b**2)
E = 100.0*gd.UNIT_GEV
hv = ham.hamiltonian_3nu_vacuum(E, **osc)

def ne(l):
    """Electron density on the chord."""
    r = np.sqrt((l - half)**2 + b**2)
    return ne_sun(r)

def H(l):
    """Hamiltonian at l along the chord."""
    h = np.array(hv, dtype=complex)
    h[0, 0] += matter.VCC_func(l, ne)
    return h

info = {}
adiabatic.find_nonadiabatic_windows(
    H, 0.0, 2*half, n_probe=20000,
    info=info)
info['gamma_max']        # 13.7580
\end{pysnip}
Here, {\tt ne\_sun} is the B16-GS98 table, interpolated in its logarithm, {\tt ne} evaluates it along the chord, and {\tt hv} is the vacuum Hamiltonian at this energy. The value shown is for a $100$-GeV neutrino crossing at an impact parameter of $b = 0.3\,R_\odot$.

The parameter $\gamma_{\rm max}$ grows with energy because the vacuum splitting $\Delta m^2/2E$ shrinks while the matter potential stays fixed. The largest value comes from the 1-2 pair, near the point where its two eigenvalues come closest: the 1-2 resonance, where the matter potential equals $\Delta m^2_{21}\cos 2\theta_{12}/2E$. This resonance occurs at a single radius in the Sun, set by the energy alone, so every chord reaches its $\gamma_{\rm max}$ at that radius. As the energy grows, that density is reached further out: $\gamma_{\rm max}$ sits at $0.97\,R_\odot$ at $100$~GeV, at $0.990\,R_\odot$ at 1~TeV, and at $0.998\,R_\odot$ at 100~TeV. By 1~PeV, it reaches $5\times10^6$ on the diameter.

\begin{figure*}[t!]
 \centering
 \includegraphics[width=\textwidth]{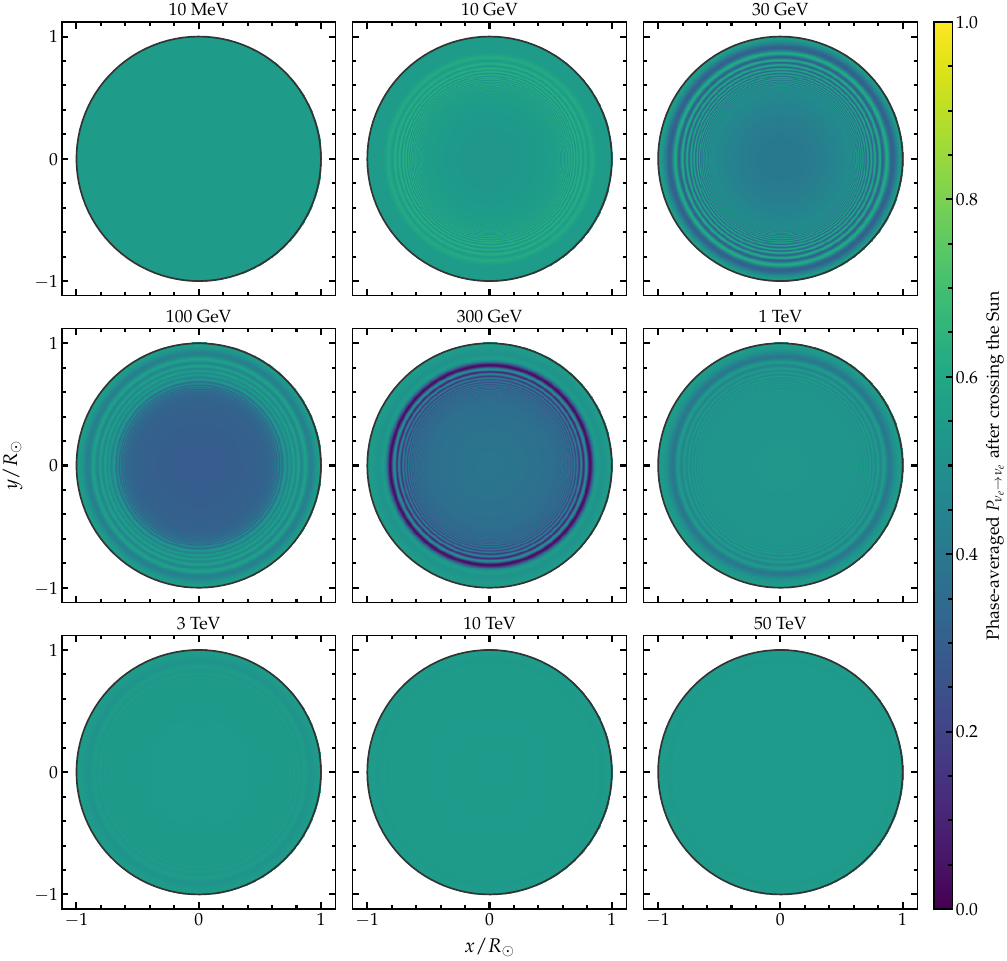}
 \caption{\label{fig:solar_tomography}\textbf{The Sun in the electron-neutrino channel.} The Sun seen face-on in the $\nu_e$ survival channel, computed with \magnus\ on the B16-GS98 solar density profile\ \cite{Vinyoles:2016djt}. The line of sight runs into the page, so each point of the disk is an impact parameter and the neutrino crosses the whole Sun along it. The color is the phase-averaged $P_{\nu_e \to \nu_e}$ at exit, for a relative energy spread of $10\%$, averaged over the area of each pixel. See notebook \#28 and Sec.~\ref{sec:solar_tomography} for details.}
\end{figure*}

Figure~\ref{fig:solar_tomography} shows the averaged survival probability across the face of the Sun, at selected energies. The line of sight runs into the page, so each point in the disk fixes the impact parameter of one trajectory, along which the neutrino crosses the whole Sun. For one impact parameter, the probability is computed via
\begin{pysnip}
import magnus.oscprob as oscprob

# The chord above: b = 0.3 R_sun
P = oscprob.osc_prob_matter_std_potential(
    3, ne, E, 2*half, average=True,
    average_initial_state='decohered',
    osc_params=osc, L0=0.0, nu_i=gd.NUE,
    nu_f=gd.NUE,
    density_is_of_number_of_electrons=True)
P                        # 0.283618
\end{pysnip}
Here, {\tt average=True} returns the phase average along the chord, and {\tt average\_initial\_state='decohered'} tells \magnus\ that the neutrino reaches the Sun decohered, \ie, as $\nu_1$, $\nu_2$, or $\nu_3$ (Sec.~\ref{sec:average_keyword}). The default, {\tt 'flavor'}, would instead treat it as a $\nu_e$ produced at the edge of the Sun, which keeps the interference between $\nu_1$ and $\nu_2$; their phase across the Sun is only about 1~rad at 100~GeV. The two choices give very different results on this chord:
\begin{pysnip}
kw = dict(
    osc_params=osc, L0=0.0, nu_i=gd.NUE,
    nu_f=gd.NUE, average=True,
    density_is_of_number_of_electrons=True)
f = oscprob.osc_prob_matter_std_potential
for start in ('flavor', 'decohered'):
    P = f(3, ne, E, 2*half, **kw,
          average_initial_state=start)
    print(start, P)  # 0.535944, 0.283618
\end{pysnip}
A neutrino from a distant source arrives without that interference, which is why \figu{solar_tomography} uses {\tt 'decohered'}. \magnus\ computes the phase-averaged probability over the default relative energy spread of 10\%, without sampling energies. Averaging a scan of probabilities over energy instead does not, in general, give the same result (Sec.~\ref{sec:averaging_is_not_estimating}). On this chord, a little interference survives the 10\% spread, so the phase-averaged probability depends slightly on the spread, by about 0.01 per $e$-fold; \magnus\ issues a {\tt PhaseAveragingWarning} saying so.

One phase survives the average: the matter phase, $\int V_{\rm CC}\,dl$, which is $8\,699$~rad across the solar diameter. Because it does not depend on energy, no energy spread averages it out. Each chord, however, crosses a different amount of matter, so the matter phase varies with the impact parameter. As $b$ changes, the phase cycles through $2\pi$ over and over, and $P$ oscillates with it, drawing fine rings in the core of the disk. The rings are $1.8\times10^{-4}\,R_\odot$ apart at $b = 0.11\,R_\odot$ and $5.1\times10^{-4}\,R_\odot$ apart at $b = 0.3\,R_\odot$, and at 1~TeV they change $P$ by up to about $\pm 0.03$. The probability computed in the snippets above, $P = 0.2836$ at $b = 0.3\,R_\odot$, lies on one of these rings: as $b$ moves by one ring spacing, $5.1\times10^{-4}\,R_\odot$, around that value, $P$ changes appreciably, from 0.282 to 0.308. These rings are far finer than any realistic experimental angular resolution, so an observation would average over many of them. Each pixel of \figu{solar_tomography} is averaged over its area, which also keeps the rings from aliasing into false patterns. Each panel from 30~GeV to 3~TeV uses $7\,791$ impact parameters. Most of them, $6\,790$, lie inside $b = 0.5\,R_\odot$, where the rings are densest.

At $10$~MeV, the disk is uniform at $\sum_i \lvert \mathbb{R}_{ei} \rvert^4 \approx 0.55$: the Sun is transparent, as $P^{\rm cross} = \mathbb{1}$ requires. At intermediate energies, where \figu{solar_adiabaticity} shows the passage to be non-adiabatic, rings appear: there, the level-crossing probabilities depart from the identity. The phase average also keeps part of the interference between eigenstates that \equ{averaged_varying} drops, and this changes the depth of the rings: at 300~GeV, the deepest ring reaches $P = 0.03$ at $b = 0.81\,R_\odot$, against $0.22$ from \equ{averaged_varying}. At higher energies, the deepest ring moves outward, to $b = 0.88\,R_\odot$ at 1~TeV and $0.91\,R_\odot$ at 10 and 50~TeV. At 50~TeV, the disk is uniform again, but for a different reason than at 10~MeV: inside the Sun, the matter potential dominates, so a $\nu_e$ that enters the Sun leaves it as a $\nu_e$, at every impact parameter, having only gained a phase, which does not affect the survival probability.

No neutrino telescope can resolve this structure. Super-Kamiokande has imaged the Sun in MeV neutrinos~\cite{Super-Kamiokande:2005wtt}, but the image is blurred to about $25^\circ$ at 10~MeV, because the recoil electron in neutrino-electron scattering follows the neutrino direction only loosely; the solar disk spans half a degree. At TeV energies, neutrino telescopes reconstruct muon tracks to within about $1^\circ$~\cite{IceCube:2019cia}, comparable to the size of the disk, and particle showers only to within $10^\circ$--$40^\circ$~\cite{IceCube:2019lzm}. The rings of \figu{solar_tomography}, by contrast, are a few to a few tens of arcseconds wide. Moreover, at GeV energies, where the rings first appear, no astrophysical neutrino flux has been identified above the atmospheric background. Finally, both \figu{solar_adiabaticity} and \figu{solar_tomography} include only oscillations: absorption in the Sun, which becomes important above about 100~GeV, is outside what \magnus\ models (Sec.~\ref{sec:limitations}).


\subsubsection{Astrophysical relativistic jet}
\label{sec:jet}

Gamma-ray bursts are the relativistic jets that some massive stars launch as they collapse~\cite{Woosley:1993wj, MacFadyen:1998vz, Piran:2004ba, Meszaros:2006rc}. Shells of plasma ejected by the central engine collide with one another along the jet~\cite{Rees:1994nw}. Reference~\cite{Paczynski:1994uv} first proposed these collisions as a source of neutrinos. Protons accelerated at those collisions can make TeV--PeV neutrinos~\cite{Waxman:1997ti, Guetta:2003wi, Murase:2005hy, Hummer:2011ms, Bustamante:2014oka}. The jet itself does not change their flavor composition: at the radii where the collisions happen, $10^{13}$--$10^{15}$~cm, its density is around $10^{-11}$~g~cm$^{-3}$, nine orders of magnitude below the density at which a TeV neutrino resonates; the phase that the jet matter adds to the evolution is below $10^{-4}$. Matter effects arise only while the jet is still inside the star, in the choked jets that never break out and in the precursor phase of the jets that do\ \cite{Mena:2006eq, Razzaque:2009kq, Sahu:2010ap, Varela:2014mma, Xiao:2015gea}. The neutrinos then cross the stellar envelope. Its density falls from about 0.3~g~cm$^{-3}$ at the head of the jet to zero at the surface, so every neutrino above about 50~GeV meets its 1-3 resonance somewhere along the way. At low energies, the crossing is adiabatic, and the neutrino leaves the star in a mass eigenstate; at high energies, it is not, and the neutrino keeps its flavor content. The flavor ratios at Earth depend on the neutrino energy, in a way that reflects the density profile of the star\ \cite{Razzaque:2009kq, Xiao:2015gea}.

As an example, we take a jet whose head is at $r_0 = 6.3 \cdot 10^{10}$~cm, inside a blue supergiant of radius $R_\star = 3 \cdot 10^{12}$~cm, with the hydrogen envelope of \Refs\ \cite{Mena:2006eq, Razzaque:2009kq}, $\rho(r) = 3.3 \cdot 10^{-6}\,(R_\star/r - 1)^3$~g~cm$^{-3}$, and an electron fraction of $Y_e = 1$:
\begin{pysnip}
import numpy as np
import magnus.globaldefs as gd
import magnus.hamiltonians as ham
import magnus.oscprob as oscprob

OSC = gd.load_nufit_params('NuFIT 6.1')
CM = 1.0e-5*gd.UNIT_KM   # One cm in eV^-1
RSTAR = 3.0e12           # Star's radius, cm
R0 = 6.3e10              # Jet head, cm

def rho(l):
    """Density, g/cm3, at l in eV^-1."""
    return 3.3e-6*(RSTAR/(l/CM) - 1.0)**3
\end{pysnip}
The density at the head is $0.33$~g~cm$^{-3}$. The oscillation length there is set by the potential, $2\pi/V_{\rm CC} \approx 5 \cdot 10^{9}~{\rm cm} \approx 0.1 r_0$.

A neutrino born as $\nu_\alpha$ at the jet head leaves the star in the state $\mathbb{U}(R_\star, r_0)\,\lvert\nu_\alpha\rangle$, where $\mathbb{U}(R_\star, r_0)$ is the evolution operator across the envelope. Its weight in the mass eigenstate $\nu_i$ is
\begin{equation}
 w_{\alpha i} = \bigl\lvert [\mathbb{V}^\dagger(R_\star)\,\mathbb{U}(R_\star, r_0)]_{i\alpha} \bigr\rvert^2 \;,
\end{equation}
where $\mathbb{V}(R_\star)$ is the vacuum mixing matrix. On its way to Earth, over a cosmological distance, the mass eigenstates lose coherence\ \cite{Razzaque:2009kq}. The probability at Earth is then
\begin{equation}
 \langle P_{\nu_\alpha \to \nu_\beta} \rangle = \sum_i w_{\alpha i}\, \lvert \mathbb{V}_{\beta i}(R_\star) \rvert^2 \;.
\end{equation}
This is \equ{averaged_varying}, with the contribution of the star computed directly as $w_{\alpha i}$ and the vacuum eigenbasis at detection.

The weights $w_{\alpha i}$ depend on the relative phases of the flavor amplitudes at the surface of the star. The flavor probabilities, $\lvert \mathbb{U}_{\beta\alpha}(R_\star, r_0) \rvert^2$, discard those phases, so the weights must be computed from the evolution operator. The scenario functions return the operator alongside the probabilities when called with {\tt return\_evolution\_operator=True}. As usual, the refinement ladder chooses the slabs:
\begin{pysnip}
E = 1.0*gd.UNIT_TEV  # 1 TeV
# Evolve from the jet head (L0) to the
# surface (L); also return the operator
out = oscprob.osc_prob_matter_std_potential(
    3, rho, E, RSTAR*CM, OSC, L0=R0*CM,
    electron_fraction=1.0,  # hydrogen
    rtol=1.0e-6, atol=1.0e-6,
    density_matter_is_in_g_per_cm3=True,
    return_evolution_operator=True)
# Probabilities and evolution operator
P, U = out
\end{pysnip}
Here, {\tt P} holds the flavor probabilities at the surface of the star, and {\tt U} is $\mathbb{U}(R_\star, r_0)$ in the flavor basis.

\begin{figure}[t!]
 \centering
 \includegraphics[width=\columnwidth]{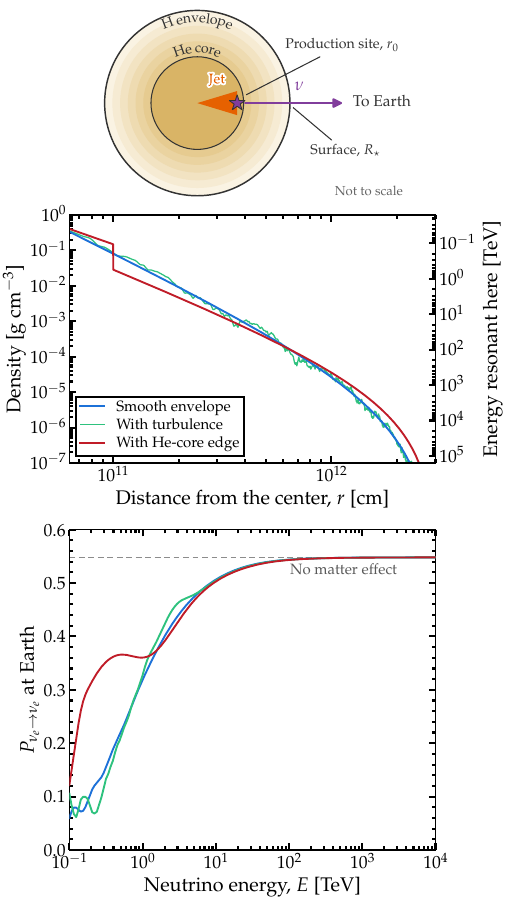}
 \caption{\label{fig:jet}\textbf{Jet neutrinos in a stellar envelope.} Neutrinos from a relativistic jet inside a collapsing star. \textit{Top:} the setup. Neutrinos are made at the head of the jet, at $r_0 = 6.3 \cdot 10^{10}$~cm from the center, and leave the star at $R_\star = 3 \cdot 10^{12}$~cm. \textit{Middle:} matter density along their path, for a smooth envelope, the same envelope with turbulence overlaid, and the same envelope with a drop in density at the edge of the helium core. The right axis gives the energy whose 1-3 resonance lies at each density. \textit{Bottom:} probability that a $\nu_e$ produced at the jet head is detected as $\nu_e$ at Earth, against its energy, for the three envelopes. See notebook \#28 and Sec.~\ref{sec:jet} for details.}
\end{figure}

From {\tt U}, the weights and the probability at Earth follow:
\begin{pysnip}
# Vacuum mixing matrix, V(R_star)
hv = np.array(ham.hamiltonian_3nu_vacuum(
    E, **OSC), dtype=complex)
_, R = np.linalg.eigh(hv)
# Weights, w[i, alpha]
w = abs(R.conj().T @ U)**2
# Probability at Earth, P[beta, alpha]
P_earth = abs(R)**2 @ w
P_earth[gd.NUE, gd.NUE]    # 0.315
\end{pysnip}
Without matter effects, the same calculation gives $0.55$.

Figure~\ref{fig:jet} shows this probability from 0.1~TeV to 10~PeV for three density envelopes. The first envelope is the smooth profile above. The second adds turbulence to it, as one realization of forty Kolmogorov modes with wavelengths from $10^{10}$ to $10^{12}$~cm and a root-mean-square amplitude of 20\%. The third is model C of \Ref\ \cite{Mena:2006eq}, in which the density drops by a factor of five at the edge of the helium core, at $10^{11}$~cm. The code receives this edge as a breakpoint. The neutrinos are made across a region about one oscillation length wide, so each curve averages over eight production points spread across that length. At the lowest energies, the resonance lies only a few oscillation lengths from the head. A single production point would then show an interference pattern, which the average removes. 

Listing~\ref{lst:jet} contains the calculation behind the figure. Each envelope takes well under a minute, over 161 energies and eight production points. At the requested tolerance of $10^{-6}$, the ladder uses a few hundred slabs per energy at PeV energies and a few thousand at 0.1~TeV.

\begin{lstlisting}[language=Python, caption={\textbf{Jet neutrinos in a stellar envelope.} Computing the data in \figu{jet}: the averaged $\nu_e$ survival probability at Earth for neutrinos from a jet head at $r_0$, through the three envelopes. Each call returns the evolution operator from the refinement ladder alongside the probabilities; the projection onto the vacuum mass states and the sum over them are \equ{averaged_varying}, with the eigenbasis at detection the vacuum one. The eight production points span one oscillation length at the jet head, $4.9 \cdot 10^{9}$~cm. The turbulent envelope is one realization of forty Kolmogorov modes between $10^{10}$ and $10^{12}$~cm at 20\% rms; the stepped one is model C of \Ref\ \cite{Mena:2006eq}, with the drop at the helium-core edge declared through {\tt t\_breakpoints}. See notebook \#28 and Sec.~\ref{sec:jet} for details.}, label={lst:jet}, float=t]
import numpy as np
import magnus.globaldefs as gd
import magnus.hamiltonians as ham
import magnus.oscprob as oscprob

OSC = gd.load_nufit_params('NuFIT 6.1')
CM = 1.0e-5*gd.UNIT_KM               # One cm in eV^-1
RSTAR, R0, RHE = 3.0e12, 6.3e10, 1.0e11      # cm
E = np.geomspace(0.1, 1.0e4, 161)*gd.UNIT_TEV

# The three envelopes, g/cm3, r in cm
def smooth(r):
    return 3.3e-6*(RSTAR/r - 1.0)**3

rng = np.random.default_rng(7)        
lam = np.geomspace(1.0e12, 1.0e10, 40)  # Modes, cm
k = 2*np.pi/lam
amp = k**(-1/3)                         # Kolmogorov
amp *= 0.20/np.sqrt(0.5*np.sum(amp**2)) # 20% rms
phi = rng.uniform(0.0, 2*np.pi, 40)
def turbulent(r):
    d = np.sum(amp*np.cos(k*r + phi))
    return smooth(r)*(1.0 + d)

def stepped(r):            # Mena et al., model C
    x = RSTAR/r - 1.0
    return 6.3e-6*(20.0*x**2.1 if r < RHE else x**2.5)

# Production points spread over one oscillation
# length at the jet head
r0s = R0 + 4.9e9*np.arange(8)/8.0

def p_earth(rho, **extra):
    """P at Earth, phase-averaged, and averaged
    over the production points."""
    P_out = np.zeros((len(E), 3, 3))
    for r0 in r0s:
        P, U = oscprob.osc_prob_matter_std_potential(
            3, lambda l: rho(l/CM), E, RSTAR*CM, OSC,
            L0=r0*CM, electron_fraction=1.0,
            density_matter_is_in_g_per_cm3=True,
            rtol=1.0e-6, atol=1.0e-6,
            return_evolution_operator=True, **extra)
        for i, e in enumerate(E):
            hv = np.array(ham.hamiltonian_3nu_vacuum(
                e, **OSC), dtype=complex)
            w, R = np.linalg.eigh(hv)
            content = abs(R.conj().T @ U[i])**2
            P_out[i] += abs(R)**2 @ content/len(r0s)
    return P_out

P_smooth = p_earth(smooth)
P_turb = p_earth(turbulent)
P_step = p_earth(stepped, t_breakpoints=[RHE*CM])
# Drawn: P_smooth[:, gd.NUE, gd.NUE], and the others
\end{lstlisting}

For the smooth envelope, the $\nu_e$ survival probability at Earth rises with energy. At 0.1~TeV, a $\nu_e$ leaves the star mostly as $\nu_3$, the mass eigenstate with the smallest electron content, so it is rarely detected as $\nu_e$. At 100~TeV, the probability reaches its vacuum value. The probability rises because the resonance crossing becomes less adiabatic as the energy grows. The resonance density falls as $1/E$. Deep in the envelope, $\rho \propto r^{-3}$, so the resonance moves outward as $r_{\rm res} \propto E^{1/3}$. There, the density changes over a distance $\rho/\lvert d\rho/dr \rvert = r_{\rm res}/3 \propto E^{1/3}$, while the oscillation length at the resonance grows as $E$. Therefore, their ratio, which sets how adiabatic the crossing is, falls as $E^{-2/3}$. At high energies, the crossing is non-adiabatic, and the neutrino keeps its flavor. At the PeV energies that IceCube detects, the star leaves the flavor content unchanged.

Turbulence shifts the probability by up to 0.05 below a few TeV. At these energies, the turbulent modes near the resonance have wavelengths comparable to the oscillation length (as in Sec.~\ref{sec:turbulence}). At the drop, the density falls by a factor of five within a small fraction of an oscillation length. Neutrinos between 0.1 and 0.55~TeV have their 1-3 resonance density inside the drop. They skip over their resonance, rather than crossing it gradually, so the crossing is non-adiabatic. This produces the plateau at about 0.36 in \figu{jet}, which extends to about 1~TeV. Above about 1.5~TeV, the resonance lies well outside the helium core, and the drop no longer matters: the stepped and smooth envelopes give the same probability to within 0.01.


\subsubsection{Long-range interactions in the Sun}
\label{sec:lri_sun}

\begin{figure}[t!]
 \centering
 \includegraphics[width=\columnwidth]{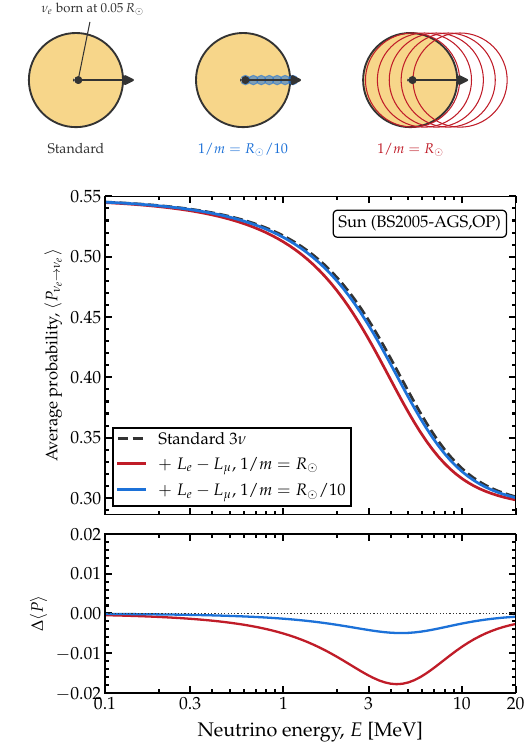}
 \caption{\label{fig:solar_lri}\textbf{A long-range force in the Sun.} Averaged $\nu_e$ survival probability of solar neutrinos under a long-range $L_e - L_\mu$ interaction sourced by solar electrons. \textit{Top:} the electrons the neutrino feels at each mediator range. A $\nu_e$ born at $0.05\,R_\odot$ leaves along the drawn ray; the discs enclose the electrons within $1/m$ of it. At $1/m = R_\odot/10$, it feels a narrow tube of electrons; at $1/m = R_\odot$, most of the Sun, even beyond the surface. The charged-current potential is local, so it has no disc. \textit{Middle:} $\langle P_{\nu_e \to \nu_e}\rangle$ against energy in the B16-GS98 model\ \cite{Vinyoles:2016djt}, without the new potential and with it at both ranges. The two ranges share one coupling, set so that the longer-ranged potential is $0.1\,V_{\rm CC}$ at birth. The probability is read at $20\,R_\odot$, where it has faded. \textit{Bottom:} the difference from the standard case. See notebooks \#19 and \#28, and Sec.~\ref{sec:lri_sun} for details.}
\end{figure}

This example adds a new term to the Hamiltonian, one that \magnus\ does not ship. We write it as described in Sec.~\ref{sec:building_new_hamiltonian}, and average the probability as in the standard solar case of Sec.~\ref{sec:sun}. The new term comes from gauging $L_e - L_\mu$, the difference between the electron and muon lepton numbers, which is anomaly-free. Its gauge boson, $Z'$, is a new light mediator\ \cite{He:1990pn, Bustamante:2018mzu, Agarwalla:2024ylc}. The charge $L_e - L_\mu$ is $+1$ for the electron and $\nu_e$, $-1$ for the muon and $\nu_\mu$, and zero for every other field of the Standard Model. The electrons of the Sun source a potential that $\nu_e$ and $\nu_\mu$ feel with opposite signs, and that $\nu_\tau$ does not feel. The Hamiltonian is
\begin{equation}
 \label{equ:h3_lri}
 \mathbb{H}^{\rm LRI}(E, l)
 =
 \mathbb{H}^{\rm vac}(E) + V_{\rm CC}(l)\,\mathbb{P}
 + V_{e\mu}(l)\,{\rm diag}\left(1, -1, 0\right) \;,
\end{equation}
where the first two terms are the standard vacuum and charged-current terms of Sec.~\ref{sec:recap}, and the third is the new one.

Unlike $V_{\rm CC}$, the new potential is non-local. Each electron sources a Yukawa potential around it, with a range set by the mass $m$ of the $Z'$. A neutrino at $\mathbf{r}$ feels the electrons within about $1/m$ of it:
\begin{equation}
 \label{equ:yukawa}
 V_{e\mu}(\mathbf{r})
 =
 \frac{g^{\prime 2}}{4\pi}
 \int d^3r^\prime \, n_e(\mathbf{r}^\prime) \,
 \frac{e^{-m \lvert \mathbf{r} - \mathbf{r}^\prime \rvert}}
      {\lvert \mathbf{r} - \mathbf{r}^\prime \rvert} \;,
\end{equation}
where $g^\prime$ is the new gauge coupling. 

When the interaction range, $1/m$, is larger than the body, the potential depends mainly on how many electrons the body holds, and only weakly on where they are\ \cite{Wise:2018rnb, Bustamante:2018mzu, Agarwalla:2023sng}. When $1/m$ is much shorter than the distance over which the density changes, only nearby electrons contribute, and $V_{e\mu} \to g^{\prime 2} n_e(\mathbf{r})/m^2$. In this limit, the potential is local and proportional to $V_{\rm CC}$, like a non-standard interaction with constant couplings~\cite{Coloma:2020gfv}. Between the two limits, the potential depends on the shape of the density profile. For a spherical body, like the Sun, the angular integral can be done analytically. The potential then splits into the contributions of the electrons inside and outside radius $r$:
\begin{equation}
 \label{equ:yukawa_1d}
 \begin{split}
 V_{e\mu}(r)
 = {} &
 \frac{g^{\prime 2}}{r}\, e^{-m r}
 \int_0^{r} dr^\prime\, r^{\prime 2}\, n_e(r^\prime)\, {\rm shc}(m r^\prime) \\
 & + g^{\prime 2}\, {\rm shc}(m r)
 \int_r^{R_\odot} dr^\prime\, r^\prime\, n_e(r^\prime)\, e^{-m r^\prime} \;,
 \end{split}
\end{equation}
where ${\rm shc}(x) \equiv \sinh(x)/x$. Both are running integrals over the density profile, so one pass over the table of a solar model gives the potential at every radius. For a massless mediator, ${\rm shc}$ and the exponentials tend to 1, and \equ{yukawa_1d} becomes the Coulomb potential of the electrons. Notebook \#19 derives \equ{yukawa_1d} and checks it against the closed form for a uniform ball.

Figure~\ref{fig:solar_lri} shows the averaged $\nu_e$ survival probability from 0.1 to 20~MeV. The density profile is from the B16-GS98 solar model\ \cite{Vinyoles:2016djt}. The neutrino is born at $0.05\,R_\odot$, near where $^8$B production peaks in that model. The figure compares the standard case with two mediator ranges, $1/m = R_\odot$ and $R_\odot/10$, corresponding to $m \approx 3 \cdot 10^{-16}$ and $3 \cdot 10^{-15}$~eV, respectively. Both ranges share one coupling. We fix it so that the longer-ranged potential is a tenth of $V_{\rm CC}$ where the neutrino is born. The longer-ranged potential lowers the probability by up to 0.018, near 4~MeV. The shorter-ranged one lowers it by at most 0.005. Listing~\ref{lst:arbitrary} contains the calculation behind the figure.

At the same coupling, a longer range lets the neutrino feel more of the Sun's electrons. Where the neutrino is born, the longer-ranged potential is $0.10\,V_{\rm CC}$, about four times the shorter-ranged one, $0.027\,V_{\rm CC}$. This difference at birth sets the difference between the two curves. The neutrino leaves the Sun adiabatically, so its averaged probability depends only on the Hamiltonian where it is born and where it is detected.

Along the path, both new potentials fall more slowly than $V_{\rm CC}$. The electron density of the Sun falls by about five orders of magnitude from the center to $0.98\,R_\odot$, and $V_{\rm CC}$ falls with it. In contrast, $V_{e\mu}$ is an integral over the electrons within about $1/m$, which for the longer range include those of the dense core. In Listing~\ref{lst:arbitrary}, the ratio $V_{e\mu}/V_{\rm CC}$ at half a solar radius is
\begin{pysnip}
half = np.searchsorted(r, 0.5*R_SUN)
v_emu[1.0][half]/vcc[half]     # 2.39
v_emu[0.1][half]/vcc[half]     # 0.098
\end{pysnip}
At $0.98\,R_\odot$, the two ratios are about 700 and 1.2. The top panel of \figu{solar_lri} illustrates this. With $1/m = R_\odot$, the neutrino feels most of the Sun wherever it is, and the potential extends beyond the surface. We read out the probability at $20\,R_\odot$ away from the solar center, where the new potential has faded.

In Listing~\ref{lst:arbitrary}, the Hamiltonian accepts an array of positions, so the engine can evaluate it at many positions in one call (Sec.~\ref{sec:cost}). The averaged probability comes from {\tt osc\_prob\_energy\_baseline} with {\tt average=True}. The standard case uses the same call, with $V_{e\mu} = 0$. At 10~MeV,
\begin{pysnip}
i = np.argmin(abs(Es - 10.0*gd.UNIT_MEV))
P_std[i]                      # 0.325
P_lri[1.0][i], P_lri[0.1][i]  # 0.317, 0.323
\end{pysnip}

\begin{figure}[t!]
 \centering
 \includegraphics[width=\columnwidth]{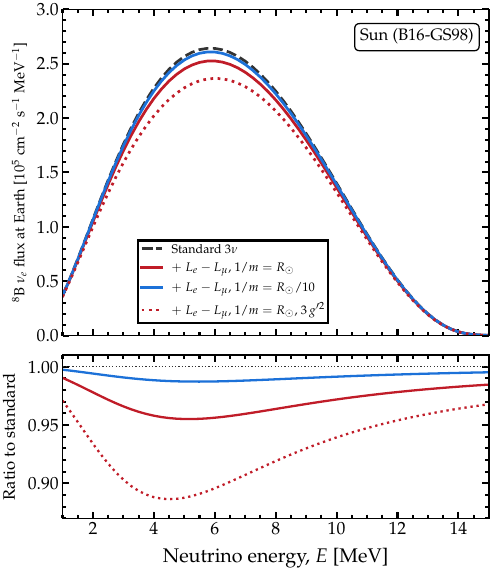}
 \caption{\label{fig:solar_8b_flux}\textbf{The $^8$B neutrino flux at Earth under a long-range force.} \textit{Top:} flux of $\nu_e$ from $^8$B decay at Earth, against energy, for the standard case and for the two mediator ranges of \figu{solar_lri}, with the same coupling. The dotted curve has $1/m = R_\odot$ and a coupling $g^{\prime 2}$ three times larger. The fluxes are the $^8$B spectrum of Ref.~\cite{Winter:2004kf} times the total flux of the B16-GS98 model, $5.46 \cdot 10^6$~cm$^{-2}$~s$^{-1}$\ \cite{Vinyoles:2016djt}, times the survival probability averaged over where $^8$B neutrinos are made in the model. \textit{Bottom:} each case divided by the standard one. See notebook \#28 and Sec.~\ref{sec:lri_sun} for details.}
\end{figure}

Figure~\ref{fig:solar_8b_flux} shows the flux of $\nu_e$ from $^8$B decay at Earth, from 1 to 15~MeV. Since $^8$B neutrinos are produced over a range of radii, 80\% of them between $0.025\,R_\odot$ and $0.081\,R_\odot$ in the B16-GS98 model, we average the survival probability over the production distribution:
\begin{equation}
 \label{equ:b8_flux}
 \frac{d\Phi_{\nu_e}}{dE}
 =
 \Phi_{^8{\rm B}}\, \lambda(E)
 \int_0^{R_\odot} dr\, f(r)\,
 \langle P_{\nu_e \to \nu_e}(E; r) \rangle \;,
\end{equation}
where $\Phi_{^8{\rm B}} = 5.46 \cdot 10^6$~cm$^{-2}$~s$^{-1}$ is the total flux of the model\ \cite{Vinyoles:2016djt}, $\lambda(E)$ is the $^8$B spectrum normalized to unity\ \cite{Winter:2004kf}, $f(r)$ is the radial distribution of production, normalized to unity, and $\langle P_{\nu_e \to \nu_e}(E; r) \rangle$ is the averaged survival probability of a neutrino born at radius $r$. We evaluate the integral on 28 radii between $0.005\,R_\odot$ and $0.14\,R_\odot$, which contain 99.9\% of production.

(We compute $f(r)$ from the temperature, density, and composition of B16-GS98, as the rate of $^7$Be$(p,\gamma)^8$B. In this rate, $^7$Be is in equilibrium between its production by $^3$He$(\alpha,\gamma)^7$Be and its destruction by electron and proton capture, with the reaction rates of Ref.~\cite{Adelberger:2010qa}. To test this calculation, we apply it to an older solar model, BS2005-AGS,OP, whose $^8$B production distribution is published\ \cite{Bahcall:2004pz}. The two distributions agree within 1\% of the peak. Notebook \#28 contains the calculations.)

Extending Listing~\ref{lst:arbitrary}, \equ{b8_flux} for $1/m = R_\odot$ is
\begin{pysnip}
# Birth radii, weighted by where 8B is made
r_b = np.linspace(0.005, 0.14, 28)*R_SUN
w = f(r_b)*np.gradient(r_b)
w = w/w.sum()
# Average over them, then scale by the 8B
# spectrum and total flux, cm^-2 s^-1 MeV^-1
P = sum(wk*averaged(v_emu[1.0], rk)
        for rk, wk in zip(r_b, w))
flux = 5.46e6*spectrum(Es)*P
\end{pysnip}
where {\tt f} and {\tt spectrum} interpolate $f(r)$ and $\lambda(E)$. The longer-ranged potential lowers the flux by up to 4.5\%, near 5~MeV, and by 3.6\% in total above 1~MeV. The shorter-ranged one lowers it by at most 1.3\%. Tripling $g^{\prime 2}$ for $1/m = R_\odot$ lowers it by up to 11\%, near 4.5~MeV.

\begin{lstlisting}[language=Python, caption={\textbf{A long-range force in the Sun.} Computing \figu{solar_lri}. The electron density comes from the B16-GS98 model\ \cite{Vinyoles:2016djt}. The potential $V_{e\mu}$ is \equ{yukawa_1d} on that profile. The neutrino is born at $0.05\,R_\odot$, and the probability, \equ{averaged_varying}, is read at $20\,R_\odot$. See notebooks \#19 and \#28, and Sec.~\ref{sec:lri_sun} for details.}, label={lst:arbitrary}, float=tbp]
import numpy as np
import magnus.globaldefs as gd
import magnus.hamiltonians as ham
import magnus.matter as matter
import magnus.oscprob as oscprob
import magnus.solarmodels as solarmodels

osc = gd.load_nufit_params('NuFIT 6.1')

# The B16-GS98 model that ships with Magnus,
# on its tabulated radii and on to 20 R_sun
model = 'B16-GS98'
tab = solarmodels.load_solar_model(model)
r_tab = tab['r_over_r_sun']*gd.SUN_RADIUS*gd.UNIT_KM
R_SUN = r_tab[-1]                   # The surface
R0, R_END = 0.05*R_SUN, 20.0*R_SUN  # Birth, readout
r = np.concatenate([r_tab, np.geomspace(
    1.002*R_SUN, R_END, 800)])
ne_func = solarmodels.electron_density_profile(model)
n_e = ne_func(r)                    # Fades past R_SUN
per_ne = matter.VCC_func(0.0, lambda l: 1.0)
vcc = per_ne*n_e

def vcc_sun(l):
    """V_CC at l."""
    return per_ne*ne_func(l)

def cumulative(y, x):
    """Running trapezoidal integral, zero at x[0]."""
    steps = 0.5*(y[1:] + y[:-1])*np.diff(x)
    return np.concatenate([[0.0], np.cumsum(steps)])

def shc(x):
    """sinh(x)/x, equal to 1 at the origin."""
    x = np.asarray(x, dtype=float)
    return np.divide(np.sinh(x), x, out=np.ones_like(x),
                     where=(x != 0.0))

def long_range_potential(r, n_e, m):
    """V_emu(r)/g'^2 for a spherical n_e."""
    I_in = cumulative(r**2*n_e*shc(m*r), r)
    I_out = cumulative(r*n_e*np.exp(-m*r), r)
    I_out = I_out[-1] - I_out      # From r outward
    r_safe = np.where(r > 0.0, r, 1.0e-30)
    return np.exp(-m*r)*I_in/r_safe + shc(m*r)*I_out

# Two mediator ranges, one coupling: g'^2 makes V_emu
# a tenth of V_CC at R0 when 1/m = R_sun
v_emu = {}
for frac in (1.0, 0.1):            # 1/m in R_sun
    v_emu[frac] = long_range_potential(
        r, n_e, 1.0/(frac*R_SUN))
g2 = 0.1*vcc_sun(R0)/np.interp(R0, r, v_emu[1.0])
v_emu = {frac: g2*v for frac, v in v_emu.items()}

# The standard part from the shipped builders; the
# new term is the only line written here
h_vac = ham.\
    hamiltonian_3nu_vacuum_energy_independent(**osc)
q = np.diag([1.0, -1.0, 0.0])     # L_e-L_mu charges

def H_lri(v):
    """H(E, l) of the text, with V_emu = v."""
    def H(E, l):
        h_matt = ham.hamiltonian_3nu_matter_td(
            l, vcc_sun)
        return (h_vac/E + h_matt
                + np.interp(l, r, v)[..., None, None]*q)
    return H

Es = np.logspace(-1.0, np.log10(20.0), 70)*gd.UNIT_MEV

def averaged(v, r0=R0):
    """<P_ee> from r0 to R_END, per E."""
    return oscprob.osc_prob_energy_baseline(
        H_lri(v), Es, R_END, r0, nu_i=gd.NUE,
        nu_f=gd.NUE, average=True)

P_std = averaged(0.0*vcc)
P_lri = {frac: averaged(v_emu[frac]) for frac in v_emu}
\end{lstlisting}


\subsubsection{Turbulent matter profile}
\label{sec:turbulence}

\begin{figure}[t!]
 \centering
 \includegraphics[width=\columnwidth]{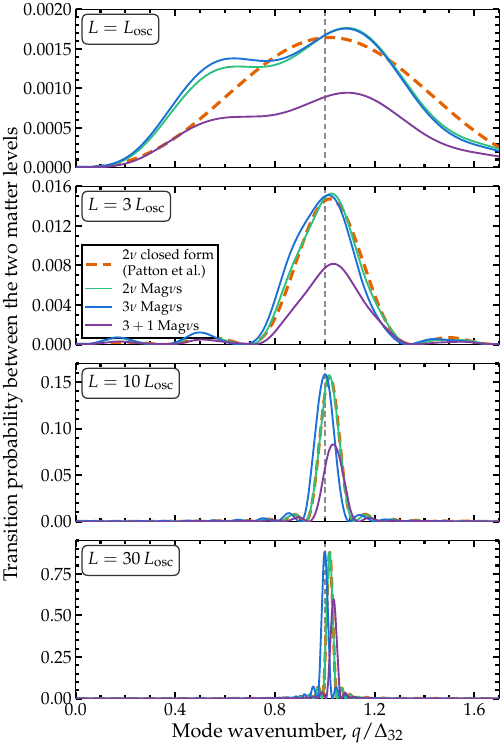}
 \caption{\label{fig:turbulence_rabi}\textbf{Turbulence: numerical against theory.} Transition probability between two matter levels, against the wavenumber $q$ of a single density mode of amplitude $C = 0.03$, for a 5-GeV neutrino crossing a region of mean density $\rho_0 = 4$~g~cm$^{-3}$. The levels are those of the mean density, between which no transition occurs without the mode. The panels are region lengths of $1$, $3$, $10$, and $30\,L_{\rm osc}$, with $L_{\rm osc} = 2\pi/\Delta_{32} = 11\,030$~km. The closed form of \Refs\ \cite{Patton:2013dba, Patton:2014lza}, \equ{turb_rabi}, is evaluated for the two-flavor reduction of the 1--3 sector. \magnus\ is run at two, three, and four flavors, the last with $\sin^2\theta_{14} = \sin^2\theta_{24} = 0.10$ and $\Delta m^2_{41} = 1$~eV$^2$. See notebook \#28 and Sec.~\ref{sec:turbulence} for details.}
\end{figure}

\begin{figure}[t!]
 \centering
 \includegraphics[width=\columnwidth]{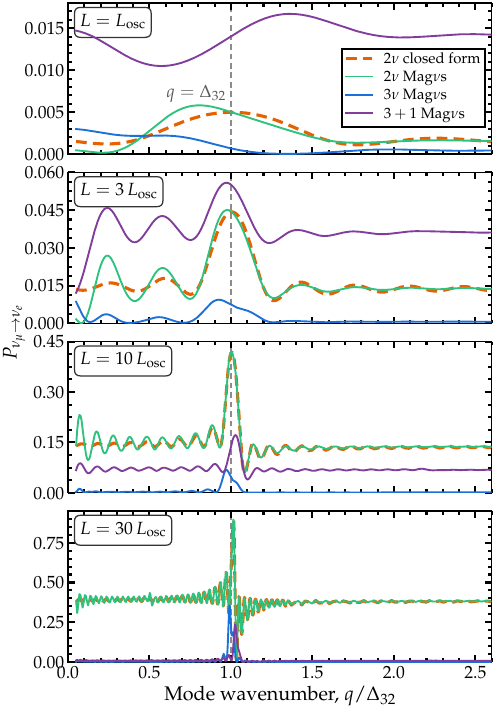}
 \caption{\label{fig:turbulence}\textbf{The same mode in the flavor channel.} Probability $P_{\nu_\mu \to \nu_e}$ for the same neutrino, medium, mode, and region lengths as Fig.~\ref{fig:turbulence_rabi}, at two, three, and four flavors. The line labeled $q = \Delta_{32}$ marks the mode that matches the three-flavor gap; the two- and four-flavor gaps lie 2\% and 3.5\% above it. The curves sit at different heights because, even without the mode, the three flavor counts give different probabilities. The closed-form curve is \equ{turb_rw_flavor}, for the same two-flavor system as in Fig.~\ref{fig:turbulence_rabi}. See notebook \#28 and Sec.~\ref{sec:turbulence} for details.}
\end{figure}

Matter behind a supernova shock does not settle, because convection keeps stirring it. Its density fluctuates about its mean by tens of percent, on length scales from the size of the star to far below it\ \cite{Loreti:1995ae, Fogli:2006xy, Patton:2013dba, Patton:2014lza}. Thus, the density along the path of an escaping neutrino has two components: a smooth profile that decreases with radius, and random fluctuations on top of it. The fluctuations can be decomposed into Fourier modes, each a periodic density variation of wavenumber $q$. A mode moves neutrinos between two eigenstates of $\mathbb{H}$ when $q$ matches the gap between their eigenvalues, $\Delta_{jk} = \lambda_j - \lambda_k$. This is the parametric resonance of Sec.~\ref{sec:order_is_the_slab_product}.

References\ \cite{Patton:2013dba, Patton:2014lza} solved this problem in closed form. They treated the random component as a perturbation of the regular one and kept only its resonant term. For a density mode $\rho(l) = \rho_0 \left[1 + C \cos(q l)\right]$, with $\rho_0$ the mean density and $C$ the relative amplitude of the mode, they obtained a Rabi formula for the transition between the two levels,
\begin{equation}
 P_{jk}
 =
 \frac{\kappa^2}{p^2 + \kappa^2}\,
 \sin^2\!\left(\sqrt{p^2 + \kappa^2}\; L\right) \;,
 \label{equ:turb_rabi}
\end{equation}
where $L$ is the length of the turbulent region and $p \equiv \tfrac{1}{2} (\Delta_{jk} - q )$ is the detuning of the mode from the gap. The coupling the mode induces between the levels is $\kappa \equiv \tfrac{1}{2} C V_{\rm CC} \lvert U^m_{ej} U^m_{ek}\rvert$, with $U^m$ the mixing matrix in matter at $\rho_0$. On resonance, where $q = \Delta_{jk}$, the probability is $\sin^2(\kappa L)$. Complete conversion requires a region of length $\pi/(2\kappa)$, which grows as $C$ shrinks.

As an example, we consider a 5-GeV neutrino crossing a region of mean density $\rho_0 = 4$~g~cm$^{-3}$ that carries one mode of amplitude $C = 0.03$:
\begin{pysnip}
import numpy as np
import magnus.globaldefs as gd
import magnus.matter as matter
import magnus.hamiltonians as ham

OSC = gd.load_nufit_params('NuFIT 6.1')
E = 5.0*gd.UNIT_GEV    # Neutrino energy
RHO0 = 4.0             # Mean density, g/cm3
C = 0.03               # Mode amplitude

def one_mode(q):
    """Density along the ray: the mean, plus
    one mode of wavenumber q."""
    def rho(l):
        x = np.asarray(l, dtype=float)
        return RHO0*(1.0 + C*np.cos(q*x))
    return rho
\end{pysnip}
The gaps of $\mathbb{H}$ at $\rho_0$ set the range of $q$ to scan:
\begin{pysnip}
# n_e and V_CC of the smooth medium
ne0 = matter.num_density_e_func(
    0.0, lambda l: RHO0,
    density_matter_is_in_g_per_cm3=True)
V0 = matter.VCC_func(0.0, lambda l: ne0)

# H at the mean density: vacuum plus matter
H0 = np.array(ham.hamiltonian_3nu_vacuum(
    E, **OSC), dtype=complex)
H0 += ham.hamiltonian_3nu_matter(V0)

# Eigenvalues w, matter eigenvectors Vm, and
# the gap the mode has to match
w, Vm = np.linalg.eigh(H0)
d32 = abs(w[2] - w[1])
LOSC = 2*np.pi/d32     # 11 030 km

# Coupling of levels 2 and 3, and the length
# of complete conversion, in units of LOSC
kap = 0.5*C*V0*abs(Vm[0, 1]*Vm[0, 2])
np.pi/(2*kap)/LOSC     # 37.4
\end{pysnip}
The mode that matches $\Delta_{32}$ has wavelength $L_{\rm osc} \equiv 2\pi/\Delta_{32} = 11\,030$~km, the oscillation length of levels 2 and 3. Complete conversion between them takes $37\,L_{\rm osc}$. Only these two levels carry appreciable electron flavor, since $\lvert U^m_{e1} \rvert^2 = 5.5 \times 10^{-4}$. Hence, the modes that match the other two gaps, $\Delta_{21}$ and $\Delta_{31}$, couple their levels about 15 and 40 times more weakly.

\magnus\ computes the same transition probability without the approximations behind \equ{turb_rabi}, from the evolution operator $\mathbb{U}$ of \equ{probability}. Written in the basis of the eigenstates at $\rho_0$, its element $\lvert [\mathbb{U}]_{jk} \rvert^2$ is the probability of a transition from level $k$ to level $j$. In a medium of constant density, $\mathbb{H}$ is diagonal in that basis at every point. Each level then evolves on its own, and $[\mathbb{U}]_{jk}$ vanishes for $j \neq k$. The mode adds $C V_{\rm CC} \cos(ql)\, U^{m*}_{ej} U^m_{ek}$ to each element of $\mathbb{H}$ in that basis. The off-diagonal elements mix the levels as the neutrino advances, and $\lvert [\mathbb{U}]_{jk} \rvert^2$ measures that mixing over the whole region. The diagonal elements raise and lower the levels with the density. \magnus\ returns $\mathbb{U}$ in the flavor basis; rotating it to the matter basis takes one line:
\begin{pysnip}
from magnus.oscprob import (
  compute_evolution_operator_multiple_slabs
  as chain)

L = LOSC             # The region, one L_osc
rho = one_mode(d32)  # Mode tuned to the gap

def H(l):
    """H at position l, with the mode on."""
    h = np.array(ham.hamiltonian_3nu_vacuum(
        E, **OSC), dtype=complex)
    # V_CC follows the density
    return h + ham.hamiltonian_3nu_matter(
        V0*rho(l)/RHO0)

# One operator per slab, from 9 nodes at
# Magnus order 6
edges = np.linspace(0.0, L, 501)
slabs = np.stack([edges[:-1], edges[1:]], 1)
Us = chain(H, slabs, 9, 6)

# Ordered product, earliest slab first
U = Us[0]
for u in Us[1:]:
    U = u @ U

# Rotate to the matter basis, then read the
# transition between levels 2 and 3
P32 = abs((Vm.conj().T @ U @ Vm)[2, 1])**2
P32                    # 0.001758
\end{pysnip}

Figure~\ref{fig:turbulence_rabi} shows this probability against $q$ for four lengths of the region, computed with \magnus\ and with \equ{turb_rabi}. We evaluate the closed form for the two-flavor 1--3 sector, the system for which it was derived. The closed form relies on the rotating-wave approximation, which neglects a rapidly oscillating part of the coupling; that part averages out only over many oscillation lengths. Hence, it is accurate only in long regions: it differs from the two-flavor \magnus\ result by 28\% of the peak at $L = L_{\rm osc}$, but by only 0.4\% at $L = 30\,L_{\rm osc}$. The resonance narrows as the region lengthens, because a longer region requires a closer match between $q$ and the gap. Adding flavors changes the eigenvalues of $\mathbb{H}$, and with them the gap that the mode must match. Hence, the resonance lies at $q/\Delta_{32} = 1.020$, $1$, and $1.035$ for two, three, and four flavors. The sterile state also draws 10\% of the electron flavor into the fourth level. This lowers $\lvert U^m_{e2} U^m_{e3} \rvert$ from 0.33 to 0.25, so the coupling $\kappa$ is weaker and the four-flavor peak is lower.

Figure~\ref{fig:turbulence} shows the flavor probability $P_{\nu_\mu \to \nu_e}$ for the same sweep, computed via
\begin{pysnip}
import magnus.oscprob as oscprob

kw = dict(
    nu_i=gd.NUMU, nu_f=gd.NUE,
    density_matter_is_in_g_per_cm3=True)

# Two flavors: the 1-3 sector
OSC2 = dict(sth=OSC['s13'], Dm2=OSC['D31'])
P2 = oscprob.osc_prob_matter_std_potential(
    2, one_mode(d32), E, L, OSC2, **kw)

# Three flavors
P3 = oscprob.osc_prob_matter_std_potential(
    3, one_mode(d32), E, L, OSC, **kw)

# The same call with a sterile state added
OSC4 = dict(OSC, s14=np.sqrt(0.10), D41=1.0,
            s24=np.sqrt(0.10), s34=0.0,
            d14=0.0, d24=0.0)
P4 = oscprob.osc_prob_matter_std_potential(
    4, one_mode(d32), E, L, OSC4, **kw)
\end{pysnip}
From $10\,L_{\rm osc}$ on, the probability changes most near $q = \Delta_{32}$. This is the resonance of \figu{turbulence_rabi}, seen in the flavor channel; as before, it narrows as the region lengthens. In shorter regions, modes with $q < \Delta_{32}$ change the probability by as much or more; we explain why below. For $q > \Delta_{32}$, the curves return to their smooth-medium values. The three-flavor curve shows no resonance at the other gaps, $\Delta_{21} = 1.26\,\Delta_{32}$ and $\Delta_{31} = 2.26\,\Delta_{32}$, because the matching modes couple their levels weakly.

Figure~\ref{fig:turbulence} also shows a flavor probability computed from the closed form. Equation~(\ref{equ:turb_rabi}) follows from the propagator of the two-level system in the rotating-wave approximation\ \cite{Patton:2013dba, Patton:2014lza}. Rotated to the flavor basis, it reproduces the resonance computed by \magnus, but not the effect of modes slower than the gap. In the matter basis at $\rho_0$, the propagator is
\begin{equation}
 \tilde{\mathbb{U}}(L)
 =
 {\rm diag}\!\left(e^{iqL/2}, e^{-iqL/2}\right)
 \exp\!\left[
 -i L
 \begin{pmatrix}
  \bar\lambda - p & \kappa_c \\
  \kappa_c^* & \bar\lambda + p
 \end{pmatrix}
 \right] ,
 \label{equ:turb_rw_u}
\end{equation}
where $\bar\lambda = \tfrac{1}{2}(\lambda_1 + \lambda_2)$ is the mean of the two levels, $p = \tfrac{1}{2}(\lambda_2 - \lambda_1 - q)$ is the detuning, and $\kappa_c = \tfrac{1}{2} C V_{\rm CC}\, U^{m*}_{e1} U^m_{e2}$ is the coupling, with $\lvert \kappa_c \rvert = \kappa$. The matrix in the exponent is the Hamiltonian in a frame that rotates with the mode. In that frame, the Hamiltonian does not depend on position, so the propagator is a single matrix exponential. The diagonal prefactor transforms the result back to the matter basis. In the flavor basis, the probability is
\begin{equation}
 P_{\nu_\mu \to \nu_e}
 =
 \left\lvert \left[ U^m\, \tilde{\mathbb{U}}(L)\, U^{m\dagger} \right]_{e\mu} \right\rvert^2 .
 \label{equ:turb_rw_flavor}
\end{equation}

Of the terms that the mode adds to $\mathbb{H}$, the rotating-wave approximation keeps only the off-diagonal part that oscillates in step with the transition between the two levels. It discards the rest, including the diagonal terms, which shift the levels up and down with the density. Below the resonance, the off-diagonal terms drive no transition, as \figu{turbulence_rabi} shows. The diagonal terms still matter there: the levels follow the slowly varying density, which changes the phases they accumulate and, with them, $P_{\nu_\mu \to \nu_e}$. Equation~(\ref{equ:turb_rw_u}) lacks these terms, so its curve stays at the value for the smooth medium.


\subsection{Running a scan in parallel}
\label{sec:njobs}

A scan of many probabilities can be shared among processes. Every wrapper accepts the argument {\tt n\_jobs} for that, the number of processes to use, with {\tt n\_jobs = 1} as default and {\tt n\_jobs = -1} for all cores:
\begin{pysnip}
import numpy as np
import magnus.oscprob as oscprob
import magnus.globaldefs as gd

N = 5000
E = np.logspace(-0.3, 1.3, N)*gd.UNIT_GEV
L = np.linspace(2e3, 11467.8, N)*gd.UNIT_KM
kw = dict(costhz=-0.9, rtol=1e-6, atol=1e-8,
          nu_i=gd.NUMU, nu_f=gd.NUMU)

# Every point has its own baseline, so no
# batched engine takes this scan
P = oscprob.osc_prob_3nu_earth(
    E, L=L, n_jobs=4, **kw)
\end{pysnip}

The first point of the scan runs by itself, in the main process, and fixes two numbers: how many slabs the trajectory needs, $N_{\rm slabs}$, and how many sample points each slab needs. Those two numbers, set two rungs below where the first point landed, become the starting point of the refinement ladder for every later point, so each worker begins close to the answer. The rest of the scan goes to the workers, never more of them than there are points left or cores. Each worker takes a chunk of consecutive points and computes them one after another; {\tt joblib} sizes the chunks to run for between $0.2$ and $2$~s, so cheap points are gathered in bulk and an expensive point is handed over on its own. If the first point shows that the rest of the scan would take less than about 1~s in a single process, \magnus\ finishes it there and starts no workers, since starting them would cost more.

Two conditions decide whether using {\tt n\_jobs} larger than 1 helps. The first is the shape of the scan. A batched engine answers only at {\tt n\_jobs} $= 1$, so asking for more processes sends the scan down the per-point path instead, which for a batchable scan is the slower of the two (Sec.~\ref{sec:batching}). A scan whose points share one baseline should keep the default of {\tt n\_jobs} $= 1$ and let the batching of Sec.~\ref{sec:batched_calls} do the work:
\begin{pysnip}
# One baseline for all energies: the batched
# engine takes it, at the default n_jobs = 1
P = oscprob.osc_prob_3nu_earth(
    E, L=11467.8*gd.UNIT_KM, **kw)
\end{pysnip}
Parallelism is for scans that no batched engine accepts, such as the first one above, where every energy has a different baseline.

The second condition is length. Starting the workers costs time, paid once per call, so the scan has to be long enough to recover it. On the scan above, four workers more than double the speed at $2\,000$ points or more. At a few hundred points, they bring no gain, and can even slow the scan down. Section~\ref{sec:cost} and \figu{njobs_scaling} show how the gain grows with the size of the scan.

Parallelism changes more than the speed. A serial scan starts each point from where its neighbor converged; a parallel scan starts every point from where the first converged. A ladder that starts on a different rung can stop on a different rung, so two runs that differ in {\tt n\_jobs} agree to the requested tolerance, but not always to the last digit. On the chord above at {\tt rtol} $= 10^{-6}$, a scan of 40 points, short enough to stay in one process, agrees exactly, while 500 points differ by about $10^{-7}$. The user should hold {\tt n\_jobs} fixed when two runs must match to the last digit.

{\tt n\_jobs} affects only a scan of several probabilities. A call for a single probability accepts it and ignores it; its slabs go to the batched kernel in a single call, and no worker is started:
\begin{pysnip}
# One point: n_jobs is accepted and does
# nothing; the call runs in one process
P = oscprob.osc_prob_3nu_earth(
    10.0*gd.UNIT_GEV, L=11467.8*gd.UNIT_KM,
    n_jobs=4, **kw)
\end{pysnip}


\subsection{Using \magnus\ from the command line}
\label{sec:cli}

\magnus\ installs a console script, {\tt magnus}, that computes one probability, or one probability matrix, without writing Python. It chooses one of the wrappers of Sec.~\ref{sec:scope_interface} through three options, {\tt -{}-flavors}, {\tt -{}-environment}, and {\tt -{}-scenario}. The physical inputs are options named after the arguments of the wrapper, \eg, {\tt -{}-energy}, {\tt -{}-baseline}, and {\tt -{}-rho}, and the oscillation parameters default to the NuFIT 6.1 values. Listing~\ref{lst:cli} shows two calls: the probability matrix of Listing~\ref{lst:constant}, in matter of constant density, and the $\nu_\mu \to \nu_e$ probability along the chord from Fermilab to Homestake. {\tt magnus -{}-help} and the documentation site\ \cite{MagnusDocs} list every option.

\begin{lstlisting}[language=bash, caption={\textbf{The command-line interface.} Two probabilities computed from the command line. The first call returns the probability matrix of Listing~\ref{lst:constant}, with a row for each initial flavor and a column for each final flavor. The second returns one channel along the chord from Fermilab to Homestake, whose length is computed from the two named locations. The oscillation parameters take their NuFIT 6.1 defaults. See Sec.~\ref{sec:cli} for details.}, label={lst:cli}, float=tbp]
$ magnus --flavors 3 --environment matter \
    --density-profile constant --rho 3.0 \
    --energy 1 --baseline 1000
Magνs 1.2.0 -- osc_prob_3nu_matter_constant_density
E = 1 GeV, L = 1000 km

            nu_e   nu_mu  nu_tau
nu_e      0.9853  0.0135  0.0012
nu_mu     0.0136  0.9863  0.0001
nu_tau    0.0011  0.0002  0.9987

$ magnus --flavors 3 --environment earth \
    --loc-ini fermilab --loc-fin homestake \
    --energy 2.5 --nu-i mu --nu-f e
Magνs 1.2.0 -- osc_prob_3nu_earth
E = 2.5 GeV

P = 0.0724
\end{lstlisting}

Every environment except the Earth needs {\tt -{}-baseline}. Through the Earth, the pair {\tt -{}-loc-ini} and {\tt -{}-loc-fin} fixes both the chord and its length, as in the second call of Listing~\ref{lst:cli}. The option {\tt -{}-costhz} fixes only the direction of the chord, so it needs {\tt -{}-baseline} beside it, or {\tt -{}-detector-depth} or {\tt -{}-source-depth}, from which the length is computed.

The numerical settings are options too: {\tt -{}-rtol} and {\tt -{}-atol} set the tolerances of Sec.~\ref{sec:refinement}, {\tt -{}-magnus-exp-order} and {\tt -{}-integration-method} the order and the quadrature rule of Sec.~\ref{sec:quadrature}, and {\tt -{}-strategy} the {\tt strategy} keyword of Sec.~\ref{sec:engines}. The tolerances default to $10^{-3}$, as in the Python functions, whereas most figures of this paper use ${\tt rtol} = 10^{-8}$ and ${\tt atol} = 10^{-10}$. For the second call of Listing~\ref{lst:cli}, tightening them to those values changes the probability by $2 \cdot 10^{-10}$.

Adding {\tt -{}-json} prints the result as a JSON object instead, for use in a pipeline (Listing~\ref{lst:cli_json}). The object names the function that computed the result and the options that selected it, and gives the energy and the baseline in natural units.

\begin{lstlisting}[language=bash, caption={\textbf{Command-line output as JSON.} The $\nu_\mu \to \nu_e$ entry of the matrix of Listing~\ref{lst:cli}, printed as JSON: the function that computed it, the options that selected that function, the energy and the baseline in natural units, and the probability. See Sec.~\ref{sec:cli} for details.}, label={lst:cli_json}, float=tbp]
$ magnus --flavors 3 --environment matter \
    --density-profile constant --rho 3.0 \
    --energy 1 --baseline 1000 \
    --nu-i mu --nu-f e --json
{
  "function": "osc_prob_3nu_matter_constant_density",
  "flavors": 3,
  "environment": "matter",
  "scenario": "std",
  "nubar": false,
  "energy_eV": 1000000000.0,
  "baseline_eV-1": 5067730000000.0,
  "probability": 0.013589091782265706
}
\end{lstlisting}

The script computes one probability at a time, at one energy and one baseline. It offers no scans, no averaged probability, no density profile read from a file, and no Hamiltonian of one's own; those are done in Python, with the calls of the preceding sections. The script is meant for a quick number, a check against another code, or a probability inside a shell pipeline. (Where the console script is not on the path, {\tt python -m magnus} runs the same program.)


\section{Performance, precision, and reach}
\label{sec:performance}


\subsection{Timing measurement protocol}
\label{sec:timing_protocol}

Every timing in this paper---this section, Sec.~\ref{sec:comparison}, and every other cost quoted---comes from one machine model and one software stack: a 13th-generation Intel Core i5-1334U with 16~GB of memory, running Ubuntu 24.04.5 LTS on Linux kernel 7.0.0, with Python 3.12.7, {\tt NumPy} 1.26.4, and \magnus\ 1.2.0. The timings of the other codes in Figs.~\ref{fig:prem_plane} and \ref{fig:shock_cost} come from the benchmark suite of {\tt NuOscProbExact}~\cite{Bustamante:2019ggq, NuOscProbExact}, run on the same machine model under that suite's own protocol.

Absolute times mean little on their own; we quote ratios whenever possible. A comparison of two code paths in \magnus---or two codes---runs on a harness that interleaves them round-robin and carries a control workload that the change cannot touch. A factor of two or more in run times between two codes reflects the codes and persists on other machines; a difference of a few percent may not.

Two features of the protocol keep one-off costs out of the measurement; without them, \magnus\ would look slow for reasons unrelated to how it computes. The first is a warm-up pass. The first call compiles the {\tt Numba} kernels, which takes about 2~s, far longer than the computation being timed. Compiled kernels are cached on disk, but each later session still spends a fraction of a second loading them on its first call. We discard the first call and start timing from the second.

The second is an adaptive repeat count. We time a setting by calling it many times in one block and dividing by their number; too few calls leave the block so short that the timer's resolution and background jitter dominate. The number of calls grows until the block lasts at least 50~ms, and we report the fastest of a few blocks. When checked, the coefficient of variation across blocks---the standard deviation over the mean---stays below 0.10. Figure~\ref{fig:prem_plane} instead follows the protocol of the {\tt NuOscProbExact} suite, which reports the mean over at least thirty blocks of at least 0.25~s, each a scan of 25 values of $\delta_{\rm CP}$.


\subsection{Slab width follows the profile, not the phase}
\label{sec:phase_vs_profile}

\begin{figure*}[t!]
 \centering
 \includegraphics[width=\textwidth]{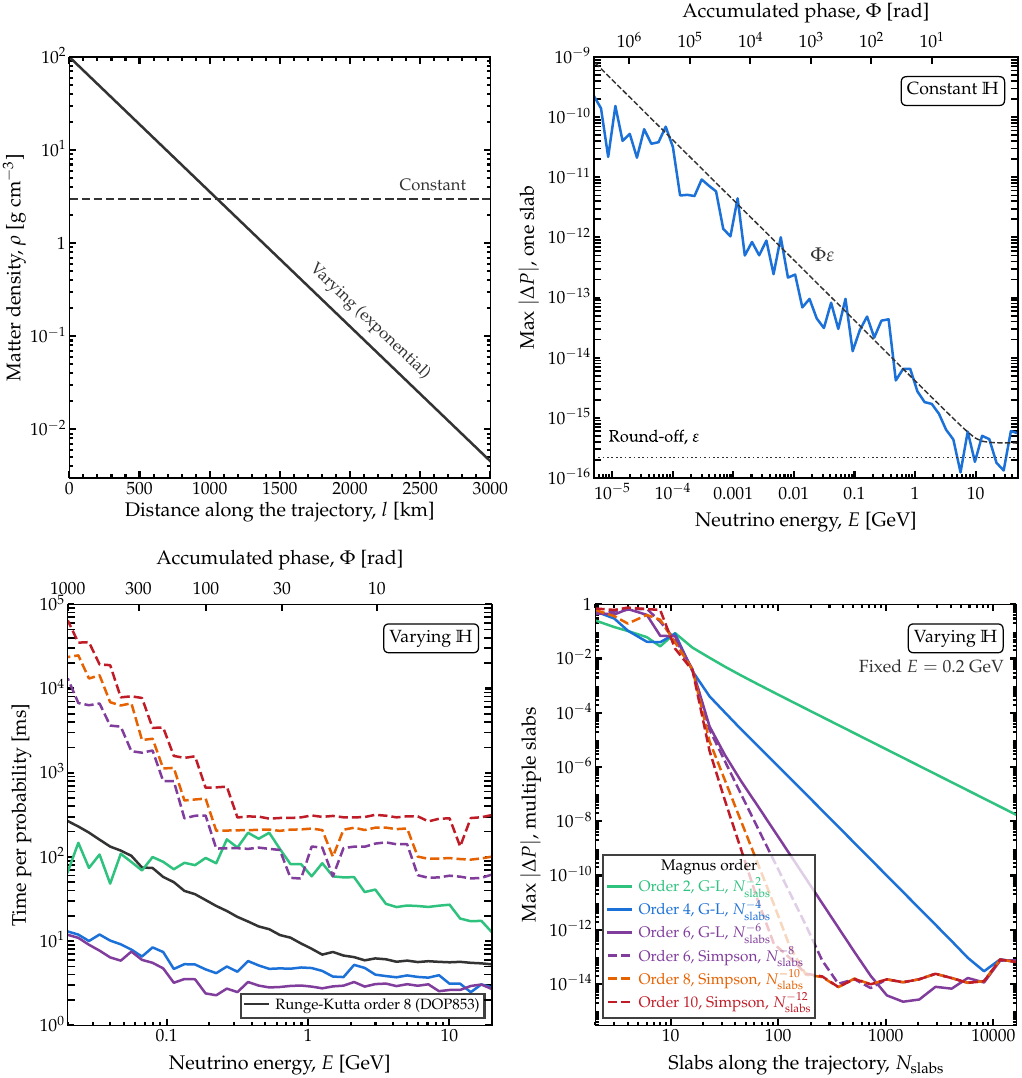}
 \caption{\label{fig:phase_vs_profile}\textbf{Cost and accuracy of a Magnus slab, at three flavors.} \emph{Top left}: the two matter-density profiles, both at an electron fraction of 0.5. The constant 3~g~cm$^{-3}$ is used in the top-right panel; the exponential, of central density 100~g~cm$^{-3}$ and scale height 300~km, in the bottom panels. \emph{Top right}: deviation of one slab from an exact exponentiation of the same constant $\mathbb{H}$, against energy, with the accumulated phase $\Phi$ of \equ{phase} on the upper axis. The series terminates at $\Omega_1$, so only arithmetic remains: the deviation bottoms out near machine epsilon, $\varepsilon = 2.2 \cdot 10^{-16}$, and rises along $\Phi\varepsilon$. \emph{Bottom left}: time per probability for the six Magnus configurations and for {\tt DOP853}, which is timed alongside them and is not the reference. All run at a tolerance of $10^{-8}$, with the slab count and the samples per slab chosen by the package. Cost climbs in steps because the code refines in discrete stages. The worst deviations from the reference range from $2.5 \cdot 10^{-7}$ for {\tt DOP853} to $1.2 \cdot 10^{-11}$ for order 10. \emph{Bottom right}: deviation from the reference against the number of slabs, at 0.2~GeV. Every reference is Richardson-extrapolated three times in extended precision: at fifty decimal digits for the right-hand panels, converged to about $2 \cdot 10^{-19}$, and at thirty digits, one solution per energy, converged past $10^{-11}$, for the bottom-left panel. See Sec.~\ref{sec:phase_vs_profile} for details.}
\end{figure*}

Figure~\ref{fig:phase_vs_profile} measures what sets the width of a slab: the density profile, not the oscillation phase. \magnus\ makes each slab narrow enough that the density changes little across it, however many oscillations fall inside. So its cost grows far more slowly than the phase, while that of a step-by-step solver grows almost in proportion to it, since it must resolve every oscillation.

The top-left panel of \figu{phase_vs_profile} shows the two density profiles used in this benchmarking: a constant 3~g~cm$^{-3}$ and an exponential of central density 100~g~cm$^{-3}$ with a 300-km scale height. The top-right panel holds the Hamiltonian constant and measures the error of a single slab as the phase across it grows. The bottom-left panel sweeps the energy along the exponential profile and times one probability at a fixed tolerance. Finally, the bottom-right panel fixes the energy and follows the error down against the slab count. We elaborate on these below.

\smallskip

\textit{Reference probability.---}We score the probabilities computed by \magnus\ against a reference that is far more accurate than any of them. The reference cuts the path into equal-width slabs, reads the density at the midpoint of each, and multiplies the exponentials together. We compute the reference in extended precision with {\tt mpmath}, so rounding errors do not affect it. The reference still depends on the number of slabs. We remove that dependence by Richardson extrapolation. Doubling the slab count shrinks the leading error term by a known factor, so combining two runs cancels that term. Three successive combinations cancel the three largest terms. We keep doubling until the last two extrapolated values agree within a set threshold; their difference is the reference's error. The right-hand panels use fifty digits, because their deviations reach the double-precision round-off near $10^{-16}$; the reference there converges to about $10^{-19}$. The bottom-left panel needs only $10^{-11}$ but a separate reference at each energy of the sweep, so we use thirty digits there.

\smallskip

\textit{Why the sweep is in energy.---}The accumulated phase, \equ{phase}, is set by the eigenvalue gap of $\mathbb{H}$. In \figu{phase_vs_profile}, its vacuum part scales as $1/E$ and its matter part does not (see Sec.~\ref{sec:recap}), so lowering the energy raises the phase while leaving the density profile untouched. That is what makes an energy sweep a test of the two against each other: one quantity moves and the other does not. The phase does not scale purely $\propto 1/E$: across the bottom-left panel, from 20 to 0.02~GeV, a factor of a thousand in energy carries $\Phi$ from 6 to 952 radians, a factor of about 150, because the energy-independent matter term dominates the gap at the high-energy end. Each panel in \figu{phase_vs_profile} carries $\Phi$ on a second axis above the energy.

\smallskip

\textit{A constant Hamiltonian costs one slab.---}If $\mathbb{H}$ does not depend on position, $l$, then every commutator in \equ{recursion} vanishes identically, $\Omega = -i\mathbb{H}\Delta l$ exactly, so the expansion terminates at its first term. The evolution operator over the whole interval is one exponential, whatever the accumulated phase. The upper-right panel of \figu{phase_vs_profile} tests this over six decades of phase. What grows is $\Phi\varepsilon$, with $\varepsilon = 2.2 \cdot 10^{-16}$ the relative precision of a double-precision number, which reaches about $10^{-9}$. That is the accuracy with which the phase is representable. It has nothing to do with the Magnus expansion: a closed-form probability~\cite{Bustamante:2019ggq} evaluated in double precision at the same phase carries the same accuracy. At the top of that span, a double-precision exponentiation is itself in error by about $10^{-10}$, the same size as the deviation being measured, so a double-precision reference could not tell the two apart. This is why we carry the reference at fifty decimal digits.

\smallskip

\textit{A varying Hamiltonian costs more.---}The corrections $\Omega_2, \Omega_3, \ldots$ in \equ{recursion} are built from commutators of $\mathbb{A}$ at two different positions. Each commutator is small when $\mathbb{A}$ changes little between those positions. Ideally, a slab is narrow enough that the \emph{Hamiltonian}---and not necessarily the neutrino state---is nearly constant across it.

Inside a slab, the neutrino state and the Hamiltonian change at different rates. The state oscillates, rotating through $\Phi$ radians, while the Hamiltonian changes only as fast as the matter density, which can be far slower. A traditional solver that advances the state in steps (like {\tt DOP853} below) must keep each step shorter than an oscillation period, so its cost grows with the phase. \magnus\ instead integrates the Hamiltonian over the slab, which involves no oscillation; the exponential of the result produces all $\Phi$ radians at once. Hence, the slab width is set by how fast $\mathbb{H}$ varies, not by how many oscillations fall inside the slab. Both codes are timed returning the full evolution matrix. For {\tt DOP853}, that means integrating all initial flavors together in a single run, which is faster than integrating them one at a time.

The bottom-left panel of \figu{phase_vs_profile} shows what that difference costs. We compare \magnus\ against a stand-alone Runge-Kutta order-8 solver, the \texttt{DOP853} algorithm as implemented in \texttt{scipy.integrate}. We score both against the extended-precision reference, over the energy sweep along the exponential profile, both described above. Every configuration, {\tt DOP853} included, runs at a tolerance of $10^{-8}$. \magnus\ chooses the slab count and the samples per slab from that tolerance, as in normal use. Its refinement proceeds in discrete stages, so its cost rises in steps; each jump marks a new stage. The panel shows:
\begin{itemize}
\item \textit{Cost follows the profile, not the phase.} Over the sweep, the phase grows about 150-fold. The time per probability of {\tt DOP853} grows 50-fold, from about 5 to 250~ms. That of \magnus\ at orders 4 and 6 on Gauss-Legendre (G-L) grows only about 3-fold, from about 2 to 4--7~ms. The gap between the two widens from 2--3 to 40--60.
\item \textit{\magnus\ is also more accurate.} Neither code guarantees the tolerance it is asked for. Asked for $10^{-8}$, {\tt DOP853} returns up to $2.5 \cdot 10^{-7}$. Every Magnus configuration stays below $10^{-8}$, order 10 by nearly three orders of magnitude. At the low-phase end, the highest orders fall below the reference's own error, so we can say only that they match it.
\item \textit{The slab count is set by the profile.} At low phase, the slab count of \magnus\ stops responding to the phase: on the exponential profile, at order 6, it stays at about 160 while the phase grows sixfold, up to 36 radians. A different profile changes it at the same energy: a constant density needs only a few slabs, and an exponential ten times flatter needs a third fewer than the steep one. {\tt DOP853} responds the opposite way, though only slightly: a steeper profile holds less matter along the path, so it carries less phase, and {\tt DOP853} takes \emph{fewer} steps.
\item \textit{Low order pays at resonances.} Order-2 Magnus is the slowest G-L configuration (Sec.~\ref{sec:quadrature}). It converges only as $N_{\rm slabs}^{-2}$, so it needs thousands of slabs at any phase. The hump in its curve is the $\Delta m^2_{31}$ MSW resonance, which the path crosses at intermediate energies. Below about 0.3~GeV, the resonant density exceeds the profile's central 100~g~cm$^{-3}$, so the resonance disappears. Near it, $\mathbb{H}$ changes fastest and the slabs must narrow. Magnus orders 4 and 6 absorb that change with their higher-order terms, so they feel it far less, yielding shallower humps.
\item \textit{Simpson's rule needs many more samples.} The Simpson configurations of Sec.~\ref{sec:quadrature}, at orders 6, 8, and 10, cost the most. Their slab counts track the phase like the G-L counts, but their samples per slab grow from about 60 to a ceiling of 500, while G-L keeps 2. At the high-phase end, order 10 on Simpson evaluates $\mathbb{H}$ some $4 \cdot 10^{5}$ times, taking nearly a minute per probability. Order 6 on Gauss-Legendre evaluates it about 3\,000 times, in about 4~ms.
\end{itemize}

\smallskip

\textit{The round-off floor.---}In the bottom-right panel of \figu{phase_vs_profile}, the error falls as slabs are added, then stops falling. Two errors compete. The truncation error falls as $N_{\rm slabs}^{-p}$ at order $p$. The round-off error grows with $N_{\rm slabs}$, because every slab adds another double-precision matrix product. Their sum is smallest, between a few times $10^{-15}$ and a few times $10^{-14}$, at a few hundred slabs at order 10 and several thousand at order 4. Order 2 never gets there within the slab counts shown. Past that minimum, more slabs make the answer worse, and the error climbs toward $10^{-13}$. The refinement ladder cannot detect the floor: it stops once refining no longer changes the probability, and near the floor refining no longer changes it. Consequently, a tolerance below the floor is reported as met without being reached. Section~\ref{sec:smooth_comparison} meets the same limit in a cross-code comparison.


\subsection{What a probability costs}
\label{sec:cost}

\begin{figure}[t!]
 \centering
 \includegraphics[width=\columnwidth]{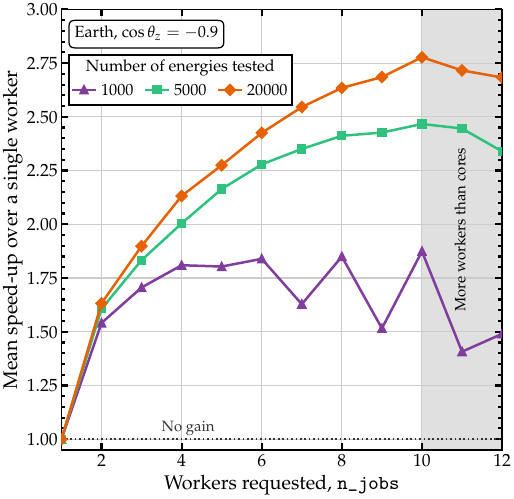}
 \caption{\label{fig:njobs_scaling}\textbf{Parallel speed-up of an energy scan.} Speed-up against the number of worker processes, {\tt n\_jobs}, for scans of 1\,000, 5\,000, and 20\,000 energies from 0.5 to 20~GeV along an Earth chord at $\cos\theta_z = -0.9$. Each energy has its own baseline, which no batched engine accepts, so every run takes the same per-point path and the speed-up reflects the parallelism alone. Each point is the mean wall-clock time over interleaved rounds---30 at 1\,000 energies, 5 at the larger sizes---with one thread per process and the worker pool started before timing. The shaded region marks requests beyond the machine's ten physical cores (two performance, eight efficiency). Ten workers give speed-ups of $1.9\times$, $2.5\times$, and $2.8\times$ as the scan grows. The two larger sizes were measured with the worker counts in a fixed order; at 1\,000 energies, that order raised the ten-worker speed-up from $1.87\times$ to $2.23\times$, so the larger sizes may be overestimated by a similar margin. See Sec.~\ref{sec:cost} for details.}
\end{figure}

A single three-flavor probability through the Earth takes about 2~ms at the default tolerance of {\tt rtol} = {\tt atol} = $10^{-3}$. Across 164 Earth and solar configurations---two to five flavors, standard and non-standard Hamiltonians, neutrinos and antineutrinos---the median call takes 2~ms and the slowest under a second. A scan of many points usually costs far less than the same points computed one at a time.

Four features set that cost. The user controls three of them; \magnus\ applies the fourth automatically:
\begin{enumerate}
\item \emph{Pass arrays.} (User's decision; \magnus\ implements it.) Every wrapper accepts an array of energies, an array of baselines, or both. \magnus\ then shares work across the points instead of starting afresh at each. For many energies at one baseline, the energy-batched engine of Sec.~\ref{sec:engines} samples the density profile once for the whole scan. For many baselines at one energy, the cumulative engine computes all of them in a single pass. Where the potential does not vary along the path, the constant-Hamiltonian engine computes the whole scan as one batch of matrix exponentials. Against the same points computed one at a time, this is worth one to two orders of magnitude, at every number of flavors from two to five. Batched answers agree with point-by-point ones to about the requested tolerance.
\smallskip
\item \emph{Write the Hamiltonian to accept an array of positions.} (User's decision.) A user who supplies $\mathbb{H}(l)$ decides how it is evaluated. Written to take a single position, it is called in a Python loop, once at every quadrature node of every slab; \magnus\ then issues a {\tt ScalarHamiltonianWarning}. Written to take an array of positions and return a stack of matrices---a \textit{vectorized} function, in Python terms---it is called once per refinement stage, for all slabs together. On a three-flavor exponential profile, the second form is several times faster, with identical output. It is the largest speed-up under the user's control. Section~\ref{sec:lri_sun} shows a Hamiltonian written this way.
\smallskip
\item \emph{Symmetry of Earth chords.} (Automatic.) A neutrino crossing a spherically symmetric Earth, such as the PREM, meets every radius twice: once on the way in and once on the way out. Thus, the Hamiltonian along the second half of the chord repeats the first half in reverse. \magnus\ evaluates it on the first half only and mirrors it for the second. This saves only the evaluations of the Hamiltonian; the exponentials and the commutators are computed as before, since their order matters. For a plain PREM density lookup, which is cheap, the saving is negligible. For an expensive Hamiltonian, it can make the call about 1.5 times faster. \magnus\ infers the symmetry from the geometry of the chord rather than detecting it, since detecting it would require the very evaluations it saves.
\smallskip
\item \emph{Ask for worker processes where no batched engine applies.} (User's decision; \magnus\ implements it.) The {\tt n\_jobs} argument distributes the requested points over several processes, as described in Sec.~\ref{sec:batching}.

Figure~\ref{fig:njobs_scaling} shows the resulting speed-up against the number of processes, at three scan sizes, along an Earth chord on which every energy has its own baseline. No batched engine accepts that request, so the single-process and multi-process runs take the same per-point path; their ratio isolates the effect of parallelism. With {\tt n\_jobs} $=$ 10, the run is 2--3 times faster, more so for larger scans, but far from 10 times. Asking for more processes than the benchmarking machine's ten physical cores brings no further gain.

More often, the whole scan shares one baseline. There, the batched engine answers only with {\tt n\_jobs} $=$ 1; any larger value sends the request to the per-point path, computed in parallel. On a scan of 5\,000 energies, that parallel run was several times slower than the batched call in one process. Thus, one process is the right default.
\end{enumerate}

The refinement ladder works against these savings: it computes every slab count below the one that converges, then discards them. On an Earth chord, this makes a call about three to four times slower than one given the right slab count in advance (Sec.~\ref{sec:prem_planes}). Section~\ref{sec:refinement} shows how to pass that count directly.


\subsection{Precision, accuracy, and tolerance}
\label{sec:diagnostics}

\textit{Precision.---}Precision is how closely \magnus\ agrees with itself. Calling \magnus\ twice with the same inputs returns the same number bit for bit. A parallel run agrees with a serial one only to the requested tolerance, because the two start their refinement from different slab counts. With the slab grid fixed, a batched scan and the same points computed one at a time agree to about $10^{-14}$ at three flavors and $10^{-13}$ at four. Left to refine adaptively, the two stop at different slab counts and differ at the level of the requested tolerance. The two matrix-exponential backends of Sec.~\ref{sec:exponential} agree to about $10^{-15}$ for a single exponential. On a solar profile that chains about 34\,000 of them, they differ by about $10^{-12}$. That is within the $N_{\rm slabs}\,\varepsilon \approx 7 \cdot 10^{-12}$ that rounding can accumulate over that many products, with $\varepsilon = 2.2 \cdot 10^{-16}$ the relative precision of a double-precision number.

\smallskip

\textit{Accuracy.---}Accuracy is how close \magnus\ comes to the true probability. Section~\ref{sec:validation} measures it against external references. The actual error can differ from the requested tolerance in either direction. At the default {\tt rtol} $=$ {\tt atol} $= 10^{-3}$, a probability through the Earth is usually much more accurate than $10^{-3}$; in the rare cases where it is not, the error stays within about twice the tolerance. We tested this on eight Earth chords, from grazing to core-crossing, at six energies between 0.5 and 20~GeV, comparing each default result with the same call at a tolerance of $10^{-7}$. The median difference is about $10^{-6}$; nine cases in ten differ by less than $10^{-4}$. The largest difference, roughly $10^{-3}$, is on the core-crossing chord at 0.5~GeV.

That accuracy is already far better than a typical oscillation analysis needs. Such an analysis evaluates the probability on a grid---\eg, in energy and arrival direction for an atmospheric experiment, in energy alone for an accelerator experiment---interpolates between the nodes, then folds the result with a flux, a cross section, and a detector response. The error from the grid and its interpolation is orders of magnitude larger than anything the oscillation code contributes\ \cite{IceCube:2018ikn}. A code accurate to $10^{-13}$ and one accurate to $10^{-6}$ return the same answer once both have passed through that chain.

Regardless, higher accuracy is still useful. It makes \magnus\ a reference against which another code, or a coarser setting of \magnus\ itself, can be checked. Where no closed-form probability exists, it is often the only reference available. It matters for quantities computed once rather than averaged over bins, as in Secs.~\ref{sec:sun} and \ref{sec:astro}. It also separates the error of a method from the error of a model, which Sec.~\ref{sec:comparison} relies on throughout. We report the accuracy \magnus\ can reach for these uses, not because a fit to oscillation data requires it.

\smallskip

\textit{Tolerance.---}The tolerance is a stopping rule: the refinement stops once successive answers agree within it. It is not a guarantee that the answer is that close to the true value. To judge whether an answer can be trusted, the user should check the warnings \magnus\ issues. (They are standard Python warnings. By default, each message is shown once per session; {\tt warnings.simplefilter('always')} shows every occurrence, and filtering on {\tt ToleranceNotAchievedWarning} catches every warning that an answer may be outside tolerance.)

The warnings err on the side of caution: they often fire on answers that prove accurate, because most of them flag a property of the input, such as a density jump, rather than predict the error. Less often, an answer is inaccurate and no warning fires. Two successive slab counts can be wrong by the same amount and still agree, so the ladder stops with an answer outside the tolerance and no warning. Section~\ref{sec:validation} calls this a silent miss. On randomly generated smooth profiles---the hardest family we tested---about one answer in twenty-five was a silent miss.


\subsection{Accuracy and speed: smooth density profile}
\label{sec:smooth_comparison}

\begin{figure*}[t!]
 \centering
 \includegraphics[width=\textwidth]{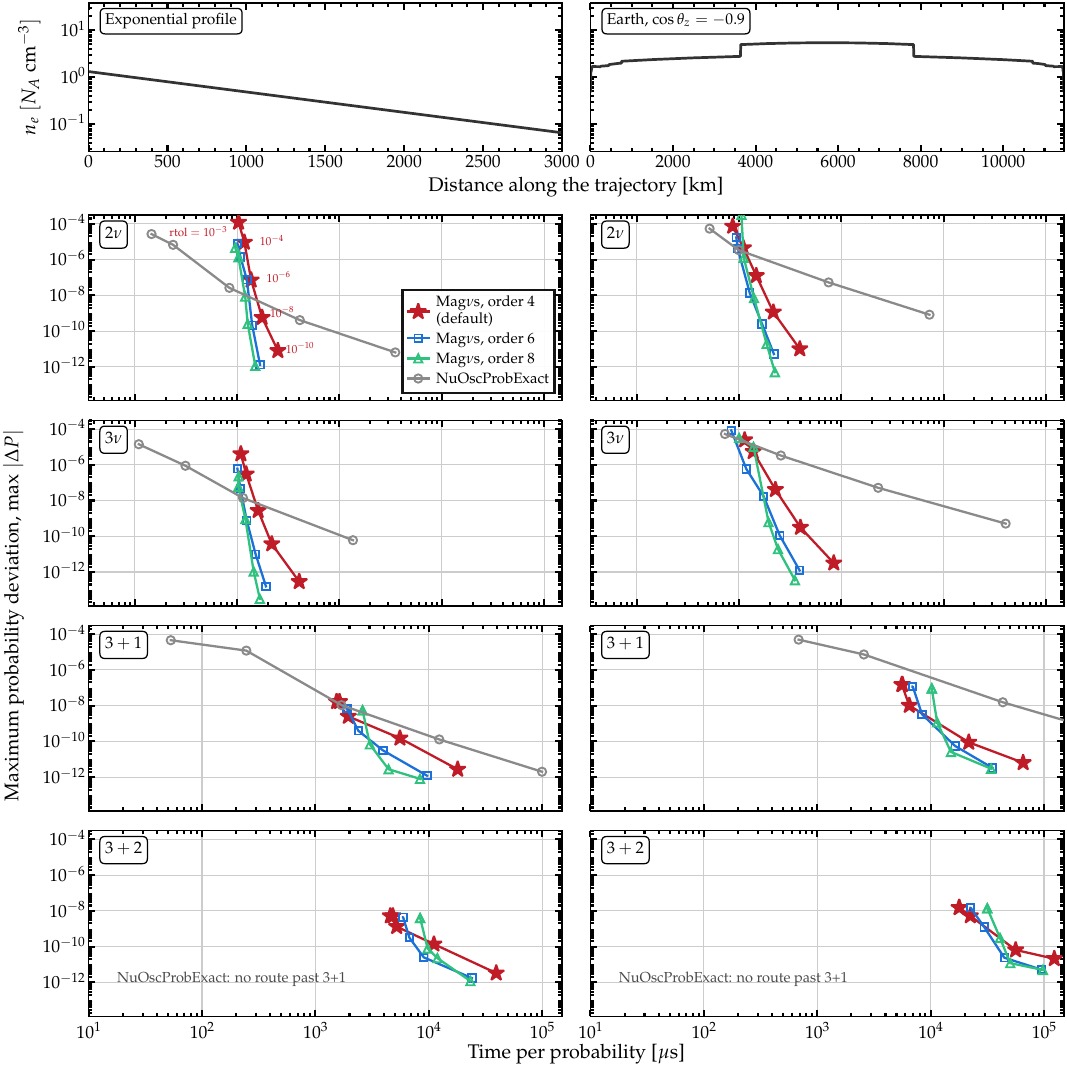}
 \caption{\label{fig:smooth_reach}\textbf{Accuracy against cost, compared with {\tt NuOscProbExact}.} Deviation of the computed probability from a reference, $\lvert \Delta P \rvert$, against the time per probability, at two to five flavors, on two density profiles. \emph{Left}: a smooth exponential profile, as in the Sun. \emph{Right}: a core-crossing Earth chord at $\cos\theta_z = -0.9$, along which the PREM profile crosses sixteen layer boundaries; each boundary is passed to \magnus\ as a slab edge. The top row shows the two profiles. Both codes are scored against the same reference, computed in extended precision with {\tt mpmath} and Richardson-extrapolated. Each curve is dialed by a requested tolerance, {\tt rtol}: \magnus\ at orders 4, 6, and 8, and {\tt NuOscProbExact}\ \cite{NuOscProbExact} through its own adaptive refinement. Both codes were timed in one process on one machine (Sec.~\ref{sec:timing_protocol}). Both codes and the reference use a uniform electron fraction $Y_e = 0.5$, the default of {\tt NuOscProbExact}, in place of \magnus's layered one inside the Earth. The bottom row has no {\tt NuOscProbExact} curve, since that code stops at four flavors. See notebook \#28 and Sec.~\ref{sec:smooth_comparison} for details.}
\end{figure*}

Figure~\ref{fig:smooth_reach} compares \magnus\ with {\tt NuOscProbExact}\ \cite{NuOscProbExact, Bustamante:2019ggq} at two to five flavors, plotting the deviation from the reference against the time per probability. We compare against {\tt NuOscProbExact} because, among the codes of Sec.~\ref{sec:relation}, it is the one that accepts the same arbitrary Hamiltonians at two to four flavors and, like \magnus, composes the evolution from slabs; Sec.~\ref{sec:comparison} compares a wider set of codes on the Earth.

\magnus\ is called as
\begin{verbatim}
 P = oscprob.osc_prob_matter_std_potential(
    d, ne_of, E, baseline, osc, L0=0.0,
    nu_i=gd.NUMU, nu_f=gd.NUMU,
    density_is_of_number_of_electrons=True,
    magnus_exp_order=4,
    rtol=r, atol=r*1.0e-2)
\end{verbatim}
with {\tt d} the number of flavors, {\tt ne\_of} the electron density along the trajectory, and {\tt r} the requested tolerance, swept from $10^{-3}$ to $10^{-10}$. Orders 6 and 8 differ only in {\tt magnus\_exp\_order}. The closed form of {\tt NuOscProbExact} is the cheaper code at loose tolerances. At tight tolerances it runs into an accuracy floor that more slabs cannot lower; \magnus\ reaches past it.

(We call the general matter \magnus\ scenario function rather than the Earth wrapper, so that the same call serves both profiles and, on each, \magnus, {\tt NuOscProbExact}, and the reference read the same electron-density function. On the Earth chord, the call passes the sixteen layer crossings as {\tt t\_breakpoints}, as the Earth wrapper would do by itself. Calling the Earth wrapper instead, with {\tt electron\_fraction=0.5}, returns the same probabilities to $10^{-15}$ at the same cost.)

\smallskip

\textit{The floor.---}As in Sec.~\ref{sec:phase_vs_profile}, composing many slab operators accumulates round-off, so past some tens of thousands of slabs the error stops falling. At three flavors on the exponential profile, {\tt NuOscProbExact} levels off at a few times $10^{-11}$ and cannot confirm convergence at a requested $10^{-10}$. This happens in any long product of matrices in floating point; it is not a defect of {\tt NuOscProbExact}. \magnus\ meets the same floor a hundred times lower or more, depending on the order, because a higher-order rule reaches a given accuracy with far fewer factors.

\smallskip

\textit{The ceiling.---}Adding flavors limits the two codes differently. The SU($n$) construction of {\tt NuOscProbExact}, with $n$ the number of flavors, needs the eigenvalues of the Hamiltonian from its characteristic polynomial, which is solvable in radicals only up to degree four. {\tt NuOscProbExact} has no five-flavor route, so the bottom row of \figu{smooth_reach} has no closed-form curve. \magnus\ has no such limit. At four and five flavors, its curve is flat between requested tolerances of $10^{-3}$ and $10^{-4}$: both requests return the same answer, already orders of magnitude more accurate than either asks for. From $10^{-6}$ on, each tighter request returns a smaller error, as at two and three flavors.

\smallskip

\textit{Where \magnus\ becomes faster.---}{\tt NuOscProbExact} is faster at loose tolerances, \magnus\ at tight ones. At three flavors, the curves cross at a requested tolerance between $10^{-6}$ and $10^{-8}$ on the exponential profile, and between $10^{-3}$ and $10^{-4}$ on the Earth chord. \magnus\ wins by needing fewer slabs, not cheaper ones: at two flavors, one of its order-4 slabs, run in compiled kernels, costs about four times a closed-form slab~\cite{Bustamante:2019ggq}, and its refinement discards every level below the one that converges. That overhead makes the closed form cheaper at the loose end of every panel. At tight tolerances, the slab count dominates instead: at a requested $10^{-8}$ on the chord at two flavors, \magnus\ needs a few hundred slabs against well over a hundred thousand, and is more than 30 times faster. At $10^{-6}$, it is about 6.5 times faster.

Adding flavors raises the cost of both codes. On the chord, a closed-form slab costs some 70 to 90 times more at four flavors than at two. The crossing moves little: on the chord, it sits near a requested $10^{-3}$ at two and three flavors and between $10^{-4}$ and $10^{-6}$ at four; on the exponential profile, it sits between $10^{-6}$ and $10^{-8}$ at every number of flavors.

\smallskip

\textit{The cost of the Hamiltonian.---}Every curve in \figu{smooth_reach} uses a Hamiltonian that is cheap to evaluate, because its density is an exponential function or a PREM table lookup. \magnus\ evaluates the Hamiltonian twice per slab and the closed form once, but the closed form needs so many more slabs that it makes far more evaluations in total. Counting the discarded refinement levels of both codes, the closed form makes about thirty times more evaluations at a requested $10^{-6}$ and a hundred times more at $10^{-8}$. With a cheap density, those extra evaluations add almost no time, so \figu{smooth_reach} does not show them. For a costlier Hamiltonian, the difference would show. Once one evaluation of $\mathbb{A}$ takes more than a few tens of nanoseconds---as it does for the tabulated BS2005-AGS,OP solar model of \figu{solar}---\magnus\ is faster at every requested tolerance of $10^{-4}$ or tighter.


\subsection{Accuracy and speed: a supernova shock front}
\label{sec:shock_comparison}

\begin{figure}[t!]
 \centering
 \includegraphics[width=\columnwidth]{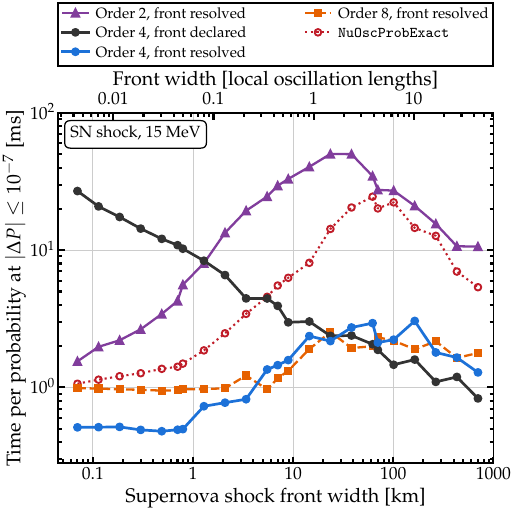}
 \caption{\label{fig:shock_cost}\textbf{Cost of crossing a supernova shock front.} Time to compute one probability to within $\lvert \Delta P \rvert \leq 10^{-7}$ on the supernova ray of \figu{shock}, against the width of its two fronts. The upper axis divides each width by the local oscillation length, that of the pair of eigenstates split by $\Delta m^2_{21}$ in the shocked shell, about 16~km. For each width and each curve, we find the smallest slab count that reaches the target and time that setting. Each time is the best of three blocks after a discarded warm-up, divided by 61, the number of baselines one call returns, so it gives the cost of one probability within a scan. \textit{Front declared} passes the two edges of each front through {\tt t\_breakpoints}; \textit{front resolved} also marks points inside each front. See notebook \#28 and Sec.~\ref{sec:shock_comparison} for details.}
\end{figure}

Figure~\ref{fig:shock_cost} shows the time to compute one probability to within $\lvert \Delta P \rvert \leq 10^{-7}$ on the supernova ray of \figu{shock}, for fronts from 0.07 to 701~km wide, using \magnus\ and {\tt NuOscProbExact}. \magnus\ is called as
\begin{verbatim}
 P = oscprob.osc_prob_matter_std_potential(
    3, ne_shock(width), E, Ls, osc, L0=L0,
    density_is_of_number_of_electrons=True,
    t_breakpoints=front_edges(width, n,
                              resolve),
    magnus_exp_order=order, cumulative=True,
    n_slabs=n, max_n_slabs=n,
    rtol=1.0e-12, atol=1.0e-14)
\end{verbatim}
with {\tt E} the neutrino energy, 15~MeV, and {\tt Ls} an array of 61 baselines along the ray, so that one call returns that many probabilities. The function {\tt ne\_shock(width)} gives the electron density along the ray. Beyond the forward shock, the star is undisturbed and the density falls as $r^{-2.4}$. The shock compresses it about tenfold at the front. Behind the front, a rarefaction lowers it to about a fifth of the undisturbed value by the contact discontinuity, across which it rises again by a factor of 2.5. The two fronts sit at 30\,323 and 12\,348~km; {\tt width} sets the distance over which each makes its transition. The function {\tt front\_edges} returns the two edges of each front, plus points inside them when {\tt resolve} is {\tt True}; {\tt t\_breakpoints} places a slab edge at each of those positions. Setting {\tt n\_slabs} and {\tt max\_n\_slabs} to the same value fixes the slab count at exactly {\tt n}. We dial slab count rather than tolerance because on this ray a tighter tolerance does not reliably give a better answer: the error comes from the slab spanning the front, which refining the rest of the ray does not change. For each width and each curve, we take the smallest {\tt n} that reaches the target and time the call at that value.

The front width sets the cost, differently for each code. \magnus\ becomes expensive when the front is narrower than its slabs and only two edges are declared: a single slab then spans the whole front. Marking points inside the front---\textit{resolving} it---removes that cost. {\tt NuOscProbExact}, and \magnus\ at order 2, which works the same way, become expensive at intermediate widths. There, the density changes too fast to be treated as constant within a slab, but too slowly to be treated as one jump.

\smallskip

\textit{Declaring a front vs.~resolving it.---}Declaring the two edges of a front with {\tt t\_breakpoints} places a slab edge on each, bracketing the front at every level of the refinement. Elsewhere, the slabs are uniform, with a spacing equal to the length of the ray divided by the slab count. On the 70\,000-km ray, that spacing stays wider than a 0.07-km front until the slab count approaches a million. Until then, the front is spanned by a single slab between the two declared edges. Raising the slab count narrows every other slab on the ray, but not that one, and that one slab dominates the error, since the Hamiltonian changes sharply only there. At Magnus order 4, the error $\lvert \Delta P \rvert$ stays between $2.5 \cdot 10^{-6}$ and $4 \cdot 10^{-6}$ from 10\,000 to 250\,000 slabs. (Figure~\ref{fig:shock_cost} does not show this, since it plots cost rather than error against slab count.) Resolving the front, by also marking points inside it, removes that plateau: eight slabs per front are enough to reach $\lvert \Delta P \rvert \leq 10^{-7}$. Figure~\ref{fig:shock_cost} shows both cases, as the \textit{front declared} and \textit{front resolved} curves.

Resolving the front can reduce the cost substantially. At 0.07~km and Magnus order 4---the default---reaching $\lvert \Delta P \rvert \leq 10^{-7}$ takes 27~ms per probability with the fronts declared, and 0.5~ms with them resolved. The saving depends on how many slabs the uniform grid would place inside the front by itself. Where it would place fewer than one, resolving saves a factor of tens. At a 9-km front, where it places about nine, resolving saves a factor under two. The declared and resolved curves at order 4 cross near 30~km. Beyond that, resolving only adds work: at 700~km, where the grid already places about 190 slabs inside each front, the extra edges add about half to the cost. Hence, the user should resolve a front only when the uniform grid would not reach inside it.

\smallskip

\textit{Why a front of middling width is expensive.---}The costs of \magnus\ at order 2 and of {\tt NuOscProbExact} rise as the front widens, peak between about 20 and 60~km, then fall; \magnus\ at orders 4 and above shows at most a mild rise. The reason lies in how a neutrino crosses a front. The states that propagate unchanged in matter are combinations of flavor states set by the local density, so a front rotates them. Across a wide front, the rotation is slow compared with an oscillation, and the neutrino follows it: the crossing is adiabatic, and the front needs no resolving. Across a thin front, the neutrino has no time to follow and crosses unchanged: the crossing is sudden, and a declared breakpoint represents it exactly. A front of middling width is in neither limit, so the slabs must resolve it.

Two measures locate this range. The first is the front width in local oscillation lengths, about 16~km, shown on the upper axis of \figu{shock_cost}: the cost peaks between about 1.5 and 4. The second can be computed without running the solver: the non-adiabaticity $\gamma_{jk}$ of \equ{hf}, evaluated across the front, taking the largest over all pairs of instantaneous eigenstates. It falls from about 730 at the thinnest front to about 0.08 at the widest, and crosses 1 between 38 and 62~km, where the costs peak.

Higher Magnus orders largely avoid the peak because of their commutators. A constant-density slab, as in {\tt NuOscProbExact} or \magnus\ at order 2, holds the Hamiltonian fixed across its length. At a middling front, the propagating states rotate inside each slab, so every slab is slightly wrong, and the only remedy is to narrow them: hence the peak. At orders 4 and above, \magnus\ samples $\mathbb{A}$ at several points inside each slab and keeps the commutators between those samples. These capture the change of the Hamiltonian within a slab, so fewer, wider slabs suffice and the run time falls.


\subsection{Accuracy and speed: the Sun}
\label{sec:solar_comparison}

\begin{figure}[t!]
 \includegraphics[width=\columnwidth]{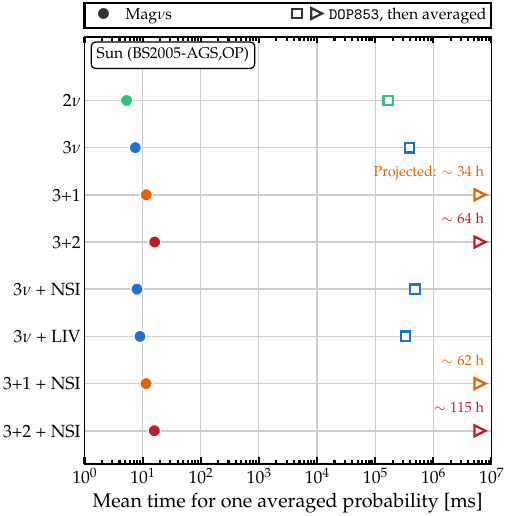}
 \caption{\label{fig:solar_average_cost}\textbf{Cost of an averaged solar probability.} Time to compute one averaged $\nu_e$ survival probability through the Sun at 5~MeV, for eight configurations, on the tabulated BS2005-AGS,OP model, interpolated in its logarithm. (The default exponential fit of {\tt osc\_prob\_*\_sun} is a different problem, with a different cost.) The evolution is adiabatic and every pair of eigenstates has decohered on arrival, so \equ{averaged_varying} needs only one eigendecomposition at production and one at detection, and is exact to about $10^{-15}$. The comparison is a {\tt DOP853} integration at ${\tt rtol} = 10^{-8}$, which agrees with \magnus\ to better than $10^{-4}$, followed by the mean of the oscillating probability over the last 1\% of the ray; the time covers both steps. The sterile parameters are $\sin^2\theta_{14} = \sin^2\theta_{24} = 0.1$ and $\Delta m^2_{41} = 1$~eV$^2$ for 3+1, plus $\sin^2\theta_{15} = \sin^2\theta_{25} = 0.05$ and $\Delta m^2_{51} = 2$~eV$^2$ for 3+2. With eV-scale splittings, the ray holds $10^{8}$ oscillations, so the integration time is extrapolated from a segment. See notebook \#28 and Sec.~\ref{sec:solar_comparison} for details.}
\end{figure}

Figure~\ref{fig:solar_average_cost} shows the cost of the averaged survival probability, \equ{averaged_varying}, of a 5-MeV solar neutrino for two to five flavors, with and without non-standard interactions and Lorentz violation. \magnus\ is called as
\begin{verbatim}
 P = oscprob.osc_prob_3nu_sun(E,
    solarmodels.table_edge('BS05-AGS-OP'),
    0.0,
    density_profile='BS05-AGS-OP',
    average=True, nu_i=gd.NUE, nu_f=gd.NUE)
\end{verbatim}
for the standard three-flavor case, and similarly for the others. The option {\tt density\_profile} selects the tabulated BS2005-AGS,OP solar model, and {\tt solarmodels.table\_edge} returns the distance to its last tabulated radius, where the path ends ($0.983\,R_\odot$). \magnus\ is four or more orders of magnitude faster than the {\tt DOP853} integration.

The two approaches differ in what they compute:
\begin{itemize}
\item \textit{{\tt DOP853}} integrates the evolution along the whole path, resolving some $10^{5}$ oscillation cycles at three flavors and some $10^{8}$ with an eV-scale sterile state, and only then averages them away.
\item \textit{\magnus} computes no oscillation at all. It diagonalizes the Hamiltonian at production and at detection, then evaluates \equ{averaged_varying} from the two sets of eigenvectors. This two-endpoint formula holds only if the neutrino stays in the same instantaneous eigenstate all the way out, so \magnus\ first searches the path for non-adiabatic crossings (Sec.~\ref{sec:hybrid}); on this profile it finds none.
\end{itemize}
The search takes about four fifths of \magnus's time; the two diagonalizations, a fraction of a percent. Both depend on the size of the Hamiltonian, set by the number of flavors, and hardly on its entries. The cost rises from 5~ms at two flavors to 16~ms at five, while non-standard interactions and Lorentz violation, which change only the entries, add little.


\section{Comparison to other codes}
\label{sec:comparison}


\subsection{Relation to other oscillation codes}
\label{sec:relation}

We compare \magnus\ against several mature oscillation codes:
\begin{itemize}
\item {\tt GLoBES}\ \cite{GLoBES, Huber:2007ji} is an experiment-simulation framework in which probabilities are one layer inside a package built for fluxes, cross sections, and detector response.
\item {\tt Prob3++}\ \cite{Prob3pp}, long used in atmospheric analyses, evaluates the analytic three-flavor probability in matter\ \cite{Barger:1980tf}.
\item {\tt nuCraft}\ \cite{nuCraft} integrates numerically along a chord through the Earth.
\item {\tt NuFast-LBL}\ \cite{Denton:2024pzc} and {\tt NuFast-Earth}\ \cite{Denton:2025oxf} are fast specializations of the three-flavor case.
\item {\tt NuOscProbExact}\ \cite{NuOscProbExact, Bustamante:2019ggq} evaluates closed-form SU($n$) expansions of the evolution operator for an arbitrary Hermitian Hamiltonian at two, three, and four flavors, and composes them across slabs of constant density.
\item {\tt nuSQuIDS}\ \cite{Delgado:2014kpa, Arguelles:2021twb, NuSQuIDS} is the most general: it propagates a density matrix rather than a state, which admits interactions, attenuation, and open-system terms that no closed form reaches, at the cost of solving a differential equation.
\end{itemize}
Structurally, \magnus\ sits between {\tt NuOscProbExact} and {\tt nuSQuIDS}: like the first, it composes slab exponentials; like the second, it accounts for how the Hamiltonian varies along the path, which constant-density slabs do not.


\subsection{Speed-accuracy comparison}
\label{sec:prem_planes}

\begin{figure*}[t!]
 \centering
 \includegraphics[width=\textwidth]{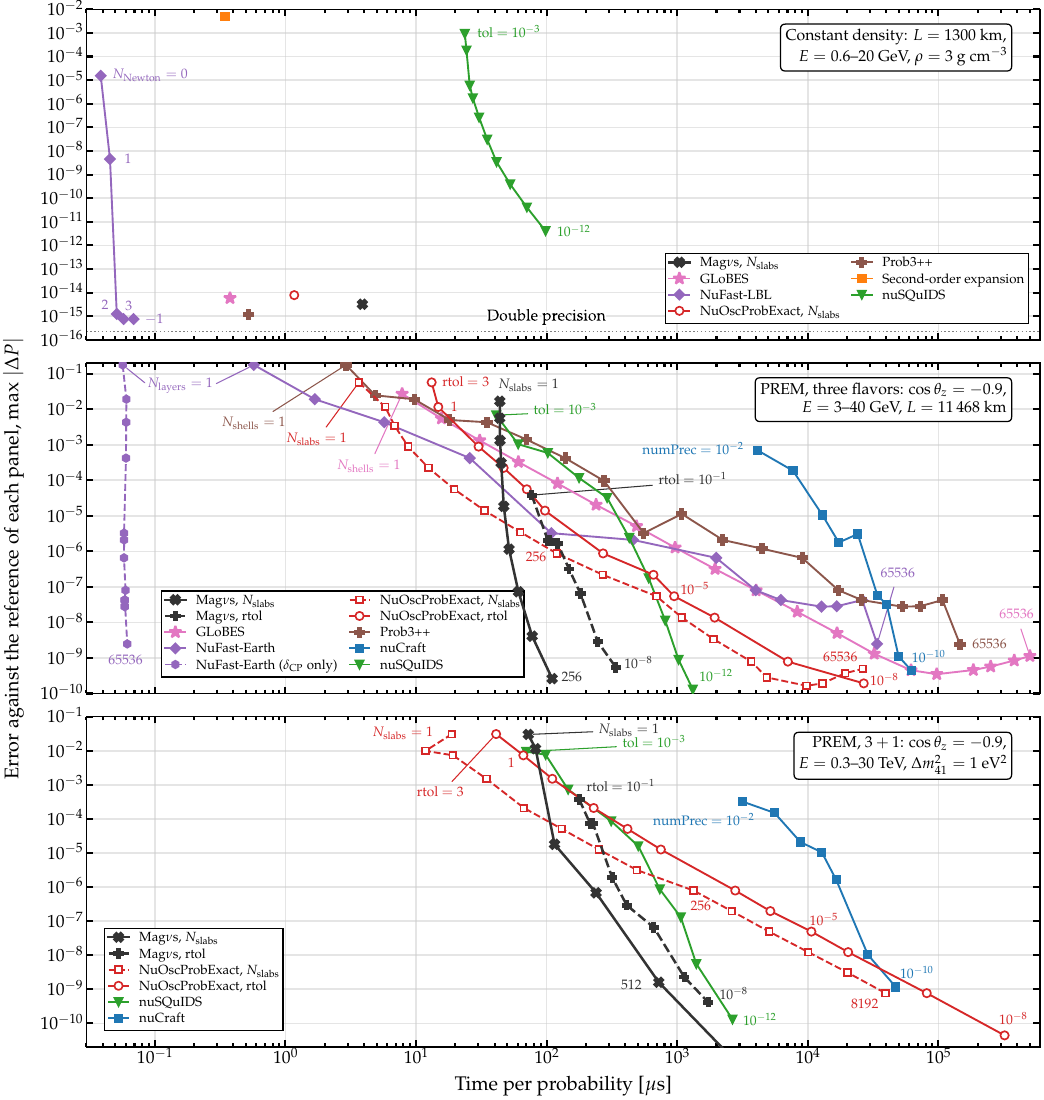}
 \caption{\label{fig:prem_plane}\textbf{Speed against accuracy, eight codes.} Error in the probability, measured against each code's own reference, plotted against the time per probability, on three setups: constant density, over a 1300-km baseline at 3~g~cm$^{-3}$ ({\it top}); a core-crossing PREM chord at $\cos\theta_z = -0.9$, with a baseline of $11\,468$~km, at three flavors ({\it middle}); and the same chord with one sterile state ({\it bottom}). \magnus\ and {\tt NuOscProbExact} appear twice in the Earth panels because each has two dials, a slab count and a tolerance; the tolerance curve of \magnus\ is drawn down to ${\tt rtol} = 10^{-8}$, and tighter settings fall beneath the panels. {\tt NuFast-Earth} appears twice because its cost depends on which oscillation parameter the scan varies. For {\tt nuCraft} at $3+1$, the reference uses the ratio of the sterile to the charged-current potential that {\tt nuCraft} assumes, 0.5016, 0.32\% above the value of 0.5 for isoscalar matter. See notebook \#28 and Sec.~\ref{sec:prem_planes} for details.}
\end{figure*}

Figure~\ref{fig:prem_plane} compares the eight codes above on three setups: constant density at three flavors, a core-crossing Earth chord at three flavors, and the same chord at $3+1$ flavors. Each point is one setting of a code's dial, placed by its time per probability and its error. Each code is scored against an extended-precision reference consistent with its own constants, chiefly the mass per nucleon that converts a density into $V_{\rm CC}$, which some codes take as the free-nucleon mass and others as the atomic mass unit, 0.8\% apart. Thus, two codes can both lie at the bottom of a panel and still disagree with each other by more than $10^{-4}$.

Every code evaluates the probability on a fixed logarithmic grid of energies; the error plotted in \figu{prem_plane} is the largest deviation over that grid. The constant-density panel uses sixty energies from $0.6$ to $20$~GeV; the Earth chord at three flavors, twelve from $3$ to $40$~GeV; and the same chord at $3+1$, twelve from $300$ to $30\,000$~GeV, three decades higher, because an eV-scale $\Delta m^2_{41}$ places its resonance there. The mixing parameters are the same for every code: $\delta_{\rm CP} = 217^\circ$, $\Delta m^2_{21} = 7.39 \cdot 10^{-5}$~eV$^2$, and $\Delta m^2_{31} = 2.525 \cdot 10^{-3}$~eV$^2$. For the timing, $\delta_{\rm CP}$ is stepped 25 times, with the whole energy grid evaluated at each step.

\smallskip

\textit{Constant density.---}At constant density (\figu{prem_plane}, top panel), the evolution operator is a single exponential, so there is no profile to discretize. Six codes---{\tt NuFast-LBL}, {\tt NuFast-Earth}, {\tt GLoBES}, {\tt Prob3++}, {\tt NuOscProbExact}, and \magnus---compute it as accurately as double-precision arithmetic allows, with errors near $10^{-15}$. Most of them have no cheaper, less accurate setting: the refinement dials of {\tt GLoBES}, {\tt Prob3++}, and {\tt NuOscProbExact} subdivide a profile that does not vary, so they change the cost a little and the error not at all, and their points lie along a short horizontal line. The closed-form codes are the cheapest, from about $0.06$~$\mu$s per probability for {\tt NuFast} to 1~$\mu$s for {\tt NuOscProbExact}. \magnus, built for profiles that vary, takes $3.9$~$\mu$s; that is about a sixth of what {\tt nuSQuIDS} needs even at its loosest setting, and about 4~s for a million probabilities.

In contrast, the dial of {\tt NuFast} affects its accuracy: it sets the number of Newton iterations used to refine the eigenvalues. Two iterations reduce the error of {\tt NuFast-LBL} from about $10^{-5}$ to the double-precision limit, at an added cost of only $0.01$~$\mu$s, so its points lie along a nearly vertical line. {\tt NuFast-LBL} is the cheapest code on this panel because it evaluates a three-flavor closed form for constant-density matter using only scalar arithmetic. {\tt nuSQuIDS}, in turn, trades cost for accuracy through its solver tolerance, from an error of about $10^{-3}$ at 24~$\mu$s to $10^{-12}$ at 100~$\mu$s, without reaching the double-precision limit. It numerically integrates the evolution of a density matrix, even though the answer here is a single matrix exponential, and so pays for a generality that constant density does not require.

Finally, \figu{prem_plane} shows the analytic approximation to $P_{\nu_\mu \to \nu_e}$ to second order in $\alpha = \Delta m^2_{21}/\Delta m^2_{31}$ and $s_{13}$\ \cite{Cervera:2000kp, Freund:2001pn}. It has no adjustable setting: its error, $5 \cdot 10^{-3}$, comes from truncating the series, not from discretization, so it cannot be reduced. At $0.35$~$\mu$s per probability, it is cheaper than every code except {\tt NuFast}, but it is the least accurate.

\smallskip

\textit{Earth chord, three flavors.---}Through the Earth (\figu{prem_plane}, middle panel), the density changes along the chord. Each code approximates the profile differently, and its error shows how well it resolves the profile. At the accuracy that an analysis typically needs, around $10^{-3}$, the closed-form codes are the cheapest: {\tt NuFast-Earth}, by far, when only $\delta_{\rm CP}$ varies, and {\tt NuOscProbExact} otherwise. At high accuracy, \magnus\ is the cheapest code, by an order of magnitude over {\tt nuSQuIDS}, the next cheapest, and by more over the rest.

{\tt GLoBES}, {\tt Prob3++}, and {\tt NuFast-Earth} cut the Earth into constant-density layers, and their cost is proportional to the number of layers. They are among the cheapest at low accuracy and among the most expensive at high accuracy. {\tt nuCraft} is accurate even at its loosest setting, but expensive at every setting, because it passes the equation of motion to {\tt SciPy}'s {\tt zvode} solver, which rebuilds the Hamiltonian at every internal step.

Against {\tt NuOscProbExact}, \magnus\ starts more expensive, but becomes the cheaper code at high accuracy. The difference lies in how each code treats a slab. {\tt NuOscProbExact} holds the density constant across it, which is the traditional slab approximation of Sec.~\ref{sec:order_is_the_slab_product}, so its error falls as $h^2$, with $h$ the slab width. \magnus, at its default order 4, also resolves how the density varies across the slab, from two Gauss-Legendre samples (Sec.~\ref{sec:quadrature}), so its error falls as $h^4$ (Table~\ref{tab:orders}). Thus, as slabs are added, the error of \magnus\ falls faster, and \magnus\ eventually overtakes {\tt NuOscProbExact}.

Figure~\ref{fig:prem_plane} shows two \magnus\ curves: one varying its slab count, {\tt n\_slabs}, and one varying its tolerance, {\tt rtol}. With {\tt n\_slabs}, \magnus\ evaluates once, at that count. Adding slabs lowers the error by orders of magnitude at little extra cost, because most of the cost is a fixed overhead, not the slabs. With {\tt rtol}, \magnus\ climbs the refinement ladder and pays for every level it evaluates and discards, so it costs a few times more at the same accuracy; in exchange, it returns evidence that the result has converged. Hence, when the slab count needed is known, \eg, from an earlier call, passing {\tt n\_slabs} is the cheaper choice.

{\tt NuFast-Earth} appears twice because its cost depends on which parameter a scan varies. When only $\delta_{\rm CP}$ varies, as in every timed scan of \figu{prem_plane}, it reuses its layer solution and is very cheap at any layer count. When $\Delta m^2_{31}$ varies, it must redo the solution, and its cost rises to that of the other codes, whose cost barely depends on which parameter varies.

\smallskip

\textit{Earth chord, $3+1$ flavors.---}The bottom panel of \figu{prem_plane} adds a sterile state, which {\tt NuFast-Earth}, {\tt Prob3++}, and {\tt GLoBES} (without extensions) cannot handle. Of the codes that can, \magnus\ is again the cheapest at high accuracy: it costs about half as much as {\tt nuSQuIDS}, and far less than {\tt NuOscProbExact} and {\tt nuCraft}. Beneath the panel, it reaches $10^{-11}$ in 3~ms.

With a sterile state, {\tt NuOscProbExact} keeps its advantage only at the coarsest accuracies, above an error of about $10^{-4}$. Its closed form owes its speed to finding the eigenvalues algebraically, which works at three flavors\ \cite{Bustamante:2019ggq}; at $3+1$, it must find them numerically, so a slab costs it about 10 times as much as at three flavors, against about 5 times for \magnus.


\subsection{When to use \magnus}
\label{sec:when_magnus}

\magnus\ is the code to reach for when a problem lies outside what a composition of constant-density slabs does well:
\begin{itemize}
\item \textbf{The density varies continuously.} A slab code holds the density constant across each slab, so it needs far more slabs than \magnus, which resolves the variation inside each. The costlier the density is to evaluate, \eg, a tabulated solar model, the larger the advantage of \magnus\ (Sec.~\ref{sec:smooth_comparison}).
\item \textbf{The profile has structure at a known place.} A shock front, a layer boundary, or a kink is passed as {\tt t\_breakpoints}, which puts a slab edge on it at every refinement level (Sec.~\ref{sec:declaring_edges}).
\item \textbf{The problem has more than three flavors.} At $3+1$, {\tt NuOscProbExact} must find its eigenvalues numerically, at about 10 times the cost per slab; past four flavors, the SU($n$) closed forms stop altogether.
\item \textbf{The Hamiltonian has no closed form of its own.} Nothing in \magnus\ assumes a form for $\mathbb{H}(l)$ beyond Hermiticity, so a non-standard interaction, a Lorentz-violating background, or an entirely new Hamiltonian is supplied as a matrix, with no change to the solver.
\item \textbf{The observable is an average.} Phase-averaged probabilities and flavor compositions are returned directly, rather than reconstructed from a scan (Sec.~\ref{sec:averaged}).
\end{itemize}

For narrower problems, a specialized code may be cheaper. On constant density, the three-flavor specializations {\tt NuFast-LBL} and {\tt NuFast-Earth} are faster than \magnus. Through the Earth, at the accuracy that an analysis typically needs, around $10^{-3}$, {\tt NuOscProbExact} is cheaper, and {\tt NuFast-Earth} is cheaper still when only $\delta_{\rm CP}$ varies. Decay, decoherence, and collective effects lie outside every unitary solver here; they need the density-matrix treatment of {\tt nuSQuIDS} (Sec.~\ref{sec:limitations}). \textbf{No other code, however, combines what \magnus\ offers: any number of flavors, any Hermitian Hamiltonian, and any density profile, computed to near double-precision accuracy, at a cost of microseconds to a few milliseconds per probability, and with evidence that the result has converged.}


\section{Conclusions}
\label{sec:conclusions}

We have presented \magnus\ \cite{Magnus}, a general-purpose code for computing neutrino oscillation probabilities, in the Standard Model and beyond. It accepts any Hamiltonian, position-dependent or not, at any number of flavors, through any density profile. The restriction is that the Hamiltonian be Hermitian. \magnus\ is designed to take numerical work off the user: given a Hamiltonian, it chooses the method, sets the resolution needed to meet the requested tolerance, and returns the probabilities, so that the user need not write or validate a solver of their own. \magnus\ is robust, flexible, and fast even when the neutrino oscillates many times along its path.

At the core of \magnus\ is the Magnus expansion, which replaces the solution of the differential equation for the evolution operator by integrals of the Hamiltonian and its commutators along the neutrino trajectory. \magnus\ places no restriction on the size of the Hamiltonian, on what it contains, or on why it varies along the trajectory. A changing matter profile, a Lorentz-violating background, and active-sterile mixing enter only through the Hamiltonian; the same method computes the probabilities for all of them.

To apply the expansion, \magnus\ cuts the trajectory into slabs, computes the evolution operator across each, and composes them time-ordered, determining the number of slabs needed to meet the requested tolerance. The evolution is unitary at any number of slabs and any order of the expansion, so the probabilities always sum to one over all flavors. To boost accuracy, slab edges can be pinned to the discontinuities of a profile, as inside the Earth or across a supernova shock front.

Where a request can be answered exactly or more cheaply than with the expansion, \magnus\ does so automatically: with a single exponential when the Hamiltonian does not vary, by following the slowly changing eigenstates of a smooth profile, or by computing many energies or baselines together. Besides oscillating probabilities, \magnus\ returns averaged ones, which describe, \eg, solar and high-energy astrophysical neutrinos. Against existing oscillation codes, \magnus\ matches their accuracy, and where high accuracy is needed on a varying density profile, it is the fastest of those we compared; at coarse accuracy or constant density, simpler codes are cheaper (Sec.~\ref{sec:comparison}).

We have shown \magnus\ at work on examples across the field: layered matter, core-crossing Earth chords, the Sun, a supernova shock front, the flavor composition of astrophysical neutrinos, geoneutrinos, a jet inside a collapsing star, turbulent matter, and several scenarios of new physics, including a Hamiltonian supplied by the user. Each is a variation on one call, with the Hamiltonian, the profile, or the observable changed. A command-line tool computes the same probabilities without writing any Python code. The repository~\cite{Magnus} holds 29 executable notebooks, from first steps to advanced uses; one of them reproduces every figure in this paper.

Two classes of problem lie beyond \magnus. First, decoherence, coupling to a bath, and decay to invisible states are not unitary and need a density-matrix treatment. Second, collective oscillations in a dense neutrino gas are nonlinear: self-interaction makes the Hamiltonian change as they evolve.

\magnus\ is available from the Python Package Index with {\tt pip install magnuspy} and from its repository\ \cite{Magnus}.

\section*{Declaration of generative AI and AI-assisted technologies in the manuscript preparation process}

During the preparation of this work, the author used Claude for research assistance in code development, numerical auditing, workflow documentation, repository maintenance, and editorial revision. All scientific judgments, validation decisions, and responsibility for the manuscript remain with the author.


\section*{Acknowledgments}

MB is supported by the {\sc Villum Fonden} project no.~29388.




\appendix

\renewcommand{\thefigure}{\Alph{section}.\arabic{figure}}
\renewcommand{\thetable}{\Alph{section}.\arabic{table}}
\newcommand{\appresetfloats}{\setcounter{figure}{0}\setcounter{table}{0}}


\section{Magnus expansion at any order}
\label{app:terms}
\appresetfloats

Section \ref{sec:expansion} gave the definition of the Magnus expansion via recursion. This appendix gives its structure, which allows a compact implementation. The maximum expansion order in \magnus\ is set to {\tt MAGNUS\_EXP\_ORDER\_MAX=10}. Orders nine and ten of the expansion need either Simpson's rule or the trapezoidal rule to integrate, since the Gauss-Legendre collocation integrators of \equ{gl} stop at order eight ({\tt MAGNUS\_EXP\_ORDER\_MAX\_GL=8}).

\subsection{Computing Magnus terms}

Unrolling the recursion equation, \equ{recursion}, every term of $\Omega_n$ is a right-nested chain
\begin{equation}
 \label{equ:nested}
 \left[\Omega_{m_1}, \left[\Omega_{m_2}, \ldots
 \left[\Omega_{m_j}, \mathbb{A}\right] \ldots \right]\right] \;,
 \quad
 \sum_{i=1}^{j} m_i = n-1 \;,
\end{equation}
carrying the coefficient $B_j/j!$. Therefore, the terms of the $j$-th group are indexed by the compositions of $n-1$ into $j$ positive parts, of which there are $\binom{n-2}{j-1}$. Since $B_j$ vanishes for odd $j \geq 3$, the number of terms in $\Omega_n$ is
\begin{equation}
 \label{equ:term_count}
 \#\,\Omega_n = \sum_{j\,:\,B_j \neq 0} \binom{n-2}{j-1} \;.
\end{equation}
An implementation to order 10 needs five nonzero coefficients:
\begin{eqnarray}
 \label{equ:bernoulli}
 && \frac{B_1}{1!} = -\frac{1}{2} , \;
 \frac{B_2}{2!} = \frac{1}{12} , \;
 \frac{B_4}{4!} = -\frac{1}{720} , \; \nonumber \\
 && \frac{B_6}{6!} = \frac{1}{30240} , \;
 \frac{B_8}{8!} = -\frac{1}{1\,209\,600} \;.
\end{eqnarray}
A term with $j$ nested commutators requires $j \leq n-1$, so $B_8$ first appears at order 9 (and $B_{10}$ not until order 11). Equation (\ref{equ:term_count}) gives $1, 1, 2, 3, 5, 9, 17, 33, 65, 129$ terms at orders 1 to 10, the count doubling from order 5 onward, which is why \magnus\ implements an order ceiling: the terms remain easy to generate far beyond it, but the work per slab grows with their number.

The module {\tt magnus.expansionterms} evaluates all of this in exact rational arithmetic at any order, with coefficients returned as fractions so that they can be compared exactly, not at a tolerance. This module shares no code with the numerical core of \magnus: the test suite regenerates the coefficients that the numerical core hard-codes for the low orders and compares them against the derived ones, and separately evaluates the generated terms numerically on a sampled $\mathbb{A}(l)$ and compares them order by order against what the core produces internally. Agreement is at machine precision for every order from one to ten, covering the hard-coded low orders and the generated high ones.

\subsection{Why the collocation orders are even}

Two facts fix the leading power of $h$ in each $\Omega_k$. The first is a parity: traversing a slab backwards inverts the evolution, so $\Omega$ changes sign; it does so termwise, since reversal cannot change how many factors of $\mathbb{A}$ a term carries. Measured from the slab midpoint, reversal replaces $h$ by $-h$, so each $\Omega_k$ carries only odd powers of $h$. The second is the effect of the commutators. Above $\Omega_1$, every term holds at least one commutator of $\mathbb{A}$ at two points of the slab. That commutator vanishes for constant $\mathbb{A}$, so its magnitude follows the variation of $\mathbb{A}$ across the slab, $\mathcal{O}(h)$ for smooth $\mathbb{A}$. Each term above $\Omega_1$ gains one power of $h$ beyond its $k$ integrations, placing $\Omega_k$ at $h^{k+1}$.

The two facts meet at odd $k$, where $h^{k+1}$ is an even power. The parity forbids it, so the term enters at $h^{k+2}$. Hence $\Omega_1$ enters at $h$ and $\Omega_2$ at $h^3$; from $\Omega_3$ upward the terms pair off, $\Omega_{2j-1}$ and $\Omega_{2j}$ both entering at $h^{2j+1}$ for $j \geq 2$. Both statements of Sec.~\ref{sec:quadrature} follow. A scheme of order $p$ carries a local error $\mathcal{O}(h^{p+1})$, so it retains every term entering below that power: $\Omega_1$ at $p = 2$, $\Omega_1$ through $\Omega_{p-2}$ for $p \geq 4$. The leading term it drops sets its error on one slab, at an odd power of $h$. A trajectory of $N$ slabs accumulates those errors, so $h^q$ on each leaves $h^{q-1}$ over the path. Odd $q$ makes $q-1$ even: the schemes exist at even order.


\section{Predefined oscillation-parameter sets}
\label{app:params}
\appresetfloats

Section \ref{sec:params} showed how to select a set of standard oscillation parameters by name. \tabl{params} lists the fifty-two sets \magnus\ recognizes: one per mass ordering for each NuFIT release\ \cite{Gonzalez-Garcia:2012hef, Gonzalez-Garcia:2014bfa, Esteban:2016qun, Esteban:2018azc, Esteban:2020cvm, Esteban:2024eli, NuFIT61}, and, from NuFIT 4.0 on, one each with and without Super-Kamiokande atmospheric data.

{\def\arraystretch{1.08}
\begin{table*}[t!]
{\small
\begin{tabular}{@{}p{2.1cm}p{1.0cm}p{2.6cm}p{10.0cm}@{}}
 Release & Year & Fit & How to select it \\
 \hline
 NuFIT 1.0 & 2012 & free fluxes & {\tt OSC\_PARAMS\_NU\_FIT\_1\_0} \\
  &  & Huber fluxes & {\tt category='huber\_fluxes\_no\_rsbl'} \\
 NuFIT 1.1 & 2013 & free fluxes & {\tt OSC\_PARAMS\_NU\_FIT\_1\_1} \\
  &  & Huber fluxes & {\tt category='huber\_fluxes\_no\_rsbl'} \\
 NuFIT 1.2 & 2013 & free fluxes & {\tt OSC\_PARAMS\_NU\_FIT\_1\_2} \\
  &  & Huber fluxes & {\tt category='huber\_fluxes\_no\_rsbl'} \\
 NuFIT 1.3 & 2014 & free fluxes & {\tt OSC\_PARAMS\_NU\_FIT\_1\_3} \\
  &  & Huber fluxes & {\tt category='huber\_fluxes\_no\_rsbl'} \\
 NuFIT 2.0 & 2014 & single fit & {\tt OSC\_PARAMS\_NU\_FIT\_2\_0} \\
 NuFIT 2.1 & 2016 & LEM analysis & {\tt OSC\_PARAMS\_NU\_FIT\_2\_1} \\
  &  & LID analysis & {\tt category='LID'} \\
 NuFIT 2.2 & 2016 & single fit & {\tt OSC\_PARAMS\_NU\_FIT\_2\_2} \\
 NuFIT 3.0 & 2016 & single fit & {\tt OSC\_PARAMS\_NU\_FIT\_3\_0} \\
 NuFIT 3.1 & 2017 & single fit & {\tt OSC\_PARAMS\_NU\_FIT\_3\_1} \\
 NuFIT 3.2 & 2018 & single fit & {\tt OSC\_PARAMS\_NU\_FIT\_3\_2} \\
 NuFIT 4.0 & 2018 & with SK & {\tt OSC\_PARAMS\_NU\_FIT\_4\_0\_SK} \\
  &  & without SK & {\tt OSC\_PARAMS\_NU\_FIT\_4\_0\_NOSK} \\
 NuFIT 4.1 & 2019 & with SK & {\tt OSC\_PARAMS\_NU\_FIT\_4\_1\_SK} \\
  &  & without SK & {\tt OSC\_PARAMS\_NU\_FIT\_4\_1\_NOSK} \\
 NuFIT 5.0 & 2020 & with SK & {\tt OSC\_PARAMS\_NU\_FIT\_5\_0\_SK} \\
  &  & without SK & {\tt OSC\_PARAMS\_NU\_FIT\_5\_0\_NOSK} \\
 NuFIT 5.1 & 2021 & with SK & {\tt OSC\_PARAMS\_NU\_FIT\_5\_1\_SK} \\
  &  & without SK & {\tt OSC\_PARAMS\_NU\_FIT\_5\_1\_NOSK} \\
 NuFIT 5.2 & 2022 & with SK & {\tt OSC\_PARAMS\_NU\_FIT\_5\_2\_SK} \\
  &  & without SK & {\tt OSC\_PARAMS\_NU\_FIT\_5\_2\_NOSK} \\
 NuFIT 5.3 & 2024 & with SK & {\tt OSC\_PARAMS\_NU\_FIT\_5\_3\_SK} \\
  &  & without SK & {\tt OSC\_PARAMS\_NU\_FIT\_5\_3\_NOSK} \\
 NuFIT 6.0 & 2024 & with SK & {\tt OSC\_PARAMS\_NU\_FIT\_6\_0\_SK} \\
  &  & without SK & {\tt OSC\_PARAMS\_NU\_FIT\_6\_0\_NOSK} \\
 NuFIT 6.1 & 2025 & with SK & {\tt OSC\_PARAMS\_NU\_FIT\_6\_1\_SK} \\
  &  & without SK & {\tt OSC\_PARAMS\_NU\_FIT\_6\_1\_NOSK} \\
\end{tabular}}
\caption{\label{tab:params}\textbf{Oscillation parameter sets.} Every set of standard oscillation parameters \magnus\ recognizes. A name in the fourth column is completed with {\tt \_NO} for normal mass ordering or {\tt \_IO} for inverted, then passed as {\tt default\_osc\_params\_set\_name} to any {\tt osc\_prob\_3nu\_*} wrapper. {\tt OSC\_PARAMS\_DEFAULT} is a further name for {\tt OSC\_PARAMS\_NU\_FIT\_6\_1\_SK\_NO}, which is what an unset parameter falls back on. Some rows give a {\tt category} instead of a name. Those variants have no constant of their own; reach them with {\tt load\_nufit\_params(version, ordering, category)}, which is also how any named set can be loaded. The third column names the secondary split each release reports: whether the Super-Kamiokande atmospheric $\chi^2$ map is folded in, whether the reactor fluxes are left free or fixed to the Huber model, and, for 2.1, which of its two analyses is taken.}
\end{table*}}


\section{The tutorial notebooks}
\label{app:notebooks}
\appresetfloats

The {\tt notebooks/} folder contains 29 notebooks, numbered in suggested reading order. They are the long form of several of the examples in Sec.~\ref{sec:examples}: what is a paragraph and a figure here is, in a notebook, the same calculation with the reasoning around it---why a convention is what it is, what happens at the edges of the parameter range, and what the numbers were checked against. Each is stored with its outputs, so it can be read without being run. Each ends with a pointer to its neighbors. The figures that cost hours to produce are cached in the repository as well: with {\tt MAGNUS\_PAPER\_CACHE\_ONLY} set, as continuous integration sets it, notebook \#28 redraws them from stored data rather than recomputing them, and fails rather than silently recomputing if a cache entry is missing.

The notebooks are generated: one script builds them, executes them, and refuses to finish if any came back without stored outputs. Because a notebook is documentation that claims to work, continuous integration executes them to ensure this. It does so on one interpreter, splitting the load of the 29 notebooks across four parallel shards. It keys a cache on the package source and on the notebook inputs: a notebook whose inputs have not moved is served from that cache rather than run again, to save time. A change to the package re-executes all of the notebooks, and a change to one notebook re-executes that one alone. The unit tests run across the four supported versions of Python; the notebooks are not.

{\def\arraystretch{1.15}
\begin{table*}[t!]
{\small
\begin{tabular}{@{}p{0.6cm}p{4.3cm}p{9.2cm}p{2.9cm}@{}}
 \# & Notebook & What it covers & Sections \\
 \hline
 01 & {\tt introduction} & What the package computes, shown on a first probability. & \ref{sec:synopsis} \\
 02 & {\tt 2nu\_vacuum\_matter} & Two-neutrino probabilities, in vacuum and in matter. & \ref{sec:minimal}, \ref{sec:constant_density} \\
 03 & {\tt 3nu\_vacuum\_matter} & Three-neutrino probabilities, in vacuum and in matter. & \ref{sec:minimal}, \ref{sec:constant_density} \\
 04 & {\tt long\_baseline} & Long-baseline accelerator configurations. & \ref{sec:prem} \\
 05 & {\tt biprobability} & Biprobability plots. & \ref{sec:biprobability} \\
 06 & {\tt oscillograms} & Oscillograms in energy and arrival direction. & \ref{sec:prem} \\
 07 & {\tt bsm\_sterile\_nu} & Sterile neutrinos, at four and five flavors. & \ref{sec:bsm} \\
 08 & {\tt bsm\_nsi} & Non-standard interactions. & \ref{sec:bsm} \\
 09 & {\tt bsm\_liv} & Lorentz-invariance violation. & \ref{sec:bsm} \\
 10 & {\tt averaged\_probability} & Phase-averaged probabilities: what {\tt average=True} returns. & \ref{sec:averaged}, \ref{sec:sun} \\
 11 & {\tt matrix\_exponential} & The matrix exponential, with its two backends. & \ref{sec:exponential} \\
 12 & {\tt adiabatic\_hybrid\_strategy} & The {\tt strategy} keyword: {\tt 'auto'}, {\tt 'hybrid'} and {\tt 'magnus'}. & \ref{sec:hybrid}, \ref{sec:engines} \\
 13 & {\tt tabulated\_solar\_model} & The BS2005-AGS,OP solar model. & \ref{sec:sun}, \ref{sec:solar_comparison} \\
 14 & {\tt supernova\_shock} & A supernova shock front, at two widths. & \ref{sec:shock}, \ref{sec:shock_comparison} \\
 15 & {\tt antineutrinos} & Antineutrinos, including the conventions they require. & \ref{sec:recap}, \ref{sec:nu_nubar} \\
 16 & {\tt exact\_vs\_approximations} & Exact probabilities against the textbook approximations. & \ref{sec:recap} \\
 17 & {\tt ordering\_and\_octant} & Mass ordering and the $\theta_{23}$ octant. & \ref{sec:prem} \\
 18 & {\tt unusual\_density\_profiles} & Unusual density profiles: the arrangement matters, not only the mean. & \ref{sec:phase_vs_profile}, \ref{sec:shock} \\
 19 & {\tt custom\_hamiltonian} & Supplying a Hamiltonian of your own. & \ref{sec:lri_sun} \\
 20 & {\tt numerical\_edge\_cases} & Numerical edge cases: the meaning of each warning. & \ref{sec:limitations}, \ref{sec:implementation_limits} \\
 21 & {\tt what\_tolerance\_means} & What {\tt rtol} and {\tt atol} promise, against what they do not. & \ref{sec:refinement}, \ref{sec:diagnostics} \\
 22 & {\tt which\_engine\_answered} & Which of the seven engines answered a call, with the reason. & \ref{sec:engines} \\
 23 & {\tt when\_averaging\_helps} & Where averaging rescues a calculation, where it fails. & \ref{sec:averaging_is_not_estimating}, \ref{sec:astro} \\
 24 & {\tt performance} & Which optimizations are worth applying. & \ref{sec:cost}, \ref{sec:timing_protocol} \\
 25 & {\tt against\_other\_codes}$^{\dagger}$ & Every comparison against another code, in full. & \ref{sec:performance}, \ref{sec:comparison} \\
 26 & {\tt nufit\_evolution} & Fourteen years of NuFIT, traced through the probabilities. & \ref{sec:recap} \\
 27 & {\tt animations} & Animated scenes. & \ref{sec:documentation} \\
 28 & {\tt paper\_figures} & Every figure in this paper. & All figures \\
 29 & {\tt pseudo\_dirac} & Pseudo-Dirac neutrinos. & \ref{sec:bsm} \\
\end{tabular}}
\caption{\label{tab:notebooks}\textbf{The tutorial notebooks.} The twenty-nine notebooks of \magnus, in reading order. They are in the source repository, not in the PyPI package. Each file is {\tt notebooks/NN\_magnus\_$\langle$name$\rangle$.ipynb}, with {\tt NN} and $\langle$name$\rangle$ from the first two columns. Each is stored with its outputs, so it can be read without being run, and ends with a pointer to its neighbors. Continuous integration re-executes a notebook whenever the package or that notebook changes, so a broken one fails the build. The last column gives the sections of this paper that each notebook covers in long form; notebook \#28 produces every figure. $^{\dagger}$Notebook \#25 is the only one that needs another oscillation code installed.}
\end{table*}}

\tabl{notebooks} lists the notebooks. Three notebooks are especially important. Notebook \#28 produces every figure in the paper: the probabilities are computed as the notebook runs, except in the figures comparing this code against another, where every code's numbers---including this one's, where the timing has to be comparable---are read from data frozen in the repository. Notebook \#25 is the whole of Secs.~\ref{sec:performance} and \ref{sec:comparison}: every comparison against another code is made and discussed there. Notebook \#13 is the tabulated solar model of Sec.~\ref{sec:sun}, including the window-mean sweep that drifts further from the answer the wider the window is.


\section{The test suite}
\label{app:tests}
\appresetfloats

The test suite of Sec.~\ref{sec:validation} runs on every push and every pull request. Coverage is a gate rather than a report: the suite fails below 90\%. A release runs the suite again, on a fresh install of the package, before anything reaches the Python Package Index. The tests are not packaged with the library; to run them, the user clones the repository and runs
\begin{verbatim}
pip install -e ".[test]"
pytest
\end{verbatim}

The test suite is not only a regression net. It checks five different kinds of thing.
\begin{enumerate}
\item \emph{The mathematics, vs.~an external implementation.} The Magnus recursion coded independently, closed-form vacuum and constant-density expressions, a tight {\tt DOP853} integration against probabilities computed by \magnus, an exact {\tt expm} where the profile makes it exact.
\smallskip
\item \emph{The properties a probability must have.} Unitarity, the behavior of reversed channels, agreement between the scalar, vectorized, and compiled paths.
\smallskip
\item \emph{Agreement between code paths.} The batched scan against the per-point path on the same slab grid, to round-off, and a parallel run against a serial one, to the tolerance, so that an optimization which changed an answer fails rather than passes quietly.
\smallskip
\item \emph{The distribution of the error over random profiles}, rather than individual cases.
\smallskip
\item \emph{The repository against its own documentation.} The documented file tree is generated from a table and checked against what is tracked, the notebooks are checked against the generator that writes them, and the notebooks' README is checked against the notebooks it describes. Adding a file fails the suite until the table follows it.

\end{enumerate}


\end{document}